\pdfoutput=1
\documentclass[showinstructions,final,faculty=fw,department=nat,phddegree=fys]{adsphd}

\title{De Sitter space, complexity, and the double-scaled SYK model}

\author{Sergio Ernesto}{Aguilar Gutierrez}

\president{Prof.~dr.~Gerda Neyens}{}
\supervisor{Prof.~dr.~Thomas Van Riet}{}
\jurymember{Prof.~dr.~Nikolay Bobev}{}
\jurymember{Prof.~dr.~Thomas Hertog}{}
\jurymember{Prof.~dr.~Christian Maes}{}
\externaljurymember{Prof.~dr.~Damián Galante}{King's College London}
\externaljurymember{Prof.~dr.~Michal P. Heller}{Ghent University}

\researchgroup{Institute for Theoretical Physics}
\website{https://fys.kuleuven.be/itf}
\email{sergioernesto.aguilargutierrez@kuleuven.be}

\address{Celestijnenlaan 200D box 2415}

\date{June 2024}
\copyyear{2024}
\usepackage{nomencl}
\renewcommand{\nomname}{List of Symbols}
\newcommand{\myprintnomenclature}{
  \cleardoublepage
  \printnomenclature
  \chaptermark{\nomname}
  \addcontentsline{toc}{chapter}{\nomname}
}
\makenomenclature

\usepackage{glossaries}
\newcommand{\glossname}{List of Abbreviations}
\newcommand{\myprintglossary}{
  \renewcommand{\glossaryname}{\glossname}
  \cleardoublepage%
  \printglossary[title=\glossname]
  \chaptermark{\glossname}
  \addcontentsline{toc}{chapter}{\glossname}
}
\makeglossaries

\usepackage{textcomp,gensymb,cite}
\usepackage{pifont}
\usepackage{booktabs,comment,physics,bbm,subcaption,epigraph}
\usepackage{amssymb,amsthm,amsmath}

\newtheorem{thm}{Theorem}[section] 
\newtheorem{deff}[thm]{Prescription} 
\newtheorem{conv}[thm]{Convention}

\definecolor{darkpastelgreen}{rgb}{0.01, 0.75, 0.24}

\usepackage{mathrsfs}
\newcommand{\eps}{\epsilon}
\newcommand{\rme}{\mathrm e}
\newcommand{\rmd}{\mathrm d}
\newcommand{\rmi}{\mathrm i}
\newcommand{\beq}{\begin{equation}}
\newcommand{\eeq}{\end{equation}}
\newcommand{\bea}{\begin{eqnarray}}
\newcommand{\eea}{\end{eqnarray}}
\newcommand{\Fig}[1]{Fig.~\ref{#1}}
\def\le{\left(}
\def\ri{\right)}
\newcommand{\eg}{{\it e.g.,}\ }
\newcommand{\ie}{{\it i.e.,}\ }
\newcommand{\TT}{\text{T}\overline{\text{T}}}
\def\le{\left(}
\def\ri{\right)}
\definecolor{linkcolour}{rgb}{0,0.2,0.9}\hypersetup{colorlinks=true,citecolor=red,urlcolor=linkcolour, linkcolor=linkcolour}
\usepackage{blindtext,etoolbox}%
\patchcmd{\thebibliography}{\chapter*}{\par\let\clearpage\relax\chapter*}{\typeout{success}}{\typeout{failure}}
\usepackage{array}
\newcommand{\PreserveBackslash}[1]{\let\temp=\\#1\let\\=\temp}
\newcolumntype{C}[1]{>{\PreserveBackslash\centering}p{#1}}
\newcolumntype{R}[1]{>{\PreserveBackslash\raggedleft}p{#1}}
\newcolumntype{L}[1]{>{\PreserveBackslash\raggedright}p{#1}}

\begin{document}

\makefrontcoverXII

\maketitle

\frontmatter
\vspace{17cm}$ $

\epigraph{``\textit{What do you think is the biggest opponent of a researcher in your opinion? \dots It's when you give up in the middle of research. It's yourself.}''}{Naoki Urasawa}

\cleardoublepage

\includepreface{preface}
\includeabstract{abstract}
\includeabstractnl{abstractnl}
\includepublications{publications}

  \graphicspath{{\chaptersdir/Nomenclature/image/}}%
  \newglossaryentry{Ch.}{name={Ch.},description={Chapter}}
\newglossaryentry{Sec.}{name={Sec.},description={Section}}
\newglossaryentry{App.}{name={App.},description={Appendix}}
\newglossaryentry{dS}{name={dS},description={de Sitter}}
\newglossaryentry{AdS}{name={AdS},description={anti-de Sitter}}
\newglossaryentry{QFT}{name={QFT},description={Quantum field theory}}
\newglossaryentry{RMTs}{name={RMTs},description={Random matrix theory}}
\newglossaryentry{CFT}{name={CFT},description={Conformal field theory}}
\newglossaryentry{GR}{name={GR},description={General relativity}}
\newglossaryentry{QM}{name={QM},description={Quantum mechanics}}
\newglossaryentry{DSSYK}{name={DSSYK},description={doubled scaled Sachdev–Ye–Kitaev}}
\newglossaryentry{JT}{name={JT},description={Jackiw–Teitelboim}}
\newglossaryentry{OTOCs}{name={OTOCs},description={Out-of-time-order correlators}}
\newglossaryentry{NEC}{name={NEC},description={Null energy condition}}
\newglossaryentry{EF}{name={EF},description={Eddington–Finkelstein}}
\newglossaryentry{TFD}{name={TFD},description={Thermofield double}}
\newglossaryentry{RT}{name={RT},description={Ryu-Takayanagi}}
\newglossaryentry{SdS}{name={SdS},description={Schwarzschild-de Sitter}}
\newglossaryentry{SAdS}{name={SAdS},description={Schwarzschild-anti-de Sitter}}
\newglossaryentry{CV}{name={CV},description={Complexity=Volume}}
\newglossaryentry{CV2.0}{name={CV2.0},description={Complexity=Spacetime Volume}}
\newglossaryentry{WDW}{name={WDW},description={Wheeler-DeWitt}}
\newglossaryentry{CA}{name={CA},description={Complexity=Action}}
\newglossaryentry{EOM}{name={EOM},description={Equation of motion}}
\newglossaryentry{CAny}{name={CAny},description={Complexity=Anything}}
\newglossaryentry{CMC}{name={CMC},description={constant mean curvature}}
\newglossaryentry{FLRW}{name={FLRW},description={Friedmann–Lemaître–Robertson–Walker}}
\newglossaryentry{CMB}{name={CMB},description={Cosmic microwave background}}
\newglossaryentry{ETW}{name={ETW},description={End-of-the-world}}
\newglossaryentry{UV}{name={UV},description={Ultraviolet}}
\newglossaryentry{IR}{name={IR},description={Infrared}}
\newglossaryentry{LdS$_2$}{name={LdS$_2$},description={Liouville-de Sitter CFT$_2$}}
\newglossaryentry{L/R}{name={L/R},description={Left and right}}

\nomenclature{$\Lambda$}{Cosmological constant}
\nomenclature{$\dot{}=\dv{\sigma}$}{Derivative with respect to a general parameter $\sigma$}
\nomenclature{$G_N$}{Newton's constant}
\nomenclature{$A$}{Area}
\nomenclature{$S$}{Entropy}
\nomenclature{$I$}{Action}
\nomenclature{$k$}{Spatial curvature}
\nomenclature{$M$}{Mass}
\nomenclature{$E$}{Energy}
\nomenclature{S$^n$}{$n$-sphere}
\nomenclature{$f(r)$}{Blackening factor}
\nomenclature{$r_{\rm st}$}{Stretched horizon radius}
\nomenclature{$r_{h}$}{Event horizon radius}
\nomenclature{$r_{c}$}{Cosmological horizon radius}
\nomenclature{$r_{t}$}{Turning point radius}
\nomenclature{$r_{\rm bdy}$}{Asymptotic boundary regulator}
\nomenclature{$\Omega_d$}{Solid angle of a $d$-sphere}
\nomenclature{$a(r)$}{General function of $r$}
\nomenclature{$\mathcal{M}$}{Bulk manifold}
\nomenclature{$\Sigma_{\pm}$}{Future and past boundary space-like slices of $\mathcal{M}$}
\nomenclature{$\mathcal{C}_V$}{Volume complexity}
\nomenclature{$\mathcal{C}_A$}{Action complexity}
\nomenclature{$\mathcal{C}_{V2.0}$}{Spacetime volume complexity}
\nomenclature{$\mathcal{B}$}{Codimension-one bulk hypersurface}
\nomenclature{$\mathcal{V}_x$}{Overal factor $\int\rmd^{d-1}x$}
\nomenclature{$\ell$}{Arbitrary length scale}
\nomenclature{$V$}{Volume}
\nomenclature{$\mathcal{C}_{2.0V}$}{Spacetime volume complexity}
\nomenclature{$\mathcal{C}_A$}{Action complexity}
\nomenclature{$\mathcal{C}^\epsilon$}{Codimension-one complexity=Anything}
\nomenclature{$\Sigma$}{Boundary slice}
\nomenclature{$\mathcal{C}_{\rm C}$}{Computational complexity}
\nomenclature{$\mathcal{C}_{\rm N}$}{Nielsen complexity}
\nomenclature{$\mathcal{C}_{\rm Q}$}{Query complexity}
\nomenclature{$\mathcal{C}_{\rm S}$}{Spread complexity}
\nomenclature{$\mathcal{C}_{\rm K}$}{Krylov complexity}
\nomenclature{$q=\rme^{-\lambda}$, $\lambda>0$}{Number of interceptions of Hamiltonian chords}
\nomenclature{$[B,~C]_q=BC-qCB$}{q-deformed commutator}
\nomenclature{$I\equiv\qty{i_1,\dots i_p}$}{Collective subindex of length $p$}
\nomenclature{$(a;q)_n=\prod_{k=1}^n(1-aq^{k-1})$}{q-Pochhammer symbol}
\nomenclature{$(a_1,\dots,a_k;q)_n=\prod_{i=1}^k(a_i;q)$}{}
\nomenclature{$\mathcal{H}$}{Hilbert space}
\nomenclature{$\Psi_I=\psi_{i_1\dots i_p}$}{String of p-fermion operators}
\nomenclature{$\ket{\mathbb{E}_0}$}{Maximal entropy state}
\nomenclature{$\ket{\chi_{\rm phys}}$}{Physical states}
\nomenclature{$\mathcal{O}_{\rm phys}$}{Physical operators}
\nomenclature{$H$}{Hamiltonian}
\nomenclature{$N$}{Number of particles}
\nomenclature{$p$}{Number of all to all body interactions}
\nomenclature{$\ket{\theta}$}{Energy eigenbasis}
\nomenclature{$\ket{n}$}{Chord number eigenbasis}
\nomenclature{$H_n(\cos\theta|q)=\bra{\theta}\ket{n}$}{q-Hermite polynomial of order n.
}
\nomenclature{$\mu(\theta)=(\rme^{\pm2\rmi\theta},q;q)_\infty=(\rme^{2\rmi\theta};q)_\infty(\rme^{-2\rmi\theta};q)_\infty(q;q)_\infty$}{Measure for the inner product in energy basis.}
\nomenclature{$A^\dagger$ and $A$}{Creation and annihilation operators, respectively.}%

\myprintglossary
\myprintnomenclature
\tableofcontents
\listoffigures
\listoftables

\mainmatter

\instructionschapters\cleardoublepage

\part{Prologue}\label{part:1}
  \graphicspath{{\chaptersdir/Introduction/image/}}%
  \chapter{Introduction}\label{ch:introduction}

\instructionsintroduction

\section{Motivation}
Our currently best-tested theory describing very massive objects, general relativity (\gls{GR}), is a classical field theory of gravity. In contrast, a successfully tested theory describing systems of very small spatial extent, including the other known fundamental forces in nature (strong and electroweak), the standard model of particle physics, employs the framework of quantum field theory (\gls{QFT}). If we are interested in describing our universe, should we worry about gravity and quantum fields being very relevant at the same time? Arguably, the most prominent examples appear in black hole physics and cosmology. As noticed by Hawking, black holes are expected to evaporate due to the emission and absorption of pairs of particles from quantum vacuum fluctuations \cite{Hawking:1974rv,Hawking:1975vcx}, and the rate of emission is greatly influenced by the size of the black hole. Meanwhile, during the early universe, according to the theory of cosmic inflation proposed in \cite{Guth:1980zm}, just before the Big Bang, our universe experienced an era of accelerated expansion, known as cosmic inflation, during which it extended dramatically its spatial size in a very short time scale, in comparison with our current rate of expansion. While this has not yet been confirmed experimentally, it is expected that the energy scales of particles during the early universe were much higher than those we have access to in present-day experiments according to the standard model of Big Bang cosmology \cite{Dodelson:2003ft}. Lastly, on the cosmological distance scale, the current era of accelerated expansion of the universe is thought to be driven by vacuum fluctuations of quantum particles \cite{Carroll:2000fy}, which have an associated constant energy density. We will provide further discussion about this point in section (\gls{Sec.}) \ref{ssec:dS space}.

Thus, a better understanding of quantum effects at different lengths and energy scales is expected to be very relevant to describing the nature and evolution of our universe. 
In this endeavor, one might want to ease the assumptions about the background as much as possible to have some analytic control over these complex problems. Maximally symmetric spacetimes, including flat, de Sitter (\gls{dS}), and anti-de Sitter (\gls{AdS}) space are particularly useful arenas to test our understanding of quantum gravity. These spacetimes naturally arise when solving Einstein equations in the presence of a cosmological constant, $\Lambda$, resulting from the variation of the Einstein-Hilbert action:\footnote{Here and throughout the thesis, we adopt natural units with $c = \hbar = 1$; and the Lorentzian-signature metric convention with signature $(-, +,\dots,+)$ corresponding to the ordering of time and spatial coordinates.}
\begin{equation}\label{eq:Einstein H action}
    I_{\rm EH}=\frac{1}{16\pi G_N}\int\rmd^{d+1} x~\sqrt{-g}(\mathcal{R}-2\Lambda)+I_{\rm bdry}+I_{\rm matter}~,
\end{equation}
with $\mathcal{R}$ the Ricci scalar, and we considered a spacetime of $d+1$-spacetime dimensions, with a metric $g_{\mu\nu}$, and $G_N$ is the Newton constant; $I_{\rm bdry}$ is the Gibbons-Hawking-York boundary action (required for the action to have a well-defined variation); and $I_{\rm matter}$ is a matter action.  Flat, dS, and AdS space correspond to $\Lambda=0$, $\Lambda>0$, and $\Lambda<0$ respectively.

When using the variational principle on this action, one finds the Einstein equations,
\begin{equation}\label{eqe:EEqs}
    \mathcal{R}_{\mu\nu}-\frac{1}{2}g_{\mu\nu}\mathcal{R}+\Lambda g_{\mu\nu}=8\pi G_N T_{\mu\nu}~,
\end{equation}
where $\mathcal{R}_{\mu\nu}$ is the Ricci tensor, and $T_{\mu\nu}$ is an energy momentum tensor originating from $I_{\rm matter}$. A spherically symmetric and static solution to vacuum Einstein equations is the following
 \begin{equation}\label{eq:spherically symmetric metric}
 \rmd s^2=-f(r)\rmd t^2+\frac{\rmd r^2}{f(r)}+r^2\rmd \Omega_{d-1}^2    ~,
 \end{equation}
with $\Omega_{d-1}$ the solid angle of a $(d-1)$-dimensional sphere. In this equation, $f(r)$ is the blackening factor that will determine the structure of the manifold depending on the cosmological constant and symmetries.

On the theoretical side, there has been a lot of interest in asymptotically AdS space. Much of the formal development in this area sparked from the pioneering work of \cite{Maldacena:1997re}, and soon followed by \cite{Witten:1998qj,Gubser:1998bc}.

On the phenomenological side, dS space plays an important role in understanding the early and late time history of our universe. Despite important advances in this field, there is no good understanding, for instance, of the statistical interpretation of the entropy of dS space, as counting a number of microstates associated with its cosmological horizon. 

On the other hand, the techniques of quantum information theory have successfully added new insights about the evolution of asymptotically AdS space geometries, as discussed in more detail in the section below. However, they remain vastly less understood in dS space. Given the importance of dS space for the history of the universe, it is urgent to improve our understanding of this framework in the presence of $\Lambda>0$. This thesis focuses on addressing the quantum nature of dS space from the perspective of a particular set of observables inspired by quantum information for quantum gravity, known as holographic complexity. Below, we summarize some relations between these areas, some lessons from AdS holography that have been incorporated in dS holography, some of the puzzles one encounters in the process, and how to make sense of them. This will lead to the research question, stated in Sec. \ref{ssec:overview}.

\section{Holography in AdS/CFT}
In the pioneering work \cite{Maldacena:1997re}, it was found that several observables of type IIB string theory on an AdS$_5\times $S$^5$ background form by $N$ number of parallel D3-branes\footnote{Dp-branes are ($1+p$)-dimensional objects embedded in a higher dimensional spacetime with Dirichlet boundary conditions upon which open strings can end.} can be reproduced from those of a special unitary group of degree $N$ ($SU(N)$) supersymmetric Yang-Mills theory\footnote{We will provide more explanations in chapter (\gls{Ch.}) \ref{ch:Geometry} and more background material.} in a ($1+3$)-dimensional flat spacetime. This result is an example of a holographic duality, a correspondence between a gravitational theory and a lower-dimensional non-gravitational QFT with special symmetries, named conformal field theory (\gls{CFT}), which lives in the asymptotic boundary of the AdS space. Meanwhile, the S$^5$ is an example of a compact geometry emerging from a so-called Kaluza-Klein dimensional reduction of type IIB string theory, which results in different matter fields (called KK modes) in the bulk geometry. Many other examples of the correspondence were found in string and M-theory (see \cite{Nastase:2007kj} for an introduction to this subject), and several non-trivial tests of the conjecture followed.

In recent years, it has become more common to employ this correspondence by considering a semiclassical gravity approximation, where one takes Newton's constant to be very small, $G_N\rightarrow0$, while keeping the number of quantum degrees of freedom to be very large (sometimes called the large $N\gg1$ expansion, introduced in \cite{tHooft:1973alw}).
The best-understood limits of the correspondence appear when we consider weak coupling on one side of the correspondence and deduce the strong coupling regime observables on the other side. For instance, in the semiclassical regime, one can solve Einstein's equations with matter sources to learn about strong coupling regimes in the dual quantum system.

In the so-called bottom-up holography, one simplifies technical evaluations even further by considering the AdS$_{d+1}$ space part of the bulk geometry while ignoring the compact manifold (and the associated KK fields) as being dual to a CFT$_{d}$. One hopes that the lessons from this approach can be generalized to the complete description of the bulk geometry coming from string theory, where the holographic duality is better established. In the bottom-up approximation, a deep connection between the geometry of the gravitational system and quantum information on the field theory side has become more manifest. One of the first examples of this connection can be found in the Ryu-Takayanagi (\gls{RT}) formula \cite{Ryu:2006bv,Ryu:2006ef} and its generalizations \cite{Hubeny:2007xt,Dong:2013qoa,Camps:2013zua,Faulkner:2013ana,Engelhardt:2014gca}. It relates the von Neumann entropy of a subregion in the asymptotic boundary with the area of a minimal surface in an asymptotically AdS space which is homologous to the boundary subregion. This provides an example of how geometry emerges from quantum entanglement \cite{Lashkari:2013koa,Faulkner:2013ica,Swingle:2014uza,Faulkner:2017tkh,Haehl:2017sot,Agon:2020mvu,Agon:2021tia}. Non-trivial checks of the holographic correspondence in the bottom-up approach have been performed \cite{Ryu:2006bv,Ryu:2006ef,Hubeny:2007xt,Dong:2013qoa,Camps:2013zua,Faulkner:2013ana,Engelhardt:2014gca} and some extensions to top-down holography have been performed in recent developments \cite{Uhlemann:2021nhu,Uhlemann:2021itz,Deddo:2022wxj,Demulder:2022aij,Deddo:2023oxn}.

Entanglement, however, falls short in describing all aspects of gravitational physics, because there are particular bulk regions that entanglement surfaces cannot reach \cite{Engelhardt:2013tra}. In particular, it fails to capture the late time growth of the wormhole inside an eternal black hole \cite{Susskind:2014moa} (which in this context refers to a codimension-one extremal surface connecting the asymptotic boundaries).

A complementary effort in this direction includes the development of a holographic dual for computational complexity. We will explain this concept and its importance in more detail below.

\section{De Sitter space}\label{ssec:dS space}
As remarked at the beginning of the chapter, there has been a lot of observational evidence that there are at least two periods where our universe experiences accelerated expansion, which is approximately described by dS space.

The first of these, cosmic inflation \cite{Guth:1980zm,Linde:1981mu,Albrecht:1982wi}, is expected to have occurred right before the Big Bang. It has received substantial evidence from the scale invariance of the power spectrum of the cosmic microwave background (\gls{CMB}). The second one concerns the late time regimes, including our current era. The current expansion is driven by a source known as dark energy, which might be explained through the presence of $\Lambda>0$. In fact, measurements provide compelling evidence that the cosmological constant adopts a value $\Lambda_{\rm Obs} \sim  5.06 \times 10^{-84}$ GeV$^2$ \cite{Planck:2015fie}. This explicit value assumes a Friedmann–Lemaître–Robertson–Walker (\gls{FLRW}) metric for the late universe, which can be expressed in polar coordinates as
\begin{equation}
    \rmd s^2=-\rmd t^2+a(t)^2\qty(\frac{\rmd r^2}{1-kr^2}+r^2\rmd\Omega_{d-1})~,
\end{equation}
where $\Omega_{d-1}$ indicates the solid angle of a S$^{d-1}$, and $k$ denotes the spatial curvature, and $a(t)$ is called the scale factor. This type of metric assumes that the large-scale structure can be described according to the cosmological principle. This in turn assumes that spacetime becomes homogeneous and isotropic. Observations point out that the universe is approximately well described by a spatially flat metric ($k=0$) with a substantial amount of dark energy (approximately 68.3\% of the energy density in the universe \cite{Abdullah:2022eqa}), driving the acceleration of the universe.

Famously, $\Lambda_{\rm Obs}$ is 122 orders of magnitude smaller than the one predicted by effective methods in QFT \cite{Adler:1995vd}, known as the cosmological constant problem. As emphasized in \cite{Galante:2023uyf}, given that $\Lambda$ and $S_{\rm GH}$ are directly related to one another through dS length scale, a better understanding of the degrees of freedom entering the Gibbons-Hawking entropy might provide new insights in the cosmological constant problem. This shows the urgency to properly study quantum effects in dS space.

Moreover, the geometry of our universe might continue approaching that dS space as matter dilutes and the cosmological constant dominates the rate of expansion of the universe. This can be more rigorously stated in terms of the cosmic no-hair theorem by Wald \cite{Wald:1983ky}. Thus, if this situation persists, dS space holds great importance in the understanding of future observers confined by their cosmological horizon, detecting its thermal radiation, and asking for its associated statistical interpretation.

\section{Holography for non-AdS spacetimes}
The holographic principle asserts that the physical properties of a gravitational system are encoded in its asymptotic boundary\footnote{The asymptotic boundary refers to a codimension-one surface at an infinitely large distance from some region of spacetime. The unfamiliar reader is referred to Figs. \ref{fig:Schwarzschild}, \ref{fig:AdS_BH} for examples.} \cite{tHooft:1993dmi,Susskind:1994vu}. It has garnered substantial support from tools in quantum information, including the RT formula. This has led to the formulation of \textit{central dogma} in holography, which states that black holes can be described in terms of unitary quantum systems carrying $\exp \le A/4 G_N \ri$ degrees of freedom\footnote{As a side comment, one of the major successes in string theory has been the recovery of the Bekenstein-Hawking formula, $A/4 G_N$, for certain supersymmetric black holes from its microscopic degrees of freedom. The reader is referred to \cite{Strominger:1996sh} for more details.}, as observed from the outside of an event horizon with area $A$ \cite{Almheiri:2020cfm}.

Similarly, the cosmological horizon of dS space also carries information about its thermodynamics, as observed by Gibbons and Hawking \cite{PhysRevD.15.2738}. This includes the celebrated result for the entropy of dS space
\begin{equation}\label{eq:GH entropy}
    S=\frac{A(r_c)}{4G_N}~,
\end{equation}
where $r_c$ is the radius of the cosmological horizon. In contrast to the black hole horizon, the location of the cosmological horizon is observer-dependent, as the observer will always located at its center. The result (\ref{eq:GH entropy}) is quite mysterious as there is no clear interpretation connected to the microscopic structure encoded by the horizon. One might interpret this entropy as measuring our ignorance about the information that is contained outside of our cosmological horizon, which cannot reach us due to the accelerated expansion of spacetime. It can also be interpreted as the entanglement entropy between the static patches, described by thermal density matrices.

The Gibbons-Hawking entropy has motivated extending holography to dS space. However, in contrast to AdS space, there is no asymptotic timelike boundary where gravity can be switched off to locate the holographic dual and construct diffeomorphism\footnote{Diffeomorphism is defined as a smooth, differentiable, and invertible coordinate transformation between manifolds \cite{Diffeomorphism}.}-invariant observables anchored to this region. 
Regardless, it was conjectured that regions delimited by a cosmological horizon can be described from their inside as unitary quantum systems with a finite number of degrees of freedom \cite{Bousso:1999dw,Bousso:2000nf,Banks:2006rx,Anninos:2011af,Banks:2018ypk, Aalsma:2021kle} including the \textit{static patch} of dS space, which is the region in causal contact with an observer in this geometry (see Fig. \ref{fig:StrechedHorizon}). Although a closed system, such as the universe as a whole, evolves unitarily, the static patch of dS space is only a subregion of spacetime, and therefore, it is in principle an open system. While the cosmological central dogma cannot be the full story, in this thesis we will make simplifying assumptions to treat dS space as a unitary quantum system (expected to be a valid approximation for time-scales below the dS radius \cite{Bros:2010rku,Rajantie:2017ajw}), since the tools of quantum information have not been well developed enough to treat even this case.\footnote{Other simplifying assumptions are working with Dirichlet boundaries in SdS$_{d+1}$ space for Chs. \ref{ch:ComplexitydS}, \ref{ch:ComplexityDSSYK}, which is known \cite{Anninos:2024wpy} to be not a well-posed initial boundary value problem for $d=3$, and it is a thermodynamically unstable configuration in $d=2$.} We leave non-Hermitian effects for an improved treatment of static patch holography in the future directions (see Ch. \ref{ch:Conclusion}).

\section{Static patch holography}
\begin{figure}
    \centering
    \includegraphics[width=0.5\textwidth]{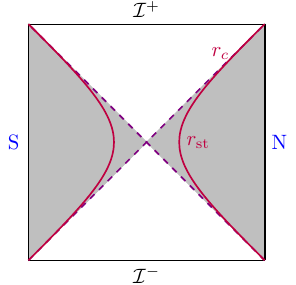}
    \caption{Penrose diagram of dS space. The black vertical lines indicate the origin of the static patch ($r=0$) where a worldline observer is placed on the north pole or south pole (blue). Future and past-like infinity are denoted by $\mathcal{I}^+$ ($r\rightarrow\infty$) and $\mathcal{I}^-$ ($r\rightarrow-\infty$). The stretched horizon (dark red) is a timelike surface at a constant radial location $r=r_{\rm st}$ in the static patch very close to the cosmological horizon, denoted by $r=r_c$ (purple). The interior of each static patch (gray) corresponds to $r<r_c$, while the exterior to $r>r_c$.}
    \label{fig:StrechedHorizon}
\end{figure}

The so-called \textit{static patch holography} postulates that the putative dual theory describing the degrees of freedom responsible for the dS entropy lives somewhere in the static patch. The two main proposals in this area are known as worldline holography \cite{Anninos:2011af,Leuven:2018ejp}; and stretched horizon
holography \cite{Susskind:2021esx} (see Fig. \ref{fig:StrechedHorizon}). 

In the former approach, the holographic dual is postulated to exist in the vicinity of the static patch observer, who is modeled as a word line particle at the center of the cosmological horizon in causally disconnected static patches of the double-sided geometry, denoted as the south and north pole\footnote{The terminology is related to the Euclidean signature geometry, where dS$_{d+1}\rightarrow$S$^{d+1}$, and the location of the worldline observers correspond to poles of the sphere.} ($\gamma_S$ and $\gamma_N$ respectively in Fig. \ref{fig:StrechedHorizon}); while in the latter, the quantum mechanical (\gls{QM}) dual theory resides in the \textit{stretched horizon}, a timelike codimension-one surface somewhere in the static patch very close to the cosmological horizon (possibly a few Planck's lengths away) \cite{Susskind:2021esx,Susskind:2022dfz,Lin:2022nss,Susskind:2022bia,Susskind:2023hnj,Susskind:2023rxm,Rahman:2022jsf,Rahman:2023pgt,Rahman:2024vyg}.\footnote{This timelike surface is a natural boundary of the static patch with respect to the worldline observer \cite{Dyson:2002pf}. Some evidence for the proposal appeared in \cite{Susskind:2011ap,Shaghoulian:2022fop} (and further developed in \cite{Franken:2023jas,Franken:2023pni}), where a modification of the RT formula \cite{Ryu:2006bv,Ryu:2006ef} for the static patch of dS space were proposed. These approaches recover the Gibbons-Hawking entropy for pure dS space, and other consistency checks have been performed.} Both the worldline observers and the stretched horizons are shown in Fig. \ref{fig:StrechedHorizon}, where we describe dS$_{d+1}$ space with static patch coordinates using (\ref{eq:spherically symmetric metric}) with 
\begin{equation}
    f(r)=1-r^2/L^2~,\label{eq:static patch dS}
\end{equation}
$f(r)=1-r^2/L^2$, where $L$ is the dS length scale, which is defined through the following relation with cosmological constant $\Lambda$:
\begin{equation}\label{eq:Lambda dS}
    \Lambda=\frac{d(d-1)}{2L^2}~.
\end{equation}
In the coordinate system (\ref{eq:spherically symmetric metric}), the location of the stretched horizon is postulated to be at a $r=r_{\rm st}$ constant surface. We will review how to write Penrose diagrams in Ch. \ref{ch:Geometry} for the reader unfamiliar with this topic.

A concrete microscopic proposal was put forward by Susskind \cite{Susskind:2021esx}. It has received some support from connections between the doubled scaled Sachdev–Ye–Kitaev (\gls{DSSYK}) model, dS Jackiw–Teitelboim (\gls{JT}) gravity, and Schwarzschild-dS (\gls{SdS}) black holes recently made more precise in
\cite{Blommaert:2023opb,Narovlansky:2023lfz,Blommaert:2023wad,Verlinde:2024znh,Verlinde:2024zrh}. In either of these approaches, one performs gravitational dressing of observables in dS space by anchoring them to the region that contains the quantum theory dual.

There are multiple reasons to believe that static patch holography may provide a well-defined framework.
First, it has been recently understood that timelike boundaries play an important role in dS space.
For instance, in three bulk dimensions, it is possible to provide a dual interpretation in terms of irrelevant $T\bar{T}$ deformations\footnote{We will not explain this topic in the thesis; however, the interested reader is referred to \cite{Zamolodchikov:2004ce} for starting work in this area and \cite{Jiang:2019epa} for a pedagogical review, and references therein.} of two-dimensional CFTs, followed by the inclusion of an additional term that introduces a positive cosmological constant in the geometry \cite{Shyam:2021ciy}.
In this way, one can perform a microstate counting inside the static patch \cite{Lewkowycz:2019xse,Coleman:2021nor}, which captures the leading and the logarithmic contributions to the entropy \cite{Anninos:2020hfj}.
A further hint of the role played by timelike boundaries is that they define sensible thermodynamics \cite{Banihashemi:2022htw,Banihashemi:2022jys}.
Second, the role of the observer in the presence of gravity has recently been stressed in \cite{Chandrasekaran:2022cip,Witten:2023qsv,Witten:2023xze,Mirbabayi:2023vgl}. 
In particular, the static patch of dS space provides a setting where an observer needs to be included to define a sensible algebra of observables.

Alternative proposals for a holographic dual of dS space exist. However, in this thesis, we will focus on static patch holography.\footnote{It is worth mentioning one of the first appearing in the literature is the dS/CFT correspondence \cite{Strominger:2001pn,Witten:2001kn,Maldacena:2002vr}, where the holographic theory is postulated to exist in the far future and far past of dS space. Evidence was presented in \cite{Strominger:2001pn,Spradlin:2001pw,Bousso:2001mw} that correlation functions of a scalar operator inserted on $\mathcal{I}^\pm$ are dual to those of an Euclidean CFT on a lower dimensional sphere. A renormalization group flow in the Euclidean CFT at $\mathcal{I}^+$ can be used to evolve backward in time in the bulk \cite{Strominger:2001gp,Godet:2024ich}. Concrete realizations of this correspondence have been made sharper in several works, including models with higher spin gravity \cite{Vasiliev:1999ba,Vasiliev:1990en} in dS$_4$ space \cite{Anninos:2011ui,Ng:2012xp,Anninos:2012ft}; an SU($2$) Wess-Zumino-Witten model dual to dS$_3$ space \cite{Hikida:2021ese,Hikida:2022ltr}, as well as through a transition from Euclidean AdS space to dS$_4$ space \cite{Hartle:2012tv,Hertog:2011ky,Bobev:2022lcc}. On the other hand, in two spacetime dimensions, there are interpolating geometries between asymptotically AdS and dS spaces, known as centaur geometries \cite{Anninos:2017hhn}, which allows one to use the tools of AdS holography to probe the interior dS space.} See \cite{Galante:2023uyf} and references therein for a review of other approaches to dS holography.

\section{Computational and holographic complexity}
An exciting advantage within the framework of static patch holography is that many lessons from the AdS/CFT correspondence carry over. This has allowed several developments in quantum information to be applied to dS space, including holographic complexity.

Computational complexity can be defined in terms of states or operators (see \cite{Chapman:2021jbh} for a review). In the state definition, it measures the difficulty of building a target state from a reference state by applying a given set of operations. Meanwhile, for operators, it is defined as the minimal number of elementary gates (a discrete set of unitaries), from a universal gate set, that is needed to model a particular unitary operator to a given precision \cite{nielsen_2010}. Despite the ambiguities, one can derive lower and upper bounds for computational complexity using a geometric approach where circuit complexity is approximated by geodesics distances in a Lie group manifold that replaces the discrete gates approximating unitary operators, as shown by Nielsen \cite{Nielsen1}.

Complexity in quantum information theory plays a crucial role in establishing the advantages of quantum over classical computation; in classifying computational problems for algorithm optimization; and in measuring quantum chaos in many-body systems (See Sec. \ref{sec:computational complexity} for a detailed explanation). Computational complexity has been further developed for several applications in QM, QFT, and CFT \cite{Caputa:2017yrh,Jefferson:2017sdb,Chapman:2017rqy,Bhattacharyya:2018wym,Chapman:2018hou,Camargo:2018eof,Ge:2019mjt,Brown:2019whu,Balasubramanian:2019wgd,Chapman:2019clq,Caputa:2018kdj,Auzzi:2020idm,Caginalp:2020tzw,Flory:2020eot,Flory:2020dja,Chagnet:2021uvi,Basteiro:2021ene,Brown:2021uov,Balasubramanian:2021mxo,Brown:2022phc,Erdmenger:2022lov,Baiguera:2023bhm}.

Although computational complexity has several practical uses, it also suffers from several ambiguities in its definition due to the dependence on the details about reference and the type of elementary operations to reach the target state in the state definition; or related to the type of gate sets and the precision to approximate a given unitary in the operator definition.

In the holographic context, several proposals have been motivated to match the computational complexity of a dual state in the quantum theory. We briefly summarize the holographic conjectures below. A detailed explanation is presented in Sec. \ref{ch:ComplexityAdS}, including figures of the different proposals.

Consider a $(d+1)$-dimensional bulk geometry with a codimension-one extremal volume surface $\mathcal{B}$ anchored at a time slice $\partial\Sigma$ where a dual quantum state is defined.
Complexity=volume (\gls{CV}) \cite{Susskind:2014rva,Stanford:2014jda} relates the evolution of the complexity of the state at $\partial\Sigma$ to the maximal volume $V$ of a codimension-one hypersurface $\mathcal{B}$ anchored at the $(d-1)$-dimensional boundary slice $\Sigma$ that solves the extremization problem (see Fig. \ref{fig:Planar_BH_Cod1})
\begin{equation}\label{eq:CV 1st}
    \mathcal{C}_V (\Sigma) =\max_{\Sigma=\partial \mathcal{B}}\frac{{V}(\mathcal{B})}{G_N\ell} \, ,
\end{equation}
where $G_N$ is Newton's constant and $\ell$ is an arbitrary length scale, usually identified as the (A)dS radius.

On the other hand, complexity=spacetime volume (\gls{CV2.0}) associates complexity to the spacetime volume $V_{\rm WDW}$ of the Wheeler-DeWitt (\gls{WDW}) patch, \ie the bulk domain of dependence of a spacelike surface anchored at the boundary slice $\Sigma$ \cite{Couch:2016exn} (see Fig. \ref{fig:WdWAdS})
\begin{equation}
    \mathcal{C}_{\rm 2.0V}(\Sigma)=\frac{V_{\rm WDW}}{G_N\ell^2}~.
\end{equation}
Meanwhile, complexity=action (\gls{CA}) is defined in terms of the on-shell gravitational action $I_{\rm WDW}$ of the WDW patch \cite{Brown:2015bva, Brown:2015lvg}
\begin{equation}
    \mathcal{C}_A (\Sigma)=\frac{I_{\rm WDW}}{\pi} \, .
\end{equation}
Finally, there is a plethora of additional holographic conjectures, the ``Complexity=anything" (\gls{CAny}) \cite{Belin:2021bga,Belin:2022xmt}, which unify the previous proposals. The motivation behind their definition is the expected behavior of computational complexity in the quantum circuit model of the field theory dual. The Hamiltonian time evolution in these models is known to display late-time linear growth of circuit complexity as well as the switchback effect  \cite{Stanford:2014jda}, which heuristically describes a characteristic delay in the growth of complexity as a consequence of inserting a perturbation in the system. We will explain this concept in detail during Ch. \ref{ch:ComplexityAdS}. The holographic complexity observables are defined to display a linear growth at late times, as well as the switchback effect in the presence of shock waves in AdS black hole geometries \cite{Belin:2021bga, Belin:2022xmt,Jorstad:2023kmq}.
These observables are evaluated in a spacetime region identified by the extremization of a certain functional. The precise definition is quite elaborated in this case; the technical details are provided in Sec. \ref{ch:ComplexityAdS}.

These defining features are motivated by the characteristics of complexity evolution in quantum circuits described above. In this thesis, we will be interested in a particular class
of codimension-one CAny observables defined on constant-mean curvature (\gls{CMC}) slices, introduced in \cite{Belin:2022xmt}.
Depending on the extrinsic curvature on the CMC slices, they will generally admit a different behavior compared to other complexity proposals.

Several intuitions and a plethora of examples involving all the holographic proposals have been considered in asymptotically AdS space.
In parallel, several investigations have been performed on the quantum side of AdS/CFT duality, \eg in the form of tensor networks with renormalization group-inspired techniques, or as continuous circuits in QM and QFT.\footnote{We refer the reader to the references in the review \cite{Chapman:2021jbh} for more details.}

Regardless of its precise dual, complexity has also shown to give rise to bulk gravitational dynamics \cite{Czech:2017ryf,Caputa:2018kdj,Susskind:2019ddc,Pedraza:2021mkh,Pedraza:2021fgp}, including for instance, the emergence of Einstein equations from optimization in holographic complexity \cite{Pedraza:2022dqi,Carrasco:2023fcj}, and, thus, is expected to be able to capture aspects of bulk physics beyond entanglement, and it has found recent applications in asymptotically AdS cosmologies \cite{Narayan:2024fcp}.

\section{Holographic complexity in de Sitter space}
The similarities between the AdS/CFT correspondence and static patch holography have motivated different studies about holographic complexity in asymptotically dS space. 
In this case, the prescriptions introduced above require all the geometric objects to be anchored to the timelike boundary of the static patch, \ie the stretched horizon \cite{Susskind:2021esx}. 
The time coordinate of a putative boundary theory is taken to run along the stretched horizon itself.
Arguably, the most striking feature of some complexity proposals in asymptotically dS space is the existence of a divergent (also called \textit{hyperfast}) growth of complexity \cite{Jorstad:2022mls,Auzzi:2023qbm,Anegawa:2023wrk} at some finite time scale. Moreover, beyond the time scale where this occurs, holographic complexity can no longer be defined unless one adopts an analytic continuation with an arbitrary regulator \cite{Jorstad:2022mls}.
This behavior is in sharp contrast with the eternal evolution of gravitational quantities in asymptotically AdS space, approaching a linear increase at late times.
However, this behavior is displayed only in the most commonly studied holographic complexity proposals, including CV, CA, and CV2.0. As we will present in the main text, the ambiguity in holographic complexity allows us to derive a subset of codimension-one observables inside the class of CAny proposal that can grow forever \cite{Aguilar-Gutierrez:2023zqm,Aguilar-Gutierrez:2023tic}. 
Once we incorporate black holes in dS space, \ie we consider SdS space, holographic complexity in the inflating region of the geometry does not qualitatively change.
Despite the different outcomes described so far, all the proposals display a switchback similar to black holes once shock waves are introduced in the background \cite{Anegawa:2023dad,Baiguera:2023tpt,Aguilar-Gutierrez:2023pnn}.
The realization of the switchback effect for all the complexity conjectures may be understood as a consequence of the universal feature that the insertion of a perturbation consistent with the null energy condition creates causal contact between regions that were originally separated in the geometry \cite{Gao:2000ga}.\footnote{This ultimately allows for the transfer of information in SdS space \cite{Aalsma:2021kle,Aguilar-Gutierrez:2023ymx}.} We will study this effect in detail in Sec. \ref{sec:CAny CMC}.

Perhaps the reader is skeptical about combining static patch holography and holographic complexity, as these two approaches are not yet on a rigorous footing on the likes of other aspects of AdS holography. The status of complexity in asymptotically dS space is less clear since holography is less understood, and it is more difficult to find a dual quantum circuit reproducing the expected features, despite the existence of different toy models \cite{Bao:2017iye,Bao:2017qmt,Niermann:2021wco,Cao:2023gkw}. 
However, as long as one can define diffeomorphism-invariant observables in a covariant manner, it is likely that some will have a holographic dual realization. 
For this reason, one might speculate that the holographic complexity observables can have a holographic realization in asymptotically dS space too, based on the central dogma. In contrast to the asymptotically AdS cases, one can probe regions inside a black hole or beyond a cosmological horizon.
Furthermore, by construction, all the holographic complexity proposals have very similar properties in AdS holography. They are built in AdS to recover a late-time linear growth and the switchback effect. However, it is not obvious a priori that the same properties should hold in other geometries. Applying them outside of such a setting would allow us to identify the differences between various proposals, as well as their universal aspects. In fact, the various settings studied in this thesis show different behavior of the various conjectures and may be used to distinguish which one could be more trustworthy in general backgrounds.

{Thus, our goal is to properly understand how can the different complexity measures be used to learn about the characteristics of the holographic dual theory (for instance, how fast they scramble quantum information in the sense of holographic complexity growth, see Ch. \ref{ch:ComplexitydS}; or the operator growth in case of Krylov complexity, see Ch. \ref{ch:ComplexityDSSYK}), and how these observables differ from those in asymptotically AdS spacetimes.}

\section{Overview}\label{ssec:overview}
In this thesis, we will explore different approaches for addressing the question
\begin{quote}
\emph{What are the principles for properly defining quantum information-theoretic notions of complexity in dS space?}
\end{quote}

In the remainder of Part \ref{part:1}, we present background material on complexity in quantum mechanics, conformal field theories, and the conjectures of holographic complexity for asymptotically AdS spacetimes. In Ch. \ref{ch:Geometry}, we provide an overview of the geometry of AdS and dS spacetimes, and explain how to describe spherical symmetric black holes in the presence of $\Lambda<0$ and $\Lambda>0$. In Ch. \ref{ch:QComplexity}, we present definitions of complexity for general quantum systems. In Ch. \ref{ch:ComplexityAdS}, we explain in more detail the  CV, CV2.0, CA, and CAny holographic complexity proposals that we have introduced, and we study their defining properties in the presence of AdS black holes.

Part \ref{part:2}, which contains  Ch. \ref{ch:ComplexitydS}, is dedicated to studying the holographic complexity proposals in a n-copy analytic extension of the SdS spacetime. In this setting, the extended SdS space has $n$ black hole and inflating patches. We summarize some of the previous findings in the literature regarding the hyperfast growth of complexity in SdS space in the CV, CV2.0, and CA proposals. We also introduce a family of codimension-one CAny proposals, which need to satisfy specific constraints allowing for late time growth. We also propose a covariant protocol in which the CAny proposals will display a switchback effect in SdS space, in close similarity to the behavior of the complexity proposals in asymptotically AdS spacetimes. We study how the different holographic complexity proposals are modified in the presence of both the black hole and the inflating patches. One of the main results is that the codimension-zero proposals (CV2.0 and CA) do not display any time evolution. In contrast, the codimension-one proposals can have non-trivial evolution if an equal number of black holes and inflating patches are probed with the extremal complexity surfaces. Our results show that different locations of the stretched horizon give rise to quite different behaviors of the complexity conjectures in asymptotically dS space.

Part \ref{part:3} is dedicated to defining quantum information-theoretic definitions of complexity in SdS$_3$ space from a putative microscopic theory holographic to it, which we refer to as the doubled double-scaled Sachdev–Ye–Kitaev (DSSYK) model. The holographic correspondence between these theories has been recently proposed in a series of works \cite{Susskind:2021esx,Narovlansky:2023lfz,Milekhin:2023bjv,Verlinde:2024znh,Verlinde:2024zrh} and non-trivial consistency checks have been performed. In Ch. \ref{ch:DSSYK}, we review the DSSYK model and the evidence about its correspondence with SdS$_3$ space. We will summarize the evidence for the correspondence, although, we caution the reader that this is a very recent proposal that requires more checks; including matching higher than two-point correlation functions. Our work in holographic complexity serves a role in this direction. In Sec. \ref{ch:ComplexityDSSYK}, we study four different definitions of quantum complexity in the DSSYK model and its bulk manifestations according to the elements in the dS holographic dictionary in \cite{Verlinde:2024znh}.

In part \ref{part:4}, we provide a summary of the main results of the thesis, as well as a unified discussion of the different lessons extracted from the studies of holographic complexity in asymptotically dS backgrounds, and in connection with the DSSYK model, which we have used to define complexity from microscopic principles in dS space holography. We conclude with some avenues for future research. Appendix (\gls{App.}) \ref{app:details_CA} includes details in the evaluation of the CA proposal, mostly focused on the extended SdS black hole spacetime. App. \ref{app:MERA} provides a brief introduction to a particular type of tensor network related to our analysis of the DSSYK model.

\cleardoublepage%

  \graphicspath{{\chaptersdir/Geometry/image/}}%
  \chapter{Spacetime geometry of (anti)-de Sitter space}\label{ch:Geometry}
This chapter serves as an introduction to the spacetime geometries that will appear in most of the thesis. To set the stage, we introduce the Schwarzschild black hole solution to Einstein's equations without a cosmological constant. This is useful in order to introduce the different coordinate systems that will be employed throughout the thesis and to explain the general methods for constructing Penrose diagrams, which play an important role in describing the causal structure of the spacetime of interest. We will then focus on the most relevant geometries for the rest of the thesis, namely eternal AdS black holes; then dS space and its black hole generalization, SdS space, and lastly, its periodic analytic continuation, named the extended SdS$^n$ space. We dedicate a subsection to discuss how the AdS black hole and SdS geometries are distorted due to the insertion of high energy pulses, called shockwaves, which will be relevant in our exploration of holographic complexity in later Secs. \ref{sec:CAny CMCAdS} and \ref{sec:CAny CMC}.

\section{Asymptotically flat: The Schwarzschild black hole}
 For simplicity, we will start studying black hole solutions to the vacuum Einstein equations with zero cosmological constant $\Lambda = 0$, corresponding to the Schwarzschild black hole\footnote{This is the most general spherically symmetric solution to vacuum Einstein equations. The unfamiliar reader might consult standard textbooks, e.g. \cite{Wald:1984rg,Carroll:2004st}, for more information about the Birkhoff theorem.} whose metric appears in (\ref{eq:spherically symmetric metric}) and its blackening factor is given by,
 \begin{equation}\label{eq:Schwarzschild}
     f(r)=1-\frac{2G_NM}{r^{d-2}}~.
 \end{equation}
 where $M$ is the mass, and the black hole event horizon is situated at $r_h=(2G_NM)^{1/(d-2)}$, where $f(r_h) = 0$ such that $t$ and $r$ are timelike and spacelike coordinates respectively when $r>r_h$, otherwise $r$ is timelike, and $t$ spacelike. Notice that in the limit $M\rightarrow0$ or $r\rightarrow\infty$ (for $d\geq3$) the metric becomes that of Minkowski space. The latter limit is known as asymptotic flatness.

\subsection{Eddington-Finkelstein coordinates}
Let us consider an arbitrary point $P$ in a spacetime manifold $\mathcal{M}$ and the Schwarzschild metric (\ref{eq:spherically symmetric metric}), (\ref{eq:Schwarzschild}). All timelike curves and null rays passing through $P$ are said to be causally related. The boundary of the point that can be reached by an observer at $P$ is determined by their future and past lightcones since nothing moves faster than the speed of light. One can describe the null geodesics passing through $P$ by solving $\rmd s^2 = 0$ with $\rmd\Omega^2_{d-1} = 0$, which means
\begin{equation}
    \dv{t}{r}=\pm\frac{1}{f(r)}~.
\end{equation}
We see that when $r = r_h$, then $\dv{t}{r}\rightarrow\pm\infty$, so the null geodesic seems never to reach the black hole horizon. One can suspect that the problem is the particular coordinate system we are considering, which is adapted to an observer far from the black hole.

In order to describe null rays, and to have a region where geometric observables can extend behind the event horizon, it is convenient to introduce (right) \textit{lightcone coordinates}: 
\beq
u_{\rm R} = t_{\rm R} - r^* (r) \, , \qquad
v_{\rm R} = t_{\rm R} + r^* (r) \, ,
\label{eq:general_null_coordinates}
\eeq
In this way, outgoing null rays correspond to $u_{\rm R}$ held constant, while ingoing null rays by $v_{\rm R}$. Here we have introduced the \textit{tortoise coordinate}, which is defined as
\beq
r^*(r) = \int_{r_{0}}^r \frac{d r'}{f(r')} \, ,
\label{eq:general_tortoise_coordinate}
\eeq
where $r_0$ is an integration constant, chosen such that $r^*(r \rightarrow \infty)=0$.
We can now bring the metric into the ingoing Eddington-Finkelstein (\gls{EF}) form 
\begin{align}
    \rmd s^2 &= - f(r) \rmd u_{\rm R}^2 - 2 \rmd u_{\rm R} \rmd r + r^2 \rmd\Omega_{d-1}^2 \label{eq:generic_null_metric_noshock2}\\
&= - f(r) \rmd v_{\rm R}^2 + 2 \rmd v_{\rm R} \rmd r + r^2 \rmd\Omega_{d-1}^2\,, \label{eq:generic_null_metric_noshock}
\end{align}
which behaves as
\begin{equation}
    \dv{v_{\rm R}}{r}=\begin{cases}
        0~,&\text{ingoing}\\
        \frac{2}{f(r)}~,&\text{outgoing}
    \end{cases}~,\quad \dv{u_{\rm R}}{r}=\begin{cases}
        0~,&\text{outgoing}\\
        \frac{2}{f(r)}~,&\text{ingoing}
    \end{cases}~.
\end{equation}
One can also rewrite the metric using both $u_{\rm R}$ and $v_{\rm R}$ at the same time,
 \begin{equation}
     \rmd s^2=f(r)\rmd u_{\rm R}\rmd v_{\rm R}+r^2\rmd\Omega_{d-1}^2~,
 \end{equation}
such that the event horizon lies at either $v_{\rm R} \rightarrow-\infty$ or $u_{\rm R} \rightarrow+\infty$. 

\begin{figure}
    \centering
    \subfloat[]{\includegraphics[width=0.45\textwidth]{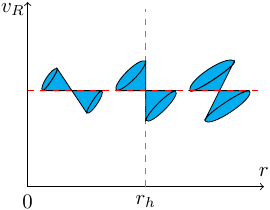}\label{sfig:crossingA}}\quad\subfloat[]{\includegraphics[width=0.45\textwidth]{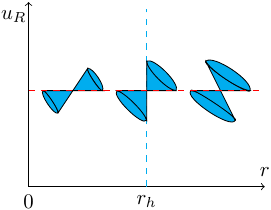}\label{sfig:crossingB}}
    \caption{Diagram depicting the lightcone of an observer (cyan) who: (a) Moves from outside a black hole on a $v_R=$constant trajectory (dashed red line); then crosses the event horizon ($r=r_h$, dashed cyan line); falls inside the black hole, and eventually falls into the singularity at $r=0$. (b) Moves from inside a white hole on a $u_R=$constant trajectory (dashed red line); then crosses the event horizon (dashed cyan line), and continues moving outside.}
    \label{fig:CrossingHorizon}
\end{figure}

It is important to observe that whenever an event horizon is crossed, either the null coordinate $u_{\rm R}$ or $v_{\rm R}$ ends its range of validity. This means that the light cone coordinate needs to be defined again using \eqref{eq:general_null_coordinates} in the new patch. One can then use the coordinate system (\ref{eq:generic_null_metric_noshock}) to enter the black hole along $v_R=$constant. Notice that, since the $r$ and $t$ become time-like and spacelike beyond $r=r_h$ respectively, then everything that enters the black hole will inevitably reach $r = 0$ (the singularity), as illustrated in Fig. \ref{sfig:crossingA}. Similarly, in the past direction, one might cross $r=r_h$ along $u_R=$constant in (\ref{eq:generic_null_metric_noshock}), where again, $r$ and $t$ change roles. As seen in Fig. \ref{sfig:crossingB} everything inside the white hole interior will eventually be emitted to the outside.

\subsection{Kruskal coordinates}
Since $v_R\rightarrow-\infty$, and $u_R\rightarrow\infty$ for the event horizon, it is convenient to define coordinates for which the event horizon location takes a finite value, such as the Kruskal coordinates, defined as
\begin{equation}\label{eq:Kruskal R Sch}
    U=-\rme^{-\frac{2\pi}{\beta}u_{\rm R}}~,\quad V=\rme^{\frac{2\pi}{\beta}u_{\rm R}}~,
\end{equation}
where $\beta=\frac{4\pi}{f'(r_h)}$ is the inverse temperature\footnote{This can be easily deduced by doing an expansion $r=r_h+\epsilon^2$ for $\epsilon\ll1$ for the metric (\ref{eq:spherically symmetric metric}) with (\ref{eq:Schwarzschild}), then performing a Wick rotation of the time-like coordinate $t=\rmi t_E$, and recognizing that one needs to demand $t_E\sim t_E+\frac{4\pi}{f'(r_h)}$ in order for the Euclidean-signature metric to have no conical singularities in the limit $\epsilon\rightarrow0$.}. The metric now takes the form
 \begin{equation}\label{eq:kruskal metric}
     \rmd s^2=\frac{f(r)}{UV}\rmd U~\rmd V+r^2\rmd\Omega_{d-1}^2~.
 \end{equation} 
At this point, the map (\ref{eq:Kruskal R Sch}) only produces $U < 0$ and $V > 0$. However, one can define an additional pair of (left) EF coordinates $u_{\rm L}$,$v_{\rm L}$, given by
\begin{equation}
    u_{\rm L}=-t_L+r^*(r)~,\quad v_{\rm L}=-t_L-r^*(r)~,
\end{equation}
which will be mapped to Kruskal coordinates as
\begin{equation}\label{eq:Kruskal L Sch}
    U=\rme^{-\frac{2\pi}{\beta}u_{\rm L}}~,\quad V=-\rme^{\frac{2\pi}{\beta}v_{\rm L}}~,
\end{equation}
allowing for $U > 0$ and $V < 0$. Notice that the combination of left and right lightcone coordinates, resulting in the coordinate range $U,~V \in(-\infty,~\infty)$ allows us to cover the whole spacetime, which we refer to as the \textit{maximal analytic continuation} of the black hole geometry.
  
\subsection{Penrose diagrams}
\begin{figure}
    \centering
    \includegraphics[width=0.8\textwidth]{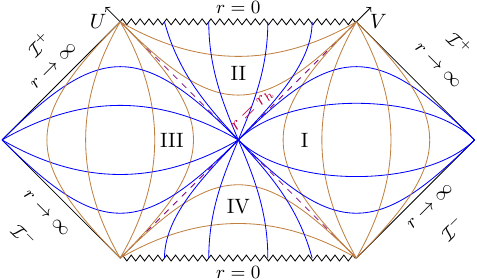}
    \caption{Penrose diagram of a Schwarzschild black hole. The different quadrants are separated by the black hole horizon ($r=r_h$, purple dashed line), where each point in the diagram represents a ($d-1$)-sphere. The $r=$constant and $t=$constant curves are shown in brown and blue respectively. Here, the Kruskal coordinates are represented by black arrows. As indicated at the start of the section, the Schwarzschild black hole approaches Minkowski space as $r\rightarrow\infty$, so its asymptotic boundary is represented by a null hypersurface (denoted by $\mathcal{I}^+$ in the future, $\mathcal{I}^-$ in the past).}
    \label{fig:Schwarzschild}
\end{figure}

We want to compactify the above-defined Kruskal coordinates in order to draw spacetime diagrams using finite coordinate ranges. This means that we need to look for a function whose image is $(-\infty,~\infty)$ for a finite domain; such as the $\tan(x)$, with $x\in\qty[-\frac{\pi}{2},~\frac{\pi}{2}]$. Then, we can define the following set of compact range coordinates
 \begin{equation}\label{eq:compact coord}
     \tilde{U}=\arctan(U)~,\quad \tilde{V}=\arctan(V)~,
 \end{equation}
to draw a diagram capturing the causal relations between spacetime points, which we refer to as a \emph{Penrose diagram}. In the (\ref{eq:compact coord}) coordinates, null rays are oriented at 45$^\degree$ angles. For an example using the Schwarzschild black hole, see Fig. \ref{fig:Schwarzschild}.

\section{Anti-de Sitter space black holes}

In this section, we consider general black hole solutions  in ($d + 1$)-dimensions for an AdS universe ($\Lambda<0$) as the vacuum solution to Einstein equations, resulting in the metric
\begin{equation}\label{eq:metric SAdS}
\begin{aligned}
        \rmd s^2&=-f(r)\rmd t^2+\frac{\rmd r^2}{f(r)}+r^2\rmd \Omega_{k,~d-1}^2~,\\
    f(r)&=\qty(\frac{r}{L})^2\qty(1-\frac{C}{r^d})+k~;
\end{aligned}
\end{equation}
where $k \in \qty{0,~\pm1}$ indicates the curvature of the ($d-1$)-dimensional line element; $C$ is a constant (related with the event horizon $r_h$ through $C=L^2r_h^{d-2}\qty(\frac{r_h^2}{L^2}+k)$ \cite{Carmi:2017jqz}); and $L$ the AdS radius, related to $\Lambda$ similar to (\ref{eq:Lambda dS})
\begin{equation}\label{eq:Lambda AdS}
    \Lambda=-\frac{d(d-1)}{2L^2}~.
\end{equation}
The case $k = 1$ in (\ref{eq:metric SAdS}) is called a Schwarschild-AdS (SAdS) black hole; while the $k = 0$ case is called a planar AdS black hole; and $k=-1$ is a hyperbolic black hole. Some examples of types of black holes in top-down constructions can be found in e.g. \cite{Gubser:1996de,Klebanov:1996un,Witten:1998zw,Emparan:1999gf}, and references within.

\begin{figure}
    \centering
    \includegraphics[width=0.6\textwidth]{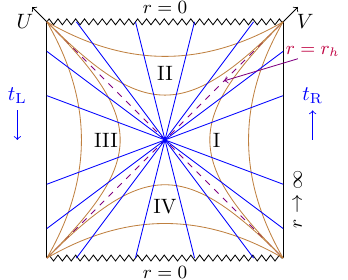}
    \caption{Penrose diagram for a two-sided eternal AdS$_{d+1}$ black hole (\ref{eq:metric SAdS}). The black hole event horizon is located at $r=r_h$ (purple dashed line), with a singularity at $r=0$ (zigzag line). Each point represents a compact $(d-1)$-surface of curvature $k=0,~\pm1$. The curves of constant tortoise coordinate $r^*(r)$ are indicated in brown, and constant time $t_{\rm L/R}$ in blue. The time arrows (blue) run along both boundaries (at $r\rightarrow\infty$). The Kruskal coordinate axes correspond to the black arrows.}
    \label{fig:AdS_BH}
\end{figure}

The Penrose diagram of a two-sided AdS black hole is shown in Fig. \ref{fig:AdS_BH}. Every point in the Penrose diagram represents a $(d - 1)$-dimensional surface with curvature $k=\pm1$ or $0$. Region II and IV indicate the black hole and white hole interiors respectively. Lastly, in this convention, the boundary time along quadrants I and III runs upwards or downwards respectively.

\subsection{Basics of the anti-de Sitter space/conformal field theory correspondence}
The AdS$_{d+1}$/CFT$_d$ correspondence can be summarized in terms of the partition function and the Hilbert space of quantum gravity on an asymptotically AdS$_{d+1}$ (times a compact manifold) with a CFT$_{d}$,
 \begin{equation}
 Z_{\rm gravity}=Z_{\rm CFT}~,\quad \mathcal{H}_{\rm gravity}=\mathcal{H}_{\rm CFT}~.    
 \end{equation}
Many efforts have been carried out to properly define the gravity side of these relations, in particular, in the $G_N\rightarrow0$ regime where the bulk is expected to be described in terms of a sum over semi-classical geometries (such as black holes with excitations of bulk fields).

As we commented in the introduction, the correspondence was introduced in the context of string theory in \cite{Maldacena:1997re}. Here, quantum particles are described as vibrations of (1+1)-dimensional fundamental objects, called strings, embedded in our higher-dimensional spacetime. Importantly, massless particles with spin-two are part of the spectrum of string vibrations, which is a promising candidate to describe gravity. It was found in \cite{Maldacena:1997re} that observables of ten-dimensional superstring theory on an AdS$_5\times$S$^5$ background (such as correlation functions) can be directly recovered from a (1+3)-dimensional $\mathcal{N} = 4$ SU($N$) super-Yang-Mills theory, a conformal theory. We will not go into detail about this topic, as it is not required for the particular topics explored in this thesis. Generically, the boundary conditions in the bulk fully specify the CFT and its particular state. We refer the reader to \cite{Aharony:1999ti} for a classic review, and \cite{Hubeny:2014bla} for a modern review on the topic.
 
According to the holographic correspondence, the eternal two-sided AdS black hole is dual to a pair of identical and decoupled CFTs living in its asymptotic boundary. Since they are decoupled, we can describe their Hilbert space as a product $\mathcal{H} = \mathcal{H}_1\otimes\mathcal{H}_2$. 
Let us focus on the SAdS case ($k=1$), whose asymptotic boundary is compact, such that the energy level $E_n$ are discrete, with corresponding eigenstates $\ket{n}_{\rm L}$ and $\ket{n}_{\rm R}$. In \cite{Maldacena:2001kr} it was realized that the CFTs that describe the eternal AdS black hole form an entangled state called the thermofield double (\gls{TFD}) state,
\begin{equation}
    \ket{\rm TFD}=\frac{1}{\sqrt{Z_\beta}}\sum_{n}\rme^{-\beta E_n/2}\ket{n}_{\rm L}\otimes\ket{n}_{\rm R}~,
\end{equation}
with $Z_\beta$ is the canonical partition function at inverse temperature $\beta$. Then, each of the boundaries is described by a thermal density matrix. Moreover, the evolution of the TFD state results from the boundary Hamiltonians as
\begin{equation}
    \rmi\partial_{t_{\rm L/R}}\ket{\rm TFD(t)}=H_{\rm L/R}\ket{\rm TFD(t)}~.
\end{equation}
Meanwhile, the total Hamiltonian $H = H_{\rm R}+H_{\rm L}$ is the generator of time translation for $t = t_{\rm R} + t_{\rm L}$, so that the TFD state can be described as
\begin{equation}
    \ket{\rm TFD(t)}=\frac{1}{\sqrt{Z_\beta}}\sum_{n}\rme^{(\rmi t-\beta/2) E_n}\ket{n}_{\rm L}\otimes\ket{n}_{\rm R}~.
\end{equation}
While the TFD evolves non-trivially under $H_{\rm R} +H_{\rm L}$; it can be seen that $(H_{\rm R} -H_{\rm L})\ket{\rm TFD(t)}=0$. The bulk interpretation is that there is a boost time isometry in the AdS black hole, seen by the following transformation in the metric in Kruskal coordinates (\ref{eq:Kruskal R Sch}, \ref{eq:kruskal metric}, \ref{eq:Kruskal L Sch}):
\begin{equation}
    U\rightarrow\rme^{\pm\frac{2\pi}{\beta}t}U~,\quad V\rightarrow\rme^{\mp\frac{2\pi}{\beta}t}V~.
\end{equation}
This isometry will be employed in Ch. \ref{ch:ComplexityAdS}, \ref{ch:ComplexitydS}, and \ref{ch:ComplexityDSSYK}.

\subsection{Shockwave geometry}\label{ssec:SWs AdS}
Let us perturb the TFD state by inserting a local unitary operator $W$ on the left or right CFT at a time $t$,
\begin{equation}\label{eq:precursor insertion}
    W(t)\ket{\rm TFD}~.
\end{equation}
We will associate an energy to the operator through the relation $W(t)=\rme^{-\rmi Ht}$, with $H$ the Hamiltonian. When this energy is of the order of the black hole temperature, we will refer to $W(t)$ as a \emph{thermal scale operator}. It noticed in \cite{Shenker:2013pqa} that the bulk dual geometry with respect to (\ref{eq:precursor insertion}) can be interpreted as a localized null energy perturbation, called a \textbf{shockwave}, acting on the eternal AdS black hole geometry.
 
\begin{figure}
    \centering
        \subfloat[]{\includegraphics[height=0.33\textwidth]{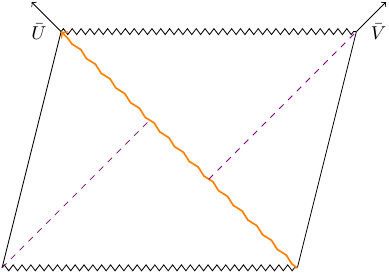}\label{sfig:SW AdS}}\hspace{0.4cm}\subfloat[]{\includegraphics[height=0.33\textwidth]{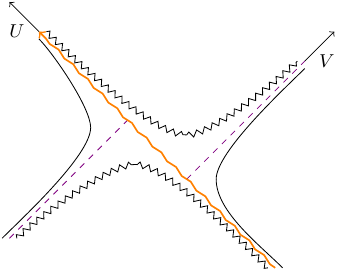}}
    \caption{Shockwave geometry for an insertion along $U=0$ (orange wavy line) in an asymptotically AdS$_{d+1\geq4}$ black hole. (a) Penrose diagrams in the $\bar{U}$, $\bar{V}$ coordinates defined in (\ref{eq:conformal coord}). (b) Diagram in Kruskal coordinates (\ref{eq:metric SW alpha}). The $U$ and $V$-axes are displayed with black arrows, and the dashed lines represent the event horizon. Figure based on \cite{Shenker:2013pqa}.}
    \label{fig:SW SAdS}
    \end{figure}

We consider that the insertion of (\ref{eq:precursor insertion}) in the CFT corresponds to a spherically symmetric shockwave inserted at time $t=t_0$ on a surface $r=r_{\mathcal{O}}$ close to the asymptotic boundary of the AdS black hole (which belong to the class of Vaidya metrics \cite{Vaidya:1999zz,Vaidya:1951fdr,Wang:1998qx}); and it propagates along the surface $U=U_0$, $\tilde{U}=\tilde{U}_0$, which we define as
\begin{equation}\label{eq:Kruskal SAdS}
    \begin{aligned}
        \tilde{U}_0&=\rme^{\frac{\tilde{f}'(\tilde{r_h})}{2}\left(\tilde{r}_{*}(r_{\mathcal{O}})-t_0\right)}~,\\
    U_0&=\rme^{\frac{f'(r_h)}{2}\left(r_{*}(r_{\mathcal{O}})-t_0\right)}\,.
    \end{aligned}
\end{equation}
See Fig. \ref{fig:SW SAdS} for an illustration of the effects of the shockwave.

We need to impose that the metric is continuous at the location where we relate $\tilde{V}$ to $V$ \cite{Israel:1966rt}, which we express as follows
\begin{equation}
\begin{aligned}
    \tilde{U}_0 \tilde{V}&=-\rme^{\tilde{f}'(\tilde{r}_{\rm b})\tilde{r}_{*}(r)}~,\\
    \quad U_0 V&=-\rme^{f'(r_h)r_{*}(r)}\,.
\end{aligned}
\end{equation}
When working in the small energy limit $E/M=\epsilon\ll1$, $\tilde{U}_0$ can be well approximated by $U_0$, allowing us to deduce that at first order the shift along the $V$ coordinate can be expressed like so
\begin{equation}\label{eq:shifts alpha}
\begin{aligned}
    \tilde{V}&=V-\frac{\epsilon}{U_o}\left(\frac{\partial}{\partial \epsilon}\rme^{\tilde{f}'(\tilde{r}_{\rm b})\tilde{r}_{*}(r)}\right)_{\epsilon=0}\\
    &\equiv V+\alpha\,.
\end{aligned}
\end{equation}
As was pointed out in \cite{Shenker:2013pqa}, $\rme^{f'(r_h) r_{*}(r)}$ can be compactly written as 
\begin{equation}
\begin{aligned}
   \rme^{f'(r_h) r_{*}(r)}\equiv (r-r_h)\,C_{\rm b}(r,r_h)\,,
\end{aligned}
\end{equation}
where $C_{\rm b}(r,r_h)$ above is a smooth and nonzero at $r=r_h$. The shift $\alpha$ becomes
\begin{equation}
\begin{aligned}
    \alpha&=\frac{G_N E \beta_{\rm b}}{ 2\pi r_h U_0}\frac{\partial}{\partial r_h}\bigg((r_h-r)\,C_{\rm b}(r,r_h)\bigg)\,,
\end{aligned}
\end{equation}
 where we used $\tilde{M}=M(1-\epsilon)$ and normalized the temperature with respect to the observer radius. Notice that Einstein equations (\ref{eqe:EEqs}) imply that
\begin{equation}
    T_{uu}=\frac{\alpha}{4\pi G_N}\delta(u)~,
\end{equation}
which means that the shockwave can be identified with localized null particles.\footnote{The null energy condition (\gls{NEC}) (see \cite{Rubakov:2014jja} for a review) states that for every future-pointing null vector field $k\mu$, the stress tensor satisfies $T_{\mu\nu}k^\mu k^\nu\geq0$, which for the shockwave source implies that $\alpha\geq0$. Classical matter satisfies NEC \cite{Penrose:1964wq}, while, commonly, quantum particles satisfy an averaged version \cite{Flanagan:1996gw}. This has implications for the causal structure of spacetime, which will not be discussed here. The reader is referred to \cite{Flanagan:1996gw} for more details.}

\begin{figure}
    \centering
   \subfloat[]{\includegraphics[width=0.8\textwidth]{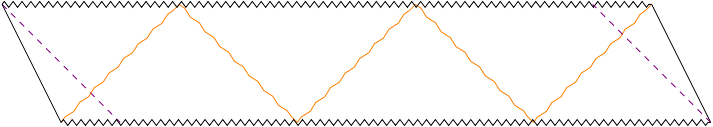}}\\\subfloat[]{\includegraphics[width=0.8\textwidth]{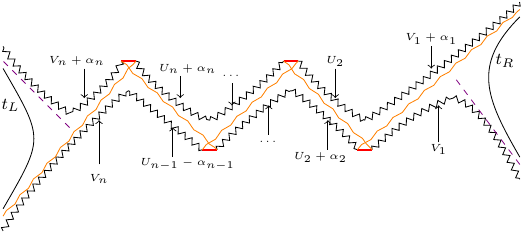}}
    \caption{Geometry of eternal AdS$_{d+1\geq1}$ black hole with multiple shockwaves (orange wavy lines) using (a) (\ref{eq:conformal coord}) coordinates, and (b) (\ref{eq:metric SW alpha AdS}) coordinates. Figure based in \cite{Shenker:2013yza}.}
    \label{fig:Kruskal_SW_AdS_BH}
\end{figure}

We now extend this result to allow insertions by considering a sequence of local unitary thermal scale operators $\qty{W_{i}(t)}$ on the TFD state, as
\begin{equation}
    W_n(t_n)\dots W_1(t_1)\ket{\rm TFD}~,
\end{equation}
where $n$ is the total number of insertions, which will be taken even, and we consider that the time separations are sufficiently large and performed in alternating order; this is defined as
\begin{equation}\label{eq:time insertions AdS}
t_{2k+1}>t_{2k},\quad t_{2k}< t_{2k-1}~,\quad t_{2k+1}-t_{2k}\gg t_*^{(\rm b)}~.
\end{equation}
where $k\in1,\dots, n/2$; and $t_*^{(\rm b)}$, called the scrambling\footnote{Heuristically, scrambling refers to the delocalization of information that cannot be probed using simple correlation functions. The reader is referred to \cite{Shenker:2013pqa} for more details.} time of the AdS black hole, is defined through the relation
\begin{equation}\label{eq:scrambling BH}
\begin{aligned}
    \alpha=2\rme^{-\frac{f'(r_b)}{2}(t_b^{(*)}\pm t_0)}~.
    \end{aligned}
\end{equation}
The bulk dual geometry represents the doubled-sided AdS black hole with $n$ alternating left- and right-moving shockwaves sent along $U=0$, as originally proposed by \cite{Shenker:2013yza}, and illustrated in Fig. \ref{fig:Kruskal_SW_AdS_BH}.
This state can be then described using the Kruskal coordinate metric (\ref{eq:kruskal metric}) with the shifts in $V$ (\ref{eq:shifts alpha}) as
\begin{align}\label{eq:metric SW alpha AdS}
        \rmd s^2=&-2A\qty(U[V+\alpha_i\Theta(U)])
    \rmd U\rmd V\\
    &+B\qty(U[V+\alpha_i\Theta(U)])\rmd \vec{x}^2~,
\end{align}
where
\begin{equation}
\begin{aligned}
    A\qty(UV)&\equiv-\frac{2}{UV}\frac{f(r)}{f'(r_h)^2}~,\quad B\qty(UV)\equiv r^2~,\\ \alpha_i&=2\rme^{-\frac{f'(r_h)}{2}(t_*^{(\rm b)}\pm t_i)}~.\label{eq:factors a b AdS}
\end{aligned}
\end{equation}
Here, $t_i$, with $i\in1,\dots n$, are the shockwave insertion times with respect to the asymptotic boundary, and the $\pm$ sign indicates the direction that the shockwaves are sent to.
 
\section{De Sitter space}
\begin{figure}
     \centering
     \includegraphics[width=0.6\textwidth]{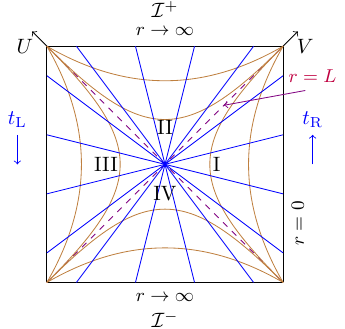}
     \caption{Penrose diagram of empty dS$_{d+1}$ space. The quadrants are separated by a cosmological horizon ($r=L$, purple dashed lines). Here, future and past time-like infinity ($\mathcal{I}^+$, $\mathcal{I}^-$ respectively) are surfaces in the inflating patch located at $r\rightarrow\infty$. The curves of constant radius are indicated in brown. In these coordinates, the time, indicated in blue, runs along the $r=0$ surface (worldline of the static patch observer). The Kruskal coordinate axes are indicated by black arrows.}
     \label{fig:dS Penrose}
 \end{figure}

Now, we search for solutions to the vacuum Einstein equations with $\Lambda >0$, starting from the simplest one, an empty dS universe, whose metric is given by (\ref{eq:spherically symmetric metric}) with (\ref{eq:static patch dS}). A detailed Penrose diagram for this spacetime is shown in Fig. \ref{fig:dS Penrose}.
 Notice that $f(L)=0$, which indicates the existence of a cosmological horizon. The spacetime region accessible to a static patch observer (corresponding to quadrant I or III)  at $r=0$ corresponds to $r<L$, while the region beyond $r=L$ (quadrant III and I) is causally disconnected with respect to them. Also, notice that there is no longer a singularity in the metric at $r=0$. Moreover, the cosmological horizon is now observer-dependent, meaning that if the wordline of the static patch observer moves somewhere, their cosmological horizon will move along, staying at the same distance $r=L$.

We will discuss the shockwave geometries of dS space in Sec. \ref{ssec:shockwaves SdS}. Below, we comment on the geometry of dS space and static patch holography.
 
 \subsection{Stretched horizons}
\label{ssec:stretched_horizons}

When comparing Fig. \ref{fig:AdS_BH} and Fig. \ref{fig:dS Penrose}, there are some similarities, the dS Penrose diagram seems to be an inside-outside version of the AdS black hole one. This has motivated the approaches to static patch holography that we mentioned in the introduction, where one introduces a timelike Dirichlet surface in the static patch, for example at $r = 0$ (worldline holography \cite{Anninos:2011af,Leuven:2018ejp}), or very near the cosmological horizon (stretched horizon holography \cite{Dyson:2002pf,Susskind:2011ap,Susskind:2021esx,Susskind:2021dfc,Shaghoulian:2021cef,Susskind:2021omt,Shaghoulian:2022fop}). Both of these approaches can realized by setting
\beq
r_{\rm st} = \rho L \, , \qquad
\rho \in [0,~1) \, ,
\label{eq:stretched_horizon_dS}
\eeq
where $r=r_{\rm st}$ is the location of the timelike Dirichlet surface, which is illustrated in Fig. \ref{fig:StrechedHorizon}. This type of ansatz for the location of the holographic dual theory to dS space will be considered in the holographic complexity proposals studies in Ch. \ref{ch:ComplexitydS}.
 
\section{Schwarzschild-de Sitter space}
\subsection{Geometry}
We consider the Schwarzschild-dS (SdS) spacetime in $d+1$ dimensions, whose metric takes the form of (\ref{eq:spherically symmetric metric}) with $\Lambda = \frac{d(d-1)}{2 L^2}$ and a blackening factor:
\beq
f (r) = 1 - \frac{2 \mu}{r^{d-2}} - \frac{r^2}{L^2} \, ,
\label{eq:asympt_dS}
\eeq
where $\mu$ a parameter related to the mass of the black hole $M$ as \cite{PhysRevD.15.2738,Balasubramanian:2001nb,Ghezelbash:2001vs}
\begin{equation}
    \mu=\frac{8\pi G_NM}{(d-1)\Omega_{d-1}}~,
\end{equation}
where $\Omega_{d-1}=2\pi^{d/2}/\Gamma(d/2)$ is the volume of a unit ($d-1$)-sphere.

When $\mu=0$, the geometry reduces to empty dS space. The case $d=2$ is special because the term proportional to $r^{-(d-2)}$ inside the blackening factor in \eqref{eq:asympt_dS} becomes constant.
In this special case, the black hole can be obtained as a discrete quotient of empty dS space with a conical defect at $r=0$.

\begin{figure}[t!]
    \centering
    \includegraphics[width=0.8\textwidth]{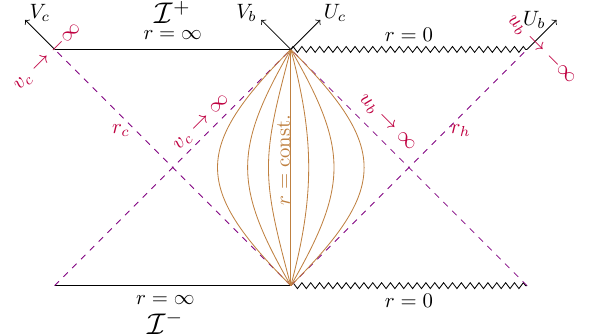}
    \caption{Penrose diagram of SdS$_{d+1}$ space in dimensions $d\geq 3,$ in the regime where the mass parameter satisfies $ \mu \in (0, \mu_{N}).$ 
    $r_h$ denotes the black hole horizon and $r_c$ the cosmological horizon (purple). We display the orientation of the null coordinates in eqs.~\eqref{eq:general_null_coordinates} and \eqref{eq:Kruskal_coord_sec2}, where the subscripts $\lbrace b,c \rbrace $ denote black hole and cosmological patches, respectively. Constant $r$ surfaces are indicated by brown lines.} 
    \label{fig:Penrose_SdS}
\end{figure}

In this work, we will only focus on the case when $d \geq 3$ and the parameter $\mu$ belongs to the range  $\mu\in (0,\,\mu_N)$, where
\beq
\mu_{N} \equiv \frac{r_{N}^{d-2}}{d}  \, , \qquad
r_{N} \equiv L \sqrt{\frac{d-2}{d}} \, .
\label{eq:critical_mass_SdS}
\eeq
In this regime, the blackening factor admits two positive roots denoted as $r_h <  r_c$, where the larger value represents a cosmological horizon and the smaller one a black hole horizon.
The Penrose diagram corresponding to the previous range of $\mu$ is depicted in \Fig{fig:Penrose_SdS}.
In the following, we will refer to the left square of the causal diagram (containing a cosmological horizon) as the \textit{cosmological} or \textit{inflating} region of the geometry, and to the right side of the causal diagram as the \textit{black hole} region.
We will also refer to the region described by the radial coordinate $r<r_h$ as the interior of the black hole, and $r>r_h$ as its exterior.
Similarly, $r<r_c$ describes the interior of the cosmological horizon, while for $r>r_c$ the exterior.
In particular, the static patch $r_h \leq r \leq r_c$ is located in the exterior of the black hole, but inside the cosmological horizon.

\subsubsection{Penrose diagram}
The Penrose diagram in \Fig{fig:Penrose_SdS}, corresponding to the maximal analytic extension of the metric, is obtained by compactifying the Kruskal coordinates $(U,V)$ defined in the central quadrant as follows\footnote{Notice that despite the existence of two sets of Kruskal coordinates in the middle quadrant in Fig.~\ref{fig:Penrose_SdS}, the null coordinates $(u,v)$ are uniquely defined in such region. Similar conventions apply to the Kruskal coordinates for Kerr black holes \cite{Heinicke:2014ipp}.  } 
\beq
 U_c =  e^{\frac{u}{L}}  \, , \quad
V_c = - e^{-\frac{v}{L}}  \, ,\qquad U_b =  -e^{-\frac{u}{L}}  \, , \quad
V_b = e^{\frac{v}{L}}  \, ,
\label{eq:Kruskal_coord_sec2} 
\eeq
Due to the existence of two horizons, we need to introduce two sets of Kruskal coordinates: $(U_c, V_c)$ cover the region with radial coordinate $r \in (r_h, \infty) $, while $(U_b, V_b)$ cover the portion with $r \in (0,r_c)$.
Both the coordinate systems are well-defined in the central static patch with $r \in (r_h, r_c)$, where one can change the coordinate chart from one system to the other.

The cosmological and black hole horizons emit thermal radiation, with associated Hawking temperature and entropy given by\footnote{ \label{foot:extremal_temp} A consistent extremal limit for the temperature of the Nariai black hole is only achieved with the Hawking-Bousso choice, corresponding to $\gamma=1/\sqrt{f(r_0)}$ with $r_0$ such that $f'(r_0)=0$. This subtlety is discussed, \eg in \cite{Bousso:1996au,Morvan:2022ybp}, but it does not play a relevant role in the present chapter.  }
\beq
     T_{h(c)} = \frac{1}{4 \pi} \left|\frac{\partial f(r)}{\partial r}\right|_{r=r_{h(c)}}  
    \qquad
    S_{h(c)} =  \frac{\Omega_{d-1} r_{h(c)}^{d-1}}{4 G} \, .
\label{eq:temp_entropy_SdS_general}  
\eeq
Notice that the Hawking temperatures in \eqref{eq:hor_cosm_temp_SdS} are defined with respect to the so-called Killing vector $\xi^\mu =\delta^\mu_t$ associated with time boots in static patch coordinates. In general, Killing-vectors are defined such that a change of coordinates ${x'}^\mu=x^\mu+\epsilon\xi^\mu$ does not alter the metric, i.e. $g_{\mu\nu}=g'_{\mu\nu}$. We will employ a notation $\xi = \partial_t$ from now on to refer to the static patch time isometry Killing vector. In contrast, there is no global timelike Killing vector of dS space.

In a general number of dimensions, it is possible to exploit the constraints $f(r_h)=f(r_c)=0$ to write all the relevant quantities defining the SdS solution in terms of the two horizons.
Therefore, we obtain the following expressions for the mass parameter and dS curvature length 
\beq
\mu = \frac{1}{2} \frac{r_c^d r_h^{d-2} - r_h^d r_c^{d-2}}{r_c^d - r_h^d} \, , \qquad
L^2 = \frac{r_c^d - r_h^d}{r_c^{d-2} - r_h^{d-2}} \, ,
\label{eq:general_relation_rhrcm}
\eeq
the blackening factor 
\beq
f(r) = \frac{1}{r_c^d - r_h^d} 
\left[ r_c^d \le 1 - \frac{r^2}{r_c^2} - \frac{r_h^{d-2}}{r^{d-2}}  \ri
- r_h^d \le 1 - \frac{r^2}{r_h^2} - \frac{r_c^{d-2}}{r^{d-2}}  \ri
\right] \, ,
\label{eq:general_blackening_SdS_rcrh}
\eeq
and the Hawking temperatures 
\beq 
T_h  = d \, \frac{r_{N}^2 - r_h^2 }{4 \pi r_h L^2}  \, , \qquad
    T_c  = d \, \frac{r_c^2 -r_{N}^2 }{4 \pi r_c L^2}  \, .
   \label{eq:hor_cosm_temp_SdS}
\eeq
For any mass parameter in the range $\mu \in (0, \mu_N)$, we get $T_h > T_c,$ which shows that the system is out of equilibrium.

In dimensions $d\geq 3,$ the Nariai geometry corresponds to the limiting case where $\mu=\mu_{N}$, such that the black hole and cosmological horizons approach each other.
However, the proper distance between the event horizons does not vanish, because the blackening factor $f(r)$ is also small in this regime.
A proper description of the spacetime in this limit is obtained by taking a near-horizon limit, \eg see Sec. 3.1 in \cite{Anninos:2012qw} or App. B in \cite{Svesko:2022txo}.
The Nariai geometry can be mapped to $\mathrm{dS}_2 \times S^{d-1}$ by a change of coordinates \cite{Maldacena:2019cbz,Anninos:2012qw}.

\paragraph{Four dimensions.}
When $d=3$, certain analytic expressions are available.
The blackening factor in \eqref{eq:asympt_dS} becomes 
\beq
f_{\rm SdS_4} (r) = 1 - \frac{2\mu}{r} - \frac{r^2}{L^2} \, ,
\label{eq:blackening_SdS4}
\eeq
and the critical mass in \eqref{eq:critical_mass_SdS} is $\mu_N^2 = L^2/27$.
In the regime $ \mu \in (0, \mu_N)$ when the blackening factor admits two roots, one can invert the identities  \eqref{eq:general_relation_rhrcm} specialized to $d=3$ and find a closed form for the event horizons in terms of the dS radius and the mass parameter \cite{Shankaranarayanan:2003ya,Choudhury:2004ph,Visser:2012wu}:
\beq
\begin{aligned}
       r_h &= r_{N} \le \cos \eta - \sqrt{3} \sin \eta \ri \, , \quad
r_c = r_{N} \le \cos \eta + \sqrt{3} \sin \eta \ri \, , \\ 
\eta &\equiv \frac{1}{3} \, \mathrm{arccos}  \le \frac{\mu}{\mu_{N}} \ri \, .
\end{aligned}
\label{eq:analytic_rh_rc_SdS4}
\eeq
The inverse of the blackening factor \eqref{eq:general_blackening_SdS_rcrh} can be integrated analytically, giving the tortoise coordinate
\beq
\begin{aligned}
r^*(r) = &
- \frac{r_c^2 + r_h r_c + r_h^2}{(r_c-r_h)(2r_c+r_h) (r_c+2r_h)} 
\left[ r_c^2 \log \left| \frac{r-r_c}{r+r_c+r_h} \right| \right. \\
& \left. + 2 r_h r_c \log \left| \frac{r-r_c}{r-r_h} \right|
-  r_h^2 \log \left| \frac{r-r_h}{r+r_c+r_h} \right|
\right] \, .
\end{aligned}
\label{eq:tortoise_coord_SdS4}
\eeq
These expressions will be used for any numerical computation involving the SdS$_4$ background in the remainder of the thesis.

\subsection{Shockwaves}\label{ssec:shockwaves SdS}
We will be interested in describing shockwaves in the geometry in Sec. \ref{sec:CAny CMC}. It is convenient to use Kruskal coordinates (\ref{eq:Kruskal_coord_sec2}) on the black hole and inflating patches. These patches are centered at the horizons $r_{h, \rm c}$, and cover the range $0\leq r< r_{\mathcal{O}}$ and $r_{\mathcal{O}} \leq r < \infty$ respectively, with $r_{\mathcal{O}}$ a reference point, which we take as the location of the static patch observer. In the present chapter, we will focus on the inflating region, as this has more relevance for our discussion in Sec. \ref{sec:CAny CMC}. We replace $U_c$, $V_c\rightarrow$ $U$, $V$ in what follows. We can then express (\ref{eq:asympt_dS}) as
\begin{equation}
    \rmd s^2=-\frac{4 f(r)}{f'(r_c)}\rme^{-f'(r_c)r_{*}(r)}\rmd U\,\rmd V +r^2 \rmd \Omega_{d-1}^2
\end{equation}
where $r_{\mathcal{O}} \leq r < \infty$.

\begin{figure}
    \centering
        \subfloat[]{\includegraphics[height=0.33\textwidth]{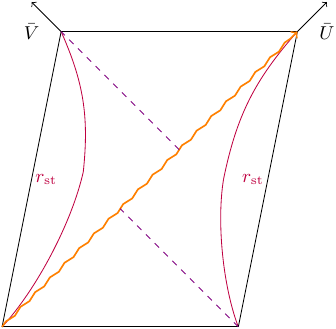}}\hspace{0.45cm}\subfloat[]{\includegraphics[height=0.35\textwidth]{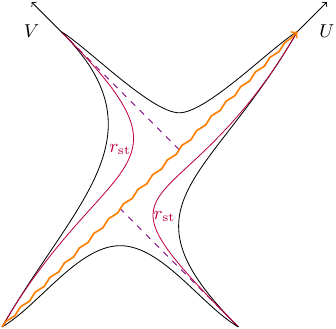}}\vspace{0.5cm}\\
        \subfloat[]{\includegraphics[height=0.35\textwidth]{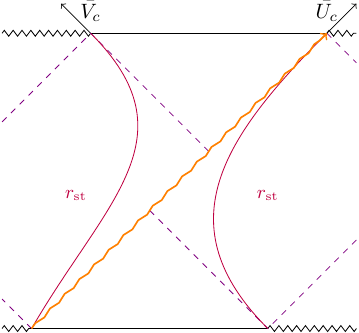}}\hspace{0.5cm}\subfloat[]{\includegraphics[height=0.36\textwidth]{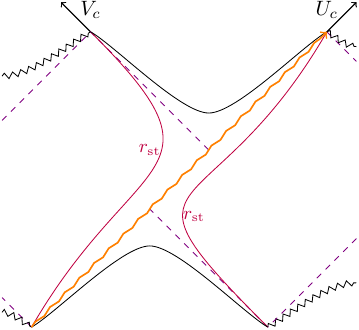}}
    \caption{Shockwave geometry for an insertion along $U=0$ (orange wavy line) in: dS$_{d+1}$ space (and SdS$_3$ space) (a, b); and SdS$_{d+1\geq4}$ space (c, d). \textit{Left column}: Penrose diagrams in the $\bar{U}$, $\bar{V}$ coordinates defined in (\ref{eq:conformal coord}). \textit{Right column}: Diagrams in Kruskal coordinates (\ref{eq:metric SW alpha}). The stretched horizon (fuchsia) is shown at a fixed location $r=r_{\rm st}$. The cosmological and black hole horizons are shown with the dashed (purple) lines. In all cases, the $U$, $V$-axis are displayed with black arrows.}
    \label{fig:SW SdS}
    \end{figure}

Once we add shockwave perturbations, the geometry becomes distorted, see Fig. \ref{fig:SW SdS}. Notice that, the location of the stretched horizon in the static patch can be modified due to the shift displayed in the diagrams. Although stretched horizon holography fixes the location of the dual theory to be located at a constant $r$ surface in the static patch, it remains unknown how the theory should behave under shockwave perturbations. We will be considering the case where the shockwaves are sent through $U=0$ and the stretched horizon remains fixed at a constant $r=r_{\rm st}$ coordinate\footnote{Similar considerations have been carried out in \cite{Baiguera:2023tpt,Anegawa:2023dad}, who have also analyzed alternative proposals.}, for which the evolution along the stretched horizon is continuous.
We are mainly interested in the metric under shockwave perturbations in the weak gravitational coupling regime to define the notions of energy below. We consider an SdS black hole with mass $M$ absorbs a shell of matter with mass $E\ll M$ along the surface 
\begin{equation}
    U=U_0=\rme^{\frac{f'(r_c)}{2}\left(r_{*}(r_{\mathcal{O}})-t_0\right)}~,
\end{equation}
with $t_0$ the static time shockwave insertion with respect to $r_{\mathcal{O}}$.

The SdS black hole after the shockwave has mass $M - E$ in (\ref{eq:asympt_dS}) for matter obeying the NEC \cite{Baiguera:2023tpt,Aguilar-Gutierrez:2023ymx}. We glue the coordinates along a shell $U,\, V$ to the past of the shell with those to the future, denoted by $\tilde{U},\, \tilde{V}$. The resulting cosmological line element for SdS black holes \cite{Aguilar-Gutierrez:2023ymx,Shenker:2013pqa}:
\begin{equation}\label{eq:tilde metric}
    \rmd s^2=-\frac{4 \tilde{f}(r)}{\tilde{f}'(r_c)}\rme^{-\tilde{f}'(\tilde{r}_{\rm c})\tilde{r}_{*}(r)}\rmd \tilde{U}\rmd \tilde{V} +r^2 \rmd \Omega_{d-1}^2
\end{equation}
where tilded quantities are given by the replacement of $M\rightarrow M-E$ in the untilded ones. In the inflating patch, the shift along the $V$ coordinate can be described by a shift in the coordinate
\begin{equation}
    \tilde{V}=V-\alpha~.
\end{equation}
The NEC also imposes that $\alpha\geq0$ \cite{Aalsma:2021kle}; while $E\ll M$ guarantees we work in the $\alpha\ll1$ limit.\footnote{See \cite{Aguilar-Gutierrez:2023ymx} for remarks on the approximation for the extremal SdS limit.} Importantly, this allows for the static patches in Fig. \ref{fig:SW SdS} to become causally connected, as it has been shown rigorously by Gao and Wald \cite{Gao:2000ga},\footnote{As discussed in \cite{Almheiri:2024xtw}, the shockwave induces a time advance that allows signals to exchanged between static patches.} in contrast to crunching geometries where the shift in $\alpha$ takes the opposite sign for matter satisfying the NEC.

We will express the shift parameter as
\begin{equation}\label{eq:scrambling cosmo}
\begin{aligned}
    \alpha=2\rme^{-\frac{f'(r_c)}{2}(t_c^{(*)}\pm t_0)}~.
    \end{aligned}
\end{equation}
Here the $\pm$ sign depends on whether the shockwave is left or right moving, and the cosmological scrambling time $t_c^{(*)}$ will be defined through this relation.

\begin{figure}
    \centering
        \subfloat[]{\includegraphics[width=0.45\textwidth]{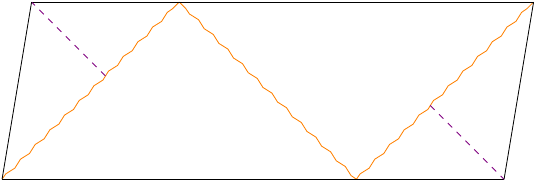}}\hspace{0.3cm}\subfloat[]{\includegraphics[width=0.45\textwidth]{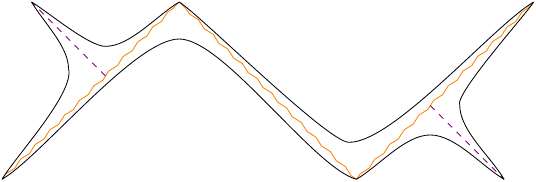}}\\
        \subfloat[]{\includegraphics[width=0.45\textwidth]{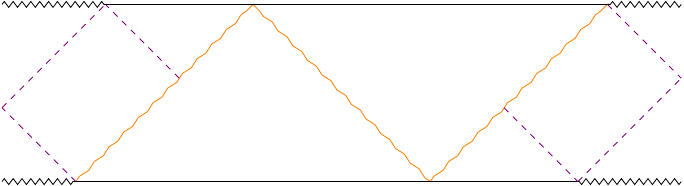}}\hspace{0.3cm}\subfloat[]{\includegraphics[width=0.45\textwidth]{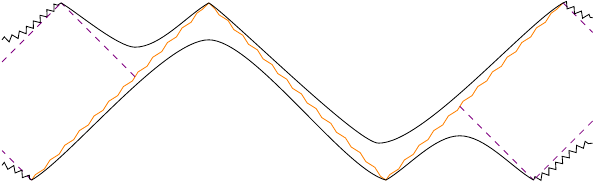}}
    \caption{Multiple shockwave geometry in SdS$_{3}$ space (above) SdS$_{d+1\geq4}$ space (below). \textit{Left column}: (\ref{eq:conformal coord}) coordinates, and \textit{right column}: (\ref{eq:Kruskal_coord_sec2}) coordinates. The figure shows two forward-evolving and one backward-evolving pulse, producing the corresponding shift (\ref{eq:scrambling cosmo}).}
    \label{fig:many SWs}
\end{figure}

As in Sec. \ref{ssec:SWs AdS}, we can further generalize this construction by performing a sequence of an even number of shockwaves, $n$, in the inflating patch (i.e. $r\in[r_{\mathcal{O}},\,\infty]$). Let us denote ${t}_1$, ${t}_2$, \dots, ${t}_n$ as the insertion static patch times with respect to the stretched horizon in alternating insertion order, i.e. ${t}_{2k+1}>{t}_{2k}$, and ${t}_{2k}<{t}_{2k-1}$, restricted to $\abs{{t}_{i+1}-{t}_i}\gg {t}_*$. {Fig. \ref{fig:many SWs} illustrates the multiple shockwave configuration in SdS space. The reader is referred to \cite{Shenker:2013pqa, Stanford:2014jda} for the asymptotic AdS black hole counterpart.}

We will take the shockwave close to the cosmological horizon and set $U_0=0$ in the following. (\ref{eq:tilde metric}) transforms into
\begin{align}\label{eq:metric SW alpha}
        &\rmd s^2=-2A\qty(U[V-\alpha_i\Theta(U)])
    \rmd U\rmd V+B\qty(U[V-\alpha_i\Theta(U)])\rmd \Omega_{d-1}^2~,
\end{align}
where $A(UV)$ and $B(UV)$ are given in (\ref{eq:factors a b AdS}), and considering the definition of $\alpha_i$ in (\ref{eq:scrambling cosmo}). Moreover, one can consider the change of coordinates
\begin{equation}\label{eq:conformal coord}
    \bar{U}=\arctan U,\quad \bar{V}=\arctan V~,
\end{equation}
for the respective Penrose diagram which is seen in Figs. \ref{fig:SW SdS} and \ref{fig:many SWs}.

\section{Extended Schwarzschild-de Sitter background.}
\begin{figure}
    \centering
\includegraphics[width=\textwidth]{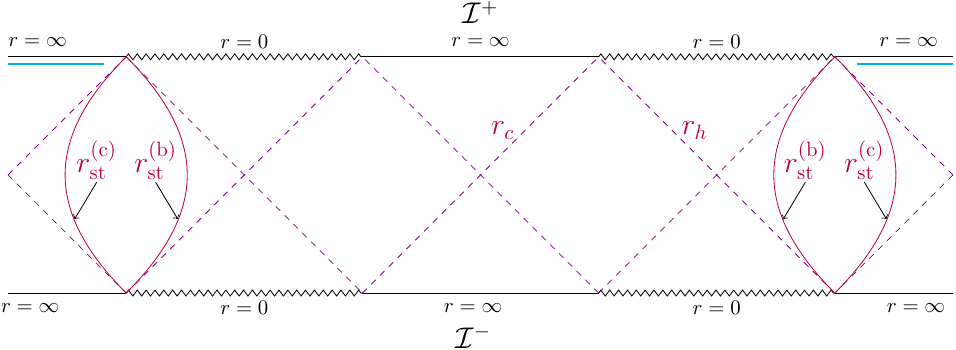}
\caption{SdS$_{d+1}^n$ space (illustrated for $n=2$) where $r_{\rm st}^{\rm (h)}$, $r_{\rm st}^{\rm (c)}$ (fuchsia solid lines) denote the stretched horizons close to the black hole and the cosmological horizon, respectively. $r_h$ and $r_c$ denote the black hole and cosmological horizon radii. The cyan line region near $\mathcal{I}^+$ indicates a spacelike slice where an observer could collect information encoded in the inflating region (as we consider in Ch. \ref{ch:ComplexitydS}).}
    \label{fig:extSdS}
\end{figure}

It is important to observe that SdS spacetime can be analytically extended even beyond the region depicted in \Fig{fig:Penrose_SdS}, \ie it is possible to build an infinite sequence of singularities 
and timelike infinities $\mathcal{I}^{\pm}$.
The Penrose diagram is composed of a periodic extension as depicted in \Fig{fig:extSdS}, where it is allowed to have $n$-pairs of black hole and cosmological patches.
We refer to this configuration as the \textit{extended SdS} spacetime, denoted with SdS$^n$.
Each of the static patches can be described by the metric introduced above:
\beq
\begin{aligned}
    \rmd s_{(i)}^2 &= - f\qty(r_{(i)}) \rmd t_{(i)}^2 + \frac{\rmd r^2}{f\qty(r_{(i)})} + r_{(i)}^2 \rmd \Omega_{d-1}^2 \, , \\  
f \qty(r_{(i)}) &= 1 - \frac{2 \mu}{r_{(i)}^{d-2}} - \frac{r_{(i)}^2}{L^2} \, .
\end{aligned}
\label{eq:multi_asympt_dS}
\eeq
The only caveat is that we need to introduce a copy of the coordinates for each copy of SdS, which here we denoted with the index $i \in \lbrace 1, \dots, n \rbrace$.
It is worth mentioning that the Euclidean continuation of SdS$^n$ geometry is singular, but this issue will not play a role in our classical gravitational analysis.\footnote{The Euclidean continuation is generically singular even when $n=1$ (\eg see \cite{Morvan:2022ybp}), unless one works in the extremal SdS limit, where the black hole and inflating patches have the same temperature. In the extended geometry, even at thermal equilibrium, there are conical singularities \cite{Aguilar-Gutierrez:2021bns,Levine:2022wos}.  \label{fn:1}}

\vskip 3mm
\noindent
Next, we define the relevant geometric quantities that will be studied in the extended SdS spacetime in this work.

\subsection{Stretched horizons}

We want to generalize the concept of stretched horizon in SdS$^n$ space. In the black hole geometry defined by \eqref{eq:asympt_dS}, the stretched horizon is defined
as a surface at the constant radial coordinate 
\beq
r_{\rm st} = (1-\rho) r_h + \rho r_c \, , \qquad
\rho \in [0,1] \, ,
\label{eq:stretched_horizon_SdS4}
\eeq
such that $r_{\rm st} \in [r_h, r_c]$ and the parameter $\rho$ interpolates between the two horizons.
In particular, we will be mainly interested in the limits $\rho \rightarrow 0$ and $ \rho \rightarrow 1$, when the stretched horizon approaches either the black hole or the cosmological horizon.\footnote{In this regard, a stationary observer in the static patch will generally experience an intermediate temperature between the cosmological and black hole one, since the system is out of equilibrium. However, if the stretched horizon is close to one of the horizons (either $\rho \rightarrow 0$ or $\rho \rightarrow 1$), then the corresponding flux of radiation will be dominant. }
There are various reasons that make this choice natural.
First, according to the central dogma for black holes and for inflationary models, the degrees of freedom and the unitary evolution of the dual quantum system should be determined from the region inside the cosmological horizon, but outside the black hole one \cite{Almheiri:2020cfm,Shaghoulian:2021cef,Aguilar-Gutierrez:2023zoi}.
This region is the static patch of the SdS background. 
Secondly, the choice \eqref{eq:stretched_horizon_SdS4} defines a timelike surface where a time coordinate for the dual quantum system can be consistently defined.
Had we chosen a surface at constant $r$ located in the intervals $[0,r_h]$ or $[r_c, \infty]$, it would have been spacelike instead.
Furthermore, holographic screens can be defined in other ways, for instance, by taking a congruence of geodesics and imposing that the expansion parameter vanishes \cite{Bousso:1999cb}.
In the case of a static and spherically-symmetric background like \eqref{eq:asympt_dS}, this requirement is equivalent to locating the stretched horizon close to one of the event horizon.
Finally, a natural static sphere observer is freely falling at fixed radius $r_{\mathcal{O}}$, \ie identified by the condition $f'(r_{\mathcal{O}})=0$. This requirement fixes $r_{\mathcal{O}} \in (r_h, r_c)$, which is the range where the stretched horizon is located. 

\begin{figure}
  	\centering
   \begin{subfigure}[t]{0.7\textwidth}
   \includegraphics[width=\textwidth]{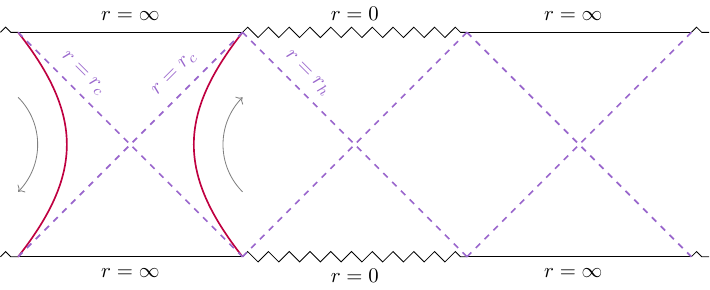}
   \caption{\textbf{Case 1}: two cosmological stretched horizons.
   }
   \label{eq:Single CH}
   \end{subfigure}
   \begin{subfigure}[t]{0.7\textwidth}
  	\includegraphics[width=\textwidth]{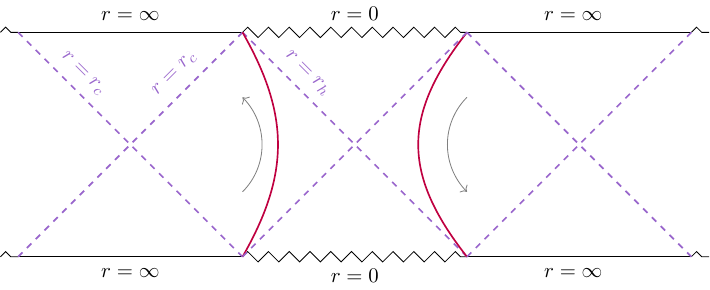}
   \caption{\textbf{Case 2}: two black hole stretched horizons.}
   \label{eq:Single BH}
   \end{subfigure}
  	\caption{ Prescriptions for the stretched horizons (in fuchsia) in a holographic setting. (a) The cosmological stretched horizons are placed in consecutive static patches. Geometric quantities anchored to the cosmological stretched horizons explore the region beyond the cosmological horizon. (b) The stretched horizons are placed in consecutive black hole patches. Geometric quantities anchored to the black hole stretched horizon explore the region behind the black hole horizon. The gray arrows represent the orientation of the Killing vector $\partial_t$. }
  	\label{fig:stretched-hor-cases12}
  \end{figure}
It turns out that the realization of the requirement \eqref{eq:stretched_horizon_SdS4} in SdS background leads to several different holographic settings:
\begin{enumerate}
\item \textbf{Two cosmological stretched horizons.} So far, all the studies of holography in asymptotically dS space in the literature focused on locating the stretched horizons just inside the cosmological horizon as depicted in Fig.~\ref{eq:Single CH}.
We refer to these surfaces at constant radii as the \textit{cosmological stretched horizons}, because they allow us to 
investigate the properties of the geometry of the cosmological region.
Indeed, it was conjectured that the Ryu-Takayanagi (RT) surface, the codimension-one extremal surfaces or the WDW patch should be anchored at the cosmological stretched horizons, such that these geometric objects lie in the inflating region of the Penrose diagram \cite{Shaghoulian:2021cef,Susskind:2021dfc,Susskind:2021esx,Shaghoulian:2022fop,Jorstad:2022mls}.
\item \textbf{Two black hole stretched horizons.} In comparison to empty dS space, the existence of a black hole event horizon in higher-dimensional SdS allows to define a so-called \textit{black hole stretched horizon} which still satisfies \eqref{eq:stretched_horizon_SdS4} but is now located in the black hole region of the Penrose diagram.
The corresponding scenario, depicted in Fig.~\ref{eq:Single BH}, implies that the geometric quantities lie behind the black hole horizon.
This configuration is very similar to the standard holographic setting considered in asymptotically AdS space. 
\item  \textbf{One cosmological and one black hole stretched horizon.}
We depict in Fig.~\ref{fig:cosmobh-stretched-hor} a novel prescription.
We take one cosmological stretched horizon and one black hole stretched horizon, both belonging to the same copy of SdS space.

\begin{figure}
    \centering
    \begin{subfigure}[t]{0.7\textwidth}
    \includegraphics[width=\textwidth]{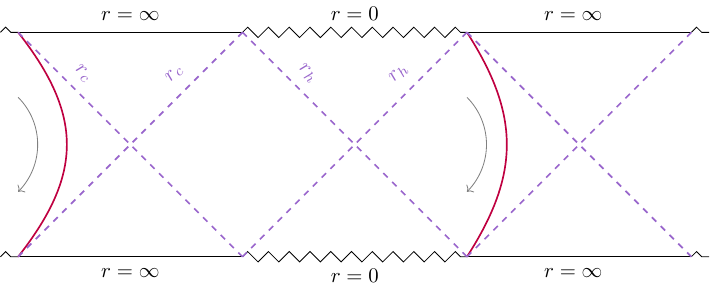}
\caption*{\textbf{Case 3}: one cosmological and one black hole stretched horizon.}
    \end{subfigure}
    \caption{New prescription for the stretched horizons in the SdS background. The left surface is a cosmological stretched horizon, while the right surface is a black hole stretched horizon. Geometric objects anchored to the stretched horizons can go behind $r_h$ and beyond $r_c$. Gray arrows represent the orientation of the Killing vector $\partial_t$. }
    \label{fig:cosmobh-stretched-hor}
\end{figure}
\item \textbf{Two cosmological stretched horizons in different copies of SdS.}
The periodicity of the SdS$^n$ geometry opens the way for more possibilities, where there are two cosmological stretched horizons belonging to different copies of the spacetime.
This setting is depicted in Fig.~\ref{fig:extended-cosmo-sec2} for the case of SdS$^2$.
\item \textbf{Two black hole stretched horizons in different copies of SdS.}
Another possibility coming from the periodicity of the Penrose diagram is to take two black hole stretched horizons in different copies, as reported in Fig.~\ref{fig:extended-bh-sec2} for the SdS$^2$ case. 
\end{enumerate}

\begin{figure}
  	\centering
   \begin{subfigure}[t]{0.7\textwidth}
   \includegraphics[width=\textwidth]{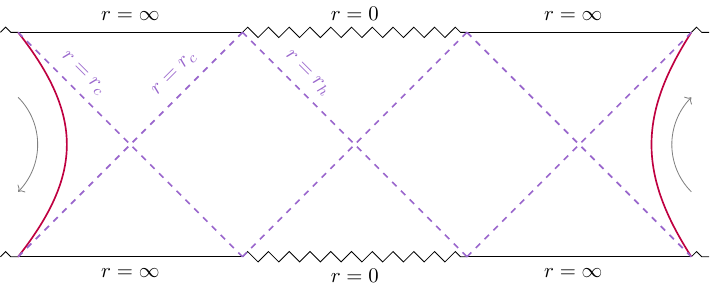}
       \caption{\textbf{Case 4}: two cosmological stretched horizons in SdS$^2$.}
       \label{fig:extended-cosmo-sec2} 
   \end{subfigure}
   \\
   \begin{subfigure}[t]{0.7\textwidth}
   \includegraphics[width=\textwidth]{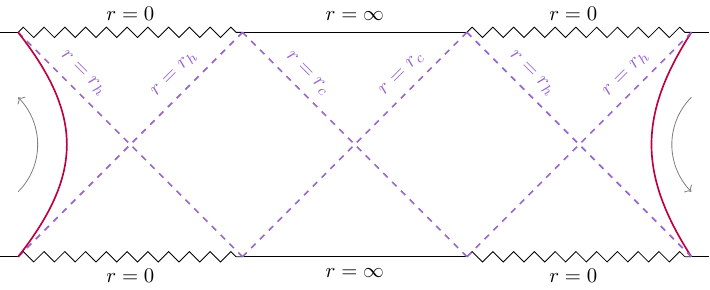}
   \caption{\textbf{Case 5}: two black hole stretched horizons in SdS$^2$.}
   \label{fig:extended-bh-sec2}
   \end{subfigure}
  	\caption{Novel prescriptions for the stretched horizons in a holographic setting with two copies of SdS space. 
   (a) The cosmological stretched horizons (in green) are located in different copies of the static patch. 
 (b) The stretched horizons (in green) are located in different copies of the black hole patch. Gray arrows represent the orientation of the Killing vector $\partial_t$.}
  	\label{fig:multiverse-hor-cases45}
  \end{figure}
In the remainder of the thesis, we will denote as $r_{\rm st}^L$ ($r_{\rm st}^R$) the left (right) stretched horizons in a certain configuration.
The common idea to all the new cases 3, 4, and 5 is that geometric objects hanging from the corresponding stretched horizons probe both the black hole interior and the exterior of the cosmological horizon, which are both of physical interest.

\subsection{Boundary times}\label{ssec:bdy_times}
We interpret the time coordinates $t_L, t_R$ running along the left or right stretched horizons of the SdS geometry as the boundary times of the corresponding dual theories.\footnote{From now on, we will refer to $t_L, t_R$ as the \textit{boundary times}, referring to the idea that they may be interpreted as the time coordinates of a putative quantum theory defined on these codimension-one surfaces.}
First of all, we define the following 
\begin{conv}
\label{conv_times}
    The boundary time $t_L (t_R)$ defined on the left (right) stretched horizon is always chosen to be oriented upward.
\end{conv}

The signs of the times $t_L, t_R$ for the left and right boundary theories, compared to the time coordinate $t_{\rm bulk}$ naturally defined in the bulk and running along the stretched horizon, are determined by the orientation of the Killing vector $\partial_t$ associated with the time-translation invariance of the metric \eqref{eq:asympt_dS}. 
In particular, this orientation flips whenever an event horizon is crossed. We emphasize again that there is no global Killing time-like vector in this geometry.

Let us clarify the convention \ref{conv_times} with an example.
In case 1, the SdS geometry is depicted in Fig.~\ref{fig:WDW_cosmo_stretched}.
The Killing vector $\partial_t$ runs upward on the right stretched horizon and downward on the left stretched horizon.
The orange lines at $\pm 45$\textdegree \, represent surfaces at constant $u (v)$ coordinates, which compose the null boundaries of the WDW patch (that will be analyzed in Sec. \ref{sec:Cod0 complexity}).
Focusing on the top-right boundary, we find that the constant value of the $u$ coordinate is given by
\beq
u_{\rm max} = t_{\rm bulk} - r^*(r^R_{\rm st}) = t_R - r^*(r^{R}_{\rm st}) \, , \qquad (\mathrm{case} \,\,\, 1)
\label{eq:identification_times_umax}
\eeq
where we used the definition \eqref{eq:general_null_coordinates} and the identification $t_{\rm bulk} = t_R$ that comes from the upward orientation of the Killing vector $\partial_t$, consistent with the upward orientation of $t_R$ according to the convention \ref{conv_times}.
On the contrary, if we evaluate the constant value of $u$ on the bottom-left boundary, we get
\beq
u_{\rm min} = t_{\rm bulk} - r^*(r^L_{\rm st}) = -t_L - r^*(r^L_{\rm st}) \, ,  \qquad (\mathrm{case} \,\,\, 1) 
\eeq
where now we identified $t_{\rm bulk} = -t_L$ because the Killing vector $\partial_t$ is oriented downward, while the convention \ref{conv_times} requires $t_L$ to be oriented upward.
A similar reasoning will be applied in the remainder of the thesis to define the boundary times in any other configuration.

Next, our goal is to introduce a time dependence into the setting, in order to probe the exterior of the cosmological horizon and the interior
of the black hole.\footnote{This guiding principle was originally introduced to study the evolution of the entanglement entropy in an eternal black hole background in \cite{Hartman:2013qma}.  }
It is well-known that the SdS background is static under a time evolution that runs along the same direction dictated by the Killing vector $\partial_t$ running along the stretched horizons.
More precisely, due to the boost symmetry associated with the time isometry in the bulk, the conjectured dual state in the boundary theory is invariant under the shift
\beq
\begin{cases}
    t_L \rightarrow t_L + \Delta t \, , \qquad
t_R \rightarrow t_R - \Delta t  & \mathrm{cases} \,\,\, 1,2,4,5 \\
t_L \rightarrow t_L + \Delta t \, , \qquad
t_R \rightarrow t_R + \Delta t & \mathrm{case} \,\,\, 3 \, ,
\end{cases}
\label{eq:time_shift_symmetry}
\eeq
where the different sign of case 3 compared to the other configurations is a consequence of the orientation of the Killing vector $\partial_t$ on the two stretched horizons.

The symmetry \eqref{eq:time_shift_symmetry} implies that we need to perform a different evolution of the boundary times in case 3 (compared to the other cases) in order to achieve a time-dependent setting.
In other words, this observation justifies the following rule

\begin{deff}
\label{rule}
   A non-trivial time evolution of the SdS background is achieved by evolving the times of the putative boundary theory upward along the stretched horizons, except on the left stretched horizon in case 3, where the time $t_L$ evolves downward.
\end{deff}

In principle, the time coordinates $t_L, t_R$ are independent, but one can synchronize them by considering an extremal codimension-one surface connecting the two stretched horizons. This geometric construction connects the time slice at time $t_R$ on the right stretched horizon with the one at time $t_L= \pm t_R$ (as specified below) on the left stretched horizon.
Throughout this work, we will refer to the special case of a symmetric (antisymmetric in case 3) time evolution whenever the following condition is met:\footnote{At the cost of being pedantic, we stress that $t$ is a boundary time, related to the bulk time $t_{\rm bulk}$ via the identification of the time coordinate on the stretched horizons dictated by the Killing vector $\partial_t$, see discussion below \eqref{eq:identification_times_umax}.  }
\beq
\begin{cases}
  \frac{t}{2} \equiv t_L = t_R  \, ,   & \mathrm{cases} \,\,\, 1,2,4,5 \\
\frac{t}{2} \equiv -t_L = t_R  \, , & \mathrm{case} \,\,\, 3 \, .
\end{cases}
\label{eq:symmetric_times}
\eeq
Using the symmetry \eqref{eq:time_shift_symmetry} of the extended black hole background, this case can always be achieved for the holographic complexity investigations, and it is not restrictive.\footnote{In principle, the time coordinates are independent among each copy of SdS. However, requiring a continuous junction between the various copies implies that the coordinates and the Killing vector $\partial_t$ are continuous. Therefore, one can synchronize the clocks of the surfaces belonging to different copies of SdS, too.}

\cleardoublepage
  \graphicspath{{\chaptersdir/QComplexity/image/}}%
  \chapter{Complexity in quantum systems}\label{ch:QComplexity}

In this chapter, we briefly review different definitions of complexity for quantum systems. We begin defining computational complexity, which has important applications for quantum circuits, and that has motivated the holographic complexity proposals presented in Ch. \ref{ch:introduction}, \ref{ch:ComplexityAdS} and \ref{ch:ComplexitydS}. Afterward, we will introduce: Nielsen complexity, originally defined in \cite{Nielsen1}; spread complexity, in \cite{Balasubramanian:2022tpr}; Krylov complexity, introduced in \cite{Parker:2018yvk}; and query complexity in \cite{Chen:2020nlj}. They have important applications to describe quantum mechanical and QFT systems\footnote{We briefly comment on the status of the complexity proposals we examine in this chapter for QFTs. For Nielsen complexity, one has to discretize the system (through lattice regularization) to avoid divergences associated with arbitrarily short length scales in the system (see e.g. \cite{Jefferson:2017sdb,Chapman:2017rqy,Bhattacharyya:2018bbv,Bueno:2019ajd,Chapman:2021jbh,Choudhury:2022xip}). In the case of spread complexity, one similarly needs a discretization procedure in order to apply the Lanczos algorithm \cite{Caputa:2022eye,Caputa:2022yju}. Meanwhile, Krylov and query complexity do not have this restriction (see Sec. \ref{ssec:query} and \ref{ssec:def Krylov} respectively). Notice that our study of complexity in Ch. \ref{ch:ComplexityDSSYK} does not require this regularization procedure, as we study these definitions in a quantum mechanical system (a pair of DSSYK models) and its presumed holographic dual theories.}, which will be examined in Ch. \ref{ch:ComplexityDSSYK} in the context of dS holography. 

We warn that this chapter contains several definitions. To improve readability, Table \ref{tab:sum complexity} contains a brief summary of the chapter; and further details can be found in the modern reviews \cite{Chen:2021lnq,Chapman:2021jbh}.
\begin{table}[t!]   
\begin{center}    
\begin{tabular}  {|C{3.5cm}|C{7.5cm}|} \hline  
\textbf{Proposal} & \textbf{Definition}\\ \hline
Computational complexity, $\mathcal{C}_{\rm C}$&For \textit{operators}: Minimal number of fundamental gates to construct a unitary (\ref{eq:computational complexity operators}). For \emph{states}: minimal complexity of all unitaries evolving a reference to a target state (\ref{eq:computational complexity states}).\\\hline
Nielsen complexity, $\mathcal{C}_{\rm N}$& For \emph{operators}: Minimal geodesic distance in the group manifold of unitary operators (\ref{eq:Nielsen complexity}).\\ \hline
Spread complexity, $\mathcal{C}_{\rm S}$&Average spread of states over an ordered basis (\ref{eq:spread complexity}).\\ \hline
Krylov complexity, $\mathcal{C}_{\rm K}$&Growth of operators in an ordered basis (\ref{eq:Krylov complexity}).\\ \hline
Query complexity, $\mathcal{C}_{\rm Q}$&Number of steps in an algorithm computing n-point correlation functions in a CFT$_2$ (\ref{eq:Query n fusions}).\\ \hline
\end{tabular}   
\caption{Summary of complexity proposals reviewed in this chapter, including computational, Nielsen, spread, Krylov, and query complexity.}
\label{tab:sum complexity}
\end{center}
\end{table}

\section{Computational complexity}\label{sec:computational complexity}
There are different notions of complexity in quantum information theory. One of the most commonly used, computational complexity, plays a crucial role in establishing the advantages of quantum over classical computation; in classifying computational problems for algorithm optimization; as a measure of quantum chaos in many-body systems\footnote{We refer to quantum chaos in the sense of quantizing classical chaotic systems. For instance, Lyapunov exponents in classical chaos have been extended to quantum mechanics and QFTs by introducing out-of-time-order correlators (\gls{OTOCs})) \cite{Rozenbaum:2016mmv} (see footnote \ref{fnt:OTOCs} for more details). The energy level spacing statistics of quantum chaotic systems have been conjectured \cite{Bohigas:1983er} to follow a Wigner-Dyson distribution \cite{wigner1993characteristic}, similar to the eigenvalue spacing of random matrices \cite{Guhr:1997ve}. This characteristic is in contrast to quantum integrable systems, whose level spacing is generically expected to follow a Poisson distribution \cite{bower1993proceedings}. We refer \cite{Jahnke:2018off} to the reader for a pedagogical review, in connection with holography.}; among different applications \cite{Caputa:2017yrh,Jefferson:2017sdb,Chapman:2017rqy,Bhattacharyya:2018wym,Chapman:2018hou,Camargo:2018eof,Ge:2019mjt,Brown:2019whu,Balasubramanian:2019wgd,Chapman:2019clq,Caputa:2018kdj,Auzzi:2020idm,Caginalp:2020tzw,Flory:2020eot,Flory:2020dja,Chagnet:2021uvi,Basteiro:2021ene,Brown:2021uov,Balasubramanian:2021mxo,Brown:2022phc,Erdmenger:2022lov,Baiguera:2023bhm,Ali:2018fcz,Bhattacharyya:2018bbv,Bhattacharyya:2019kvj,Ali:2019zcj,Bhattacharyya:2019txx,Bhattacharyya:2020iic,Bhattacharyya:2022ren,Bhattacharyya:2023sjr,Bhattacharyya:2020art,Bhattacharyya:2020rpy,Bhattacharyya:2020kgu,Bhattacharyya:2024duw}.

Computational complexity can be defined in terms of states or operators (see \cite{Chapman:2021jbh,Chen:2021lnq} for a review). To define operator computational complexity, let us first introduce one of the most basic ingredients in quantum computation, the \emph{qubit}. This is a two-level system described by a vector in a Hilbert space $\mathcal{H}$. We introduce a orthornomalized basis vectors $\ket{0},~\ket{1}\in\mathcal{H}$, to express a general qubit state as
\begin{equation}
    \ket{\psi}=a_0\ket{0}+a_1\ket{1}~.
\end{equation}
Similarly, one can define a $N$-qubit system, whose states take the form
\begin{equation}
    \ket{\psi}=\sum_{i\in\qty{0,~1}^N}\alpha_i\ket{i}~,
\end{equation}
and its Hilbert space has dimension $2^N$.

One can perform operations in this type of system using a discrete set of \emph{fundamental gates}, which are defined as a set of operations that affect only one and two-qubit states at the same time.\footnote{For example, an arbitrary unitary operator $U$ acting in the $N$-qubit system can be written as
\begin{equation}
    U=\exp(\rmi \sum_a y_ah_a)~,
\end{equation}
where $y_a\in\mathbb{R}$ and $h_a=\prod_{i=1}^N\sigma_{k_i}^{(i)}$ (e.g. $\sigma_z^{(1)}\otimes\sigma_y^{(2)}\otimes\sigma_z^{(3)}\otimes\dots$) are generators of SU$(2)^N$. One can show (see details in \cite{Chapman:2021jbh}) that any $h_a$ can be brought in the form $h=\prod_i\sigma_{z}^{(i)}$ using single-qubit operations, and that $\rme^{\rmi\alpha h}$ ($\alpha\in\mathbb{R}$) can be used as an fundamental gate.}

Fundamental gates play an important role in defining computational complexity. Consider first the norm of a generic operator $O$,
\begin{equation}
    \norm{O}=\max_{\ket{\psi}}\abs{\bra{\psi}O\ket{\psi}}~,
\end{equation}
which represents the norm of the maximal eigenvalue of $O$. It has been shown in \cite{nielsen_2010} that any unitary operator $U$ can be approximately reproduced up to arbitrary precision $\epsilon$ by a circuit made of fundamental gates that prepares an operator $V$, by making the norm of their difference\footnote{Notice that even though $U$ and $V$ are unitary, $U-V$ is not unitary, otherwise the norm would be 1.} smaller than $\epsilon$, this
\begin{equation}\label{eq:norm difference}
    \norm{U - V} <\epsilon~.
\end{equation}
Using the property (\ref{eq:norm difference}), the \textit{computational complexity} of a unitary operator $U$, $\mathcal{C}_{\rm C}(U)$, is defined as the number of fundamental gates $U_i$ (a discrete set of unitary operators) that is needed to model a particular unitary operator $U$ to a given precision:
\begin{equation}\label{eq:computational complexity operators}
    \mathcal{C}_{\rm C}(U)=\min\qty(n~,\quad \text{s.t.}\norm{U-\prod_{i=1}^nU_i}<\epsilon)~.
\end{equation}
There is a related notion, called state computational complexity, that we define as the minimization over the operator computational complexity of all unitaries that build a target state, $\ket{\psi_T}$, given a reference state, $\ket{\psi_R}$; that is:
\begin{equation}\label{eq:computational complexity states}
    \mathcal{C}_{\rm C}(\ket{\psi_T})=\min_{\ket{\psi_T}=U\ket{\psi_R}}\mathcal{C}_{\rm C}(U)~.
\end{equation}
While computational complexity has several practical uses, it also suffers from several ambiguities in its definition due to the dependence on the details about reference and the type of elementary operations to reach the target state in the state definition; or related to the type of gate sets and the precision to approximate a given operator.

\subsection{Linear late time growth}\label{ssec:linear growth CC}
 The exact value of complexity is highly sensitive to the above ambiguities. However, there are robust properties of complexity, such as its scaling with the dimensionality of the Hilbert space of the system of qubits, which are of interest to quantify the evolution of the complexity of a system.

For instance, consider a quantum circuit with $N$ qubits. We will define \emph{$k$-local gates} as unitary operators in the circuit that can act on $k \ll N$ number of qubits (e.g. the fundamental gates we defined previously in this section are 1 or 2-local) at the same time. Following \cite{Susskind:2018pmk}, we also define a \emph{$k$-local all-to-all quantum circuit} when all the $k$ qubits can interact among themselves on each gate. For convenience, we will consider that the circuit is all-to-all below.

\begin{figure}
    \centering
    \includegraphics[width=0.8\textwidth]{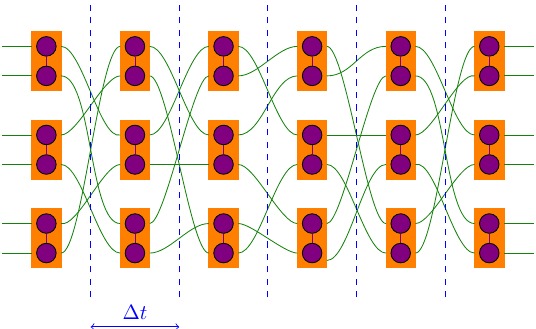}
    \caption{Representation of a $2$-local \textit{all-to-all} circuit with $N=6$ qubits (purple dots) interacting with each other in the 2-local gates (orange rectangle). Each time step ($\Delta t$, in blue) in the circuit is defined as the execution of three 2-local gates. Since the qubits move to a different gate after each $\Delta t$, the system is a fast scrambler. Moreover, since there are no overlapping qubits between the gates at a given time step, the gates commute.}
    \label{fig:2-local}
\end{figure}

One can then define the evolution of the circuit in terms of the execution of the gates acting on groups of $k$ qubits each, which we refer to as a \textit{time-step}. If after each time-step, the qubits can move to a different $k$-local gate, then eventually they will all interact with each other. This situation is referred to as \emph{fast scrambling} of information, since information contained in a subsystem (such as the state of a qubit) will quickly propagate over the rest of the system. This result indicates that computational complexity can be used intuitively as a measure of quantum chaos. See Fig. \ref{fig:2-local} for an illustration of a 2-local all-to-all fast-scrambling circuit.
 
\begin{figure}
    \centering
    \includegraphics[width=0.8\textwidth]{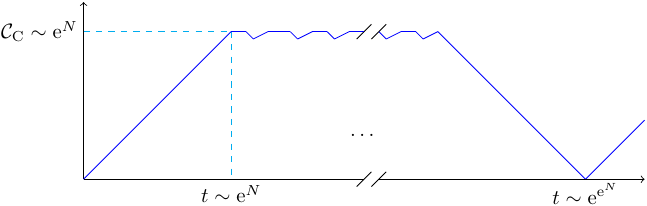}
    \caption{Evolution of computational complexity in a $k$-local all-to-all quantum circuit to very long times, conjectured by \cite{Susskind:2014rva}. It is expected that the complexity grows linearly until it reaches a maximal value ($\mathcal{C}_{\rm C}\sim\rme^N$ when $t\sim\rme^N$), after which it saturates and collisions start taking place, leading to oscillations around the saturation value. Eventually, quantum recurrence effects will lead to a decrease in complexity (see footnote \ref{fn:quantum recurrence}), expected at a time $t\sim\rme^{\rme^N}$, after which it is expected to start increasing again.}
    \label{fig:quantum_recurrence}
\end{figure}

One can notice that after each time step in this configuration, a number of $N/k$ gates will be executed, which indicates a linear increase in the number of gates with time. Given that we defined computational complexity for operators (\ref{eq:computational complexity operators}) as the minimum number of elementary gates to reproduce a given unitary operator with a given precision; it follows that their complexity will also grow linearly with time (see Fig. \ref{fig:quantum_recurrence}). 

A possible complication is when there are \textit{shortcuts} (also called \textit{collisions}) in the circuit when some of the unitary gates generated after $n$ time steps $\Delta t$ are equivalent to each other (in the sense of the norm (\ref{eq:norm difference}))  \cite{Susskind:2018pmk}, or that another circuit could generate them more efficiently \cite{Chapman:2021jbh}. It can be shown that collisions are highly unlikely during the evolution of the quantum circuit until $n\sim\rme^N$, when complexity (\ref{eq:computational complexity operators}) saturates \cite{Chapman:2021jbh}.

After a long enough time period, it is expected that \emph{quantum recurrence}\footnote{\label{fn:quantum recurrence}There exists a quantum recurrence theorem \cite{Bocchieri:1957cnr,percival1961almost,schulman1978note} (generalized to isolated quantum many-body system in \cite{Rauer_2018}), which states that non-integrable systems with a finite density of states will return almost exactly to their initial state for sufficiently long times, which we refer to the recurrence time.} will appear and lead to a decrease in computational complexity. See Fig. \ref{fig:quantum_recurrence} for an example of the expected evolution.

\subsection{Switchback effect}\label{ssec:QC switchback}
\begin{figure}
    \centering
    \subfloat[]{\includegraphics[width=0.45\textwidth]{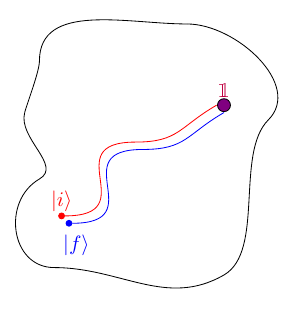}\label{sfig:Precursor1}}\quad\subfloat[]{\includegraphics[width=0.45\textwidth]{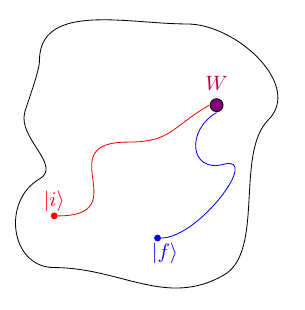}\label{sfig:Precursor2}}
    \caption{Evolution of the initial state $\ket{i}$ (red dot) to an intermediate state where an operator acts on (purple dot) and then to the final state $\ket{f}$ (blue dot) in a Hilbert space manifold (white blob) for the operation (a) $U^\dagger(t)\mathbbm{1}U(t)$ representing a backtracking (red and blue) curve; (b) $U^\dagger(t)WU(t)$ for a general $W$.}
    \label{fig:PrecursorEvol}
\end{figure}

Another of the robust properties of the evolution of computational complexity for generic quantum circuits is the scrambling of information under the presence of perturbations. We describe perturbations in a system of $N$-qubits by inserting an operator $W$ acting on a single qubit at a time $t$. Consider a \emph{precursor operator} defined as
\begin{equation}\label{eq:precursor eq}
    W(t)=\rme^{\rmi H t}W\rme^{-\rmi H t}~,
\end{equation}
in the Schrödinger picture, where we assume $H$ is an all-to-all $k$-local Hamiltonian of the $N$ qubit system. The effect of the precursor is to modify evolution in the quantum circuit, which can result in a very different final state in many body chaotic systems for a given initial state \cite{Susskind:2014rva,Stanford:2014jda}, as we illustrate in Fig. \ref{fig:PrecursorEvol}.
 
We define an \emph{infection} when a given qubit interacts with the operator $W$, or with another infected qubit. Since we consider an all-to-all circuit, all qubits will eventually become infected, similar to an epidemic model for diseases. 

Let us then define the \emph{size of the epidemic} as the number of new infected qubits per time step $\Delta t$ since $W$ is applied, and let $\Delta \tilde{s}$ represent the average number of infections. Considering an infinitesimal time step, we can express
\begin{equation}\label{eq:infection}
    \dv{\tilde{s}}{t}=\frac{(N-\tilde{s})\tilde{s}}{N-1}~,
\end{equation}
where $(N - \tilde{s})$ represents the number of uninfected qubits, and $\tilde{s}/(N - 1)$ is the probability that an uninfected qubit interacts with an infected one.

Assuming that there were no infections initially, i.e. $\tilde{s}(0)=0$, and that $N\gg1$, we have as the solution to (\ref{eq:infection}):
\begin{equation}
    \tilde{s}(t)=N\frac{\rme^{t-t_*}}{1+\rme^{t-t_*}}~,
\end{equation}
where $t_*=\log N$ is called the \emph{scrambling time}. The computational complexity of the precursor is then found by integrating over the size of the infection:
\begin{equation}\label{eq:C precursor}
    \mathcal{C}_{\rm C}(W(t))=\int_{0}^t\tilde{s}(t')~\rmd t'=N\log(1+\rme^{t-t_*})~,
\end{equation}
such that, $t_*$ can be interpreted as a time delay in the complexity growth. We can see that when $t-t_*<0$ (early times), the complexity has an exponential growth, which has been related to quantum chaos (as measured by OTOCs\footnote{\label{fnt:OTOCs}OTOCs are expectation values of the form $\Tr(\rho [W(t),~V (0)]^2)$ where $\rho$ is a thermal state and $W(0)$, $V (0)$ are local operators. The OTOC measures how much $W(t)$ and $V (0)$ fail to commute in the Heisenberg picture. Quantum chaotic systems, show exponential sensitivity to initial conditions, which can be diagnosed with the OTOC evolving as $\Tr(\rho [W(t),~V (0)]^2)\propto\rme^{\kappa_L t}$, where $\kappa_{L}$ is the Lyapunov exponent. Moreover, there is a conjectured universal bound for the Lyapunov exponent of quantum chaotic systems \cite{Maldacena:2015waa,Murthy:2019fgs}, which is further discussed in Sec. \ref{ssec:Bound OTOC}.}) in many body systems \cite{Susskind:2018pmk,Chapman:2021jbh}. On the other hand, when $t-t_*>0$, we may approximate (\ref{eq:C precursor}) as
\begin{equation}\label{eq:switchback}
    \mathcal{C}_{\rm C}(W(t))\simeq N(t-t_*)~.
\end{equation}
Thus, the effect of the precursor is to delay the growth of the operator computational complexity due to the scrambling of information. We will call this delay in the growth the \emph{switchback effect}, which is one of the main motivations for defining holographic complexity, as we will discuss in Secs. \ref{sec:CAny CMCAdS}, \ref{sec:CAny CMC}. Also notice that although we have only studied the late time growth with the precursor, it is expected that the rest of the evolution still behaves as we depicted in Fig. \ref{fig:quantum_recurrence} \cite{Susskind:2018pmk,Chapman:2021jbh}.

\begin{figure}
     \centering
     \includegraphics[width=0.3\textwidth]{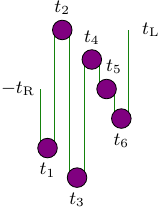}
     \caption{Representation of the operator insertions (purple dots) at times $t_1$, $t_2$\dots$t_6$ in the preparation of the state (\ref{eq:complexity time fold}) starting from the TFD state, where each switchback is represented as a folding curve after each insertion time, $t_i$; except for $t_5$, where the condition $\abs{t_i-t_{i-1}}> t_*$ (with $t_*$ being the scrambling time) is not satisfied. Figure based on \cite{Stanford:2014jda}.}
     \label{fig:timefold}
 \end{figure}

So far, we have focused on the evolution of the operator complexity of $W(t)$, but in order to motivate our study of holographic complexity, we would like to estimate how precursors could affect the state complexity of the TFD state. To do so, let us generalize the situation above by considering the TFD state after the insertion of $n$ precursors (which are not necessarily in time order) at times $t_1$, $t_2$,\dots $t_n$ on the left side of the TFD state; which we denote by $W_L(t_i)$ with $i=1,\dots,n$, and let it evolve into the state: 
\begin{equation}\label{eq:state Psi}
    \ket{\Psi(t_L,~t_R)}=U_L(t_L)U_R(t_R)W(t_n)\dots W(t_1)\ket{\rm TFD}
\end{equation}
where $U_{L/R}(t)=\rme^{-\rmi H_{L/R}t}$. Considering that $W(t)$ are thermal scale operators (defined in Sec. \ref{ssec:SWs AdS}), and that $(H_L-H_R)\ket{\rm TFD}=0$, one can re-express (\ref{eq:state Psi}) as
\begin{equation}\label{eq:state Psi2}
    \ket{\Psi(t_L,~t_R)}=U_L(t_L)W(t_1)\dots W(t_n)U_L^\dagger(-t_R)\ket{\rm TFD}~.
\end{equation}
The evolution of the TFD state due to the precursors is illustrated in Fig. \ref{fig:timefold}.

As originally noticed in \cite{Stanford:2014jda}, there is an upper bound on the state complexity of (\ref{eq:state Psi2}), given that the state complexity is related to operator complexity through (\ref{eq:computational complexity states}), and we have obtained (\ref{eq:switchback}) for each precursor $W_L(t_i-t_{i-1})$ when $\abs{t_i-t_{i-1}}> t_*$. Since the operator complexity is linear in the time difference, the total state complexity is bounded by its sum of the contributions in (\ref{eq:state Psi2}), leading to
\begin{align}\label{eq:complexity time fold}
\mathcal{C}_{\rm C}(\ket{\Psi(t_L,~t_R)})&\leq N(t_f-2n_{\rm sb}t_*)~,\\
   \text{where}~~ t_f&\equiv\abs{t_R+t_1}+\abs{t_2-t_1}+\dots+\abs{t_L-t_{n}}~.
\end{align}
Here $n_{\rm sb}$ is the number of switchbacks (i.e. the number of times that the insertions satisfy $\abs{t_i-t_{i-1}}> t_*$). This characteristic growth will play an important role in our discussion in Secs. \ref{sec:CAny CMCAdS}, \ref{sec:CAny CMC}.

\section{Quantum complexity proposals}\label{sec:QC proposals}
While computational complexity provides an intuitive proposal to quantify the difficulty in constructing a given configuration of a quantum system, in practice it becomes difficult to implement due to the above ambiguities (including the choice of fundamental gates and degree of precision in (\ref{eq:computational complexity operators}); as well as the reference and target states in (\ref{eq:computational complexity states})). Below, we will briefly review the definition of the concrete proposals that we will examine in Sec. \ref{ch:ComplexityDSSYK}, specifically: Nielsen complexity \cite{Nielsen1}, spread complexity \cite{Balasubramanian:2022tpr}; Krylov complexity \cite{Parker:2018yvk}; and query complexity \cite{Chen:2020nlj}.

\subsection{Nielsen complexity}\label{app:Nielsen}
Nielsen operator complexity was introduced in \cite{Nielsen1} to provide lower and upper bounds on the computational complexity of quantum circuits. For recent reviews, the reader is referred to \cite{Chen:2021lnq,Chapman:2021jbh}; ours will be mostly based on \cite{Yang:2018nda,Yang:2018tpo,Yang:2019iav}.

\begin{figure}
    \centering
    \subfloat[]{\includegraphics[width=0.4\textwidth]{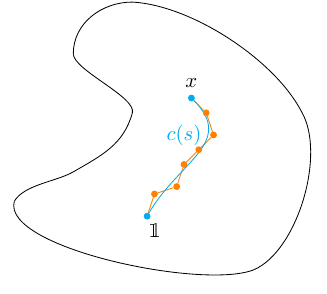}}\hspace{0.5cm}\subfloat[]{\includegraphics[width=0.4\textwidth]{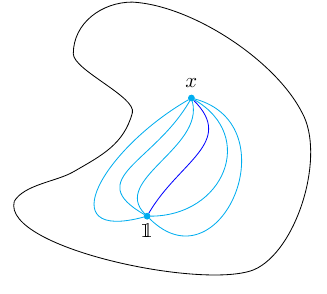}}
    \caption{Nielsen's geometric approach to operator complexity. The group manifold of unitary operators (white blob) is approximated as a smooth region. (a) A discrete set of elementary gates in a circuit (represented by orange dots) connecting the operators $\mathbbm{1}$ and $x\in\;$SU($n$) (cyan dots) is approximated through a continuous curve c($s$) (cyan). (b) Nielsen operator complexity picks the minimal length geodesic (blue) among all (cyan) of those connecting $\mathbbm{1}$ and $x$.}
    \label{fig:Nielsen approach}
\end{figure}

Consider the group manifold of unitary operators SU($n$) acting on a finite-dimensional quantum mechanical system (e.g. the Majorana fermions in the DSSYK model). In Nielsen's geometric approach, the discrete nature of this manifold is approximated by a smooth one where continuous paths connect operators.\footnote{We emphasize that the definition in \cite{Nielsen1} does not assume a particular metric in this manifold. One can ask whether some of the other complexity proposals in this section can be equivalent to Nielsen complexity by properly choosing the metric. This question was explored in \cite{Aguilar-Gutierrez:2023nyk}.} The original motivation \cite{Nielsen1} for doing this is to provide an approximation to the total number of elementary gates of the form that reproduce an arbitrary unitary operator, $x\in\;$SU($n$) to a given precision (relevant in optimization control of quantum circuits). The smooth geometric approximation becomes more accurate when the elementary gates have the form $\delta x=\rme^{-\rmi H\delta s}$, with $\delta s$ an infinitesimal parametrization (e.g. a small time step) of the gate, and $H$ is a generator of $U$ (such as the Hamiltonian). See Fig. \ref{fig:Nielsen approach} for an illustration.

We define \textit{Nielsen complexity} of an operator $x\in\;$SU($n$), $\mathcal{C}_{\rm N}(x)$, as the \textit{minimal geodesic length} between the identify operator $\mathbbm{1}$ and the operator $x$. This can be expressed as a map SU$(n)\rightarrow \mathbb{R}$, where one can impose certain axioms for it to describe a distance in the space of unitaries \cite{Yang:2018nda,Yang:2018tpo,Yang:2019iav}:
\begin{itemize}
    \item \textit{Non-negativity}:
    \begin{equation}
        \mathcal{C}_{\rm N}(x)\geq0~,~\forall x\in \text{SU($n$)}~,\label{eq:non-negative}
    \end{equation}
    where $\mathcal{C}_{\rm N}(\mathbbm{1})=0$.
    \item \textit{Triangle inequality}:
    \begin{equation}
        \mathcal{C}_{\rm N}(x)+\mathcal{C}_{\rm N}(y)\geq\mathcal{C}_{\rm N}(xy)~,~\forall x,~y\in \text{SU($n$)}
    \end{equation}
    where we will consider the definition of operator product in (\ref{eq:inner prod}).
    \item \textit{Parallel decomposition}: Let $M[x]$ denote a matrix representation for $x\in$SU($n$), then
    \begin{equation}
        \qty(\mathcal{C}_{\rm N}(M[x]\oplus M[y]))^Q=\qty(\mathcal{C}_{\rm N}(M[x]))^Q+\qty(\mathcal{C}_{\rm N}(M[y]))^Q~,
    \end{equation}
    where $Q\in\mathbb{Z}_+$.
    \item \textit{Smoothness}: Let $\delta x=\exp(\rmi H\delta s)$ represent the infinitesimal form of $x\in\;$SU($n$), where $H$ is a traceless Hermitian operator and $\delta s\geq0$. We require
\begin{equation}\label{eq:Nielsen complexity}
    \mathcal{C}_{\rm N}(\delta x)=F(H)\delta s+\mathcal{O}(\delta s^2)~,
\end{equation}
where $F[H]$, called the \emph{cost function}, is any analytic function.
\end{itemize}
Using these postulates for metric spaces, the continuous curve in the space of SU($n$) operators can be represented as
\begin{equation}\label{eq:Most general NC}
    c(s)=P_{\xi}\exp\qty(-\rmi\int_0^sH_\xi(u)\rmd u)~,\quad\xi=\qty{\rm L,~R}~,
\end{equation}
where $s\in\mathbb{R}$ represents a parametrization of this curve, $P_\xi$ refers to path ordering (since there is an evolving operator in the integral), and $\xi$ determines the orientation (R/L corresponds to building a quantum circuit from right to left, or left to right), such that under an infinitesimal displacement:
\begin{equation}
\begin{aligned}
    \delta c(s)&=-\rmi H_{\rm R}(s)c(s)\delta s=-\rmi c(s)H_{\rm L}(s)\delta s~.
\end{aligned}
\end{equation}
from which it follows that
\begin{equation}\label{eq:LR invariant H}
    H_{\rm R}(s)=c(s)H_{\rm L} c(s)^{-1}~.
\end{equation}
One can then define the length along the curve $c(s)$ starting from the identity $\mathbbm{1}$ to operator $x\in\;$SU($n$) in terms of (\ref{eq:Nielsen complexity}) as\footnote{We have used reparametrization invariance to set the limits $s\in[0,~1]$.}
\begin{equation}
    L_\xi[c]=\int\mathcal{C}[\delta x]=\int_0^1F(H_\xi(s))~\rmd s~.
\end{equation}
Nielsen operator complexity is defined as the minimum length of $c(s)$:
\begin{equation}\label{eq:Nielsen General}
    \mathcal{C}_{\rm N}^{(\xi)}(x)\equiv\min_{\qty{c(s):~c[0]=\mathbbm{1},~c[1]=x}}L_\xi[c]~.
\end{equation}
Notice that since $H_{\rm R}\neq H_{\rm L}$ (\ref{eq:LR invariant H}), the Nielsen complexity will depend on the choice of orientation in the path integral (\ref{eq:Most general NC}). 

As we mentioned in the introduction, evaluating Nielsen complexity with (\ref{eq:Nielsen General}) is quite involved and often intractable. There is, however, a great simplification by demanding the following properties on the cost function:
\begin{itemize}
    \item \textit{Unitary invariance}: Let $x\in\;$SU($n$),
    \begin{equation}\label{eq:f unitary}
    F(H_\xi)=F(x H_\xi x^\dagger)~.
    \end{equation}
    This implies that $F(H_{\rm L})=F(H_{\rm R})\equiv F(H)$ from (\ref{eq:LR invariant H}) given that $c(s)\in\;$SU($n$). The resulting metric space is said to be bi-invariant, as (\ref{eq:Nielsen General}) is invariant under transformations $c(s)\rightarrow c(s)x$ and $c(s)\rightarrow x~ c(s)~\forall x\in\;$SU($n$).
    \item \textit{Reversal invariance}: 
    \begin{equation}\label{eq:f time reverse}
        F(H)=F(-H)~.
    \end{equation}
    This property can be physically motivated when we identify $H$ as the generator of time translations, and we require that the map (\ref{eq:Nielsen General}) be time-reversal invariant.
\end{itemize}
It was shown in \cite{Yang:2018nda} (see also \cite{Yang:2018tpo,Yang:2019iav}) that (\ref{eq:non-negative}-\ref{eq:Nielsen complexity}) together with (\ref{eq:f unitary}, \ref{eq:f time reverse}) determines the form of the cost to be (up to a positive proportionality constant)\footnote{Alternatively, it is often convenient (see e.g. \cite{Nielsen1,Nielsen2,Nielsen3,Chapman:2021jbh,Baiguera:2023bhm}) to expand the Hamiltonian in a basis of operators $O_a$, as
\begin{equation}
    H(s)=\sum_aY^a(s)O_a~,
\end{equation}where $Y_a(s)$ are coefficients, which can be used to introduce a family of cost functions of the form
\begin{equation}\label{eq:more general cost}
    F_{k,\qty{p}}[Y^a]=\qty(\sum_a p_a\abs{Y^a}^k)^{1/k}~,
\end{equation}
where $p_a$ are called the penalty factors, accounting for the difficulty in executing different gates. In case all $p_a=1$, the class of cost functions in (\ref{eq:more general cost}) is directly related with (\ref{eq:Bi invariant cost}).}:
\begin{equation}\label{eq:Bi invariant cost}
    F(H(s))= \qty(\tr(H(s)H^\dagger(s))^{Q/2})^{1/Q}~.
\end{equation}
where the case $Q=2$ corresponds to a \textit{Riemannian metric} on the SU($n$) group manifold, and $Q\neq2$ to Finsler metrics (see e.g. \cite{bao2012introduction}). We will focus on the proposal (\ref{eq:Bi invariant cost}), and set $Q=2$ to make the minimization process much more tractable. 

We then study how to construct the unitary target operator of the form:
\begin{equation}\label{eq:unitaries}
    x(t)=\exp(-\rmi V)~,\quad V=H ~t+2\pi K_n~\mathbbm{1}~,\quad K_n\in\mathbb{Z}~,
\end{equation}
where $H$ is the (traceless and Hermitian) Hamiltonian. Moreover, given that $V$ is the generator of SU($n$) elements, we require $\tr(V)=0$; resulting in the constraint $\sum_{n=0}^\infty K_n=0$.

The corresponding bi-invariant Nielsen complexity, $\mathcal{C}_{\rm N}(x(t))=\mathcal{C}_{\rm N}(t)$, is then given by (\ref{eq:Nielsen General}) and (\ref{eq:Bi invariant cost}) with $Q=2$ (Riemannian case) as:
\begin{equation}\label{eq:biinvariant complexity}
\begin{aligned}
    \mathcal{C}_{\rm N}(t)&=\min_{\qty{K_n:~\sum_n K_n=0}}\sqrt{\tr(V V^\dagger)}~.
\end{aligned}
\end{equation}
The place of Nielsen complexity in the holographic dictionary is, arguably, less understood in comparison with the other complexity definitions in this section. Nevertheless, its robust features, such as the growth of circuit complexity with system size, have motivated the different holographic complexity proposals mentioned in the introduction. 

Lastly, while we have commented about Nielsen complexity of operators, there is a definition for states which follows as in (\ref{eq:computational complexity states}) replacing $\mathcal{C}_{\rm C}$ for $\mathcal{C}_{\rm N}$, which will not be relevant for our discussion in Ch. \ref{ch:ComplexityDSSYK}.

Below, we will define the notions of spread and Krylov complexity; which are sometimes referred to as Krylov complexity for states, and operators respectively. We refer the reader to ~\cite{Bhattacharjee:2023dik,Chattopadhyay:2023fob,Camargo:2024deu,Aguilar-Gutierrez:2024nau,Aguilar-Gutierrez:2023nyk,Caputa:2022eye,Afrasiar:2022efk,Caputa:2022yju,Pal:2023yik,Barbon:2019wsy,Bhattacharjee:2022vlt,Dymarsky:2019elm,Dymarsky:2021bjq,Kundu:2023hbk,Bhattacharya:2022gbz,Mohan:2023btr,Yates:2021asz,Caputa:2021ori,Patramanis:2021lkx,Trigueros:2021rwj,Rabinovici:2020ryf,Rabinovici:2021qqt,Rabinovici:2022beu,Bhattacharjee:2022qjw,Chattopadhyay:2023fob,Bhattacharjee:2023dik,Bhattacharjee:2022ave,Takahashi:2023nkt,Camargo:2022rnt,Avdoshkin:2022xuw,Erdmenger:2023wjg,Hashimoto:2023swv,Camargo:2023eev,Iizuka:2023pov,Caputa:2023vyr,Fan:2023ohh,Vasli:2023syq,Gautam:2023bcm,Iizuka:2023fba,Huh:2023jxt,Anegawa:2024wov,Caputa:2024vrn,Chen:2024imd,Caputa:2024xkp,Chattopadhyay:2024pdj,Bhattacharya:2023yec,Bhattacharya:2024hto,Basu:2024tgg,Tang:2023ocr,Banerjee:2022ime,Nandy:2023brt,Bhattacharyya:2023grv,Bhattacharyya:2023dhp} for recent developments in this are; and \cite{Nandy:2024htc} for a recent review.

\subsection{Spread complexity}\label{ssec:spread}
Starting from the Schrödinger picture for a generic pure quantum system, we would like to construct an ordered, orthonormal basis of states $\qty{\ket{B_n}}$ that minimizes $\sum_n c_n\abs{\bra{\phi(t)}\ket{B_n}}^2$ where c$_n$ is an arbitrary monotonically increasing real sequence, and
\begin{equation}
     \ket{\phi(t)}=\rme^{-\rmi Ht}\ket{\phi_0}~.
\end{equation}
It was found in \cite{Balasubramanian:2022tpr} that the solution to this problem is the so-called Krylov basis, $\ket{K_n}$, defined as follows
\begin{align}
    \ket{A_{n+1}}&\equiv (H - a_n)\ket{K_n} - b_n \ket{K_{n-1}}~, \label{eq:lanczos}\\
    \ket{K_n} &\equiv b_n^{-1}\ket{A_n}~.
\end{align}
The procedure for finding the Krylov basis above is called the \textit{Lanczos algorithm}. Here $\ket{K_0}\equiv\ket{\phi_0}$ and
\begin{equation}
    a_{n} \equiv \bra{K_n} H \ket{K_n}, \qquad b_{n} \equiv(\braket{A_n})^{1/2}~,
\end{equation}
are called the Lanczos coefficients. Using this basis, $\ket{\phi(t)}$ can be expressed as
\begin{equation}\label{eq: phi (t) spread}
    \ket{\phi(t)} = \sum^{\mathcal{K}}_{n=0} \phi_n (t) \ket{K_n}~.
\end{equation}
Here $\mathcal{K}$ denotes the Krylov space dimension, which satisfies $\mathcal{K}\leq D_{\mathcal{H}}$, with $D_{\mathcal{H}}$ the Hilbert space dimension. The Hamiltonian in this basis becomes tridiagonal, and we can express a recursive relation between the time-dependent components in (\ref{eq: phi (t) spread}) as a Schrödinger equation:
\begin{equation} \label{eq:Schro}
    \rmi\partial_t \phi_n(t) = a_n \phi_n (t) + b_{n+1}\phi_{n+1}(t) + b_n \phi_{n-1}(t)~,
\end{equation}
with $\sum_n|\phi_n(t)|^2=1$.  The spread complexity of a time-evolved pure state $\ket{\phi(t)}$ is defined as \cite{Balasubramanian:2022tpr}
\begin{equation}\label{eq:spread complexity}
    \mathcal{C}_{\rm S}(t)\equiv\sum_n n\abs{\bra{\phi(t)}\ket{K_n}}^2~,
\end{equation}
Intuitively, $\mathcal{C}_{\rm S}$ measures the average position in a one-dimensional chain generated by the Krylov basis, where each step along the chain represents an increasingly chaotic state since they roughly behave as $\ket{K_n} \approx H^n \ket{\phi_0}$.

\subsection{Krylov complexity}\label{ssec:def Krylov}
One can also define a notion of complexity in terms of an ordered Krylov basis for operators in generic quantum systems. Starting from the Heisenberg picture, we use the Choi-Jamiokowski isomorphism \cite{Jamiolkowski:1972pzh,Choi:1975nug} which maps an operator $\hat{O}$ acting on states in Hilbert space $\mathcal{H}$ to a state $\ket{O}$ in a double-copy Hilbert space $\mathcal{H}\otimes\mathcal{H}$, where $\ket{O}$ can be expanded in a complete basis of states $\qty{\ket{\chi_n}}\in\mathcal{H}$ as
\begin{equation}\label{eq:Op as state Krylov}
\begin{aligned}
  |O)&\equiv\sum_{m,\,n}O_{nm}\ket{\chi_m,~\chi_n}~,
\end{aligned}
\end{equation}
where $O_{nm}\equiv\bra{\chi_m}\hat{O}\ket{\chi_n}$. We will consider the Frobenius product\footnote{Other choices of inner products inherent related to finite temperature ensembles can be found in \cite{Parker:2018yvk,Barbon:2019wsy}.} for defining the inner product of these states as:
\begin{equation}\label{eq:inner prod}
    (X|Y)=\frac{1}{D_{\mathcal{H}}}\tr(X^\dagger Y)~,
\end{equation}
where $D_{\mathcal{H}}$ refers to the Hilbert space dimension. 

We can represent the evolution of the operator through the Heisenberg equation as
\begin{align}\label{eq: Heisenberg time evol}
\partial_t|O(t))&=\rmi\mathcal{L}|O(t))~,
\end{align}
where $\mathcal{L}$ is called the Liouvillian super-operator,
\begin{equation}
    \mathcal{L}=\qty[H,~\cdot~],\quad O(t)=\rme^{\rmi\mathcal{L}t}O~.
\end{equation}
We can then solve (\ref{eq: Heisenberg time evol}) in terms of a Krylov basis, $\qty{|O_n)}$,
\begin{equation}\label{eq:amplitudes}
\begin{aligned}
    |O (t))&=\sum_{n=0}^{\mathcal{K}-1} i^n \varphi_n(t) |O_n)~,\\
    \varphi_n(t)&=(O_n|\rme^{\rmi \mathcal{L}t}|O_n)~,\quad (O_m|O_n)=\delta_{mn}~.
\end{aligned}
\end{equation}
The Lanczos algorithm that we encountered in Sec. \ref{ssec:spread} for determining the Krylov basis, $\qty{|O_n)}$, takes the form
\begin{align}
    &|A_{n+1})=\mathcal{L}|O_{n})-b_{n}|O_{n-1})~,\label{eq:Lanczos op}\\
    &|O_{n})=b_n^{-1}|A_{n})~,\quad b_{n\geq1}=(A_{n}|A_{n})^{1/2}~,~b_0=1~.\label{eq:K amplitudes}
\end{align}
There are further implications if we assume that $O(t)$ is a Hermitian operator. In this case, the correlation function is an even function in $t$ that can be expanded as a Taylor series as
\begin{equation}\label{eq:2pnt correlator Krylov}
    \varphi_0(t)=(O(t)|O(0))=\sum_nm_{2n}\frac{(-1)^{n}t^{2n}}{(2n)!}~,
\end{equation}
where $m_{2n}$ are referred to as the ``moments". The Lanczos coefficients $b_n$ can be then determined from the moments using an equivalent algorithm to (\ref{eq:K amplitudes}) \cite{Parker:2018yvk,viswanath1994recursion,Bhattacharjee:2022ave}
\begin{align}\label{eq:Alt Lanczos}
    b_n=\sqrt{Q_{2n}^{(n)}}~,\quad Q_{2k}^{(m)}=\frac{Q_{2k}^{(m-1)}}{b_{m-1}^2}-\frac{Q_{2k-2}^{(m-2)}}{b_{m-2}^2}~,
\end{align}
where $Q_{2k}^{(0)}=m_{2k}$, and $Q_{2k}^{(-1)}=0$.

The other amplitudes can be determined through the Lanczos algorithm (\ref{eq:K amplitudes}) and the Heisenberg equation (\ref{eq: Heisenberg time evol}), resulting in the recursion relation:
\begin{equation}\label{eq:sch eq K operator}
    \partial_t\varphi_n(t)=b_n\varphi_{n-1}(t)-b_{n+1}\varphi_{n+1}(t)~.
\end{equation}
Krylov-complexity is then defined as
    \begin{equation}\label{eq:Krylov complexity}
        {\mathcal{C}_{\rm K}}(t)\equiv\sum_{n=0}^{\mathcal{K}-1}n|\varphi_n(t)|^2\,.
    \end{equation}
The definition above was originally motivated \cite{Parker:2018yvk} to describe the size of the operator under Hamiltonian evolution, as it measures the mean width of a wavepacket in the Krylov space.

\subsection{Query complexity}\label{ssec:query}
We would like a notion of state complexity for a CFT that can be naturally adapted to the Chern-Simons (CS) formulation of 3D gravity \cite{Witten:1998qj,Witten:1988hc,Witten:2007kt} so that we can define complexity in the dS$_3$ space holography in Ch. \ref{ch:ComplexityDSSYK}. A promising proposal with these characteristics was developed in \cite{Chen:2020nlj} (for global AdS$_3$ gravity), called ''\textit{query complexity}", which is based on the same concept applied in quantum algorithms\footnote{The proposal shares similarities to the initial motivation for the CV conjecture \cite{Susskind:2014rva}. The volume of a codimension-one surface in a spacetime filled with a tensor network essentially counts the total number of tensors, while query complexity counts the number of tensor contractions in a tensor network computing the expectation value of the set of operators in the network. It follows that complexity is heuristically given by the size of the network.} (see a recent review in \cite{ambainis2018understanding}). Intuitively, {query complexity} for a CFT is defined as the number of times that a subroutine in an algorithm computing multipoint correlation functions of the CFTs must be performed. In this subsection, we will briefly review the original proposal in \cite{Chen:2020nlj}. We warn the reader that in contrast to the other subsections, AdS/CFT holography is crucial for this definition.

We start by defining a map $\rho$ which translates operators in the CFT to expectation values
\begin{equation}
    \rho:\quad \mathcal{O}\rightarrow\tr(\mathcal{O}\rho)~.
\end{equation}
In the holographic context, the location of the cutoff surface in AdS space will modify the domain of the map above.\footnote{We are referring to vacuum AdS space for the present discussion, but one should in principle account for heavy and light states when the proposal is generalized to other holographic CFTs, see comments on this in \cite{Chen:2022fbg}.} Without the cutoff state complexity would be trivially infinite. Thus, we would like to implement $\rho$ as an algorithm that reads an input of local CFT operators, $\mathcal{O}(x_1)$, $\mathcal{O}(x_2)$, \dots $\mathcal{O}(x_n)$, where $x_i$ parametrizes the location on the cutoff region; and that it evaluates n-point correlation functions $\bra{0}\mathcal{O}(x_1)\mathcal{O}(x_2)\dots\mathcal{O}(x_n)\ket{0}$, with $\ket{0}$ being the ground state of the CFT. Moreover, if the cutoff surface is performed along a geodesic path, the map $\rho$ cannot take more than one input $\mathcal{O}$, otherwise, it would be outside the constant mode sector of the CFT cutoff.

To study a concrete way of implementing this algorithm, we employ topological gravity in the AdS$_3$/CFT$_2$ setting. The correlation functions above can be repeatedly evaluated through fusion rules in the CFT, corresponding to the junction of Wilson lines in the bulk. We will be using the SL$(2,~\mathbb{R})\times$SL$(2,~\mathbb{R})$ CS formulation of global AdS$_3$ space,
\begin{equation}\label{eq:CS AdS3}
    I=\frac{k}{4\pi}\qty(\int\tr(\mathcal{A}\rmd \mathcal{A}+\frac{2}{3}\mathcal{A}\wedge\mathcal{A}\wedge\mathcal{A})-\int\tr(\bar{\mathcal{A}}\rmd \bar{\mathcal{A}}+\frac{2}{3}\bar{\mathcal{A}}\wedge\bar{\mathcal{A}}\wedge\bar{\mathcal{A}}))~,
\end{equation}
where $k$ is the coupling constant; $(\mathcal{A},~\bar{\mathcal{A}})$ are 1-form gauge fields; and $\tr$ denotes contraction using the Killing forms of the algebra. The equations of motion of (\ref{eq:CS AdS3}), called flatness conditions:
\begin{equation}\label{eq:flatness}
    \rmd\mathcal{A}+\mathcal{A}\wedge\mathcal{A}=0~,\quad \rmd \bar{\mathcal{A}}+\bar{\mathcal{A}}\wedge\bar{\mathcal{A}}=0~,
\end{equation}
which are equivalent to those of Einstein gravity in AdS$_3$ space (the reader is referred to \cite{Witten:1988hc} for details). 

We study bulk Wilson lines of the form
\begin{equation} 
W_R(\gamma)=\tr_{R}\qty(P\exp(-\int_{\gamma}\mathcal{A})P\exp(-\int_{\gamma}\bar{\mathcal{A}}))
\end{equation}
where $R$ denotes a continuous series representation of SL$(2,~\mathbb{R})\times$SL$(2,~\mathbb{R})$, $P$ represents path-ordering along the curve $\gamma$. If the path is closed, one has a Wilson loop, which is trivial in global AdS$_3$, while if $\gamma$ is open, its endpoints of $\gamma$ need to end at the asymptotic boundary to define gauge invariant quantities.

Expectation values of the Wilson lines are evaluated using the representation theory of SL$(2,~\mathbb{R})\times $SL$(2,~\mathbb{R})$, where primary states are denoted by
\begin{equation}
    \ket{h,~\overline{h}}=\mathcal{O}_{h\overline{h}}(z,~\overline{z})\ket{0}~.
\end{equation}
Here $\mathcal{O}_{h\overline{h}}(z,~\overline{z})$ is a local operator, $z\in\mathbb{C}$, and $h$, $\overline{h}$ are called the \textit{conformal weights}. These are constants appearing in the \textit{operator product expansion} of $\mathcal{O}_{h\overline{h}}(z,~\overline{z})$ with the stress energy tensor components of the CFT$_2$,
\begin{align}
    T_{zz}(z)\mathcal{O}_{h\overline{h}}(0,~0)&=\frac{h}{z^2}\mathcal{O}_{h\overline{h}}(0,~0)+\frac{1}{z}\partial_z\mathcal{O}_{h\overline{h}}(0,~0)+\dots~,\\
    T_{\bar{z}\bar{z}}(\overline{z})\mathcal{O}_{h\overline{h}}(0,~0)&=\frac{\overline{h}}{\overline{z}^2}\mathcal{O}_{h\overline{h}}(0,~0)+\frac{1}{\overline{z}}\partial_{\bar{z}}\mathcal{O}_{h\overline{h}}(0,~0)+\dots~.
\end{align}
where the dots indicate non-singular terms in the regime $z,~\overline{z}\rightarrow0$.

Now, consider global time slices of AdS$_3$ gravity. Since we are interested in an algorithm computing n-point correlation functions, we study how to combine Wilson lines. We will call a junction of Wilson lines when a pair (or higher number) of Wilson lines merge. It was proposed in \cite{Chen:2020nlj} to define the rule to junction ($\mathcal{J}$) Wilson lines purely in terms of CFT operators by mapping at least two primary states (or their descendants) $\ket{h_1,~\overline{h}_1}$ and $\ket{h_2,~\overline{h}_2}$ to a new one $\ket{h_3,~\overline{h}_3}$ as:
\begin{equation}\label{eq:fusions}
\begin{aligned}
    \mathcal{J}&\qty(\mathcal{O}_{h_1\overline{h}_1}(u,~\bar{u})\ket{0},~\mathcal{O}_{h_2\overline{h}_2}(v,~\bar{v})\ket{0})\\
    &=\int\rmd^2 w\sum_{h_3,~\overline{h}_3}c^{h_3\overline{h}_3}_{h_1\overline{h}_1h_2\overline{h}_2}(u,~\bar{u},~v,~\bar{v},~w,~\bar{w})\mathcal{O}_{h_3\overline{h}_3}(w,~\bar{w})\ket{0}
\end{aligned}
\end{equation}
where the functional dependence of the coefficients $c^{h_3\overline{h}_3}_{h_1\overline{h}_1h_2\overline{h}_2}(u,~\bar{u},~v,~\bar{v},~w,~\bar{w})$ are determined by the transformation rules of SL$(2,~\mathbb{R})\times $SL$(2,~\mathbb{R})$. Namely, the coefficients are invariant under a gauge transformation that affects all the Wilson lines simultaneously, and they need to transform covariantly when the gauge transformation acts only on a single one of the Wilson lines in the junction; similar to the operator product expansion (OPE) of local CFT operators. This definition of junction, together with the flatness conditions guarantees that the Wilson line network is deformable under diffeomorphisms in the bulk \cite{Chen:2020nlj}; and thus that it computes multipoint correlation functions from the fusion algebra of the CFT (through OPE expansions) if the network is placed on the asymptotic boundary.

\begin{figure}
    \centering
    \includegraphics[width=0.5\textwidth]{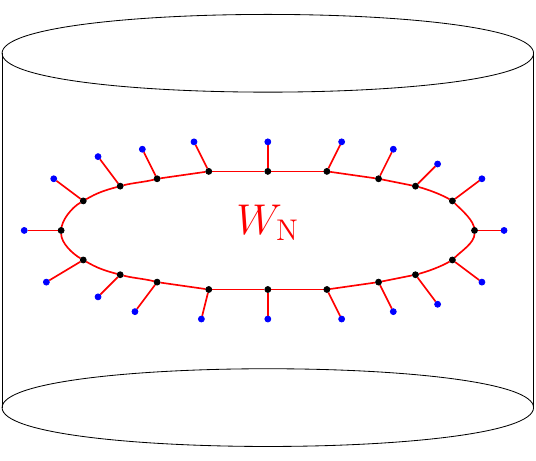}
    \caption{Wilson line network (labeled $W_N$) on a global time slice in pure AdS$_3$ space. The Wilson lines (red lines) have been junctioned together (black dots) according to the rule (\ref{eq:fusions}) and amputated (blue dots) along a cutoff surface in the bulk interior, which is not necessarily at a constant radial location.}
    \label{fig:Network_trunc}
\end{figure}

On the other hand, since maps $\rho$ are defined on a cutoff surface, one can introduce the concept of ``\emph{amputation}''. This operation removes the ends of the Wilson lines that extend to the asymptotic boundary up to a given cutoff surface of the open Wilson lines, as illustrated in Fig. \ref{fig:Network_trunc}. In terms of the CFT, this operation represents a renormalization group (RG) flow in the sense that its input is the incoming representation of the local operators from a reference scale (e.g. close to the asymptotic boundary) and coarse grains them to the cutoff scale of the network, whose output will be a number (the n-point correlation function). As such, this operation will not be gauge invariant; instead, it will depend sensitively on the choice of cutoff. {As a remark, notice that if we input a trivial representation of $R$ on any of the open lines of the amputated network, this will reduce the order of the $n$-point correlation function to $n-1$.}

After defining the characteristics of the Wilson line network, one can now define query complexity in terms of the number of times that fusion rules in (\ref{eq:fusions}) are applied to compute a n-point correlation function, or equivalently, the number of junctions in the amputated network in Fig. \ref{fig:Network_trunc}. We can represent this relation as
\begin{equation}\label{eq:Query n fusions}
    \boxed{\mathcal{C}_{\rm Q}=\text{number~of~ fusions}~.}
\end{equation}
Given that the only differential invariants on the surface introduced in the Wilson line network in a static configuration are the {proper length of the induced curve, $\lambda$; its mean curvature $K$; and torsion $\mathcal{T}$,} \footnote{\label{fn:Powers K T}Arbitrary powers of these differential invariant quantities are in principle allowed, however, they will be ill-defined given that we consider junctions of Wilson lines to form the network, instead of smooth surfaces. Thus, they will not be considered.} it follows that the density of the state complexity will be given by
\begin{equation}\label{eq:complexity dlambda}
    \dv{\mathcal{C}_{\rm Q}}{\lambda}=c_1+c_2 K+c_3\mathcal{T}~,
\end{equation}
where $c_{i}\in\mathbb{R}$ ($i\in\qty{1,~2,~3}$) are constants. 

However, as we discussed at the beginning of the subsection, the map $\rho$ should take no more than one input operator $\mathcal{O}(x)$ if the network, formed by the Wilson lines, follow geodesic trajectories (for which $K=0$, $\mathcal{T}=0$); otherwise, {there would be more than single place where the representations in $R$ could originate from within a single cutoff surface, for which we associate no query complexity to this configuration}. This implies that $c_1=0$ in (\ref{eq:complexity dlambda}). After fixing this constant, one can then integrate (\ref{eq:complexity dlambda}), and use the Gauss-Bonnet theorem at a fixed global time slice:
\begin{equation}\label{eq:Query global time slice}
    \mathcal{C}_{\rm Q}=c_2\qty(-\int \mathcal{R}~\rmd V+2\pi)+c_3\int\rmd\lambda~\mathcal{T}~.
\end{equation}
Given that $\mathcal{R}=-\ell_{\rm AdS}^{-2}$ for pure AdS$_3$ space, then (\ref{eq:Query global time slice}) indicates a direct relation between query complexity with the CV proposal if one could fix $c_3=0$, although there is no a priori reason for it.

\cleardoublepage
  \graphicspath{{\chaptersdir/ComplexityAdS/image/}}%
  \chapter{Holographic complexity in asymptotically anti-de Sitter space}\label{ch:ComplexityAdS}
This chapter is partially based on complementary material in \cite{Aguilar-Gutierrez:2023zqm}, \cite{Aguilar-Gutierrez:2023pnn} and \cite{Aguilar-Gutierrez:2024rka}. After our encounter with the complexity of quantum systems, we are interested in their bulk description. This is an ongoing task. Several holographic complexity proposals in AdS black holes have been introduced,\footnote{While in this chapter we are mostly interested in the eternal AdS black hole, the formalism can be applied in single-sided \cite{Alishahiha:2022kzc,Aguilar-Gutierrez:2023ccv}, and evaporating black holes \cite{Craps:2022ahp}.} all of them sharing the universal features of computational complexity for quantum circuits including (i) late-time linear growth, and (ii) the switchback effect. We dedicate a section to each of the most common holographic complexity conjectures, including the CV proposal in Sec. \ref{ssec:CV}; CV2.0 in Sec. \ref{ssec:CV20AdS}; CA in Sec. \ref{ssec:CAAdS} and then we present their recent unification under the CAny proposal, Sec \ref{sec:Cany AdS}.\footnote{We have selected this order in the presentation of the proposals for pedagogical purposes, while in Ch. \ref{ch:ComplexitydS} they will be presented according to the number of co-dimensions for each proposal.} Importantly, we adapt some of the results in \cite{Aguilar-Gutierrez:2023pnn} to show that the CAny proposals in AdS black hole backgrounds need further constraints to obey the switchback effect, an unnoticed point in their original formulation \cite{Belin:2021bga,Belin:2022xmt}.

\section{Complexity=volume}
\label{ssec:CV}
\begin{figure}
    \centering
    \subfloat[]{\label{subfig:Planar_BH_Cod1}\includegraphics[width=0.445\textwidth]{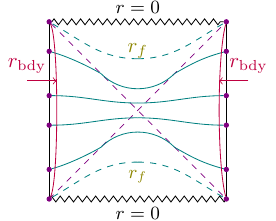}}\hfill\subfloat[]{\label{subfig:Ueff_AdS}\includegraphics[width=0.55\textwidth]{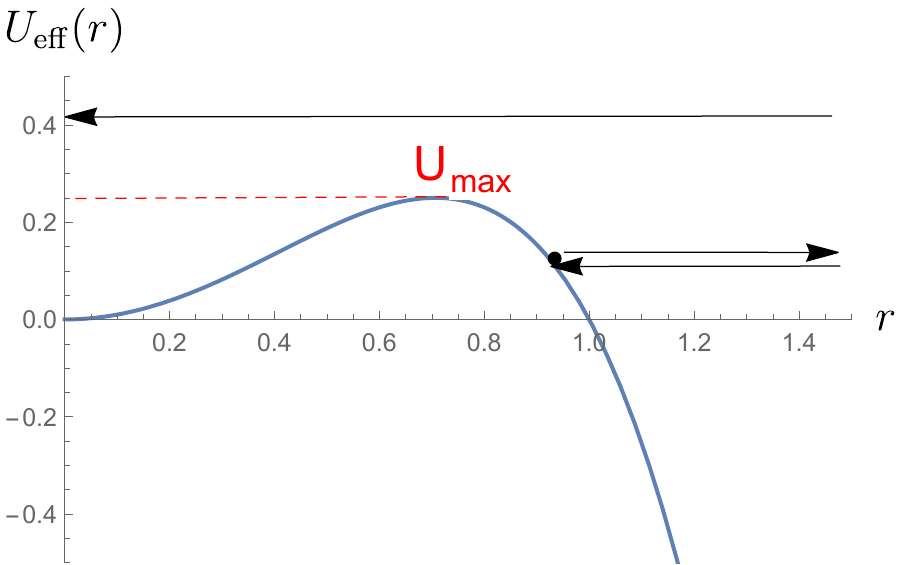}}
    \caption{(a) Extremal CV surfaces (dark aqua) anchored (purple dots) at a cutoff surface ($r_{\rm bdy}$ fuchsia curve) close to the asymptotic boundary, and subject to symmetric evolution. There is a critical value $r=r_{\rm f}$ (dashed line) when $t_{\rm L}=t_{\rm R}\xrightarrow{}\infty$. (b) Qualitative behavior of the effective potential $U_{\rm eff}(r)$ (\ref{eq:classical_motion_AdS}), which has a maximum value $U_{\rm max}$. For $P^2>U_{\rm max}$, the extremal volume surface falls into the black hole singularity. For $P^2\leq U_{\rm max}$, the extremal surface connects the left and right $r_{\rm bdy}$ regions after going through the turning point $r_t$ (\ref{eq:turning}). We have set $d=2$, $r_h= 1$, $L=1$ and $k=1$. The shape of the potential (and the appearance of a maximum value, $U_{\rm max}$) is qualitatively the same for $k=-1$, $0$, and in higher dimensions.}
    \label{fig:Planar_BH_Cod1}
\end{figure}

Historically, the first holographic complexity proposal in asymptotically AdS space, CV, was introduced in \cite{Susskind:2014rva} In this conjecture, the state complexity of a holographic CFT can be associated with the maximal volume $V$ of a codimension-one surface $\mathcal{B}$ anchored to the asymptotic boundary on a time slice $\Sigma$ (a codimension-two surface), as explained in (\ref{eq:CV 1st}), which we repeat here for the convenience of the reader,
\begin{equation}
    \mathcal{C}_V (\Sigma) =\max_{\Sigma=\partial \mathcal{B}}\frac{{V}(\mathcal{B})}{G_N\ell} \, .
\end{equation}
Assuming that the codimension-one surface preserves the symmetry of the background, we describe it in terms of a general parameter, $\sigma$, such that\footnote{Note that one must introduce an infrared (IR) regulator in the $k=-1$, $0$ cases in order for the CV (as well as other proposals we consider) to reproduce a finite value.}
\begin{equation}\label{eq:vol computation}
    \begin{aligned}
    \mathcal{C}_V = \frac{V}{G_N L} &=   \frac{\Omega_{k,d-1}}{G_N L} \int r^{d-1} \sqrt{-f(r) \dot{u}^2 - 2 \dot{u} \dot{r}} \, d\sigma \\
    &= \frac{\Omega_{k,d-1}}{G_N L} \int r^{d-1} \sqrt{-f(r) \dot{v}^2 + 2 \dot{v} \dot{r}} \, d\sigma \, ,
\end{aligned}
\end{equation}
where we introduced the coordinates \eqref{eq:general_null_coordinates}
\begin{equation}
    r=r(\sigma),\quad v=v(\sigma),\quad u=u(\sigma)~,
\end{equation}
with the notation $\cdot \equiv d/d\sigma$, and we defined $\Omega_{k,d-1}=\int\rmd \Omega_{k,d-1}$. See Fig. \ref{subfig:Planar_BH_Cod1} for an illustration of how the holographic complexity extremal surfaces in the CV proposal behave when they probe an AdS black hole background. Here, we anchor the maximal volume surfaces close to the asymptotic boundary (by introducing a radial ultraviolet (UV) regulator) $r_{\rm bdy}$. We emphasize that just as computational complexity has several ambiguities (see Sec. \ref{sec:computational complexity}), likewise the CV proposal suffers from the dependence on the UV regulator. Nevertheless, we anticipate that the choice of scheme does not affect the rate of growth of the CV proposal. Without loss of generality, we will also consider the symmetric evolution of the boundary times (i.e. $t_L=t_R\equiv t/2$).

Since the volume functional does not depend on $u~ (v)$, there is a corresponding conserved momentum
\begin{equation}
    P_u \equiv \frac{-f(r) \dot{u} - \dot{r}}{\sqrt{-f(r) \dot{u}^2 - 2 \dot{u} \dot{r}}} \, r^{d-1} =  \frac{-f(r) \dot{v} + \dot{r}}{\sqrt{-f(r) \dot{v}^2 + 2 \dot{v} \dot{r}}} \, r^{d-1} \equiv P_v \, .
    \label{eq:conserved_momentum_CV}
\end{equation}
From now on, we define $P_u = P_v \equiv P \, L^{d-1}$.
The functional \eqref{eq:vol computation} is invariant under reparametrizations.
This allows us to fix the gauge
\begin{equation}
    \sqrt{-f(r) \dot{u}^2 - 2 \dot{u} \dot{r}} = \sqrt{-f(r) \dot{v}^2 + 2 \dot{v} \dot{r}} = \left( \frac{r}{L} \right)^{d-1} \, ,
    \label{gauge}
\end{equation}
under which choice the volume reads
\begin{equation}\label{eq:Vol interm}
    V = \Omega_{k,d-1} \int \frac{r^{2(d-1)}}{L^{d-1}} d\sigma \,.
\end{equation}
We will need to derive an expression for $\dv{r}{\sigma}$ to evaluate this integral. Using the gauge choice in \eqref{gauge}, the conserved momentum \eqref{eq:conserved_momentum_CV} becomes
\begin{equation}\label{eq:P u dot}
    P = -f(r) \dot{u} - \dot{r} = -f(r) \dot{v} + \dot{r} \, .
\end{equation}
Solving for the derivatives of the null coordinates and plugging the results into \eqref{gauge}, we get
\begin{equation}\label{eq:dot u dot v}
    \dot{r}_{\pm} = \pm \sqrt{P^2 + f(r) \qty(\frac{r}{L})^{2(d-1)}} \, , \qquad
    \dot{u}_{\pm} = \frac{-P - \dot{r}_{\pm}}{f(r)} \, , 
    \qquad \dot{v}_{\pm} = \frac{-P + \dot{r}_{\pm}}{f(r)} \, .
\end{equation}
For a fixed orientation of the parameter $\sigma$ along the maximal slice, when $r$ increases in the same direction as $\sigma$ the solution is given by the $+$ branch, otherwise we must look at the $-$ branch.

The first equation in \eqref{eq:dot u dot v} can be written in the form
\begin{equation}
    \dot{r}_{\pm}^2 + U_{\rm eff}(r) = P^2 \, , \qquad
    U_{\rm eff}(r) = - f(r) \, \le \frac{r}{L} \ri^{2(d-1)} \, , 
    \label{eq:classical_motion_AdS}
\end{equation}
which describes the motion of a classical non-relativistic particle with energy $P^2$ in a potential $U_{\rm eff}(r)$ \cite{Carmi:2017jqz,Chapman:2018dem,Chapman:2018lsv,Belin:2021bga}.
The turning points $r_t$ of the extremal codimension-one surface satisfy $\dot{r}_{\pm}=0$, and are thus the solutions to the equation
\begin{equation}
 P^2 = U_{\rm eff}(r_t)  \, .
    \label{eq:turning}
\end{equation}
The qualitative behavior of the function $U_{\rm eff}(r)$ can be easily inferred from the properties of the blackening factor $f(r)$. In particular, $U_{\rm eff}(r)$ in (\ref{eq:classical_motion_AdS}) vanishes for $r= 0$, or the event horizon; and it has an opposite sign compared to $f(r)$. The shape of the effective potential $U_{\rm eff}(r)$ has crucial consequences on the time evolution of the maximal codimension-one slice. 
For a fixed value of the conserved momentum $P$, from \eqref{eq:classical_motion_AdS} it is clear that the corresponding maximal surface can only explore spacetime regions where $P^2 \geq U_{\rm eff}(r)$. When the condition is saturated, \ie $P^2 = U_{\rm eff}(r_t)$, a potential barrier is met and the maximal surface presents a turning point $r=r_t$. This is illustrated for the AdS black hole metric (\ref{eq:metric SAdS}) in Fig. \ref{subfig:Ueff_AdS}.

Thus, after leaving the right cutoff surface close to the asymptotic boundary $r_{\rm bdy}$, the maximal volume surface extends in the direction of decreasing $r$ and enters the black hole. 
However, the value of $|P|$ cannot be arbitrarily large. Let us define $r=r_f$ such that
\begin{equation}
    U'_{\rm eff}(r_f) =0 \, , \qquad U''_{\rm eff}(r_f) \leq 0 \, .
    \label{eq:def_rf_caseAdSCV}
\end{equation}
Namely, $r=r_f$ is the local maximum of the effective potential: $U_{\rm max} \equiv U_{\rm eff} (r_f)$. For instance when $k=0$, one has $r_f=2^{-1/d}r_h$; while more generically, one can find numerical solutions for $r_f$ in (\ref{eq:def_rf_caseAdSCV}).

\begin{figure}[t!]
    \centering
\subfloat[]{ \label{subfig:CVAdS}\includegraphics[width=0.45\textwidth]{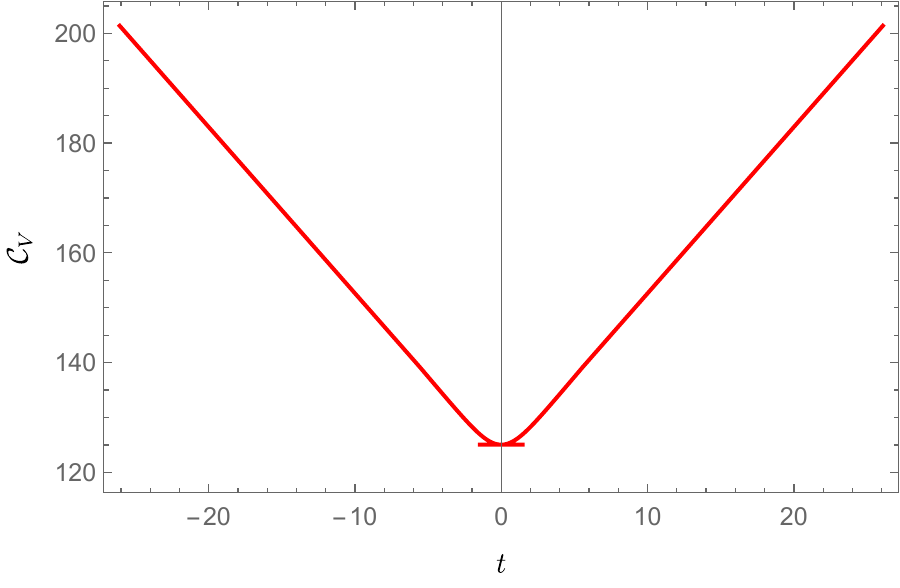}}\hfill \subfloat[]{\label{subfigb:dCVAdS} \includegraphics[width=0.48\textwidth]{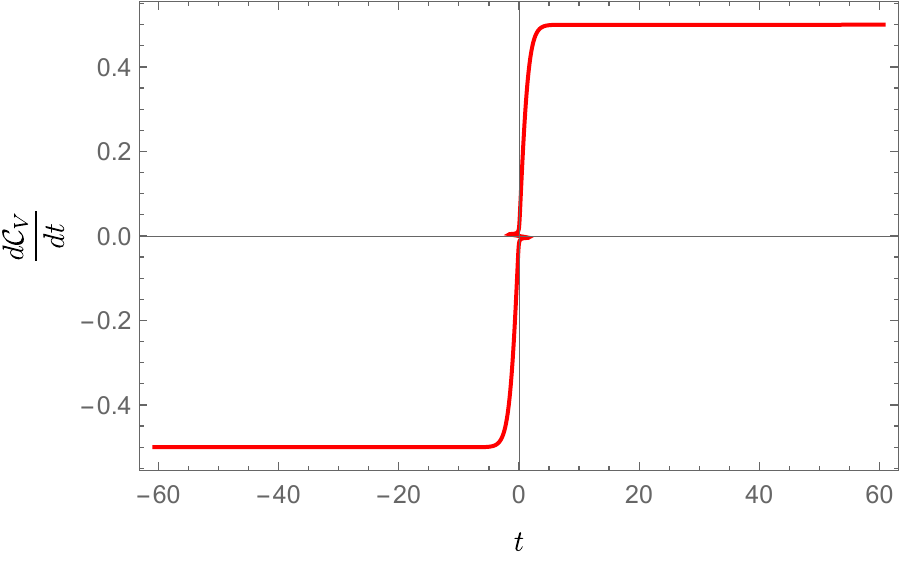}}
    \caption{Time dependence of (a): the CV proposal and (b): its rate of growth, for the prescription in Fig. \ref{fig:Planar_BH_Cod1}. We have set $d=2$, $k=1$, $r_{\rm bdy}=10$, $r_h = 1$, $L=1$, and $G_N=1$.}
    \label{fig:volume_BH_AdS}
\end{figure}

As it can be understood from Fig.~\ref{fig:Planar_BH_Cod1}
, for $P^2 > U_{\rm max}$ the codimension-one slice falls into the black hole singularity $r=0$ before connecting the left and right stretched horizons. 
Instead, for $0 \leq P^2 \leq U_{\rm max}$, the maximal surface meets the potential barrier and passes through a turning point. Then, it moves in the direction of increasing $r$ and reaches the left asymptotic boundary.
Therefore, $0 \leq P^2 \leq U_{\rm max}$ is the range of the conserved momentum $P^2$ that properly defines an extremal surface according to CV conjecture (see also footnote \ref{footnote:disconnected_solutions}).
In this case, the turning point is located at $r_{t} \leq r_h$. Thus, combining (\ref{eq:classical_motion_AdS}, \ref{eq:Vol interm}) the extremal volume and the boundary time at $r_{\rm bdy}$ read
\begin{equation}\label{eq:Vol BB AdS}
    \frac{V(P)}{\Omega_{k,d-1}} = \frac{2}{L^{d-1}} \int^{r_{\rm bdy}}_{r_{t}} \frac{r^{2(d-1)}}{\sqrt{P^2 + f(r) (r/L)^{2(d-1)}}} \rmd r \, ,
\end{equation}
\begin{equation}
    t(P) =  -2 \int_{r_{t}}^{r_{\rm bdy}} \frac{P}{f(r) \sqrt{P^2 + f(r) (r/L)^{2(d-1)}}} \rmd r \, ,
    \label{eq:t_CV_AdS}
\end{equation}
respectively. The second relation comes from expressing the boundary time in terms of the null coordinates \eqref{eq:general_null_coordinates} and our result (\ref{eq:dot u dot v}) as
\begin{align}
    r^*(r_{t}) -\frac{t}{2} - r^*(r_{\rm bdy})  &= \int_{r_{\rm bdy}}^{r_{t}} \frac{\dot{v}_-}{\dot{r}_-} dr =\int_{r_{\rm bdy}}^{r_{t}} \frac{P + \sqrt{P^2 + f(r) (r/L)^{2(d-1)}}}{f(r) \sqrt{P^2 + f(r) (r/L)^{2(d-1)}}} dr \, , \label{eq:integrations_CV_caseAdS_1}\\
    -\frac{t}{2} - r^*(r_{\rm bdy}) + r^*(r_{t}) &= \int_{r_{t}}^{r_{\rm bdy}} \frac{\dot{u}_+}{\dot{r}_+} \rmd r = \int_{r_{t}}^{r_{\rm bdy}} \frac{-P - \sqrt{P^2 + f(r) (r/L)^{2(d-1)}}}{f(r) \sqrt{P^2 + f(r) (r/L)^{2(d-1)}}} dr \, ,
    \label{eq:integrations_CV_caseAdS}
\end{align}
where we set the time at the turning point $t_{t}=0$ by symmetry.
In Fig.~\ref{fig:volume_BH_AdS} we display the volume and its growth rate as functions of the boundary time $t$. Notice that the answer depends on the UV regulator $r_{\rm bdy}$ for the volume. We emphasize that its precise value is not important, but rather its rate of growth (which we will show does not depend on the regulator below), just as in the case of computational complexity.\footnote{There is a different notion from what we are considering, called complexity of formation \cite{Chapman:2016hwi}, where one considers the difference in holographic complexity associated vacuum AdS space to that with an AdS black hole background.}

With the growth of the anchoring time $t$, the turning point approaches the black hole singularity, and the value of the conserved momentum $P$ increases. 
In the late-time limit $t \rightarrow +\infty$, the value of the conserved momentum is $P^2 = P^2_{\rm max} \equiv U_{\rm max}$, as it can be consistently checked from \eqref{eq:t_CV_AdS}. Namely, at $P = P_{\rm max}$ the turning point is $r_{t} = r_f$ and
\begin{equation}
    U_{\rm eff}(r) \sim U_{\rm eff}(r_f) + \frac{1}{2} U''_{\rm eff} (r-r_f)^2 + \dots \, ,
\end{equation}
since $U'_{\rm eff}(r_f)=0$. This condition is crucial for the stretched horizon time to diverge. Indeed,
\begin{equation}\label{eq:t_Pmax_case2}
    t(P_{\rm max}) \sim 2 \sqrt{2} \int^{r_f} \frac{P_{\rm max}}{f(r_f) \sqrt{-U''_{\rm eff}(r_f)} (r-r_f)} \, dr \sim \frac{2 \sqrt{2} \, P_{\rm max}}{f(r_f) \sqrt{-U''_{\rm eff}(r_f)}} \, \log|r-r_f| \, ,
\end{equation}
around the turning point.
At late times (i.e. $t\rightarrow\infty$), the extremal surface hugs the final slice located at constant $r=r_f$.

\paragraph{Growth rate.}
Combining (\ref{eq:Vol BB AdS}, \ref{eq:t_CV_AdS}), we write the volume as
\begin{equation}
    \frac{V}{\Omega_{k,d-1} L^{d-1}} = 
    P \, t +
    2 \int_{r_{t}}^{r_{\rm bdy}} \frac{\sqrt{P^2 + f(r) (r/L)^{2(d-1)}}}{f(r)} dr  \, .
\end{equation}
Taking the derivative with respect to the boundary time $t$, we obtain
\beq
\begin{aligned}
    \frac{1}{\Omega_{k,d-1} L^{d-1}} \frac{dV}{dt} =& P + \frac{dP}{dt} \left( t + 2 \int_{r_{t}}^{r_{\rm bdy}} \frac{P}{f(r) \sqrt{P^2 + f(r) (r/L)^{2(d-1)}}} dr \right) \\ 
    &- 2 \frac{dr_{t}}{dt} \frac{\sqrt{P^2 + f(r_{t}) (r_{t}/L)^{2(d-1)}}}{f(r_{t})} \, .
\end{aligned}
\label{eq:trick_CV_case1}
\eeq
The second term vanishes because of \eqref{eq:t_CV_AdS}, while the third term vanishes due to the definition of turning point in \eqref{eq:turning}. 
Therefore, we get
\begin{equation}\label{eq:dV dt all t AdS}
    \frac{dV}{dt} = \Omega_{k,d-1} L^{d-1} P \, .
\end{equation}
In particular, at late times we get
\begin{equation}
    \lim_{t \to +\infty} \frac{dV}{dt} = \Omega_{k,d-1} L^{d-1} \sqrt{U_{\rm max}} =
    \Omega_{k,d-1} \sqrt{- f(r_f)} \, r_f^{d-1} \, ,
    \label{eq:ratelate_CV_caseAdS}
\end{equation}
where $r_f$ is defined in \eqref{eq:def_rf_caseAdSCV}.
We notice that the final result corresponds to the volume measure evaluated on the final slice at constant $r=r_f$. Importantly, the UV regulator $r_{\rm bdy}$ does not appear in the rate of growth (\ref{eq:ratelate_CV_caseAdS}). 
This is the characteristic linear growth obtained in asymptotically AdS geometries \cite{Carmi:2017jqz} which reproduces the growth of quantum circuits in Sec. \ref{ssec:linear growth CC} and motivated the original work of \cite{Susskind:2014rva}.

\section{WDW patch}\label{sec:WdW AdS}
\begin{figure}
    \centering
    \subfloat[]{\includegraphics[width=0.49\textwidth]{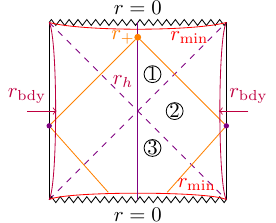}}\hfill\subfloat[]{\includegraphics[width=0.49\textwidth]{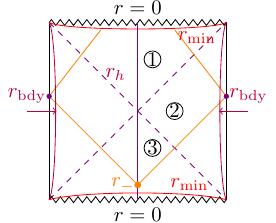}}
    \caption{The WDW patch for a double-sided AdS black hole is the causal domain of dependence (inside the orange diamond). The circled regions indicate the evaluation of (\ref{eq:CV20_case2_int_regime_AdS}). The purple dashed line indicates the event horizon ($r_h$), while the red curve ($r_{\rm min}$) is an IR cutoff surface, and the fuchsia curve indicates the regulator near the asymptotic boundary at $r=r_{\rm bdy}$ to anchor (purple dots) the WDW patch.}
    \label{fig:WdWAdS}
\end{figure}

As mentioned in Ch. \ref{ch:introduction}, the Wheeler-De Witt (WDW) patch is defined as the bulk domain of dependence of a spacelike surface anchored at the boundary slice $\Sigma_{\rm CFT}$ \cite{Couch:2016exn}. See Fig. \ref{fig:WdWAdS}.
Notice that the WDW patch covers a region behind the black hole horizon and will be anchored to the region close to the asymptotic boundary, at $r=r_{\rm bdy}$, so CA and CV2.0 will, in general, be ultraviolent divergent. We introduce a regulator surface close to the asymptotic boundary (shown in Fig. \ref{fig:WdWAdS}) in order to study their evolution.

Next, we assume without loss of generality that the boundary times satisfy $t_L=t_R=t/2$. In this configuration, we define the special positions of the WDW patch as 
\beq
\begin{aligned}
    u_{\rm min} &= -\frac{t}{2} - r^*(r_{\rm bdy}) = -r^*(r_+) \, ,\\
v_{\rm max} &= -\frac{t}{2} + r^*(r_{\rm bdy}) = r^*(r_-) \, , \qquad
r_{\pm} \leq r_h \, .
\end{aligned}
\label{eq:umax_vmin_WDWpatch_caseAdS}
\eeq
Taking derivatives, we notice that the joints satisfy
\beq
\frac{d r_{+}}{dt} =  \frac{1}{2} f(r_{+}) \, , \qquad
\frac{d r_{-}}{dt} = - \frac{1}{2} f(r_{-}) \, .
\label{eq:rates_rprmcaseAdS}
\eeq
The time evolution is symmetric. The instant $t_0$ (and its opposite $-t_0$) when the top (bottom) joint of the WDW patch reaches the future (past) singularity can be obtained from a limit of the general critical time $t_{c2}$ when the top (bottom) joint cross the future (past) cutoff surface in the black hole region, $r_{\rm min}$, \ie
\beq
t_{c2} = 2 \le  r^*(r_{\rm min}) - r^*(r_{\rm bdy})  \ri \, , \qquad
t_{0} = \lim_{r_{\rm min} \rightarrow 0} t_{c2} \, .
\label{eq:critical_time_tcAdS}
\eeq
We will use these considerations in order to evaluate the CV2.0 and CA proposals below.

\section{Complexity=volume 2.0}
\label{ssec:CV20AdS}

The CV2.0 proposal was originally defined in \cite{Couch:2016exn} after the CA proposal \cite{Brown:2015bva,Brown:2015lvg} (which we will discuss next). We will study this conjecture first, since the technical evaluation CA reduces to CV2.0 with additional terms for the class of geometries we are interested in, as we will review below. The motivation to introduce this proposal is that, compared to the CV conjecture (Sec. \ref{ssec:CV}), the CV2.0 conjecture is much easier to evaluate since the only required trajectories are null geodesics, which are, in principle, easier to obtain than maximal volume slices. It's also, arguably, more practical than the CA proposal (Sec. \ref{ssec:CAAdS}) as one does not need to evaluate an action at null surfaces.

As mentioned in Sec. \ref{ch:introduction}, the CV2.0 conjecture states that holographic complexity is proportional to the spacetime volume of the WDW patch, (see Fig. \ref{fig:WdWAdS}).

For a generic Lorentzian signature geometry, one can express this observable as
\beq 
\mathcal{C}_{2.0V} = \frac{V_{\rm WDW}}{G_N L^2} = 
\int_{\rm WDW}\rmd^{d+1}x\sqrt{-g} \, ,
\label{eq:general_CV20}
\eeq
where the limits of integration correspond to the null boundaries of the WDW patch (see Sec. \ref{ssec:CV20} for explicit evaluations in an SdS background).

Let us now evaluate CV2.0 for the AdS black hole (\ref{eq:metric SAdS}) by analyzing the time dependence of the quantity \eqref{eq:general_CV20}.
For convenience, we will always split the evaluation of CV2.0 as
\beq
\mathcal{C}_{2.0V} = \sum_{i} \mathcal{J}_i \, ,
\label{eq:decomposition_CV20_general}
\eeq
where the index $i$ refers to a subregion of the WDW patch that we specify case by case in the configurations discussed below, and $\mathcal{J}_i$ is the complexity \eqref{eq:general_CV20} computed in such subregion.

We refer to Fig.~\ref{fig:WdWAdS} for the splitting of the WDW patch into the subregions corresponding to the integrals
\begin{subequations}
\label{eq:CV20_case2_int_regime_AdS}
\beq
\mathcal{J}_{1L}=\mathcal{J}_{1R} = \frac{\Omega_{k,d-1}}{G_{N} L^2} \int_{r_+}^{r_h} dr  \,  r^{d-1} \le  \frac{t}{2} -r^*(r) +r^*(r_{\rm bdy})  \ri \, ,
\label{eq:CV20_caseAdS_int_regime_p1}
\eeq
\beq
\mathcal{J}_{2L} =
\mathcal{J}_{2R}=
\frac{\Omega_{k,d-1}}{G_{N} L^2} \int_{r_h}^{r_{\rm bdy}} dr  \,  r^{d-1} \le  2 r^*(r_{\rm bdy}) - 2 r^*(r)  \ri \, , 
\label{eq:CV20_caseAdS_int_regime_p2}
\eeq
\beq
\mathcal{J}_{3L} =
\mathcal{J}_{3R} =  \frac{\Omega_{k,d-1}}{G_{N} L^2} \int_{r_-}^{r_h} dr  \,  r^{d-1} \le - \frac{t}{2} +r^*(r_{\rm bdy}) - r^*(r)  \ri \, ,
\label{eq:CV20_caseAdS_int_regime_p3}
\eeq
\end{subequations}
where $r_{\pm}$ were defined in \eqref{eq:umax_vmin_WDWpatch_caseAdS}.
In the following discussion, we will directly perform the limit $r_{\rm min} \rightarrow 0$ for the cutoff surface in (\ref{eq:critical_time_tcAdS}) since it leads to regular results. 

Famously, holographic complexity computed in a black hole background in asymptotically AdS spacetime admits linear growth at early (late) times.
The geometric origin for this behavior is that the (bottom) top joint of the WDW patch moves behind the singularity, while the top (bottom) joint approaches a constant radial coordinate which coincides with the location of the horizon \cite{Carmi:2017jqz}.
In order to capture the linear growth, we need to study the early and late time regimes in the evolution of CV2.0. Also, notice the presence of the UV regulator $r_{\rm bdy}$ in (\ref{eq:CV20_caseAdS_int_regime_p2}). As in computational complexity, we are not concerned about the precise numerical value of the complexity itself (which is sensitive to the choice of the regulator), since complexity is only meaningful when it is compared to something else. Then, we will study its evolution, and show explicitly that its early and late rate of growth are independent of $r_{\rm bdy}$, as follows.

\begin{figure}[t!]
    \centering
    \subfloat[]{ \label{subfig:CV2AdS}
\includegraphics[width=0.475\textwidth]{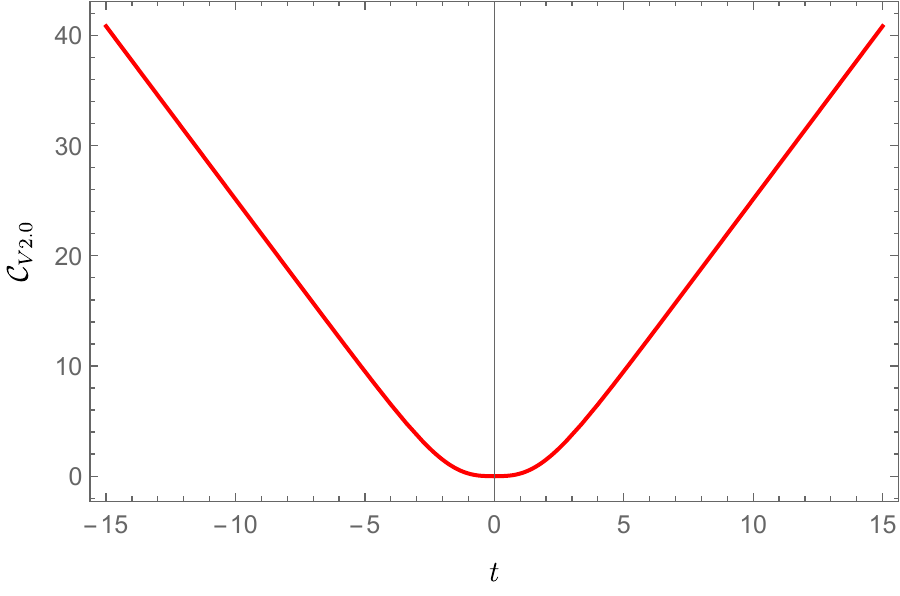}}\hfill\subfloat[]{\label{subfigb:dCV2AdS} 
\includegraphics[width=0.475\textwidth]{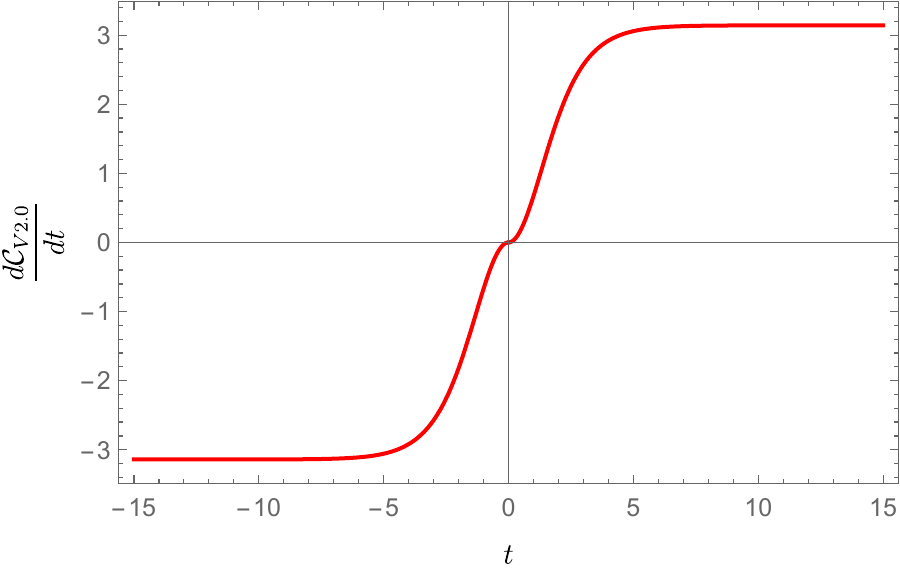}}
    \caption{Time dependence of (a): CV2.0 and (b): its rate, computed according to \eqref{eq:rateCV20_caseAdS_p2}. We set the same numerical values as in Fig. \ref{fig:volume_BH_AdS}.}
    \label{fig:CV20AdS}
\end{figure}

Consider Fig. \ref{fig:WdWAdS}. The WDW patch starts at times $t < t_0$ ($t_0 \leq 0$ by (\ref{eq:critical_time_tcAdS})) with the bottom joint behind the past singularity, while the top joint is located behind the black hole horizon.
Afterwards, both joints sit behind the respective singularities, 
and after $t > t_0$ the bottom joints move after the past singularity and stay in the interior of the black hole.
Then, taking the time derivative of (\ref{eq:CV20_case2_int_regime_AdS}), the rate of CV2.0 reads
\beq
\frac{d \mathcal{C}_{2.0V}}{dt} = 
\frac{\Omega_{d-1}}{d \, G_{N} L^2}  \times 
 \begin{cases}
 \le - r_+^d \ri  &  \mathrm{if} \,\, t < t_{0} \\
0  & \mathrm{if} \,\, t_{0} \leq t \leq -t_0 \\
  r_-^d   &  \mathrm{if} \,\, t > -t_{0} \\
\end{cases}
\label{eq:rateCV20_caseAdS_p2}
\eeq
In particular, the late (early) time behavior can be found analytically, since it corresponds to $r_- \rightarrow r_h$ ($r_+ \rightarrow r_h$).
This gives
\beq
\frac{d \mathcal{C}_{2.0V}}{dt} \,\, \underset{t \rightarrow \pm t_{0}}{{\longrightarrow}} \,\,
\pm \frac{\Omega_{k,d-1}}{d \, G_{N} L^2} \, r_h^d \, ,
\label{eq:late_times_CV20_caseAdS}
\eeq
Then, as seen in (\ref{eq:rateCV20_caseAdS_p2}), the complexity is constant in the intermediate regime between the critical times \cite{Carmi:2017jqz}, while at late and early time rate of growth approaches a finite constant, as motivated by the growth of computational complexity. This behavior is confirmed in Fig. \ref{fig:CV20AdS}.

In the following section, we will find a closely related behavior for the CV2.0 conjecture in the black hole patch of SdS space, see \ref{ssec:CV20}.

\section{Complexity=action}
\label{ssec:CAAdS}
The CA proposal was created to address some unsatisfactory aspects of the CV conjecture (\ref{eq:CV 1st}), including the dependence on the arbitrary length scale $\ell$, and the fact it is unclear why a maximal volume slice should play an important role in the holographic dictionary \cite{Brown:2015bva,Brown:2015lvg}. Nevertheless, the CA proposal still suffers from ambiguities, as we discuss below.

According to the CA proposal, holographic complexity is proportional to the on-shell gravitational action evaluated in the WDW patch, \ie
\beq
\mathcal{C}_A = \frac{I_{\rm WDW}}{\pi} \, , \qquad
I_{\rm WDW} = \sum_{\mathcal{X}}  I_{\mathcal{X}} \, , \qquad
\mathcal{X} \in \lbrace \mathcal{B}, \rm GHY, \mathcal{N}, \mathcal{J} ,\rm ct  \rbrace \, .
\label{eq:tot_grav_action}
\eeq
The terms contributing to the list labeled by $\mathcal{X}$ are collected in \cite{Lehner:2016vdi} or in App. A of \cite{Carmi:2016wjl} (see also \cite{Parattu:2015gga,Chakraborty:2016yna}).
We briefly review their expressions:
\begin{itemize}
    \item The bulk term is the Einstein-Hilbert action (\ref{eq:Einstein H action}), which we repeat here for the reader
    \beq
I_{\mathcal{B}} = \frac{1}{16 \pi G_N} \int_{\rm WDW} d^{d+1} x \, \sqrt{-g}	 \,  \le  \mathcal{R} - 2 \Lambda  \ri \, .
\label{eq:general_bulk_action}
\eeq
\item The Gibbons-Hawking-York (GHY) term is evaluated on timelike or spacelike codimension-one boundary surfaces
\beq
I_{\rm GHY} = \frac{\varepsilon_{t,s}}{8 \pi G} \int_{\mathcal{B}_{t,s}} d^d x \, \sqrt{|h|} \, K \, ,
\label{eq:GHY_term}
\eeq
where $h$ is the determinant of the induced metric and $K$ is the trace of the extrinsic curvature.
The parameter $\varepsilon_{t,s}=\pm 1$ distinguishes if the surface of interest ${\mathcal{B}_{t,s}} $ is timelike or spacelike, respectively.
\item The codimension-one null boundaries admit the contribution 
\beq
I_{\mathcal{N}} = \frac{\varepsilon_n}{8 \pi G} 
\int_{\mathcal{B}_n} d\lambda d^{d-1} x \, \, \sqrt{|\gamma|} \, \kappa (\lambda) \, ,
\label{eq:null_codim1_term}
\eeq
where $\varepsilon_n = \pm 1$ depends on the orientation of the null normal to the surface, $\lambda$ is a parameter along the congruence of geodesics, 
$\gamma$ is the induced metric along the other orthogonal directions and $\kappa(\lambda)$ is the acceleration of the null geodesics.
In the case of an affine parametrization, $\kappa=0$.
\item Codimension-two joints arise from the intersection of codimension-one surfaces (see Fig. \ref{fig:WdWAdS}). In this work, the relevant joints will always arise from intersections involving at least one null surface. Their expression reads
 \beq
I_{\mathcal{J}} =  \frac{\varepsilon_{\mathfrak{a}}}{8 \pi G} \int_{\mathcal{J}} d^{d-1} x \, 
 \sqrt{|\gamma|} \, \mathfrak{a}  \, ,
 \label{eq:joints_action}
\eeq
where the coefficient $ \varepsilon_{\mathfrak{a}} = \pm 1$ depends on the orientations of the joint, and $\gamma$ is the determinant of the induced metric along the codimension-two joint.
The integrand $\mathfrak{a}$ is 
\beq
\mathfrak{a} = \begin{cases}
\log \left| \mathbf{t} \cdot \mathbf{k}  \right|  & \mathrm{if} \,\, \mathbf{t} \,\, \mathrm{timelike} \,\, \mathrm{and} \,\, \mathbf{k} \,\, \mathrm{null}   \\
\log \left| \mathbf{n} \cdot \mathbf{k}  \right|  & \mathrm{if} \,\, \mathbf{n} \,\, \mathrm{spacelike} \,\, \mathrm{and} \,\, \mathbf{k} \,\, \mathrm{null} \\
\log \left| \frac{1}{2} \mathbf{k}_L \cdot \mathbf{k}_R  \right|  & \mathrm{if} \,\, \mathbf{k}_L, \mathbf{k}_R \,\, \mathrm{null}  \\
\end{cases}
\label{eq:integrand_a_actionjoints}
\eeq
where the argument of the logarithms is the scalar product between the outward-directed one-forms normal to the codimension-one surfaces intersecting at the joint.  
\item Since the gravitational action is not reparametrization-invariant as it stands, we need to add a counterterm on codimension-one null boundaries
\beq
I_{\rm ct} = \frac{1}{8 \pi G} \int_{\mathcal{B}_n} 
d\lambda d^{d-1} x \, \sqrt{\gamma} \,  \Theta \, \log |\ell_{\rm ct} \Theta| \, ,
\label{eq:counterterm_action}
\eeq 
where $\Theta$ is the expansion of the congruence of null geodesics and $\ell_{\rm ct}$ is an arbitrary length scale.
\end{itemize}

The asymptotically (A)dS spacetimes with metric \eqref{eq:asympt_dS} have constant Ricci scalar, thus the bulk term is simply proportional to the CV2.0 conjecture:
\beq
I_{\mathcal{B}} = 
\frac{d}{8 \pi G_N L^2} V_{\rm WDW} = \frac{d}{8 \pi} \, \mathcal{C}_{2.0V} \, .
\label{eq:bulk_term_CA}
\eeq
Therefore, the bulk contribution satisfies the same properties discovered for the CV2.0 computation performed in Sec. \ref{ssec:CV20AdS}.
In the following, we will denote the combination of boundary terms with 
\beq
I_{\rm bdy} = I_{\rm GHY} + I_{\mathcal{N}} + I_{\mathcal{J}} + I_{\rm ct} \, .
\label{eq:bdy_action}
\eeq
We will focus on a symmetric configuration of the boundary times $t_L=t_R=t/2$. As we commented about the CV2.0 proposal, the critical time obeys $t_0 <0$, as depicted in Fig.~\ref{fig:WdWAdS}. By applying the time derivatives \eqref{eq:rates_rprmcaseAdS} to the CA conjecture \eqref{eq:tot_grav_action}. We obtain
\small
\beq
\begin{aligned}
    &\frac{8 \pi^2 G_N}{\Omega_{k,d-1}} \frac{d \mathcal{C}_A}{dt} = \\
&\begin{cases}
    - \frac{Cd}{2L^2}+k\delta_{2,d} - \frac{(r_+)^d}{L^2} - \tilde{g}(r_+) \log \left| \frac{\ell_{\rm ct}^2 (d-1)^2 f(r_+) }{(r_+)^2} \right| - (r_+)^{d-1} \, \frac{f'(r_+)}{2}     &  \mathrm{if}  \,\, t < t_{0} \\
 0  & \mathrm{otherwise} \\
 \frac{Cd}{2L^2}-k\delta_{2,d}  + \frac{(r_-)^d}{L^2}  + \tilde{g}(r_-) \log \left| \frac{\ell_{\rm ct}^2 (d-1)^2 f(r_-) }{(r_-)^2} \right| + (r_-)^{d-1} \, \frac{f'(r_-)}{2}    &  \mathrm{if} \,\, t > -t_{0} \\
\end{cases}~,
\end{aligned}
\label{eq:rateCA_caseAdS_p2}
\eeq
\normalsize
where $r_\pm$ appear in (\ref{eq:umax_vmin_WDWpatch_caseAdS}); we defined $\tilde{g}(r_\pm)\equiv \frac{d-1}{2}(r_\pm)^{d-2} f(r_\pm)$; $\delta_{ij}$ is the Kronecker delta; and the results for the boundary terms were taken from appendix \ref{app:ssec:case2}; see footnote \ref{fn:AdS CA App}.

Notice the time dependence of the intermediate regime. This is a trivial consequence of the intermediate regime of CV2.0 in \eqref{eq:rateCV20_caseAdS_p2}. The vanishing rate of the boundary terms follows exactly as \eqref{eq:rate_bdy_app_case2} (with $\Omega_{d-1}\rightarrow\Omega_{k,d-1}$).
From (\ref{eq:rateCA_caseAdS_p2}), one can notice that the time derivative of CA is discontinuous at the critical time.
One can evaluate the jump at $t=-t_0$ by performing the limit $r_- \rightarrow 0$ in (\ref{eq:rateCA_caseAdS_p2}) to get
\beq
\frac{d \mathcal{C}_A}{dt}\Big|_{t= -t_0^{+}} - \frac{d \mathcal{C}_A}{dt}\Big|_{t= -t_0^-} = \frac{\Omega_{k,d-1}}{8 \pi^2 G_N} \qty(\frac{C~d}{2L^2}-k~\delta_{2,d}) \, .
\label{eq:disc_rateCA_caseAdSp2}
\eeq
An analogous computation shows that the same conclusion holds for $t=t_0$. The bulk interpretation for this discontinuity is that the WDW patch reaches either the past or future singularities (see Fig. \ref{fig:WdWAdS}), where the GHY term gives the sole contribution to (\ref{eq:disc_rateCA_caseAdSp2}), since all terms on null boundaries (either codimension-one or codimension-two) evaluated at the singularity vanish.

\begin{figure}[t!]
    \centering
    \subfloat[]{ \label{subfig:CAAdS}\includegraphics[width=0.475\textwidth]{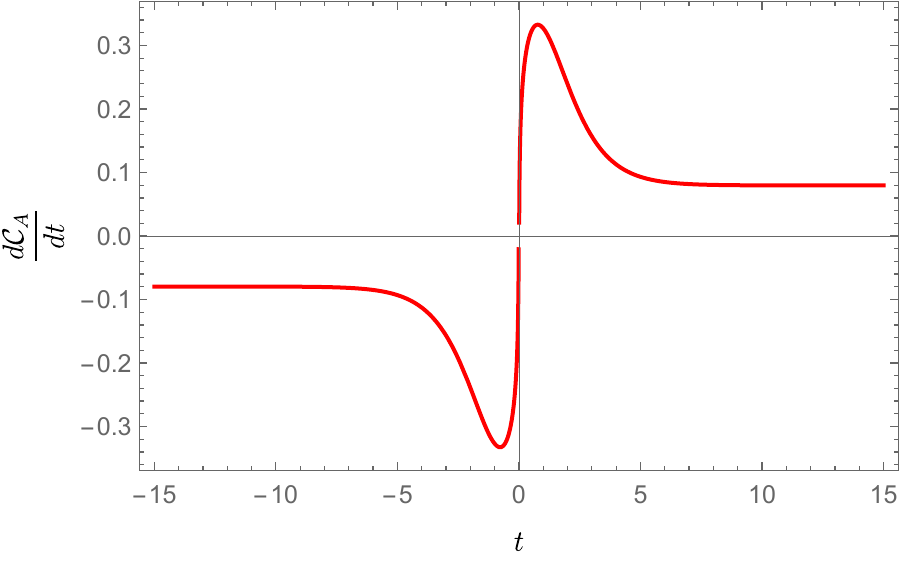}}\hfill \subfloat[]{\label{subfigb:dCAAdS} \includegraphics[width=0.475\textwidth]{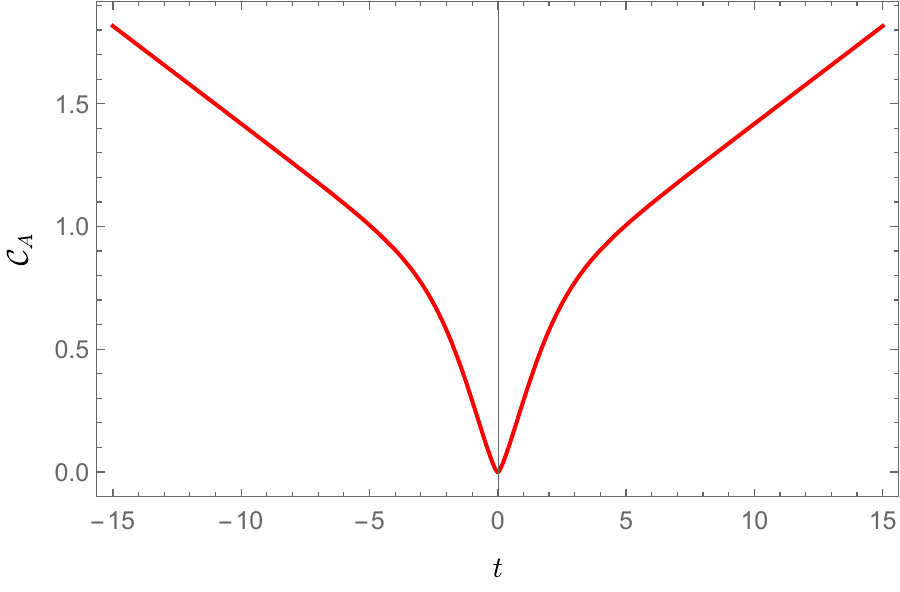}}
    \caption{Time dependence of (a): CA and (b) its rate, computed according to \eqref{eq:rateCA_caseAdS_p2}. We set the same numerical values as in Fig. \ref{fig:volume_BH_AdS} and $\ell_{\rm ct} = 10^{-4}$. }
    \label{fig:CAAdS}
\end{figure}

It is straightforward to determine the late (early) time limits, which correspond to the bottom (top) joint of the WDW patch approaching the horizon radius $r_h$.
The rate reads ($d\geq 3$)
\beq
\lim_{t \rightarrow \pm \infty} \frac{d \mathcal{C}_A}{dt} = \pm \frac{\Omega_{k,d-1}}{8 \pi^2 G_N}  \left[ \frac{2(r_h)^d-C~d}{2L^2}+k~\delta_{d,2}  + (r_h)^{d-1} \, \frac{f'(r_h)}{2}  \right] \, .
\label{eq:rate_CA_caseAdS_latelimit}
\eeq
We conclude that the CA proposal for the AdS black hole displays the celebrated late (early) time linear growth, and as in the other cases, the rate of growth remains \emph{finite}, while \ref{eq:tot_grav_action} still contains a UV regulator dependence. An example is displayed in Fig. \ref{fig:CAAdS}. The reader is also referred to \cite{Carmi:2016wjl} for several more examples of the CA proposal in AdS black hole backgrounds.

\section{Complexity=anything}\label{sec:Cany AdS}

A unification and extension of the previous proposals was developed by \cite{Belin:2021bga,Belin:2022xmt} (see also \cite{Jorstad:2023kmq,Myers:2024vve}). The new class of observables is based on the observation that while computational complexity suffers from several ambiguities, as we discussed in  Sec. \ref{sec:computational complexity}, and yet it obeys two universal properties, (\emph{i}) late time linear growth, and (\emph{ii}) the switchback effect. Likewise, the holographic complexity proposals we have encountered obey these properties\footnote{We have not discussed the holographic switchback effect in the CV, CA, and CV2.0 conjectures, as the evaluation is more involved; the reader is referred to \cite{Stanford:2014jda,Brown:2015bva,Brown:2015lvg,Belin:2022xmt} for details. We will consider the switchback effect for the AdS black hole at the end of this section for a particular class of CAny proposals.} as the UV regulator $r_{\rm bdy}$ close to the boundary in all proposals, or the scale $\ell$ in CV case. We will briefly review the definition of the new class of proposals, starting with \emph{codimension-one} observables. Similar to the CV case, consider a codimension-one surface $\Sigma_{F_2}$ which is anchored to the asymptotic boundary, which we denote as $\partial\Sigma_{F_2}=\Sigma_{\rm CFT}$, and where $F_2$ is an arbitrary scalar functional of the bulk metric $g_{\mu\nu}$, and the embedding coordinates $X^\mu(\sigma^a)$ of the codimension-one surface (parametrized by $\sigma^a$) in the bulk. The evolution of the codimension-one surface is covariantly defined through the extremization of this functional, i.e.
\begin{equation}\label{eq:CAny_cod1_1}
    \partial_X\qty(\int_\Sigma \rmd^d\sigma\sqrt{h} F_2(g_{\mu\nu},~X^\mu))=0~.
\end{equation}
The general class of codimension-one CAny observables are defined as
\begin{equation}\label{eq:CAny_cod1_2}
    O_{F_1,~\Sigma_{F_2}}(\Sigma_{\rm CFT})=\frac{1}{G_NL}\int_{\Sigma_{F_2}}\rmd^d\sigma\sqrt{\abs{h}}F_1(g_{\mu\nu},~X^\mu)~,
\end{equation}
where $F_1$ is another arbitrary scalar functional of the bulk metric $g_{\mu\nu}$, and the embedding coordinates of surface $\Sigma_{F_2}$. As an intuitive example, when $F_1=F_2=1$ one recovers the CV proposal (\ref{eq:CV 1st}). It was shown in \cite{Belin:2021bga} that (\ref{eq:CAny_cod1_1}, \ref{eq:CAny_cod1_2}) obey the same properties (\emph{i}, \emph{ii}) as computational complexity, and thus are possible holographic duals to it.
 
The infinite class of observables above was further extended to include \emph{codimension-zero} observables in \cite{Belin:2022xmt}. To define them, consider a bulk region $\mathcal{M}$ which is bounded by two codimension-one surfaces $\Sigma_{\pm}$ that do not touch or cross each other, and that is anchored to the boundary $\partial \Sigma_{\pm}= \Sigma_{\rm CFT}$. One can define covariantly define the evolution of region $\mathcal{M}$ by extremization of the functional
\begin{equation}\label{eq:CAny cod 0}
    W_{G_2,~F_{2,\pm}}=\int_{\mathcal{M}}\rmd^{d+1}x~\sqrt{\abs{g}} G_2[g_{\mu\nu}]+\sum_{\varepsilon=\pm}\int_{\Sigma_\varepsilon}\rmd^{d}\sigma~ \sqrt{\abs{h}}F_{2,\varepsilon}[g_{\mu\nu},~X_\varepsilon^\mu]~,
\end{equation}
with respect to the embedding coordinates of $\Sigma_\pm$,
\begin{equation}
    \delta_{X_\pm}\qty[W_{G_2,~F_{2,\pm}}]=0~.
\end{equation}
In (\ref{eq:CAny cod 0}), we denote $G_2[g_{\mu\nu}]$ an arbitrary scalar functional of $(d+1)$-dimensional bulk curvature invariants (which depend on the metric $g_{\mu\nu}$); similarly, $F_{2,\pm}[g_{\mu\nu},~X_\pm^\mu]$ are arbitrary local invariant functionals of the bulk metric and the embedding coordinates $X_\pm^\mu(\sigma_i)$ of $\Sigma_\pm$ respectively. Importantly, $G_2$, $F_{2,\pm}$ are independent in principle. Similar to (\ref{eq:CAny_cod1_2}), we then define new observables evaluated on $\mathcal{M}$ as
\begin{equation}\label{eq:obs CAny cod0}
\begin{aligned}
    O[{G_1,~F_{1,\pm},~G_2,~F_{2,\pm}}]=\frac{1}{G_NL^2}\int_{\mathcal{M}}\rmd^{d+1}x~\sqrt{\abs{g}} G_1[g_{\mu\nu}]&\\
    +\frac{1}{G_NL}\sum_{\varepsilon=\pm}\int_{\Sigma_\varepsilon[G_2,~F_{2,\varepsilon}]}\rmd^{d}\sigma~\sqrt{\abs{h}} F_{1,\varepsilon}[g_{\mu\nu},~X_\varepsilon^\mu]~,&
\end{aligned}
\end{equation}
where $F_{1,\pm}$ and $G_1$ are arbitrary scalar functionals of the bulk metric $g_{\mu\nu}$ and the embedding coordinates $X^\mu_\pm(\sigma^a)$ of $\Sigma_\pm$, which can be chosen to be independent of each other. By properly choosing the above functionals one can recover all the proposals we have seen so far (i.e. CV, CV2.0, CA). It was shown in \cite{Belin:2022xmt} that the new observables also display the property (\emph{i}), and some would also obey (\emph{ii}). It was later shown in \cite{Aguilar-Gutierrez:2023zqm,Aguilar-Gutierrez:2023pnn} that not all the observables in (\ref{eq:CAny cod 0}, \ref{eq:obs CAny cod0}) will display (\emph{ii}). 

We will study a particular class of the CAny observables below, and show that some of these proposals lack time-reversal invariance even in maximally symmetric backgrounds, which leads to the violation of (\emph{ii}).

\subsection{Complexity=anything: constant mean curvature slices}
\label{sec:CAny CMCAdS}
In Sec. \ref{ch:ComplexitydS}, we will study a family of codimension-one observables within the class of the CAny proposal \cite{Belin:2021bga,Belin:2022xmt} to derive an alternative to the hyperfast growth in SdS spacetimes. The late-time linear growth property for the CAny proposals in AdS black holes has been studied in \cite{Belin:2022xmt,Jorstad:2023kmq}, which we briefly review below.\footnote{Notice that \cite{Belin:2022xmt,Jorstad:2023kmq} considered a planar AdS black hole (meaning $k=0$ in \ref{eq:metric SAdS}). This is a technical convenience for obtaining explicit expressions for the rate of growth of CAny complexity proposals. We will consider this simplification until later stages of the evaluation, which we emphasize below (\ref{eq:rt locationAdS}) and (\ref{eq:switchback resultAdS}).}

\begin{figure}[t!]
    \centering
    \includegraphics[width=0.7\textwidth]{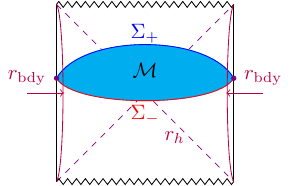}
\caption{Proposal for evaluating the holographic complexity on the CMC slices $\Sigma_\pm$, anchored to the asymptotic boundary (purple dots) of a planar AdS black hole. $\mathcal{M}$ (cyan) is the bulk region bounded by the slices $\Sigma_-$ and $\Sigma_+$ (red and blue, respectively), where $\mathcal{C}^{-}$ and $\mathcal{C}^{+}$ in \eqref{eq:Volepsilon} are evaluated. 
In the picture, we anchor the bulk region to a surface $r=r_{\rm bdy}$ (fuchsia curve) close to the black hole asymptotic boundary.
Similar considerations apply to the cosmological stretched horizons. The precise profile of the $\Sigma_\epsilon$ slices is determined by the extremization of \eqref{eq:regions CMC}.}
    \label{fig:ProposalAdS}
\end{figure}

First, we define a spacetime region by extremizing the following combination of codimension-one and codimension-zero terms with different weights
\begin{equation}\label{eq:regions CMC}
\begin{aligned}
    \mathcal{C}_{\rm CMC}&=\frac{1}{G_N L}\biggl[\alpha_+\int_{\Sigma_+}d^d\sigma\,\sqrt{h}+\alpha_-\int_{\Sigma_-}d^d\sigma\,\sqrt{h}+ \frac{\alpha_B}{L}\int_{\mathcal{M}}d^{d+1}x\sqrt{-g}\biggr] \, ,
\end{aligned}
\end{equation}
where $\mathcal{M}$ is a $(d+1)$--dimensional bulk region with future (past) boundaries $\Sigma_{\pm}$ anchored at the stretched horizons, such that $\partial\mathcal{M}=\Sigma_+\cup\Sigma_-$, as shown in Fig.~\ref{fig:ProposalAdS}.\footnote{We will label with $\epsilon=\lbrace +,- \rbrace$ the quantities defined on the codimension-one surfaces $\Sigma_{\pm}$.  }
We denote with $h$ the determinant of the induced metric on $\Sigma_{\pm}$. 
The coefficients $\alpha_{\pm}$ and $\alpha_B$ are dimensionless constants. The extremization of the functional \eqref{eq:regions CMC} defines constant mean curvature (CMC) slices, \ie the extrinsic curvature on these surfaces is given by \cite{MARSDEN1980109,Belin:2022xmt}
\begin{equation}
    K_\epsilon\equiv\eval{K}_{\Sigma_\epsilon} = -\epsilon \, \frac{\alpha_B}{\alpha_\epsilon~L}~,
    \label{eq:K_CMC}
\end{equation}
with the convention that the vectors normal to $\Sigma_{\pm}$ are future-directed.
We then define holographic complexity as the physical observable
\begin{equation}\label{eq:Volepsilon}
    \mathcal{C}^\epsilon \equiv \frac{1}{G_N L}\int_{\Sigma_\epsilon}d^d\sigma\,\sqrt{h}~F[g_{\mu\nu},\,\mathcal{R}_{\mu\nu\rho\sigma},\,\nabla_\mu]~,
\end{equation}
where $F[g_{\mu\nu},\,\mathcal{R}_{\mu\nu\rho\sigma},\,\nabla_\mu]$ is an arbitrary scalar functional composed of $(d+1)$--dimensional bulk curvature invariants built with the metric $g_{\mu\nu}$, the Riemann tensor $\mathcal{R}_{\mu\nu\rho\sigma}$ and the covariant derivative $\nabla_{\mu}$.
The quantity $\mathcal{C}^{\epsilon}$ is defined on either the future (when $\epsilon=-$) or the past (when $\epsilon=+$) codimension-one CMC slices $\Sigma_{\pm}$ determined by the extremization of the functional \eqref{eq:regions CMC} with fixed boundary conditions on the stretched horizons.

To simplify the evaluation of the CMC slices and the associated complexity observable, we employ the EF coordinates \eqref{eq:generic_null_metric_noshock}.
Using the AdS black hole background \eqref{eq:metric SAdS}, we describe the slices $\Sigma_\epsilon$ in terms of a radial parameter $\sigma$, such that the coordinates are expressed as $(v(\sigma)$, $r(\sigma))$. Without lost of generality, we can choose $\eval{K}_{\Sigma_+}=-\abs{K}$ and $\eval{K}_{\Sigma_-}=\abs{K}$ in the evaluation of \eqref{eq:regions CMC} in EF coordinates, such that we obtain
\beq
\mathcal{C}_{\rm CMC} =\frac{\Omega_{k,d-1}L^{d-2}}{G_N}\sum_{\epsilon}\alpha_\epsilon\int_{\Sigma_\epsilon}
    d\sigma\,\mathcal{L}_\epsilon~,
    \label{eq:C CMC}
\eeq
where $\Omega_{k,d-1}=\int\rmd \Omega_{k,d-1}$, and
\begin{equation}
\begin{aligned}
    \mathcal{L}_\epsilon&\equiv \qty(\frac{r}{L})^{d-1}\sqrt{-f(r)\dot{v}^2+2\dot{v}\dot{r}}-\epsilon \, \tfrac{L\abs{K}}{d}\dot{v}\qty(\frac{r}{L})^d\\
    &=\qty(\frac{r}{L})^{d-1}\sqrt{-f(r)\dot{u}^2-2\dot{u}\dot{r}}-\epsilon \, \tfrac{L\abs{K}}{d}\dot{u}\qty(\frac{r}{L})^d \, .
\end{aligned}
\end{equation}
Using the gauge choice (\ref{gauge}), the EOM associated with \eqref{eq:C CMC} can be written as 
\begin{equation}
    \dot{r}^2+\mathcal{U}(P_{\epsilon},\,r)=0~,
    \label{eq:EOM_CAny}
\end{equation}
where
\begin{subequations}
    \beq
    \begin{aligned}
    \label{eq:potential arbitrary}
P_{\epsilon}&\equiv\pdv{\mathcal{L}_\epsilon}{\dot{v}}=\dot{r}-\dot{v}\,f(r)-\epsilon \, \frac{L~\abs{K}}{d}\,\le \frac{r}{L}\ri^d\\
&=\pdv{\mathcal{L}_\epsilon}{\dot{u}}=-\dot{r}-\dot{u}\,f(r)-\epsilon \, \frac{L~\abs{K}}{d}\,\le \frac{r}{L}\ri^d~,
\end{aligned}
\eeq
\beq
    \mathcal{U}(P_{\epsilon},\,r) \equiv -f(r) \le \frac{r}{L}\ri^{2(d-1)} - \left(P_{\epsilon} +\epsilon \, \frac{L~\abs{K}}{d}  \le \frac{r}{L}\ri^d\right)^2~.\label{eq:U CAny}
\eeq
\end{subequations}
The differential equation \eqref{eq:EOM_CAny} effectively describes the motion of a particle in a potential $\mathcal{U}$.
The quantity $P_{\epsilon}$ is a conserved momentum corresponding to the fact that the variable $u(v)$ is cyclic for the Lagrangian density $\mathcal{L}_{\epsilon}$.

The codimension-one surfaces solving the variational problem in \eqref{eq:EOM_CAny} determine the CMC slices
where complexity is calculated according to \eqref{eq:Volepsilon}.
In EF coordinates, the CAny observable reads
\beq
\begin{aligned}
    \label{eq:C epsilon inter}
    \mathcal{C}^\epsilon &=\frac{\Omega_{k,d-1}L^{d-2}}{G_N}\int_{\Sigma_\epsilon}d\sigma\,\qty(\frac{r}{L})^{d-1}\sqrt{-f(r)\dot{v}^2+2\dot{v}\dot{r}}\,a(r)\\
    &=\frac{\Omega_{k,d-1}L^{d-2}}{G_N}\int_{\Sigma_\epsilon}d\sigma\,\qty(\frac{r}{L})^{d-1}\sqrt{-f(r)\dot{u}^2-2\dot{u}\dot{r}}\,a(r)\,,
\end{aligned}
\eeq
where $a(r)$ is a dimensionless scalar function corresponding to the evaluation of the functional $F$ in \eqref{eq:Volepsilon} in the geometry under consideration.
We remark that the CAny proposal evaluated on CMC slices reduces to the CV conjecture \eqref{eq:vol computation} when $K_{\epsilon}=0$ and $F=1$ in \eqref{eq:Volepsilon}.\footnote{Equivalently, it follows from \eqref{eq:K_CMC} that when $K_{\epsilon}=0$, only the $\alpha_+$ or $\alpha_-$ terms in (\ref{eq:regions CMC}) contribute to defining the region of integration, which collapses to a single extremal surface.}
Indeed, by imposing the above-mentioned conditions, we find that
the identities \eqref{eq:EOM_CAny}, \eqref{eq:potential arbitrary} and \eqref{eq:U CAny} reduce to eqs.~\eqref{eq:classical_motion_AdS} and \eqref{eq:conserved_momentum_CV}, and the observable \eqref{eq:C epsilon inter} reduces to the induced volume on the extremal surface. 

The shape of the effective potential $\mathcal{U}$ is crucial to determine the time evolution of the CMC slices.
As long as $P_{\epsilon}$ lies in a range such that the effective potential admits at least one root, then the corresponding extremal surface will present a turning point $r_t$ defined by $\mathcal{U}(P_{\epsilon}, r_t)=0$.
The major difference with the CV case is that the effective potential \eqref{eq:U CAny} is itself a function of $P_{\epsilon}$, therefore it is more difficult to determine the full time-dependence of complexity.
For this reason, we will mainly focus on the late-time behavior, which provides the most relevant and universal feature.

Next, considering symmetric time evolution $t_L=t_R=t/2$, combining (\ref{eq:C epsilon inter}) with (\ref{eq:C CMC}), we can express:
\begin{equation}\label{eq:Anything complexity cod1AdS}
\mathcal{C}^\epsilon=\frac{2\Omega_{k,d-1}L^{d-2}}{G_N} \int^{{r}_{\rm bdy}}_{{r}_{t}} \frac{a(r)~ {\qty(\frac{r}{L})}^{2(d-1)}  }{\sqrt{-\mathcal{U}(P_{\epsilon},\,{r})}} \, dr \, .
\end{equation}
where $r=r_{\rm bdy}$ is a radial cutoff value near the asymptotic boundary of the planar AdS black hole. Meanwhile, the boundary time, $t_{\rm bdy}$, can be expressed as
\begin{equation}\label{eq: t CAnyAdS}
\begin{aligned}
t&=\int_{\Sigma_\epsilon} \rmd r\frac{\dot{v}-\dot{r}/{f(r)}}{\sqrt{-\mathcal{U}(P_v^\epsilon,\,r)}}= 2  \int^{{r}_{\rm bdy}}_{{r}_{t}} \frac{ P_{\epsilon}+ \epsilon L\frac{\abs{K}}{d} \qty(\frac{r}{L})^d  }{f(r)\,\sqrt{-\mathcal{U}(P_{\epsilon}, r)}} \, dr \, .
\end{aligned}
\end{equation}
Combining (\ref{eq:Anything complexity cod1AdS}) and (\ref{eq: t CAnyAdS}), we can now express
\begin{align}\label{eq:CAny CMC AdS}
    &{\mathcal{C}^\epsilon}=\frac{\Omega_{k,d-1}L^{d-2}}{G_N}a(r_{t})\sqrt{-f(r_{t})\qty(\frac{r_{t}}{L})^{2(d-1)}}t\\
    &+\frac{2\Omega_{k,d-1}L^{d-2}}{G_N}\int^{r_{\rm bdy}}_{r_{t}}\rmd r\tfrac{a(r)f(r)\,\qty(\frac{r}{L})^{2(d-1)}+a(r_t)\sqrt{-f(r_{t})\qty(\frac{r_t}{L})^{2(d-1)}}\left(P_v^\epsilon + \frac{\epsilon L \abs{K}}{d} \qty(\frac{r}{L})^{d} \right)}{f(r)\sqrt{-\mathcal{U}(P_v^\epsilon,\,r)}}~.\nonumber
\end{align}
The time derivative is easily identified as
\begin{align}\label{eq:first term derivative complexity}
   &{\dv{\mathcal{C}^\epsilon}{t}}=\frac{\Omega_{k,d-1}L^{d-2}}{G_N} \, a(r_{t})\sqrt{-f(r_{{t}})\qty(\frac{r_{{t}}}{L})^{2(d-1)}}\\
    &+\frac{2\Omega_{k,d-1}L^{d-2}}{G_N}\dv{P_v^\epsilon}{t}\tfrac{a(r)f(r)\,\qty(\frac{r}{L})^{2(d-1)}+a(r_t)\sqrt{-f(r_{t})\qty(\frac{r_t}{L})^{2(d-1)}}\left(P_v^\epsilon + \frac{\epsilon L \abs{K}}{d} \qty(\frac{r}{L})^{d} \right)}{f(r)\sqrt{-\mathcal{U}(P_v^\epsilon,\,r)}} ~.\nonumber
\end{align}
In the $t\rightarrow\infty$ regime we then recover:
\begin{equation}\label{eq:Vol late timesAdS}
    \lim_{t\rightarrow\infty}\dv{t} \mathcal{C}^\epsilon \simeq \frac{\Omega_{k,d-1}L^{d-2}}{G_N}\sqrt{-f(r_{f}) \qty(\frac{r_f}{L})^{2(d-1)}}\,a(r_{f})~\text{ with }r_{f}\equiv\lim_{t\rightarrow \infty}r_{t}~.
\end{equation}
Here $r_f$ is a local maximum of the effective potential at late times:
\begin{equation}\label{eq:extr potentialAdS}
    \eval{\mathcal{U}}_{r_{f}}=0,\quad \eval{\partial_{{r}}\mathcal{U}}_{r_{f}}=0,\quad \eval{\partial^2_{{r}}\mathcal{U}}_{r_{f}}\leq0~.
 \end{equation}
It is easy to see that the late time singularity in the denominator of the second term in (\ref{eq:first term derivative complexity}), given by the potential term $\eval{\mathcal{U}}_{r_{f}}$, is canceled by its numerator, which then shows (\ref{eq:Vol late timesAdS}). 

The algebraic conditions in (\ref{eq:extr potentialAdS}) lead to the following relation
\begin{equation}\label{eq:rt locationAdS}
\begin{aligned}
    0= 4 r_{f}f\left(r_{f}\right)\left((d-1)f'\left(r_{f}\right)+{K}_{\epsilon}^2 r_{f}\right)+4(d-1)^2f\left(r_{f}\right){}^2+r_{f}^2f'\left(r_{f}\right){}^2~.
\end{aligned}
\end{equation}
In particular, for the blackening factor in (\ref{eq:metric SAdS}) with $k=0$, one can always find roots to the above equation in the range $r_h \leq r_f < \infty$, such that \eqref{eq:Vol late timesAdS} reproduces a late time linear growth for $\mathcal{C}^\epsilon$, as required in the CAny proposal \cite{Belin:2021bga,Belin:2022xmt}.

\subsubsection{Switchback effect}
We will examine the complexity of the AdS black hole with alternating shockwaves in (\ref{eq:time insertions AdS}), which satisfy the following property \cite{Belin:2021bga,Belin:2022xmt}:
\begin{equation}\label{eq:total SW complexity}
\begin{aligned}
    {\mathcal{C}}^\epsilon(t_L,\,t_R)=&\mathcal{C}^\epsilon(t_R,\,V_1)+\mathcal{C}^\epsilon(V_1+\alpha,\,U_2)+\dots\\
    &+\mathcal{C}^\epsilon(U_{n-1}-\alpha_{n-1},\,V_n)+\mathcal{C}^\epsilon(V_{n}+\alpha_{n},\,t_L)~,
\end{aligned}
\end{equation}
where $\mathcal{C}^\epsilon(\cdot,\,\cdot)$ denotes the contributions from $\Sigma_\epsilon$ with two fixed endpoints and all endpoints are located either on the left/right horizon or asymptotic infinity.

\begin{figure}
    \centering
    \subfloat[]{\includegraphics[width=0.32\textwidth]{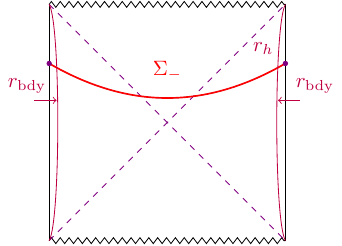}}\hfill\subfloat[]{\includegraphics[width=0.32\textwidth]{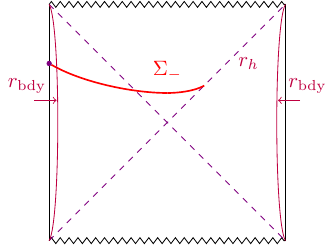}}\hfill\subfloat[]{\includegraphics[width=0.32\textwidth]{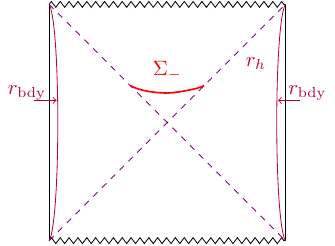}}
    \caption{Extremal complexity surfaces $\Sigma_\epsilon$ in (\ref{eq:total SW complexity}) for $\epsilon=-$ in a planar AdS black hole. (a) $\mathcal{C}^-(t_L,\,t_R)$; (b) $\mathcal{C}^-(V_R,~t_L)$; and (c) $\mathcal{C}^-(V_R,~U_L)$.}
    \label{fig:anchoringAdS}
\end{figure}

The different cases are illustrated in Fig. \ref{fig:anchoringAdS}.
To study the complexity growth evolution in the perturbed geometry, we must find the location $u_{R,\,L}$, $v_{R,\,L}$ where $\Sigma_\epsilon$ intersects with the left/right horizon $r_h$. We derive that
\begin{equation}
    v_R-v_t=\int_{r_t}^{r_h}\frac{\rmd r}{f(r)}\qty(1-\frac{P_v^\epsilon+\frac{\epsilon L\abs{K}}{d}\qty(\frac{r}{L})^d}{\sqrt{-\mathcal{U}(P_v^\epsilon,\,r)}})~,
\end{equation}
where we denote $v_t=v_R(r_t)$.

The different contributions in (\ref{eq:multiple SW}) can be then expressed with EF coordinates (\ref{eq:generic_null_metric_noshock}) as:
\begin{align}
    &\mathcal{C}^\epsilon(t_R,\,V_L)=\mathcal{C}^\epsilon(V_R,\,t_L)\\
    &=-\frac{\Omega_{k,d-1}L^{d-2}}{G_N}a(r_t)\sqrt{-f(r_t)\qty(\frac{r_t}{L})^{2(d-1)}}\qty(\int^{r_{\rm bdy}}_{r_t}+\int^{r_h}_{r_t})\frac{\left(P_v^\epsilon + \frac{\epsilon L\abs{K}}{d} \qty(\frac{r}{L})^{d} \right)\rmd r}{f(r)\sqrt{-\mathcal{U}(P_v^\epsilon, r)}}~,\nonumber\\
    &\mathcal{C}^\epsilon(V_R,\,U_L)=-\frac{2\Omega_{k,d-1}L^{d-2}}{G_N}a(r_t)\sqrt{-f(r_t)\qty(\frac{r_t}{L})^{2(d-1)}}\int^{r_h}_{r_t}\frac{\left(P_v^\epsilon + \frac{\epsilon L\abs{K}}{d} \qty(\frac{r}{L})^{d} \right)\rmd r}{f(r)\sqrt{-\mathcal{U}(P_v^\epsilon, r)}}~.\label{eq:Cepsilon VRUL}
\end{align}
Using the result in (\ref{eq:C epsilon inter}), one can express (\ref{eq:Cepsilon VRUL}) as:
\begin{align}
    {\mathcal{C}^\epsilon}&(V_R,\,U_L)=-\frac{2\Omega_{k,d-1}L^{d-2}}{G_N}a(r_t)\sqrt{-f(r_t)\qty(\frac{r_t}{L})^{2(d-1)}}\int^{r_h}_{r_t}\frac{\left(P_v^\epsilon + \frac{\epsilon L\abs{K}}{d} \qty(\frac{r}{L})^{d} \right)\rmd r}{f(r)\sqrt{-\mathcal{U}(P_v^\epsilon, r)}}\nonumber\\
    &+\frac{2\Omega_{k,d-1}L^{d-1}}{G_N}\int^{r_h}_{r_t}\rmd r\frac{a(r)f(r)\,r^{2(d-1)}+a(r_t)\sqrt{-f(r_{t})r_{t}^{2(d-1)}}\left(P_v^\epsilon + \frac{\epsilon L\abs{K}}{d} \qty(\frac{r}{L})^{d} \right)}{f(r)\sqrt{-\mathcal{U}(P_v^\epsilon,\,r)}}~.\label{eq:Late expressionCepsilon VRUL}
\end{align}
To proceed with the evaluation, we perform an expansion around the final slice where (\ref{eq:extr potentialAdS}) allows us to approximate
\begin{equation}\label{eq: Approx Potential}
\lim_{r\rightarrow r_f}U(P_v^\epsilon,\,r)\simeq \tfrac{1}{2}(r-r_f)^2\mathcal{U}''(P_v^\epsilon,\,r)+\mathcal{O}(\abs{r-r_f}^3)~.
\end{equation}
Then, the CAny proposal (\ref{eq:Late expressionCepsilon VRUL}) near the final turning point $r_f$ can be computed as follows
\begin{equation}\label{eq:P infty epsilon}
\begin{aligned}
{\mathcal{C}^\epsilon}(V_R,\,U_L)=\frac{\Omega_{k,d-1}L^{d-2}}{G_N}P_\infty^\epsilon v~,\quad P_\infty^\epsilon=a(r_f)\sqrt{-f(r_f)\qty(\frac{r_f}{L})^{2(d-1)}}~,
\end{aligned}
\end{equation}
where $r_f$ is a root of the function in (\ref{eq:rt locationAdS}).

The result above can be repeated to evaluate all the contributions in (\ref{eq:total SW complexity}) as:
\begin{align}
{\mathcal{C}^\epsilon}(V_R,~t_L)&=\frac{\Omega_{k,d-1}L^{d-1}}{G_N}P^\epsilon_{\infty}\log{\rme^{t_L/L}V_R}~,\\
{\mathcal{C}^\epsilon}(V_R,~U_L)&=\frac{\Omega_{k,d-1}L^{d-1}}{G_N}P_{\infty}^\epsilon\log{U_LV_R}~,\\
{\mathcal{C}^\epsilon}(t_R,~V_L)&=\frac{\Omega_{k,d-1}L^{d-1}}{G_N}P_{\infty}^\epsilon\log{V_L\rme^{t_R/L}}~.
\end{align}
However, there will also be an early time contribution in the shockwave geometry, given by the term
\begin{equation}
{\mathcal{C}^\epsilon}(U_R,V_L)=\frac{\Omega_{k,d-1}L^{d-1}}{G_N}P_{-\infty}^\epsilon\log{U_LV_R}~,
\end{equation}
where
\begin{equation}\label{eq:P -infty epsilon}
    P_{-\infty}^\epsilon=\lim_{t\rightarrow-\infty}a(r_I)\sqrt{-f(r_I)\qty(\frac{r_I}{L})^{2(d-1)}}~,
\end{equation}
and $r_I=\lim_{t\rightarrow-\infty}r_t$, for which there is a sign flip in $K_\epsilon\rightarrow-K_\epsilon$. This means that $r_I$ is a solution to (\ref{eq:extr potentialAdS}, \ref{eq:rt locationAdS}) with the appropriate modification of $K$. (\ref{eq:total SW complexity}) can be then expressed as
\begin{equation}\label{eq:Cepsilon totAdS}
\begin{aligned}
{\mathcal{C}^\epsilon}(t_L,\,t_R)\simeq \tfrac{\Omega_{k,d-1}L^{d-1}}{G_N}\Bigl[& P_{\infty}^\epsilon \log(V_1\rme^{t_R/L})+P_{-\infty}^\epsilon\log\qty((V_1+\alpha)U_2)\\
&+\dots+P_{\infty}^\epsilon\log((V_n+\alpha_n)\rme^{t_L/L})\Bigr]~.
\end{aligned}
\end{equation}
Extremizing (\ref{eq:Cepsilon totAdS}) with respect to an arbitrary interception point ($V_i$, $U_i$) in the multiple shockwave geometry,
\begin{equation}\label{eq:cond SW1}
\dv{{\mathcal{C}^\epsilon}(t_L,\,t_R)}{V_i}=0~,\quad \dv{{\mathcal{C}^\epsilon}(t_L,\,t_R)}{U_i}=0~,
\end{equation}
allows us to locate
\begin{align}\label{eq:cond SW2}
U_i=\frac{\alpha_i P_{-\infty}^\epsilon}{P_{-\infty}^\epsilon+P_{-\infty}^\epsilon}~,\quad V_i=-\frac{\alpha_iP_{\infty}^\epsilon}{P_{\infty}^\epsilon+P_{-\infty}^\epsilon}~.
\end{align}
Replacing the interception points into (\ref{eq:Cepsilon totAdS}) generates:
\begin{equation}\label{eq:Intermediate switchbackAdS}
{\mathcal{C}^\epsilon}(t_L,\,t_R)\simeq\frac{\Omega_{k,d-1}L^{d-2}}{G_N}\qty(P_{\infty}^\epsilon (t_R+t_L)+(P_{\infty}^\epsilon+P_{-\infty}^\epsilon)\qty(\sum_{k=1}^nt_k-nt^{(b)}_*))~,
\end{equation}
where $t^{(b)}_*$ is defined in (\ref{eq:scrambling BH}), and we have discarded subleading constant terms in $P_{\pm\infty}^\epsilon$ since we are considering the regime (\ref{eq:time insertions AdS}).

Notice that if and only if $P_{\infty}^\epsilon=P_{-\infty}^\epsilon$, we reproduce the switchback effect property:
\begin{equation}\label{eq:switchback resultAdS}
    \mathcal{C}^\epsilon\propto\abs{t_R+t_1}+\abs{t_2-t_1}+\dots+\abs{t_n-t_L}-2nt^{(b)}_*~.
\end{equation}
The original CAny proposals \cite{Belin:2021bga,Belin:2022xmt} were originally defined for the particular case $k=0$ in the metric (\ref{eq:metric SAdS}). we notice that (\ref{eq:Volepsilon}, \ref{eq:C CMC}) do not obey the constraint $P_{\infty}^\epsilon=P_{-\infty}^\epsilon$ for generic $F[\dots]$. However, this is satisfied in the special case where $F[\dots]=1$, where holographic complexity would correspond to the volume of CMC slices. 

In Sec. \ref{sec:CAny CMC}, we will propose a protocol where $P_{\infty}^\epsilon=P_{-\infty}^\epsilon$ can be satisfied for an arbitrary scalar functional $F[\dots]$ in both asymptotically AdS and dS spacetimes, and discuss more broadly its implications for the CAny proposals.

\cleardoublepage

\part{Holographic complexity in de Sitter space}\label{part:2}
  \graphicspath{{\chaptersdir/ComplexitydS/image/}}%
  \chapter{Holographic complexity in asymptotically de Sitter space}\label{ch:ComplexitydS}

This chapter is mostly a reprint of \cite{Aguilar-Gutierrez:2023zqm,Aguilar-Gutierrez:2023pnn,Aguilar-Gutierrez:2024rka}. We compute several holographic complexity conjectures (the CV, CA, CV2.0, and codimension-one CAny proposals) in the extended SdS black hole.
We consider multiple configurations of the stretched horizons to which geometric objects are anchored.
The holographic complexity proposals admit a hyperfast growth when the gravitational observables only lie in the cosmological patch, except for a class of CAny observables that admit a linear growth, and exhibit a switchback effect. 
All the complexity conjectures present a linear increase when restricted to the black hole patch, similar to the AdS case.
When both the black hole and the cosmological regions are probed, 
codimension-zero proposals are time-independent, while the codimension-one proposals can have non-trivial evolution with a linear increase at late times.
As a byproduct of our analysis, we find that codimension-one spacelike surfaces are highly constrained in SdS space. 
Therefore, different locations of the stretched horizon give rise to different behaviors of the complexity conjectures.

\subsection*{Motivations and new observables.}

\noindent As anticipated in Ch. \ref{ch:introduction}, the central dogma gives great importance to the region located outside the black hole and inside the cosmological horizons.
An observer in such a region may have the possibility to describe both the interior of a black hole and the exterior of a cosmological horizon of the SdS spacetime. 
So far, the explorations of holographic complexity in the SdS background (including of course pure dS space) only focused on the inflating patch (the region outside the cosmological horizon $r_c$ in Fig.~\ref{fig:extSdS}) \cite{Baiguera:2023tpt,Aguilar-Gutierrez:2023zqm,Aguilar-Gutierrez:2023pnn}.
Instead, no investigations have focused on the black hole patch (the region inside the black hole horizon $r_h$ in Fig.~\ref{fig:extSdS}).
In this chapter, we study the holographic complexity conjectures for several different choices of the stretched horizons to which the extremal surfaces of interest can be anchored. From the lenses of dS holography, this procedure might allow us to predict the behavior of new observables for the putative dual quantum theory. The background of interest is the periodic extension of the SdS black hole with $n$ copies (that we will denote as SdS$^n$), see Fig.~\ref{fig:extSdS}. 
{This setting allows us to generalize the previous studies, and in particular, the holographic complexity in dS space is reproduced when we restrict the complexity observables to probe only a single of the inflating patches in the geometry, as we will examine in detail in this chapter.}

The possible choices for the timelike surfaces to which geometric observables can be anchored are partially motivated by the proposal of holographic screens by Bousso \cite{Bousso:1999cb,Bousso:1999dw}, which has been previously used to compute holographic complexity in FRLW cosmologies \cite{Caginalp:2019fyt}. The hyperfast and linear growth of complexity observables in the inflating patch at late times in this geometry is a consequence of whether the extremal surfaces (where complexity is evaluated) reach timelike infinity $\mathcal{I}^{\pm}$ (e.g. see Figs. \ref{fig:WDW_evo1}, \ref{fig:WDW_evo2}, \ref{subfig:extremal_surfaces_CV_case1}), in which case 
\begin{align}\label{eq:hyperfast}
\lim_{t\rightarrow t_c}\dv{\mathcal{C}}{t}\rightarrow\infty    
\end{align}
where $\mathcal{C}$ represents the holographic complexity observable computed for a given proposal, and $t_c$ is a critical (stretched horizon) time.

Our motivations for studying complexity in the extended SdS space are two-fold. First, this type of spacetime has been used to construct multiverse models. One instance is in the context of dS space proliferation \cite{Bousso:1998bn,Bousso:1999ms}, where a near-extremal extended Nariai black hole can lead to different disconnected universes due to quantum effects. Other toy models of eternal inflation use the extended SdS background to study the von Neumann entropy for a subregion of $\mathcal{I}^+$ in lower-dimensional quantum gravity \cite{Aguilar-Gutierrez:2021bns,Shaghoulian:2022fop} ({see \cite{Langhoff:2021uct} for a higher dimensional implementation)}, which has also motivated certain recent braneworld multiverse models \cite{Yadav:2023qfg,Aguilar-Gutierrez:2023zoi}. 
In the quantum cosmology context, the typical setting involves a meta-observer living on a spacelike surface (see the horizontal cyan lines in Fig.~\ref{fig:extSdS}), from which they are able to collect coarse-grained information on the history of the spacetime. In these latter approaches, the extended SdS black hole has proven to be a useful tool (\eg see \cite{Hartle:2016tpo}).
Second, we want to accumulate evidence for universal and non-universal features of the complexity conjectures in more general asymptotically dS geometries, an effort initiated in \cite{Aguilar-Gutierrez:2024xfi}.

\subsection*{Main results}

\begin{figure}[t!]
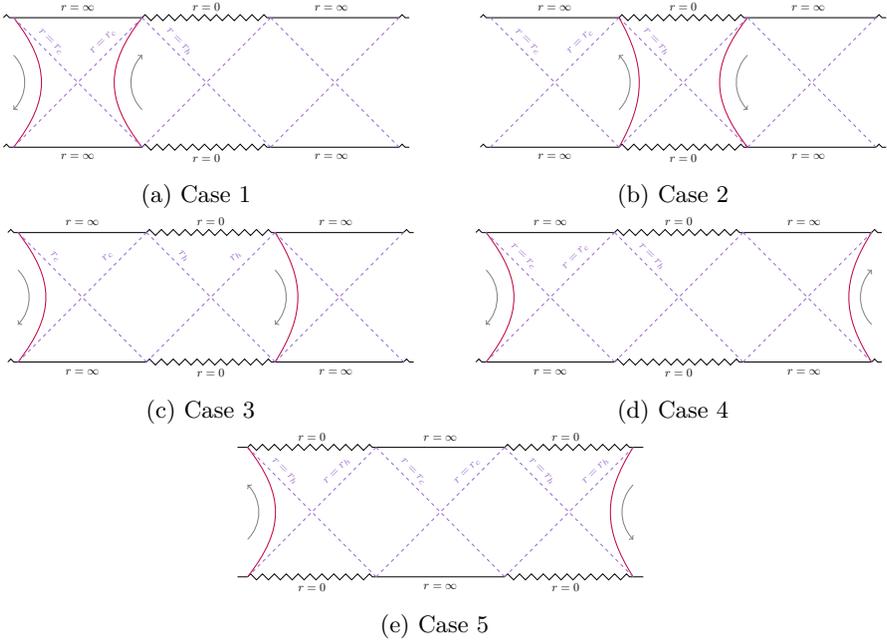

    \centering
    \subfloat[Case 1]{ \label{subfig:case1} \includegraphics[width=0.45\textwidth]{Figures/FigCase1.pdf}}
\qquad \subfloat[Case 2]{ \label{subfig:case2} \includegraphics[width=0.45\textwidth]{Figures/FigCase2}} \\
 \subfloat[Case 3]{ \label{subfig:case3} \includegraphics[width=0.45\textwidth]{Figures/FigCase3.pdf}}
\qquad \subfloat[Case 4]{\label{subfig:case4} \includegraphics[width=0.45\textwidth]{Figures/FigCase4}}\\
\subfloat[Case 5]{\label{subfig:case5} \includegraphics[width=0.45\textwidth]{Figures/FigCase5.pdf}}
    \caption{Summary of the possible configurations for the stretched horizons. Holographic observables are located between the stretched horizons.  }
    \label{fig:summary_cases}
\end{figure}

\noindent The extended SdS background allows for several possible configurations of the stretched horizons, which are summarized in Fig.~\ref{fig:summary_cases} (the details will be given in Sec. \ref{ssec:stretched_horizons}).
The setting denoted as case 1 contains a single copy of the SdS geometry where the geometric quantities only probe the cosmological patch. This is the kind of holographic complexity already studied in the literature in empty dS space for the CV, CA, and CV2.0 proposals \cite{Jorstad:2022mls}, which we extend to the case of a SdS black hole, and we propose a CAny measure of complexity that has a very different evolution with respect to the other proposals.
The configuration in case 2 focuses on the black hole region only, since it probes the interior of the corresponding event horizon.
This is similar to the investigations of holographic complexity performed in asymptotically AdS geometries.
The main novelty of this work is provided by cases 3-5, where gravitational observables probe both the interior of the black hole and the exterior of the cosmological horizon. This is possible when the stretched horizons belong to static patches in non-consecutive regions of the Penrose diagram.

\begin{table}[t!]   
\begin{center}    
\begin{tabular}  {|c|c|c|} \hline  &  \textbf{Case 1 (Fig.~\ref{subfig:case1})} & \textbf{Case 2 (Fig.~\ref{subfig:case2})}  \\ \hline
\rule{0pt}{4.9ex} $\mathcal{C}_{V2.0}$  & Hyperfast \eqref{eq:rateCV20_case1}  & Linear \eqref{eq:late_times_CV20_case2}  \\
\rule{0pt}{4.9ex} $\mathcal{C}_A$ &  Hyperfast \eqref{eq:approx_rate_CA_case1} & Linear \eqref{eq:rate_CA_case2_latelimit}  \\ 
 \rule{0pt}{4.9ex}  $\mathcal{C}_V$ &  Hyperfast \eqref{eq:ratelate_CV_case1}  & Linear \eqref{eq:ratelate_CV_case2}  \\ 
 \rule{0pt}{4.9ex} $\mathcal{C}^{\epsilon} $ &   
\eqref{eq:Vol late times} 
$ \begin{cases}
    \mathrm{Hyperfast} & \mathrm{if} \,\, |K_{\epsilon}| < K_{\rm crit} \\ 
     \mathrm{Linear} & \mathrm{if} \,\, |K_{\epsilon}| \geq K_{\rm crit}
\end{cases} $
 &  Linear \eqref{eq:Vol late times}  \\[0.2cm]
\hline
\end{tabular}   
\vskip 3mm
\begin{tabular}  {|c|c|c|} \hline  & \textbf{Case 3} (Fig.~\ref{subfig:case3}) & \textbf{Cases 4-5 (Fig.~\ref{subfig:case4}--\ref{subfig:case5})} \\ \hline
\rule{0pt}{4.9ex} $\mathcal{C}_{V2.0}$   &  Time-independent \eqref{eq:rate_CV20_case3} & Time-independent  \\
\rule{0pt}{4.9ex} $\mathcal{C}_A$ &Time-independent \eqref{eq:rate_CA_case3} & Time-independent  \\ 
 \rule{0pt}{4.9ex}  $\mathcal{C}_V$ & Linear  \eqref{CV_latetimes_case3}  &  Time-independent \\ 
 \rule{0pt}{4.9ex} $\mathcal{C}^{\epsilon} $ 
 &  Linear \eqref{eq:t inft dCdt CAny3}  &   Time-independent  \\[0.2cm]
\hline
\end{tabular}   
\caption{Late-time behavior of the holographic complexity conjectures.
$\mathcal{C}^{\epsilon}$ is a class of codimension-one observables among the CAny proposals, defined on CMC slices with extrinsic curvature $K_{\epsilon}$ ($K_{\rm crit}$ denotes a critical value).} 
\label{tab:resultsSdSn}
\end{center}
\end{table}

We anticipate the main results in table~\ref{tab:resultsSdSn}, which collects the late-time behavior of the complexity conjectures.\footnote{For visual convenience, we have split table~\ref{tab:resultsSdSn} in two parts. } 
The first set of comments comes from reading the table by columns, which allows us to scan for universal and distinguishing properties of the holographic proposals:
\begin{itemize}
    \item The first column (case 1) shows that hyperfast growth is a feature of the holographic complexity proposals evaluated in a single inflating patch of the SdS background. This generalizes previous results in empty dS space \cite{Jorstad:2022mls}. The only exception is provided by the observable $\mathcal{C}^\epsilon$, belonging to a class of codimension-one CAny observables evaluated on CMC slices, which present an eternal evolution with linear growth at late times whenever their extrinsic curvature is greater than a critical value \cite{Aguilar-Gutierrez:2023zoi}.
    \item The second column (case 2) tells us that the growth rate of complexity at late times is universally linear whenever the geometric observables are located inside the black hole patch. This is consistent with the behavior of black holes in asymptotically AdS space \cite{Carmi:2017jqz}.
    \item  From the column referring to case 3, we read that all the geometric observables extending across a single black hole and one cosmological patch do \textit{not} present hyperfast growth. However, there are some differences. The codimension-zero proposals are always time-independent, as a consequence of the WDW patch reaching timelike infinities $\mathcal{I}^{\pm}$ and the singularity at all times. 
    On the other hand, codimension-one conjectures evolve forever, approaching a linear rate at late times. 
    \item The last column (cases 4-5) refers to configurations with multiple copies of the SdS geometry, including both cosmological and black hole patches. In this setting, we find a universal behavior: all the holographic proposals are time-independent. 
    For codimension-zero observable, this is a consequence of the WDW patch reaching timelike infinities $\mathcal{I}^{\pm}$ and the singularity at all times. For codimension-one quantities, we find certain constraints that only allow for the trivial evolution generated by the time isometry of the background.  
\end{itemize}

The second interpretation of table~\ref{tab:resultsSdSn} comes from reading it by rows, to determine the common features of the various configurations of the stretched horizons:
\begin{itemize}
\item The first two lines show that codimension-zero proposals present the same features, which drastically change according to the location of the stretched horizons.
In particular, complexity is time-independent in cases 3-5.
 \item The codimension-one proposals also present similar features, except for the two possibilities mentioned above for $\mathcal{C}^{\epsilon}$ in case 1. 
 When there exists at least one inflating and one black hole patch (cases 3-5), we will show that any configuration with symmetric boundary times $t$ is forbidden, except for a maximal surface only defined at a unique $t$.
 While this leads to a time-independent volume in cases 4-5, instead the isometries in case 3 allow for a non-trivial evolution that approaches linear growth at late times.
\end{itemize}

In summary, our study shows that the holographic complexity of a black hole in asymptotically dS space presents different features depending on the location of the stretched horizons: one can find hyperfast, linear, or vanishing growth.
In case 3, which probes one cosmological and one black hole patch, there is a time regime such that CA is time-independent, while CV is not.
This also happens in asymptotically AdS black holes \cite{Carmi:2017jqz}. 
Furthermore, let us mention that a completely different behavior of CV and CA is also reminiscent of the fact that defects and boundaries show a different structure of UV divergences when evaluating CA and CV proposals \cite{Chapman:2018bqj,Sato:2019kik,Braccia:2019xxi,Auzzi:2021nrj,Baiguera:2021cba,Auzzi:2021ozb}.
The novelty is that the different behavior arises in a dynamical situation, in asymptotically dS space, and at any boundary time, including late times.

\subsection*{Outline.}

\noindent This chapter is organized as follows.
In Sec.~\ref{sec:Cod0 complexity} we study the time evolution of the WDW patch and of the codimension-zero holographic proposals (CV2.0 and CA). In Sec.~\ref{sec:Cod1 complexity} we similarly proceed with the analysis of the codimension-one proposals, including the CV and CAny proposals evaluated on CMC slices. We then conclude in Sec.~\ref{sec:ConclusionsSdSn} with a summary of our findings, and some relevant future directions.
App.~\ref{app:details_CA} contains some technical details on the computation of CA.

\section{Codimension-zero proposals}

\label{sec:Cod0 complexity}

We begin our journey across the holographic proposals in extended SdS geometry by considering the codimension-zero observables.
After defining the main properties of the WDW patch in Sec \ref{sec:WDW patch}, we explore in Secs. \ref{ssec:CV20} and \ref{ssec:CA} the time evolution of CV2.0 and CA conjectures, respectively.

\subsection{The Wheeler-DeWitt patch}\label{sec:WDW patch}
We build the WDW patch, \ie the bulk domain of dependence of a spacelike surface anchored at the stretched horizons, according to the list of proposals presented in Sec. \ref{ssec:stretched_horizons}.
In order to avoid divergences of the spacetime volume or of the gravitational action when approaching timelike infinities $\mathcal{I}^{\pm}$, we introduce a cutoff surface located at $r=r_{\rm max}$.
Although the region close to the black hole singularity does not bring any divergence, we will also introduce by analogy a cutoff surface located at $r=r_{\rm min}$.
At the end of the computation, we will send $r_{\rm max} \rightarrow \infty$ and $r_{\rm min} \rightarrow 0$. 

In all the cases described below, the WDW patch is conveniently described by the null coordinates introduced in \eqref{eq:general_null_coordinates}.
In particular, the null boundaries delimiting the shape of the WDW patch are identified by the conditions that either the $u$ or $v$ coordinate is constant, as we will specify case by case below.

\subsubsection*{Case 1}

\begin{figure}[t!]
\centering
\includegraphics[width=0.6\textwidth]{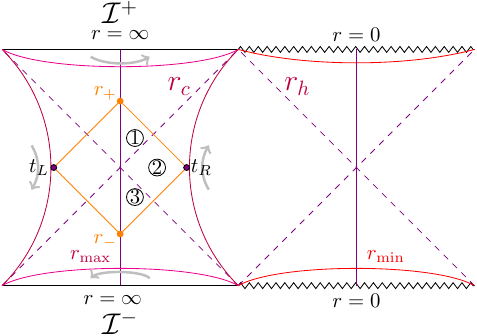}
\caption{General configuration of the WDW patch in case 1. Pink and red curves near $\mathcal{I}^\pm$ and the singularities indicate cutoff surfaces $r_{\rm max}$ for the cosmological and $r_{\rm min}$ for the black hole patches, respectively. The boundaries of the WDW patch are denoted in orange.
The vertical purple line denotes a locus of constant $t=0$.
The gray arrows indicate the orientation of the Killing vector $\partial_t$, and the circled numbers indicate the regions of integration in (\ref{eq:decomposition_CV20_general}).
}
\label{fig:WDW_cosmo_stretched}
\end{figure}

The time dependence of the WDW patch was studied in various dimensions for empty dS space (with and without perturbations) \cite{Jorstad:2022mls,Anegawa:2023wrk,Anegawa:2023dad} and for the SdS black hole in the presence of a shockwave \cite{Baiguera:2023tpt}.
We depict the relevant geometric configuration in Fig.~\ref{fig:WDW_cosmo_stretched}, focusing on the case with symmetric time evolution \eqref{eq:symmetric_times} and symmetric location of the stretched horizons, \ie $r_{\rm st} \equiv r_{\rm st}^L = r_{\rm st}^R$.

The top and bottom joints $r_{\pm}$ of the WDW patch are related to the position of the stretched horizons as
\beq
u_{\rm max} = \frac{t}{2} - r^*(r_{\rm st}) = -r^*(r_+) \, , \qquad
v_{\rm min} = \frac{t}{2} + r^*(r_{\rm st}) = r^*(r_-) \, , \qquad
r_{\pm} \geq r_c \, ,
\label{eq:umax_vmin_WDWpatch}
\eeq
where we used the symmetry of the configuration to infer that the joints are located along the vertical line at $t=0$ in the Penrose diagram.
The time dependence of the joints can be computed by taking the time derivative of the previous identities, leading to
\beq
\frac{d r_{+}}{dt} = - \frac{1}{2} f(r_{+}) \, , \qquad
\frac{d r_{-}}{dt} = \frac{1}{2} f(r_{-}) \, .
\label{eq:rates_rprmcase1}
\eeq
Due to the isometries of the spacetime, the time evolution is symmetric with respect to the boundary time $t=0.$
The critical time $t_{\infty}$ (and its opposite $-t_{\infty}$) is identified by the condition that the top (bottom) joint of the WDW patch reaches future (past) timelike infinity.
More precisely, the critical time $t_{c1}$ when the top joint crosses the future cutoff surface is given by
\beq
t_{c1} = 2 \le  r^*(r_{\rm st}) - r^*(r_{\rm max}) \ri \, , \qquad
t_{\infty} = \lim_{r_{\rm max} \rightarrow \infty} t_{c1} \, .
\label{eq:critical_time_tc1}
\eeq  

\subsubsection*{Case 2}
\begin{figure}[t!]
\centering
\includegraphics[width=0.6\textwidth]{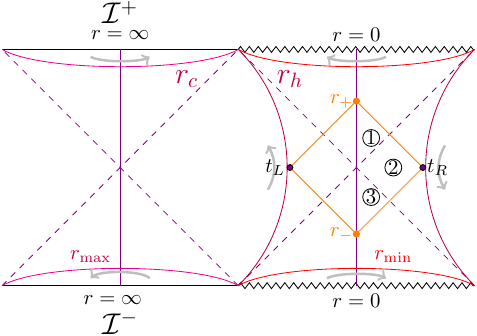}
\caption{General configuration of the WDW patch in case 2. The same coloring scheme as Fig. \ref{fig:WDW_cosmo_stretched} is used, with the circled regions indicating the evaluation of (\ref{eq:CV20_case2_int_regime}).}
\label{fig:WDW_cosmo_stretched_case2}
\end{figure}

This setting, depicted in Fig.~\ref{fig:WDW_cosmo_stretched_case2}, has several analogies with computations of holographic complexity performed in asymptotically AdS spacetimes (see Sec. \ref{sec:WdW AdS}, and \cite{Chapman:2016hwi,Carmi:2017jqz}) because the WDW patch covers a region behind the black hole horizon.
The main differences with these standard calculations are the specific blackening factor and the presence of the stretched horizons, which cut part of the geometry and forbid the WDW patch to reach any region close to $r=\infty$.
As a consequence, the corresponding holographic complexity will not be UV divergent, nevertheless, the rate of growth presents similar features compared to the AdS case.

Next, we assume without loss of generality that the boundary times satisfy \eqref{eq:symmetric_times} and that the stretched horizons are symmetric, \ie $r_{\rm st} \equiv r_{\rm st}^L = r_{\rm st}^R$.
Similar to (\ref{eq:umax_vmin_WDWpatch_caseAdS}), we define the special positions of the WDW patch as 
\beq
u_{\rm min} = -\frac{t}{2} - r^*(r_{\rm st}) = -r^*(r_+) \, , \qquad
v_{\rm max} = -\frac{t}{2} + r^*(r_{\rm st}) = r^*(r_-) \, , \qquad
r_{\pm} \leq r_h \, .
\label{eq:umax_vmin_WDWpatch_case2}
\eeq
The difference with \eqref{eq:umax_vmin_WDWpatch} is that $r_{\pm}$ lie in a different region, and the Killing vector corresponding to the time directions flows in the opposite way along the stretched horizons.
As a consequence, the derivatives of the joints now satisfy the different relations
\beq
\frac{d r_{+}}{dt} =  \frac{1}{2} f(r_{+}) \, , \qquad
\frac{d r_{-}}{dt} = - \frac{1}{2} f(r_{-}) \, .
\label{eq:rates_rprmcase2}
\eeq
Similar to (\ref{eq:critical_time_tc2}), one derives a critical time $t_{c2}$ when the top joint crosses the future cutoff surface in the black hole region, \ie
\beq
t_{c2} = 2 \le  r^*(r_{\rm min}) - r^*(r_{\rm st})  \ri \, , \qquad
t_{0} = \lim_{r_{\rm min} \rightarrow 0} t_{c2} \, .
\label{eq:critical_time_tc2}
\eeq  

\subsubsection*{Case 3}

\begin{figure}[t!]
\centering
\includegraphics[width=0.6\textwidth]{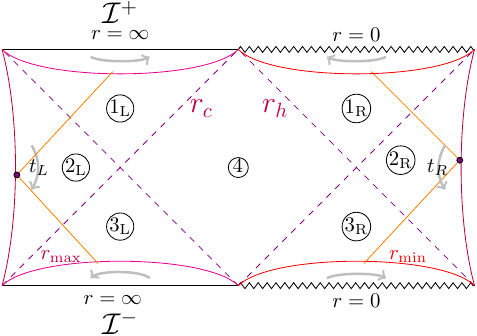}
\caption{General configuration of the WDW patch in case 3. The same coloring scheme as Fig. \ref{fig:WDW_cosmo_stretched} is used, with the circled regions indicating the evaluation of (\ref{eq:Case3_CV20}).} 
\label{fig:WDW_BH_stretched}
\end{figure}

Let us now consider the setting depicted in Fig.~\ref{fig:WDW_BH_stretched}.
We will keep $t_L, t_R$ and the locations of the stretched horizons $r_{\rm st}^L, r_{\rm st}^R$ arbitrary in the following considerations.
The joints of the WDW patch are always located behind the cutoff surfaces (either $r_{\rm max}$ or $r_{\rm min}$), even in the limit where they approach timelike infinity $\mathcal{I}^{\pm}$ and the singularity.
Regarding the time evolution of the diagram, it should be noted that the Killing vector $\partial_t$ is downward-oriented along both stretched horizons, implying that the time coordinate increases towards the bottom of the Penrose diagram.
Therefore, from the perspective of a dual boundary theory, we should denote the times on the stretched horizons as $t_{\rm bulk}= -t_L$ and $t_{\rm bulk} = -t_R$.

\subsubsection*{Cases 4-5}

\begin{figure}[t!]
\centering
\subfloat[]{\label{fig:WDW_case4} \includegraphics[width=0.9\textwidth]{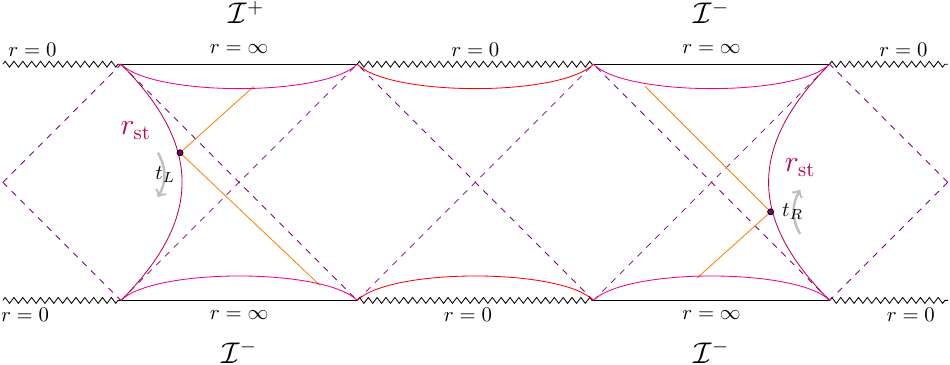}}\vspace{0.5cm}\\ 
\subfloat[]{\label{fig:WDW_case5}\includegraphics[width=0.9\textwidth]{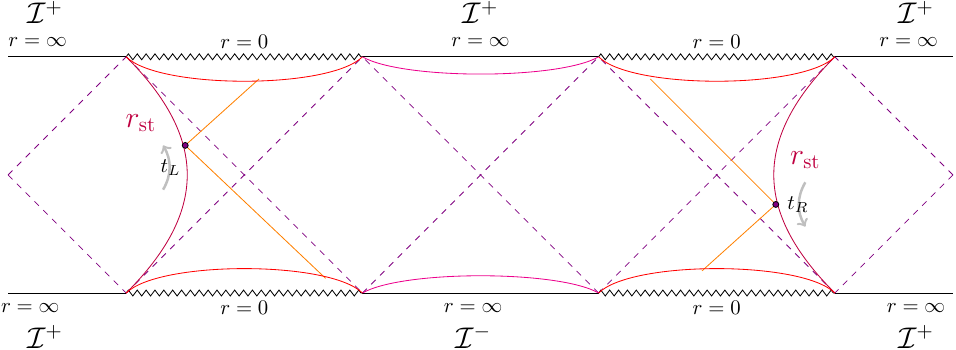}}
\caption{General configuration of the WDW patch in cases 4 (a) and 5 (b).}
\label{fig:WDW_case_45}
\end{figure}

The relevant shapes of the WDW patch in these cases are depicted in Fig.~\ref{fig:WDW_case_45}.
We observe that the diagrams are similar to case 3 in Fig.~\ref{fig:WDW_BH_stretched}, except that there is an additional inflating patch (case 4) or black hole patch (case 5).
The main feature of these configurations is that the top and bottom joints of the WDW patch always lie behind timelike infinities or singularities, independently of the precise location of the stretched horizons and of the value of the boundary times.

\subsection{Complexity=volume 2.0}
\label{ssec:CV20}

Let us analyze the time dependence of the quantity \eqref{eq:general_CV20} in the settings defined in Sec. \ref{ssec:stretched_horizons}.
We will use the split in the evaluation of CV2.0 \eqref{eq:general_CV20} shown in (\ref{eq:decomposition_CV20_general}) for each subregion.

\subsubsection*{Case 1}

The shape of the WDW patch during the time evolution is determined by the sign of the critical time $t_{\infty}$ defined in \eqref{eq:critical_time_tc1}.
One can check that by increasing the value of $\rho$ defined in \eqref{eq:stretched_horizon_SdS4} along the interval $[0,1]$, there is a transition from negative to positive values of $t_{\infty}$.
Since static patch holography proposes that the stretched horizon should be located close to the cosmological one ($\rho \rightarrow 1$), we will focus on this case and assume $t_{\infty} >0$. 
We will comment on the opposite scenario at the end of this paragraph.

The configuration of the WDW patch represented in Fig.~\ref{fig:WDW_cosmo_stretched} is the prototype for investigations of CV2.0 in this regime.
According to the decomposition into subregions reported in the picture, we compute the quantities entering \eqref{eq:decomposition_CV20_general} as
\begin{subequations}
\label{eq:CV20_case1_int_regime}
\beq
\mathcal{J}_{1L}=\mathcal{J}_{1R} = \frac{\Omega_{d-1}}{G_{N} L^2} \int_{r_c}^{r_+} dr  \,  r^{d-1} \le  \frac{t}{2} +r^*(r) -r^*(r_{\rm st})  \ri \, ,
\label{eq:CV20_case1_int_regime_p1}
\eeq
\beq
\mathcal{J}_{2L} =
\mathcal{J}_{2R}=
\frac{\Omega_{d-1}}{G_{N} L^2} \int_{r_{\rm st}}^{r_c} dr  \,  r^{d-1} \le  2 r^*(r) - 2 r^*(r_{\rm st})  \ri \, , 
\label{eq:CV20_case1_int_regime_p2}
\eeq
\beq
\mathcal{J}_{3L} =
\mathcal{J}_{3R} =  \frac{\Omega_{d-1}}{G_{N} L^2} \int_{r_c}^{r_-} dr  \,  r^{d-1} \le - \frac{t}{2} +r^*(r) - r^*(r_{\rm st})  \ri \, ,
\label{eq:CV20_case1_int_regime_p3}
\eeq
\end{subequations}
where $r_{\pm}$ were implicitly defined in \eqref{eq:umax_vmin_WDWpatch}.
The total complexity is obtained by combining the previous terms according to \eqref{eq:decomposition_CV20_general}.

\begin{figure}[t!]
    \centering
    \subfloat[]{
\includegraphics[width=0.3\textwidth]{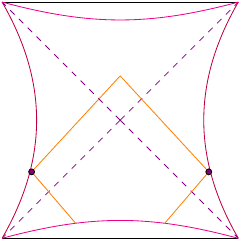} \label{subfig:WDW_evo1_regime1}}\hfill \subfloat[]{\label{subfig:WDW_evo1_regime2} \includegraphics[width=0.3\textwidth]{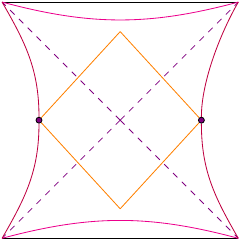}}\hfill \subfloat[]{ \label{subfig:WDW_evo1_regime3}\includegraphics[width=0.3\textwidth]{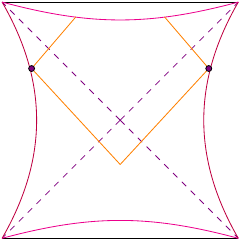}}
    \caption{Possible time evolution of the WDW patch in case 1 (and similarly for case 2). In the picture, we only report the part of the Penrose diagram included between the stretched horizons. In case 1 (2), the purple dashed lines represent the cosmological (black hole) horizon, while the magenta cutoff surfaces are located at $r=r_{\rm max}$ ($r=r_{\rm min}$). }
    \label{fig:WDW_evo1}
\end{figure}

The evolution of the WDW patch is composed of the three regimes summarized in Fig.~\ref{fig:WDW_evo1}.
At early times $t < -t_{\infty}$, the bottom joint sits behind past timelike infinity $\mathcal{I}^-$, while the top joint is located outside the cosmological horizon $r_c$.
CV2.0 is determined by eq~\eqref{eq:CV20_case1_int_regime} with the replacement $r_{-} \rightarrow r_{\rm max}$.
The intermediate regime $-t_{\infty} \leq t \leq t_{\infty}$ is described by the configuration in Fig.~\ref{fig:WDW_evo1}, whose complexity is given by the sum of the terms in \eqref{eq:CV20_case1_int_regime}.
Finally, at late times $t > t_{\infty}$ the top joint moves behind future timelike infinity $\mathcal{I}^+$, and CV2.0 is obtained by replacing $r_+ \rightarrow r_{\rm max}$ in \eqref{eq:CV20_case1_int_regime}.

\begin{figure}[t!]
    \centering
\subfloat[]{ \label{subfig:CV20rate_case1}
\includegraphics[width=0.475\textwidth]{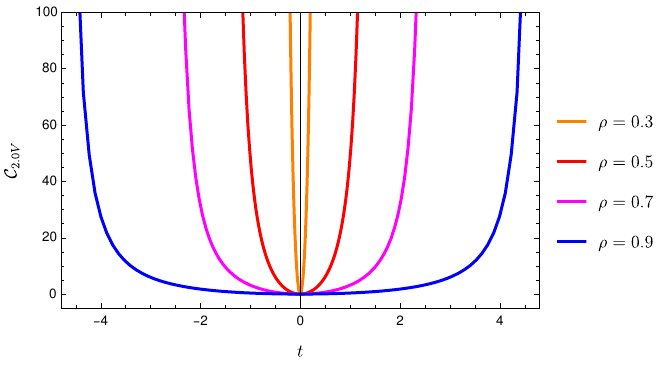}}\hfill \subfloat[]{\label{subfigb:CV20rate_case1} 
\includegraphics[width=0.475\textwidth]{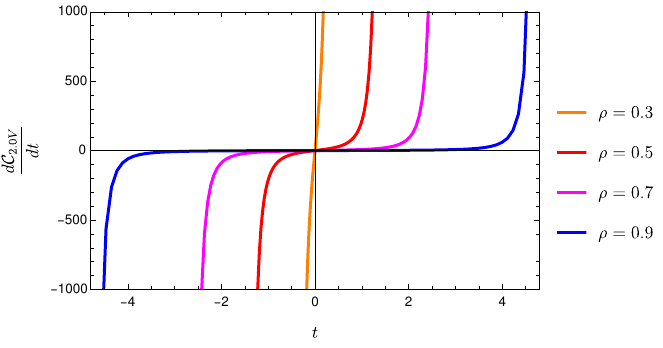}}
    \caption{Time dependence of (a): CV2.0 and (b): its rate, computed according to case 1 in \eqref{eq:rateCV20_case1} during the interval $[-t_{\infty}, t_{\infty}]$ with critical time defined in \eqref{eq:critical_time_tc1}. We set $G_N L=1, d=3, \mu=0.14$, and we vary the value of $\rho$ in \eqref{eq:stretched_horizon_SdS4}, but always keeping $t_{\infty} >0$.   }
    \label{fig:CV20case1}
\end{figure}

While there is no closed expression for the above integrals in generic dimension $d$, we can get the complexity rate by applying the fundamental theorem of integral calculus, together with \eqref{eq:rates_rprmcase1}:
\beq
\frac{d \mathcal{C}_{2.0V}}{dt} =  \frac{\Omega_{d-1}}{d \, G_{N} L^2} \times  
\begin{cases}
  r_+^d   &  \mathrm{if} \,\, t < -t_{\infty} \\
 \le r_+^d - r_-^d \ri  & \mathrm{if} \,\, -t_{\infty} \leq t \leq t_{\infty} \\
  (-r_-^d)   &  \mathrm{if} \,\, t > t_{\infty} \\
\end{cases}
\label{eq:rateCV20_case1}
\eeq
Let us focus on the intermediate regime.
At the critical time $t_{\infty} (-t_{\infty})$ defined in \eqref{eq:critical_time_tc1}, the complexity rate clearly becomes positively (negatively) divergent, since $r_+ (r_-) \rightarrow \infty$ and $r_- (r_+)$ is finite. 
This is the hyperfast growth characteristic of asymptotically dS spacetimes \cite{Susskind:2021esx}.
The plots of the time evolution of CV2.0 and its time derivative during the interval $[-t_{\infty}, t_{\infty}]$ are reported in Fig.~\ref{fig:CV20case1}, when $d=3$ and for various choices of $\rho$.
As expected, the qualitative behavior is similar to the case of empty dS space, see Fig.~5 in \cite{Jorstad:2022mls}.
One can also show that by keeping the regulator $r_{\rm max}$ finite, CV2.0 approaches a linear behavior at early and late times, in the same fashion as Fig.~7 of \cite{Jorstad:2022mls}.
This is also consistent with reference \cite{Baiguera:2023tpt} (see Fig.~24), in the limit of vanishing shockwave.

\begin{figure}[t!]
    \centering
    \subfloat[]{
\includegraphics[width=0.3\textwidth]{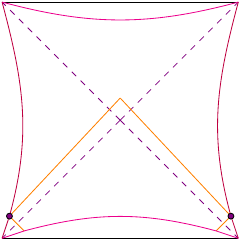} \label{subfig:WDW_evo2_regime1}}\hfill\subfloat[]{\label{subfig:WDW_evo2_regime2} \includegraphics[width=0.3\textwidth]{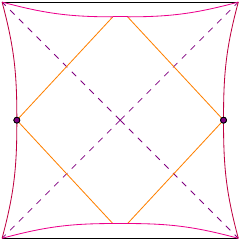}}
\hfill \subfloat[]{ \label{subfig:WDW_evo2_regime3}
\includegraphics[width=0.3\textwidth]{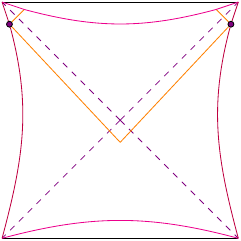}}
    \caption{Alternative time evolution of the WDW patch. Same color scheme as in Fig.~\ref{fig:WDW_evo1}. In the intermediate regime, both joints of the WDW patch lie behind the cutoff surfaces. }
    \label{fig:WDW_evo2}
\end{figure}

Let us briefly comment on the case (depicted in Fig.~\ref{fig:WDW_evo2}) when $\rho \rightarrow 0$ in \eqref{eq:stretched_horizon_SdS4}, \ie the stretched horizon approaches the black hole one and the critical time satisfies $t_{\infty} < 0$. 
When $t_{\infty} \leq t \leq -t_{\infty}$,
the top and bottom joints of the WDW patch both sit behind the respective region $\mathcal{I}^{\pm}$.
As a consequence, by direct computation one finds that the rate vanishes during the intermediate regime.
Furthermore, since at least one of the joints always reaches timelike infinity, CV2.0 is always infinite.
These results are inconsistent with the intuition coming from circuit models and from the geometry of dS space \cite{Susskind:2021esx,Shaghoulian:2022fop,Jorstad:2022mls}. 
Therefore, the case with $\rho \rightarrow 0$ does not seem the relevant regime to capture the features expected by a holographic realization of computational complexity.

\subsubsection*{Case 2}
We refer to Fig.~\ref{fig:WDW_cosmo_stretched_case2} for the splitting of the WDW patch into the subregions corresponding to the integrals (similar to (\ref{eq:CV20_case2_int_regime_AdS}))
\begin{subequations}
\label{eq:CV20_case2_int_regime}
\beq
\mathcal{J}_{1L}=\mathcal{J}_{1R} = \frac{\Omega_{d-1}}{G_{N} L^2} \int_{r_+}^{r_h} dr  \,  r^{d-1} \le  \frac{t}{2} -r^*(r) +r^*(r_{\rm st})  \ri \, ,
\label{eq:CV20_case2_int_regime_p1}
\eeq
\beq
\mathcal{J}_{2L} =
\mathcal{J}_{2R}=
\frac{\Omega_{d-1}}{G_{N} L^2} \int_{r_h}^{r_{\rm st}} dr  \,  r^{d-1} \le  2 r^*(r_{\rm st}) - 2 r^*(r)  \ri \, , 
\label{eq:CV20_case2_int_regime_p2}
\eeq
\beq
\mathcal{J}_{3L} =
\mathcal{J}_{3R} =  \frac{\Omega_{d-1}}{G_{N} L^2} \int_{r_-}^{r_h} dr  \,  r^{d-1} \le - \frac{t}{2} +r^*(r_{\rm st}) - r^*(r)  \ri \, ,
\label{eq:CV20_case2_int_regime_p3}
\eeq
\end{subequations}
where $r_{\pm}$ were defined in \eqref{eq:umax_vmin_WDWpatch_case2}.
All the other possible configurations of the WDW patch that can arise in case 2, together with the corresponding expressions for CV2.0, can be obtained by a limiting procedure from the above results.
In the following discussion, we will directly perform the limit $r_{\rm min} \rightarrow 0$.

\begin{figure}[t!]
    \centering
    \subfloat[]{ \label{subfig:CV20rate_case2}
\includegraphics[width=0.475\textwidth]{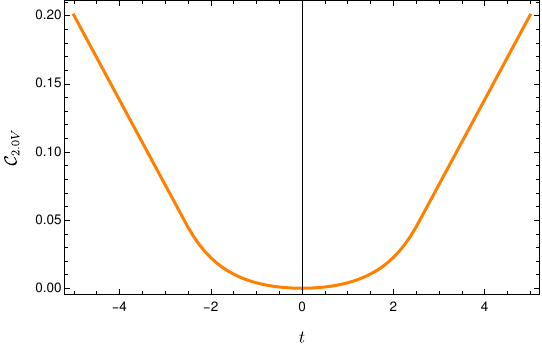}}\hfill\subfloat[]{\label{subfigb:CV20rate_case2} 
\includegraphics[width=0.475\textwidth]{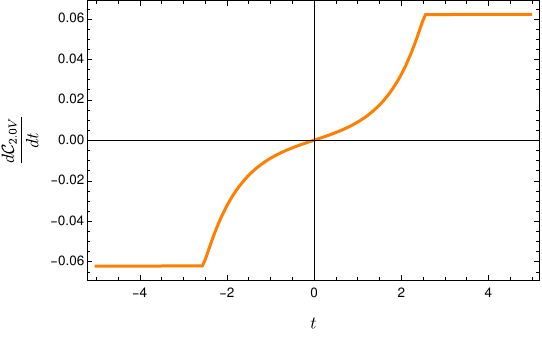}}
    \caption{Time dependence of (a): CV2.0 and (b): its rate, computed according to \eqref{eq:rateCV20_case2_p1}. We set $G_N L=1, d=3, \mu=0.14, \rho=0.01$. The intermediate interval that connects the regimes with linear growth corresponds to the times $t \in [-t_0, t_0]$, in terms of the critical value in \eqref{eq:critical_time_tc2}.  }
    \label{fig:CV20case2}
\end{figure}

The WDW patch can evolve in two different ways, distinguished by the sign of $t_0$ defined in \eqref{eq:critical_time_tc2}.
If $t_0>0$, or equivalently for a stretched horizon \eqref{eq:stretched_horizon_SdS4} close to the black hole horizon ($\rho \rightarrow 0$), the time evolution is characterized by the three regimes depicted in Fig.~\ref{fig:WDW_evo1}.
At times $t < -t_{0}$ the bottom joints are located behind the past singularity and CV2.0 is given by \eqref{eq:CV20_case2_int_regime} with the replacement $r_- \rightarrow 0$.
Next, we have an intermediate regime where both joints lie inside the black hole horizon.
When $t > t_{0}$, the top joint moves behind the future singularity, and CV2.0 is obtained by replacing $r_+ \rightarrow 0$ in \eqref{eq:CV20_case2_int_regime}. 
Using the time derivatives \eqref{eq:rates_rprmcase2}, we find the rates
\beq
\frac{d \mathcal{C}_{2.0V}}{dt} = 
\frac{\Omega_{d-1}}{d \, G_{N} L^2}  \times 
 \begin{cases}
 \le - r_+^d \ri  &  \mathrm{if} \,\, t < -t_{0} \\
 \le r_-^d - r_+^d \ri  & \mathrm{if} \,\, -t_{0} \leq t \leq t_0 \\
  r_-^d   &  \mathrm{if} \,\, t > t_{0} \\
\end{cases}
\label{eq:rateCV20_case2_p1}
\eeq
The late (early) time behavior (when $r_- \rightarrow r_h$ or $r_+ \rightarrow r_h$ respectively) results in
\beq
\frac{d \mathcal{C}_{2.0V}}{dt} \,\, \underset{t \rightarrow \pm t_{0}}{{\longrightarrow}} \,\,
\pm \frac{\Omega_{d-1}}{d \, G_{N} L^2} \, r_h^d \, ,
\label{eq:late_times_CV20_case2}
\eeq
which is the expected linear growth of complexity we also found in Sec. \ref{ssec:CV20AdS}.
The time dependence of complexity and its rate of growth are depicted in Fig.~\ref{fig:CV20case2}.
The plots show that there is an intermediate regime with non-trivial time evolution that interpolates between early and late times when the advocated linear growth is achieved.

\begin{figure}[t!]
    \centering
    \subfloat[]{ \label{subfig:CV20rate_case2r05}
\includegraphics[width=0.475\textwidth]{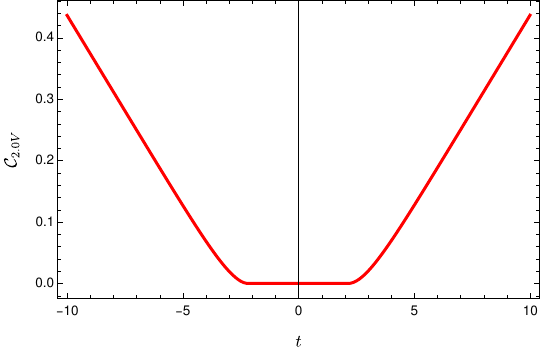}}\hfill\subfloat[]{\label{subfigb:CV20rate_case2r05} 
\includegraphics[width=0.475\textwidth]{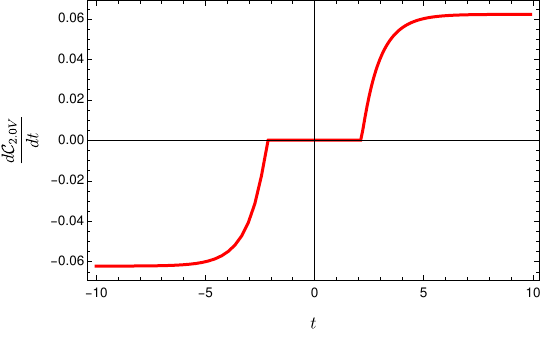}}
    \caption{Time dependence of (a): CV2.0 and (b): its rate, computed according to \eqref{eq:rateCV20_case2_p2}. We set $G_N L=1, d=3, \mu=0.14, \rho=0.5$.}
    \label{fig:CV20case2r05}
\end{figure}

The other possible evolution of the WDW patch (reported in Fig.~\ref{fig:WDW_evo2}) shows up when the critical time satisfies $t_0 \leq 0$, which always happens when $\rho \rightarrow 1$. By the same reasoning as we reported in Sec. \ref{ssec:CV20AdS}, the rate of CV2.0 in this case reads
\beq
\frac{d \mathcal{C}_{2.0V}}{dt} = 
\frac{\Omega_{d-1}}{d \, G_{N} L^2}  \times 
 \begin{cases}
 \le - r_+^d \ri  &  \mathrm{if} \,\, t < t_{0} \\
0  & \mathrm{if} \,\, t_{0} \leq t \leq -t_0 \\
  r_-^d   &  \mathrm{if} \,\, t > -t_{0} \\
\end{cases}
\label{eq:rateCV20_case2_p2}
\eeq
Again, at late (early) times the bottom (top) joint of the WDW patch approaches the black hole horizon, leading to the constant rate computed in \eqref{eq:late_times_CV20_case2}.
The results are displayed in Fig.~\ref{fig:CV20case2r05}, which are very similar to the case of black holes in asymptotically AdS spacetime (Sec. \ref{ssec:CV20AdS}, and \cite{Carmi:2017jqz}). Namely, the complexity is constant in the regime between the critical times, while at late and early times approaches a constant rate.
Therefore, this analysis suggests that the expected interior growth of a black hole in asymptotically dS spacetime is recovered when the stretched horizon is close to the cosmological one.
Notably, this choice is the same that reproduced the hyperfast growth for the inflationary patch of spacetime in case 1.

\subsubsection*{Case 3}

The configuration of the WDW patch is given by the plot depicted in fig.~\ref{fig:WDW_BH_stretched}.
Let us split the integration region into various parts, according to the labels used in the picture.
We begin by studying the right side of the Penrose diagram.
The corresponding terms in the sum \eqref{eq:decomposition_CV20_general} are defined as follows:
\begin{subequations}
\label{eq:Case3_CV20}
\beq
\mathcal{J}_{1R} = 
\frac{\Omega_{d-1}}{d G_{N} L^2} \le r_h^d - r_{\rm min}^d \ri \int_{-t_R - r^*(r_{\rm st}^R)}^{+\infty} du    \, ,
\eeq
\beq
\mathcal{J}_{2R}=
\frac{\Omega_{d-1}}{G_{N} L^2} \int_{r_h}^{r_{\rm st}^R} dr  \,  r^{d-1} \le  2 r^*(r_{\rm st}^R) - 2 r^*(r)  \ri = \mathrm{const.} 
\eeq
\beq
\mathcal{J}_{3R} = \frac{\Omega_{d-1}}{d G_{N} L^2} \le r_h^d - r_{\rm min}^d \ri \int_{-\infty}^{-t_R + r^*(r^R_{\rm st})} dv     \, , 
\eeq
\beq
\mathcal{J}_{4} = \frac{\Omega_{d-1}}{d G_{N} L^2} \le r_c^d - r_{h}^d \ri \int_{-\infty}^{\infty} du = \mathrm{const.} 
\eeq
\end{subequations}
In the previous analysis, we denoted with ``$\mathrm{const.}$'' the time-independent terms.
For convenience, the integrals referring to regions inside the black hole are performed by first integrating along the radial coordinate $r$, and then along one of the null directions $u,v$.
While some expressions are formally divergent and can be explicitly evaluated using an appropriate regularization, here we will only focus on the time dependence of the problem.
In particular, we notice that
\beq
\frac{d \mathcal{J}_{1R} }{dt_R} = - \frac{d \mathcal{J}_{3R}}{d t_R} =  \frac{\Omega_{d-1}}{d G_{N} L^2} \le r_h^d - r_{\rm min}^d \ri  \quad \Rightarrow \quad  
\frac{d}{dt_R} \le \mathcal{J}_{1R} + \mathcal{J}_{3R}  \ri = 0 \, .
\eeq
A similar computation shows that
\beq
\frac{d \mathcal{J}_{1L} }{dt_L} = - \frac{d \mathcal{J}_{3L}}{d t_L} = - \frac{\Omega_{d-1}}{d G_{N} L^2} \le r_{\rm max}^d - r_{c}^d \ri  \quad \Rightarrow \quad  
\frac{d}{dt_L} \le \mathcal{J}_{1L} + \mathcal{J}_{3L}  \ri = 0 \, .
\eeq
In conclusion, we trivially find 
\beq
\frac{d \mathcal{C}_{V2.0}}{d t_R} = 
\frac{d \mathcal{C}_{V2.0}}{d t_L} = 0 \, . 
\label{eq:rate_CV20_case3}
\eeq
Remarkably, the spacetime volumes located on the left and right sides of the WDW patch are separately time-independent.
Therefore, any choice of the relative evolution between the left and right boundary times $t_L, t_R$ leads to a vanishing complexity rate!
This behavior is very different compared to the previous cases and reflects the geometric feature that the portion of WDW patches that emerge from the past cutoff surfaces of the Penrose diagram is compensated by other regions of the same size that move behind the future cutoff surfaces.
If we aim to give a dual interpretation to the spacetime volume inside the WDW patch, we would conclude that the putative dual state has always the same computational complexity.
In particular, it is not able to give us any information about the growth of black holes, nor about the space beyond the cosmological horizon.

\subsubsection*{Cases 4-5}

Decomposing the WDW patch in a similar way as case 3, we find that CV2.0 is time-independent.

\subsection{Complexity=action}
\label{ssec:CA}
We proceed evaluating (\ref{eq:bdy_action}) for SdS$^n$ space. We report the main results, while we refer the reader to App.~\ref{app:details_CA} for more details.

\subsubsection*{Case 1}

\begin{figure}[t!]
    \centering
    \subfloat[]{ \label{subfig:CArate_case1}\includegraphics[width=0.475\textwidth]{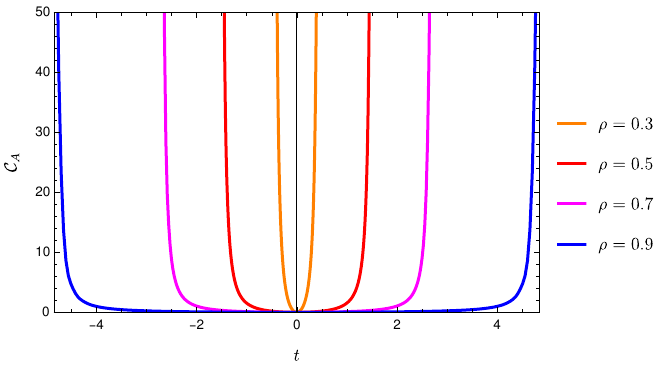}}\hfill \subfloat[]{\label{subfigb:CArate_case1} \includegraphics[width=0.475\textwidth]{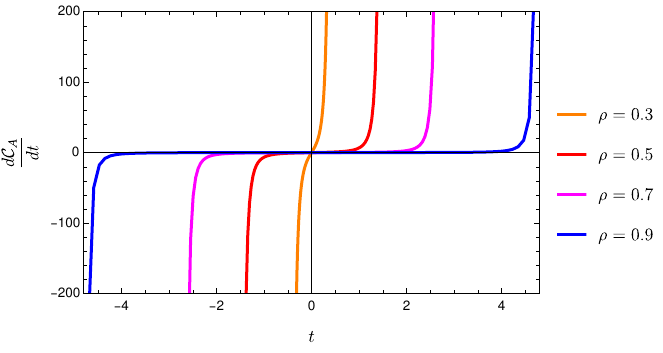}}
    \caption{Time dependence of (a): CA and (b): its rate, computed according to case 1 in \eqref{eq:rate_CA_case1} during the interval $[-t_{\infty}, t_{\infty}]$ with critical time defined in \eqref{eq:critical_time_tc1}. We set $G_N L=1, d=3, \mu=0.14, \ell_{\rm ct}=1/3$, and we vary $\rho$ in \eqref{eq:stretched_horizon_SdS4}, while keeping $t_{\infty} >0$.  }
    \label{fig:CAcase1}
\end{figure}

We study a symmetric configuration of both the time evolution \eqref{eq:symmetric_times} and the location of the stretched horizons $r_{\rm st}^R=r_{\rm st}^L$.
When the critical time satisfies $t_{\infty} >0$, the evolution of the WDW patch is described by Fig.~\ref{fig:WDW_evo1}.
During the intermediate time regime $-t_{\infty} \leq t \leq t_{\infty}$, the CA conjecture is obtained by summing the bulk term \eqref{eq:bulk_term_CA} and the boundary contribution evaluated in \eqref{eq:bdy_action_case1}.
It is more insightful to study the rate of growth of CA, obtained by applying the time derivatives \eqref{eq:rates_rprmcase1}:
\beq
\begin{aligned}
&\frac{d \mathcal{C}_A}{dt} = \frac{\Omega_{d-1}}{8 \pi^2 G_N}  \left[  \frac{(r_+)^d}{L^2}  + \frac{d-1}{2} (r_+)^{d-2} f(r_+) \log \left| \frac{\ell_{\rm ct}^2 (d-1)^2 f(r_+) }{(r_+)^2} \right| - \frac{(r_-)^d}{L^2}  \right. \\
&  \left. + \frac{(r_+)^{d-1}f'(r_+)}{2}  - \frac{d-1}{2} (r_-)^{d-2} f(r_-) \log \left| \frac{\ell_{\rm ct}^2 (d-1)^2 f(r_-) }{(r_-)^2} \right| - (r_-)^{d-1} \, \frac{f'(r_-)}{2}  \right] \, ,
\end{aligned}
\label{eq:rate_CA_case1}
\eeq
where $r_{\pm}$ were defined in \eqref{eq:umax_vmin_WDWpatch}.
We already established by inspection of \eqref{eq:rateCV20_case1} that CV2.0 displays a hyperfast growth when approaching the critical times $\pm t_{\infty}$.
To check whether the rate of CA is also divergent when the top joint of the WDW patch approaches $\mathcal{I}^+$, we perform a series expansion around $r_{+} = \infty$ of \eqref{eq:rate_CA_case1}.
Using the explicit blackening factor in \eqref{eq:asympt_dS}, we get
\beq
 \frac{d \mathcal{C}_A}{dt} \,\, \underset{t \rightarrow t_{\infty}}{{\longrightarrow}} \,\,\frac{\Omega_{d-1}}{16 \pi^2 G_N L^2} (d-1)    (r_+)^{d}  \log \left| \frac{L^2}{\ell_{\rm ct}^2 (d-1)^2} \right| \, .
\label{eq:approx_rate_CA_case1}
\eeq
The complexity rate (and therefore CA itself) is positively divergent if the counterterm length scale satisfies
\beq
\ell_{\rm ct} < \frac{L}{d-1} \, .
\label{eq:positive_CA_case1}
\eeq
This requirement is the same that fixes the positivity of complexity in empty dS space \cite{Jorstad:2022mls}.
In a similar way, when $t \rightarrow -t_{\infty}$, a series expansion around $r_-=\infty$ gives
\beq
 \frac{d \mathcal{C}_A}{dt} \,\, \underset{t \rightarrow -t_{\infty}}{{\longrightarrow}} \,\, - \frac{\Omega_{d-1}}{16 \pi^2 G_N L^2} (d-1)    (r_-)^{d}  \log \left| \frac{L^2}{\ell_{\rm ct}^2 (d-1)^2} \right| \, ,
\label{eq:approx_rate_CA_case1_past}
\eeq
and now the condition \eqref{eq:positive_CA_case1} implies that the rate is negatively divergent instead.
The time evolution of CA during the intermediate regime is depicted in Fig.~\ref{fig:CAcase1}.
In this plot, we consider various possibilities for the parameter $\rho$ defining the location of the stretched horizon, but we always keep $\ell_{\rm ct}=1/3$ to satisfy \eqref{eq:positive_CA_case1}, so that complexity is positive (negative) at late (early) times. 

The time evolution of the WDW patch when $t< - t_{\infty}$ or $t> t_{\infty}$ is different compared to the intermediate regime that we studied above because GHY terms are now non-vanishing.
The most relevant and characteristic feature of asymptotically dS spacetimes is the hyperfast growth, which we already capture with the expansion in \eqref{eq:approx_rate_CA_case1}.
At early and late times, the divergent growth of complexity can be regularized by keeping the radial coordinate $r=r_{\rm max}$ of the cutoff surface finite.
The analysis and results in the latter case are analogous to empty dS space (see Fig.~8 in \cite{Jorstad:2022mls}), and we will not report them here.

When the critical time satisfies $t_{\infty} <0$, the WDW patch follows the evolution depicted in Fig.~\ref{fig:WDW_evo2}, and  CA is always trivially divergent.
Therefore, this case does not seem to capture the expected behavior of complexity for a putative dual state.

\subsubsection*{Case 2}
\begin{figure}[t!]
    \centering
    \subfloat[]{ \label{subfig:CArate_case2}\includegraphics[width=0.475\textwidth]{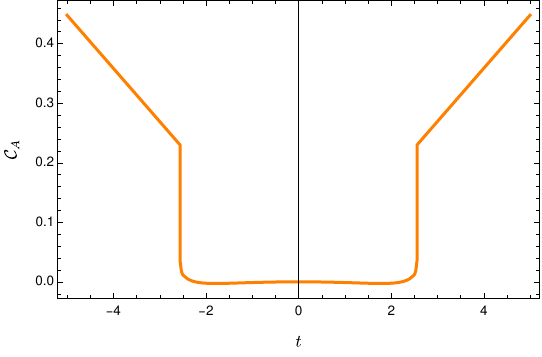}}\hfill \subfloat[]{\label{subfigb:CArate_case2} \includegraphics[width=0.475\textwidth]{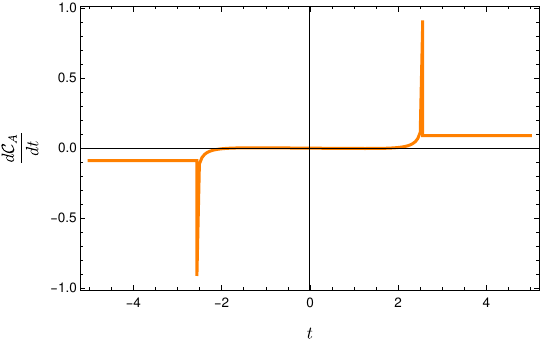}}
    \caption{Time dependence of, (a), CA and, (b), its rate, computed according to \eqref{eq:rateCA_case2_p1}. We set $G_N L=1, d=3, \mu=0.14, \rho=0.01, \ell_{\rm ct}=1/3$. }
    \label{fig:CAcase2}
\end{figure}

In the following, we focus on a symmetric configuration of the boundary times \eqref{eq:symmetric_times} and of the stretched horizons $r_{\rm st}^R=r_{\rm st}^L$.
Let us first assume that the critical time satisfies $t_0>0$, when the evolution of the WDW patch is governed by the diagrams in Fig.~\ref{fig:WDW_evo1}.
By applying the time derivatives \eqref{eq:rates_rprmcase2} to the CA conjecture \eqref{eq:tot_grav_action}, we obtain
\small
\beq
\begin{aligned}
    &\frac{8 \pi^2 G_N}{\Omega_{d-1}}  \frac{d \mathcal{C}_A}{dt} =  \\
&\begin{cases}
   - \mu d - \frac{(r_+)^d}{L^2} - \frac{d-1}{2} (r_+)^{d-2} f(r_+) \log \left| \frac{\ell_{\rm ct}^2 (d-1)^2 f(r_+) }{(r_+)^2} \right| - (r_+)^{d-1} \, \frac{f'(r_+)}{2}       &  \mathrm{if}  \,\, t < -t_{0} \\ 
 \Big(  - \frac{(r_+)^d}{L^2}  - \frac{d-1}{2} (r_+)^{d-2} f(r_+) \log \left| \frac{\ell_{\rm ct}^2 (d-1)^2 f(r_+) }{(r_+)^2} \right| - (r_+)^{d-1} \, \frac{f'(r_+)}{2}   & \\
    + \frac{(r_-)^d}{L^2}  + \frac{d-1}{2} (r_-)^{d-2} f(r_-) \log \left| \frac{\ell_{\rm ct}^2 (d-1)^2 f(r_-) }{(r_-)^2} \right| + (r_-)^{d-1} \, \frac{f'(r_-)}{2}  \Big) & \mathrm{if} \,\, t \in [-t_0, t_0] \\
 \mu d  + \frac{(r_-)^d}{L^2}  + \frac{d-1}{2} (r_-)^{d-2} f(r_-) \log \left| \frac{\ell_{\rm ct}^2 (d-1)^2 f(r_-) }{(r_-)^2} \right| + (r_-)^{d-1} \, \frac{f'(r_-)}{2}   &  \mathrm{if} \,\, t > t_{0} \\
\end{cases}
\end{aligned}
\label{eq:rateCA_case2_p1}
\eeq
\normalsize
where the results for the boundary terms were taken from App. \ref{app:ssec:case2}.
We represent the numerical evolution of CA and of its rate in Fig.~\ref{fig:CAcase2}.
It is evident that while the complexity is continuous, instead its time derivative presents a discontinuity at the critical times.
The value of the jump at $t= t_0 $ is simply obtained by comparing the rate in the last regime with the rate at intermediate times in the limit $r_{+} \rightarrow 0$, giving
\beq
\frac{d \mathcal{C}_A}{dt}\Big|_{t= t_0^{+}} - \frac{d \mathcal{C}_A}{dt}\Big|_{t=t_0^-} = \frac{\Omega_{d-1}}{8 \pi^2 G_N} \mu d \, .
\label{eq:disc_rateCA_case2p1}
\eeq
The GHY term is fully responsible for this jump since all the terms on null boundaries (either codimension-one or codimension-two) evaluated at the singularity vanish.
At $t=-t_0,$ the same conclusion (except for an overall minus sign) holds.
It is straightforward to determine the late (early) time limits, which correspond to the bottom (top) joint of the WDW patch approaching the horizon radius $r_h$.
The rate reads
\beq
\lim_{t \rightarrow \pm \infty} \frac{d \mathcal{C}_A}{dt} = \pm \frac{\Omega_{d-1}}{8 \pi^2 G_N} \left[ \mu d  + \frac{(r_h)^d}{L^2}  + (r_h)^{d-1} \, \frac{f'(r_h)}{2}  \right] \, ,
\label{eq:rate_CA_case2_latelimit}
\eeq
which is clearly a constant.
This shows that CA asymptotically reproduces the celebrated linear trend of complexity.

\begin{figure}[t!]
    \centering
    \subfloat[]{ \label{subfig:CArate_case2r05}\includegraphics[width=0.475\textwidth]{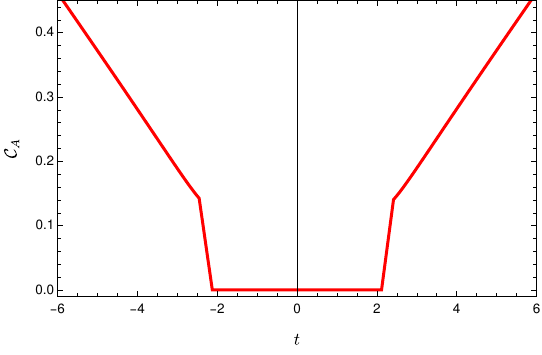}}\hfill \subfloat[]{\label{subfigb:CArate_case2r05} \includegraphics[width=0.475\textwidth]{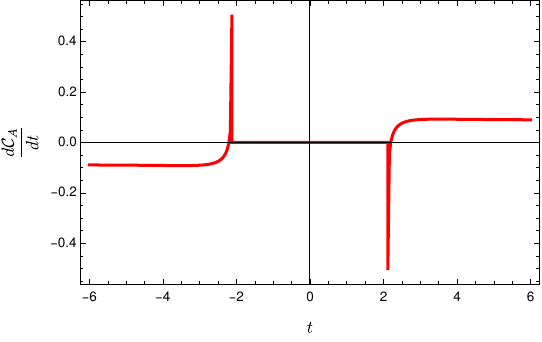}}
    \caption{Time dependence of (a): CA and (b) its rate, computed according to \eqref{eq:rateCA_case2_p2}. We set $G_N L=1, d=3, \mu=0.14, \rho=0.5, \ell_{\rm ct} = 1/3$. }
    \label{fig:CAcase2r05}
\end{figure}

Next, let us assume that the critical time is instead negative, \ie $t_0 <0$.
In this case, the evolution of the WDW patch is depicted in Fig.~\ref{fig:WDW_evo2}, and similar computations lead to the rate
\small
\beq
\begin{aligned}
    &\frac{8 \pi^2 G_N}{\Omega_{d-1}} \frac{d \mathcal{C}_A}{dt} = \\
&\begin{cases}
    - \mu d - \frac{(r_+)^d}{L^2} - \frac{d-1}{2} (r_+)^{d-2} f(r_+) \log \left| \frac{\ell_{\rm ct}^2 (d-1)^2 f(r_+) }{(r_+)^2} \right| - (r_+)^{d-1} \, \frac{f'(r_+)}{2}     &  \mathrm{if}  \,\, t < t_{0} \\
 0  & \mathrm{otherwise} \\
 \mu d  + \frac{(r_-)^d}{L^2}  + \frac{d-1}{2} (r_-)^{d-2} f(r_-) \log \left| \frac{\ell_{\rm ct}^2 (d-1)^2 f(r_-) }{(r_-)^2} \right| + (r_-)^{d-1} \, \frac{f'(r_-)}{2}    &  \mathrm{if} \,\, t > -t_{0} \\
\end{cases}
\end{aligned}
\label{eq:rateCA_case2_p2}
\eeq
\normalsize
whose main difference with the previous scenario is the time-independence of the intermediate regime.\footnote{Analogous to Sec. \ref{ssec:CAAdS}, this is a trivial consequence of the intermediate regime of CV2.0 in \eqref{eq:rateCV20_case2_p2} and the vanishing rate found in \eqref{eq:rate_bdy_app_case2}.   }
We present a numerical evaluation of CA and of its rate in Fig.~\ref{fig:CAcase2r05}.
The discontinuity in CA at $t=-t_0$ can be evaluated in the $r_- \rightarrow 0$ limit, leading to
\beq
\frac{d \mathcal{C}_A}{dt}\Big|_{t= -t_0^{+}} - \frac{d \mathcal{C}_A}{dt}\Big|_{t= -t_0^-} = \frac{\Omega_{d-1}}{8 \pi^2 G_N} \mu d \, ,
\label{eq:disc_rateCA_case2p2}
\eeq
which is the same result obtained in \eqref{eq:disc_rateCA_case2p1}. The same discontinuity occurs for $t=t_0$. Meanwhile, the early and late time limits are also given by the same expression \eqref{eq:rate_CA_case2_latelimit} determined above.
Therefore, we conclude that the location of the stretched horizons in case 2 mainly affects the intermediate regime.
When taking $\rho \rightarrow 1$, complexity shows an evolution very similar to the case of black holes in AdS (which we discussed in Sec. \ref{ssec:CV20AdS}; see also \cite{Carmi:2017jqz}).

\subsubsection*{Case 3}

The boundary terms of the gravitational action in the setting described by Fig.~\ref{fig:WDW_BH_stretched} are evaluated in App. \ref{app:ssec:case3}.
The main observation is that the rates in eqs.~\eqref{eq:rate_CV20_case3} and \eqref{eq:rate_CA_bdy_case3} vanish for any choice of the boundary times $t_L, t_R$, therefore implying that
\beq
\frac{d \mathcal{C}_A}{d t_L} = \frac{d \mathcal{C}_A}{d t_R} = 0 \, .
\label{eq:rate_CA_case3}
\eeq
In particular, the on-shell action inside the WDW patch evaluated on the right and left sides of the Penrose diagram are separately time-independent.

\subsubsection*{Cases 4-5}

In analogy with case 3, one can show that CA conjecture in cases 4 and 5 is also time-independent.
The precise steps are similar to the ones reported in App. \ref{app:ssec:case3}.

\section{Codimension-one proposals}
\label{sec:Cod1 complexity}

We continue the analysis of holographic complexity in the extended SdS background with the codimension-one proposals: the volume (section \ref{ssec:CV_dS}), and a special class of the complexity=anything conjectures (section \ref{sec:CAny CMC}).

\subsection{CV proposal}\label{ssec:CV_dS}
We proceed to evaluate the CV proposal in (\ref{eq:Vol interm}), with $\sigma$ determined from (\ref{eq:classical_motion_AdS}), for the different configurations in cases displayed in Fig. \ref{fig:summary_cases}. In all of these configurations, the sign of $P$ for a given maximal slice depends on the orientation of $\sigma$. This can be understood by noting from \eqref{eq:P u dot} that
\begin{equation}
    P = -f(r) \dot{t}  \, .
\end{equation}
Outside the cosmological horizon and inside the black hole we have $f(r) < 0$, so ${\rm sign}(P) ={\rm sign}(\dot{t})$. 
In other words, in these regions $P>0$ if $\sigma$ is oriented as the Killing vector $\partial_t$, otherwise $P<0$.\footnote{Of course, in the external region $f(r) > 0$, so ${\rm sign}(P) =- {\rm sign}(\dot{t})$ and opposite conclusions follow.} 
It can be then seen from Fig \ref{fig:Penrose_SdS} that if $\sigma$ runs from the right stretched horizon to the left one, $P>0$ in the future black hole interior and $P<0$ in the future inflating region. 
This is the convention we will employ in the rest of the work.

From (\ref{eq:turning}) we deduce that $U_{\rm eff}(r) \geq 0$ for $0 \leq r \leq r_h$ and $r \geq r_c$, whereas $U_{\rm eff}(r) < 0$ for $r_h < r < r_c$. Hence, we deduce that $U_{\rm eff}(r)$ has at least a (local) maximum in the interval $(0, r_h)$ and at least a minimum in the range $(r_h, r_c)$, while it diverges for $r \rightarrow +\infty$. A plot of the function $U_{\rm eff}(r)$ is displayed in Fig. \ref{fig:Ueff_1}, from which it can be seen that there exist only one local maximum and one local minimum.

In the parameter space we have numerically explored, the function $U_{\rm eff}(r)$ keeps the same qualitative behavior. So, in what follows we will assume the existence of a single local maximum in $U_{\rm eff}(r)$.

The following observations thus follow:
\begin{itemize}
    \item If the maximal surface extends into the inflating region $r \geq r_c$, the potential barrier prevents the surface from reaching timelike infinity $\mathcal{I}^{\pm}$. Strictly speaking, timelike infinity can only be reached by the maximal surfaces with $P \rightarrow \pm \infty$.
    \item If the maximal surface enters the black hole $r \leq r_h$, it ends up in the singularity $r=0$ unless a potential barrier is met. Therefore, the singularity is avoided when the corresponding $P^2$ is lower than the maximal value of the effective potential inside the black hole.
\end{itemize}    

These observations have important consequences for the evolution of holographic complexity.
First, the CV prescription requires computing the maximal volume of a connected surface anchored at the stretched horizons. 
Therefore, any extremal surface starting from one stretched horizon and falling into the singularity before reaching the other stretched horizon will be discarded in our analysis.\footnote{ \label{footnote:disconnected_solutions}  In principle, one can generalize the CV conjecture to include the possibility that an extremal surface is composed of two parts ending at $r=0$, and connected through the singularity. This case is considered in Sec. 2.2.2 of \cite{Belin:2022xmt} for asymptotically AdS space, and a similar prescription is employed in empty dS space in Sec. 5 of \cite{Jorstad:2022mls}. However, these configurations are either extremal but not maximal, or they appear after some critical time. 
We will not consider them in this work.  }
Second, hyperfast growth is achieved whenever the extremal surface reaches timelike infinity $\mathcal{I}^{\pm}$. 
Given the discussion above, we conclude that as long as the conserved momentum is bounded $|P| < \infty$, hyperfast growth is avoided. This is displayed in the below cases.

\subsubsection*{Case 1}

\begin{figure}[t!]
  	\centering
 \subfloat[]{ \label{subfig:extremal_surfaces_CV_case1} \includegraphics[width=0.4\textwidth]{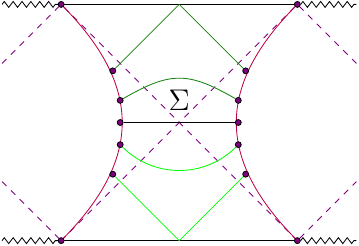} }\vspace{0.1cm}\subfloat[]{ \label{subfig:Ueff_CV_case1} \includegraphics[width=0.5\textwidth]{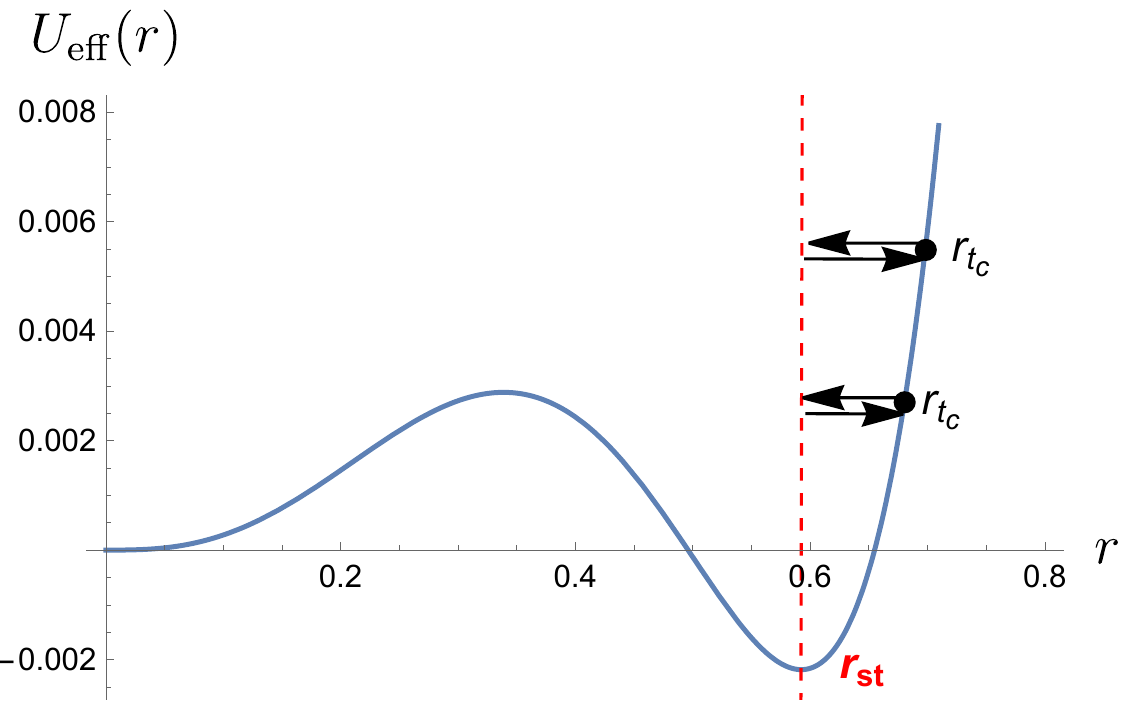} }
  	\caption{(a): The left and right stretched horizons (in fuchsia) are placed in consecutive static patches. In this prescription, the codimension-one extremal surface $\Sigma$, anchored to the stretched horizon (purple dots), explores the region outside the cosmological horizon. The extremal surfaces are characterized by $P<0$ ($P>0$) in the upper (lower) half of the Penrose diagram, see the dark (light) green curves. The black curve has $P=0$.
    (b): Qualitative behavior of the effective potential $U_{\rm eff}(r)$ in \eqref{eq:turning}. For every value of $P^2$, the surface connects the two stretched horizons after going through the turning point $r_{t_c}$.
    We have set $d=3$, $\mu= 0.187$, and $L=1$.
    }
  	\label{fig:Ueff_1}
  \end{figure}

\begin{figure}[t!]
    \centering
\subfloat[]{ \label{subfig:CV_case1}\includegraphics[width=0.45\textwidth]{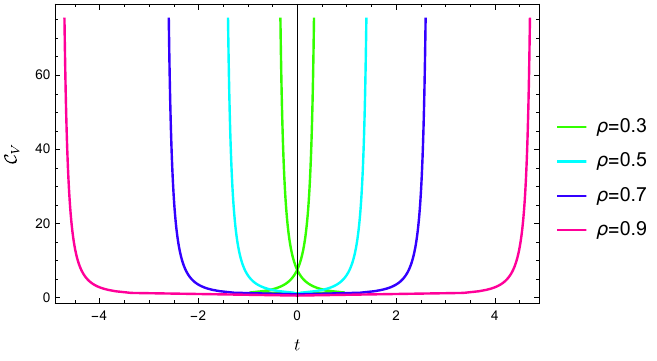}}\hfill \subfloat[]{\label{subfigb:CVrate_case1} \includegraphics[width=0.48\textwidth]{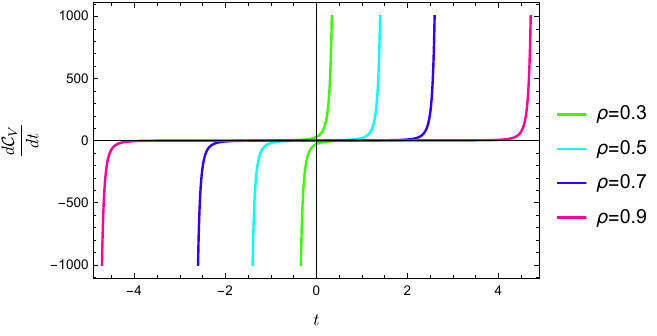}}
    \caption{Time dependence of (a): CV and (b): its rate for the prescription in Fig. \ref{fig:Ueff_1}. Different curves correspond to different values of the stretched horizon parameter $\rho$.
    We have set $d=3$, $\mu = 0.14$, and $G_N L=1$.
    When multiple extremal surfaces exist (see the green curve), it is understood that complexity corresponds to the maximal value of the volume at each fixed time $t$.  }
    \label{fig:volume_cosmo}
\end{figure}

We place the left and right stretched horizons $r_{\rm st} \equiv r_{\rm st}^L = r_{\rm st}^R$ into two consecutive static patches, see Fig.~\ref{fig:Ueff_1}, and we attach the codimension-one surface to the left and right stretched horizon at times $-t_L$ and $t_R$, respectively. 
Due to the time-translation symmetry of the configuration, it is not restrictive to consider the symmetric case \eqref{eq:symmetric_times}.  
After leaving the right stretched horizon $r_{\rm st}$, the maximal surface moves in the direction of increasing $r$ and enters the inflating region. As it can be seen in Fig. \ref{fig:Ueff_1}, here it passes through a turning point and proceeds in the direction of decreasing $r$ until it reaches the left stretched horizon $r_{\rm st}$. 
In summary, the maximal surface admits a single turning point $r_{t_c}$ satisfying $r_{t_c} \geq r_c$. 
Therefore, the extremal volume and the boundary time at the stretched horizon $r_{\rm st}$ are respectively given by
\begin{equation}
    \frac{V(P)}{\Omega_{d-1}} = \frac{2}{L^{d-1}} \int_{r_{\rm st}}^{r_{t_c}} \frac{r^{2(d-1)}}{\sqrt{P^2 + f(r) (r/L)^{2(d-1)}}} dr \, ,
    \label{eq:V_CV_1}
\end{equation}
\begin{equation}
    t(P) =   2 \int_{r_{\rm st}}^{r_{t_c}} \frac{P}{f(r) \sqrt{P^2 + f(r) (r/L)^{2(d-1)}}} dr \, ,
    \label{eq:t_CV_1}
\end{equation}
where the second expression is obtained by adding the left and right-hand sides of
\begin{align}
    - r^*(r_{t_c}) -\frac{t}{2} + r^*(r_{\rm st})  &= \int_{r_{\rm st}}^{r_{t_c}} \frac{\dot{u}_+}{\dot{r}_+} dr =\int_{r_{\rm st}}^{r_{t_c}} \frac{-P - \sqrt{P^2 + f(r) (r/L)^{2(d-1)}}}{f(r) \sqrt{P^2 + f(r) (r/L)^{2(d-1)}}} dr \, , \label{eq:equi1}\\
 -\frac{t}{2} + r^*(r_{\rm st}) - r^*(r_{t_c}) &= \int_{r_{t_c}}^{r_{\rm st}} \frac{\dot{v}_-}{\dot{r}_-} dr = \int_{r_{t_c}}^{r_{\rm st}} \frac{P + \sqrt{P^2 + f(r) (r/L)^{2(d-1)}}}{f(r) \sqrt{P^2 + f(r) (r/L)^{2(d-1)}}} dr \, , \label{eq:equi2}
\end{align}
in which we set the time at the turning point $t_{t_c}=0$ by symmetry.
In the previous equations, we exploited the fact that
\begin{equation}
    \int_a^b \frac{dr}{f(r)} = r^*(b) - r^*(a) \, .
\end{equation}
In Fig.~\ref{fig:volume_cosmo} we show numerical plots of volume complexity and its growth rate as functions of time for several values of the stretched horizon parameter $\rho$ for $d=3$. For different spacetime dimensions, we qualitatively get the same behaviors.  
In particular, the volume diverges at a finite critical time, which increases both with the spacetime dimension $d$ and with the value of $\rho$.
Restricting to $t>0$, for small values of $\rho$ there are two main regimes (see the green curve in Fig. \ref{fig:volume_cosmo}): at small times there are three extremal surfaces anchored at the same boundary time $t$, whereas at large times there are two. 
Such configurations are symmetric under time-reflection $t \rightarrow -t$.
According to the CV prescription, complexity is determined by the maximal volume among these possibilities. 
An analog case with multiple extremal surfaces happened in \cite{Chapman:2021eyy,Jorstad:2022mls,Auzzi:2023qbm}.
Following the same arguments presented in \cite{Jorstad:2022mls,Auzzi:2023qbm}, we deduce that, for fixed time $t$, the surface with larger volume is the one with larger $\left| P \right|$.
For bigger values of $\rho$, there is instead only one extremal surface, and no further maximization is required. 

The volume diverges when the turning point reaches $r_{t_c} = r_{\rm max} \rightarrow +\infty$. From \eqref{eq:turning}, we note that this happens for 
\begin{equation}
    P^2 \sim \left( \frac{r_{\rm max}}{L} \right)^{2d} \rightarrow +\infty \, . 
\end{equation}
Consequently, from \eqref{eq:t_CV_1} the critical time (for $P \rightarrow -\infty$) can be expressed as
\begin{equation}
    t_{c1} = -2 \int_{r_{\rm st}}^{r_{\rm max}} \frac{dr}{f(r)} = 2 (r^*(r_{\rm st})-r^*(r_{\rm max})) \, ,
    \qquad 
    t_{\infty} = \lim_{r_{\rm max} \rightarrow \infty} t_{c1} \, ,
\end{equation}
in accordance with the result obtained for codimension-zero proposals in \eqref{eq:critical_time_tc1}.
Since the contribution from empty dS space in the blackening factor $f(r)\sim-(r/L)^2$ is the dominant term in the previous integrals, the critical time obtained in empty dS (discussed in Sec. 5 of \cite{Jorstad:2022mls}) is valid for this case as well. Therefore, we conclude that in the SdS background, we get the following dominant contribution around the critical time
\begin{equation}
    t - t_{\infty} \sim \frac{L^2}{r_{\rm max}} \, .
    \label{t_critical_case1}
\end{equation}

\paragraph{Growth rate.}
In order to analyze the growth rate of volume complexity, it is convenient to observe that
\begin{equation}
  \frac{(r/L)^{2(d-1)}}{\sqrt{P^2 + f(r) (r/L)^{2(d-1)}}} = 
  - \frac{P^2}{f(r) \sqrt{P^2 + f(r) (r/L)^{2(d-1)}}}
  +\frac{\sqrt{P^2 + f(r) (r/L)^{2(d-1)}}}{f(r)}  \, .
  \label{eq:rewriting_CV}
\end{equation}
We can thus re-express the volume as
\begin{equation}
    \frac{V}{\Omega_{d-1} L^{d-1}} = 
   - P \, t +
    2 \int_{r_{\rm st}}^{r_{t_c}} \frac{\sqrt{P^2 + f(r) (r/L)^{2(d-1)}}}{f(r)} dr  \, .
\end{equation}
Similar to (\ref{eq:dV dt all t AdS}), we get
\begin{equation}
   \frac{dV}{dt} =-\Omega_{d-1} L^{d-1} P \, .
   \label{eq:V_rate}
\end{equation}
In particular, at the critical time, we find
\begin{equation}
    \lim_{t \to t_{\infty}} \frac{dV}{dt} = 
    \Omega_{d-1} \frac{r_{\rm max}^d}{L} 
    \sim \frac{1}{(t_{\infty} -t)^d} \, ,
    \label{eq:ratelate_CV_case1}
\end{equation}
where \eqref{t_critical_case1} has been used.
This result describes the hyperfast growth of CV in asymptotically dS geometries.

\subsubsection*{Case 2}
\begin{figure}[t!]
  	\centering
  \subfloat[]{\label{subfig:Penrose_CV_case2}  \includegraphics[width=0.4\textwidth]{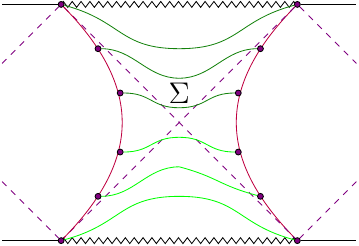}}\qquad\subfloat[]{ \label{subfig:potential_CV_case2} \includegraphics[width=0.5\textwidth]{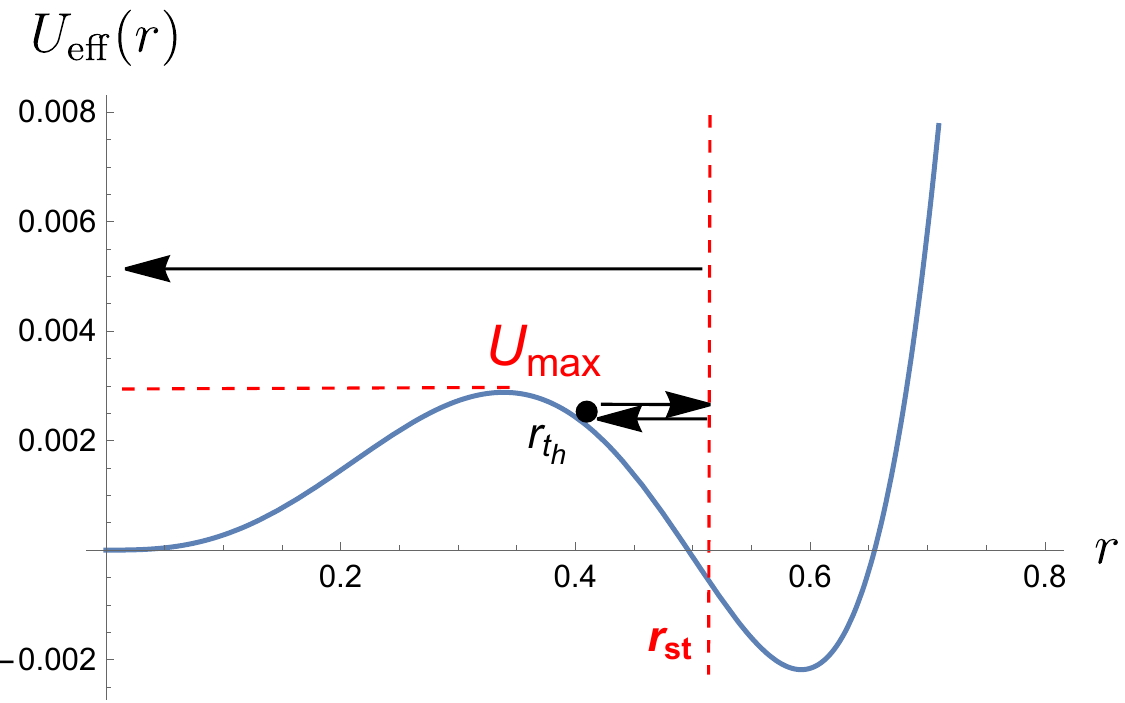}}
  	\caption{(a): The left and right stretched horizons are placed in consecutive static patches. In this prescription, the codimension-one extremal surface $\Sigma$ explores the region behind the black hole horizon.
    The extremal surfaces have $P>0$ ($P<0$) in the upper (lower) half of the Penrose diagram, see the dark (light) green curves.
    (b): Qualitative behavior of the effective potential $U_{\rm eff}(r)$. For large $P^2$, the surface falls into the black hole singularity. For small enough $P^2$, the surface connects the two stretched horizons after going through the turning point $r_{t_h}$.
    We have set $d=3$, $\mu= 0.187$, and $L=1$.
    }
  	\label{fig:Ueff_2}
  \end{figure}

One can proceed in a similar way with the configuration depicted in Fig.~\ref{fig:Ueff_2}, where we anchor the maximal volume surfaces to a pair of black hole stretched horizons $r_{\rm st} \equiv r_{\rm st}^L = r_{\rm st}^R$. Without loss of generality, we consider the symmetric choice \eqref{eq:symmetric_times} for the boundary times along the stretched horizons.
After leaving the right stretched horizon $r_{\rm st}$, the maximal surface extends in the direction of decreasing $r$ and enters the black hole. 
Contrary to case 1, the value of $|P|$ cannot be arbitrarily large, since there exists a local maximum for the effective potential (\ref{eq:def_rf_caseAdSCV}) at a final slice $r_f$, as displayed in Fig.~\ref{fig:Ueff_2} (otherwise, the codimension-one slice falls into the singularity at $r=0$). As explained in sec. \ref{ssec:CV}, the range of the conserved momentum $P^2$ most be $0 \leq P^2 \leq U_{\rm max}$ for it to be an extremal surface.
Since the turning point is located at $r_{t_h} \leq r_h$, we have that
\begin{equation}\label{eq:Vol BB}
    \frac{V(P)}{\Omega_{d-1}} = \frac{2}{L^{d-1}} \int^{r_{\rm st}}_{r_{t_h}} \frac{r^{2(d-1)}}{\sqrt{P^2 + f(r) (r/L)^{2(d-1)}}} dr \, ,
\end{equation}
\begin{equation}
    t(P) =  -2 \int_{r_{t_h}}^{r_{\rm st}} \frac{P}{f(r) \sqrt{P^2 + f(r) (r/L)^{2(d-1)}}} dr \, ,
    \label{eq:t_CV_2}
\end{equation}
by the same reasoning as (\ref{eq:Vol BB AdS} and \ref{eq:t_CV_AdS}), with $t(P)$ now representing the boundary time at the stretched horizon $r_{\rm st}$.\footnote{\label{fnt:Sign of P}Notice the opposite sign in (\ref{eq:t_CV_2}) with respect to (\ref{eq:t_CV_1}). The reason for this difference is that $P$, as defined in (\ref{eq:conserved_momentum_CV}), takes values $P\leq 0$ in the future region of the inflating patch, while $P\geq 0$ in the black hole patch, as the Killing vector corresponding to time translations adopts opposite orientations in these two cases. This determines the sign of the late time rate of growth of the CV proposal, as previously pointed out in \cite{Chapman:2018lsv}.}

\begin{figure}[t!]
    \centering
\subfloat[]{ \label{subfig:CV_case2}\includegraphics[width=0.45\textwidth]{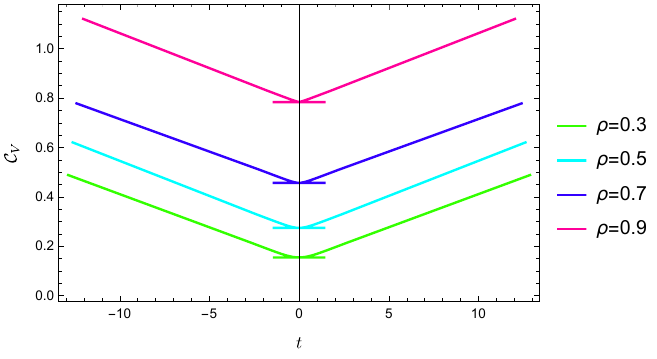}}\hfill \subfloat[]{\label{subfigb:CVrate_case2} \includegraphics[width=0.48\textwidth]{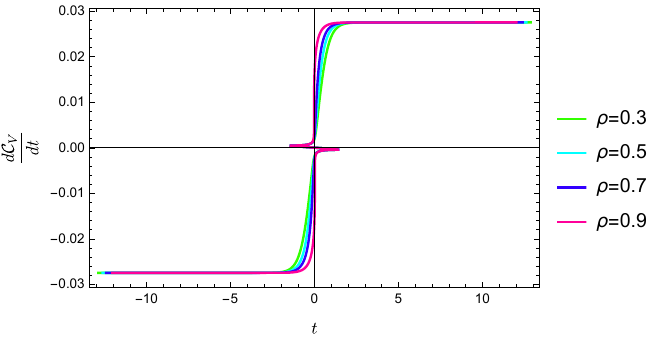}}
    \caption{Time dependence of (a): CV and (b): its rate for the prescription in Fig. \ref{fig:Ueff_2}. Different curves correspond to different values of the stretched horizon parameter.
    We have set $d=3$, $\mu = 0.14$, and $G_N L=1$. For any fixed time $t$, holographic complexity corresponds to the maximum of each colored curve.  }
    \label{fig:volume_BH}
\end{figure}

In Fig.~\ref{fig:volume_BH} we display the volume and its growth rate as functions of the boundary time $t$ for different values of the stretched horizon parameter $\rho$. For any value of $\rho$, at small times, a window exists in which for fixed $t$ there are three extremal surfaces. Again, the maximal surface is the one with larger $\left| P \right|$.
For larger times, only one extremal surface exists. One can also find that (\ref{eq:t_Pmax_case2}) still holds in this case, following the same steps as in Sec. \ref{ssec:CV}, the extremal volume surface approaches $r_f$ at late times.

\paragraph{Growth rate.}
In contrast to case 1, the extremal volume slices do not explore the inflating region. Instead, the same type of late-time growth found for asymptotically AdS black holes trivially follows, as we analyzed in Sec. \ref{ssec:CV}, namely (\ref{eq:dV dt all t AdS}) for $k=1$, which at late times leads to the characteristic late time linear growth
\begin{equation}
    \lim_{t \to +\infty} \frac{dV}{dt} = \Omega_{d-1} L^{d-1} \sqrt{U_{\rm max}} =
    \Omega_{d-1} \sqrt{- f(r_f)} \, r_f^{d-1} \, ,
    \label{eq:ratelate_CV_case2}
\end{equation}
where $r_f$ is defined in \eqref{eq:def_rf_caseAdSCV}.

\subsubsection*{Case 3}

 \begin{figure}[t!]
  	\centering
  	\subfloat[]{ \label{subfig:Penrose_CV_case3} \includegraphics[width=0.5\textwidth]{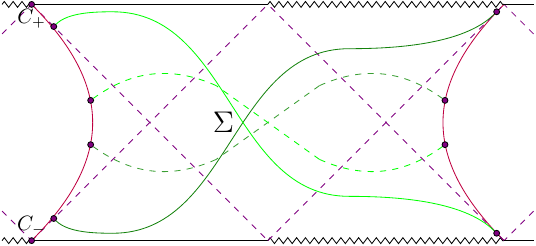}}\qquad\subfloat[]{ \label{subfig:potential_CV_case3} \includegraphics[width=0.4\textwidth]{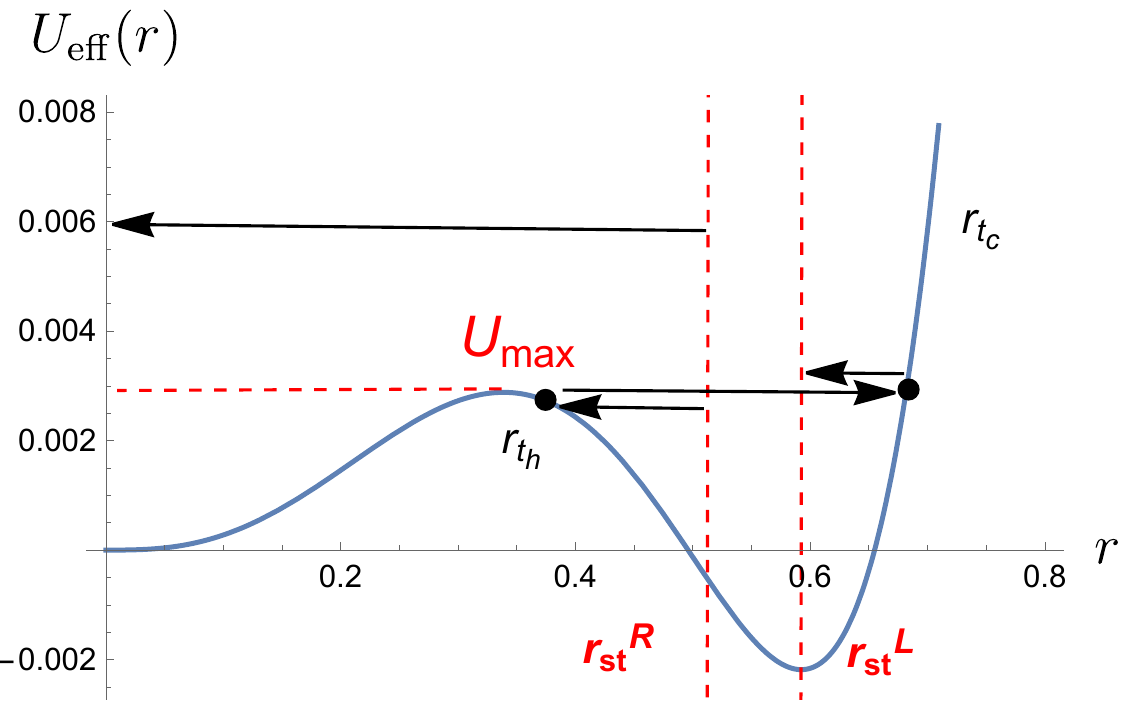}}
  	\caption{(a): The left and right stretched horizons are placed in non-consecutive static patches. According to the prescription \ref{rule}, the maximal volume surfaces $\Sigma$ can evolve to arbitrary late boundary time $t_R=-t_L$. The dark (light) green extremal surfaces have $P>0$ ($P<0$). The dashed curves correspond to finite boundary time, while the solid curves are achieved at late (early) times.
    (b): Qualitative behavior of the effective potential $U_{\rm eff}(r)$. For large $P^2$, the surface directly falls into the black hole singularity. For small enough $P^2$, the surface connects the two stretched horizons after going through two turning points $r_{t_c}$ and $r_{t_h}$.
    We have set $d=3$, $\mu= 0.187$, and $L=1$.}
  	\label{fig:Ueff_3}
  \end{figure}

We place the left and right stretched horizons at $r=r_{\rm st}^L$ and $r=r_{\rm st}^R$ into non-consecutive static patches, as in Fig.~\ref{fig:Ueff_3}.
Starting from the right stretched horizon $r_{\rm st}^R$, the maximal surface moves towards decreasing $r$ until it enters the black hole. For the surface to reach the left stretched horizon $r_{\rm st}^L$, the value of the conserved momentum $P$ must be properly chosen, see Fig.~\ref{fig:Ueff_3}.
For $P^2 > U_{\rm max}$, the maximal surface does not encounter a potential barrier inside the black hole, so it falls into the singularity.\footnote{Let us remind that $U_{\rm max}$ is the value of the effective potential $U_{\rm eff} (r)$ at the location of its local maximum. } Instead, for $0 \leq P^2 \leq U_{\rm max}$, the surface passes through a turning point $r_{t_h}$ and turns to the direction of increasing $r$.
Next, it explores the inflating region, where it meets another turning point $r_{t_c}$.
Finally, it attaches to the left stretched horizon $r_{\rm st}^L$. 
In summary, for $0 \leq P^2 \leq U_{\rm max}$ there are two turning points, one inside the inflating region and one inside the black hole. 
In this case, the volume reads\footnote{One can straightforwardly generalize the arguments in case of $n$ cosmological and black hole patches upon the replacement $\int_{r_{t_h}}^{r_{t_c}}\rightarrow n\int_{r_{t_h}}^{r_{t_c}}$. However, for clarity, we will limit the discussion to $n=1$ in case 3, and to $n=2$ in cases 4-5. \label{fnt:case3 CV}}
\begin{align}\label{eq:Vol BC}
    \frac{V(P)}{\Omega_{d-1}} \, =& \frac{1}{L^{d-1} } \qty(\int_{r_{t_h}}^{r_{\rm st}^R} +
    \int_{r_{t_h}}^{r_{t_c}}+  \int_{r_{\rm st}^L}^{r_{t_c}})\frac{r^{2(d-1)}}{\sqrt{P^2 + f(r) (r/L)^{2(d-1)}}} dr \, .
\end{align}
To determine the boundary time as a function of the conserved momentum $P$, it is convenient to introduce the function
\begin{equation}
    \tau_{\pm}(r,P) \equiv \frac{-P \pm \sqrt{P^2 + f(r) (r/L)^{2(d-1)}}}{f(r) \sqrt{P^2 + f(r) (r/L)^{2(d-1)}}} \, .
    \label{eq:def_tau}
\end{equation}
For a fixed $P$, the maximal surface explores the future (past) black hole interior and the past (future) inflating region.
We define the anchoring time at the left and right stretched horizon as $-t_L$ and $-t_R$ respectively, according to the convention \ref{conv_times} and the orientation of the Killing vector $\partial_t$ depicted in Fig.~\ref{fig:WDW_BH_stretched}. 
In this way, we have
\begin{subequations}
\label{eq:identities_CV_case3}
\begin{equation}\label{eq:time contribution4}
    t_{t_h} -r^*(r_{t_h}) +t_R + r^*(r_{\rm st}^R) =
   \int^{r_{t_h}}_{r_{\rm st}^R} \frac{\dot{u}_-}{\dot{r}_-} dr =
   -\int^{r_{t_h}}_{r_{\rm st}^R} \tau_+(r,P) dr \, ,
\end{equation}
\begin{equation}
    t_{t_c} +r^*(r_{t_c}) - t_{t_h} - r^*(r_{t_h}) =
   \int^{r_{t_c}}_{r_{t_h}} \frac{\dot{v}_+}{\dot{r}_+} dr =
   \int^{r_{t_c}}_{r_{t_h}} \tau_+(r,P) dr \, ,
   \label{eq:identity2_CV_case3}
\end{equation}
\begin{equation}\label{eq:time contribution1}
   -t_L- r^*(r_{\rm st}^L) - t_{t_c} + r^*(r_{t_c}) =
   \int^{r_{\rm st}^L}_{r_{t_c}} \frac{\dot{u}_-}{\dot{r}_-} dr =
   -\int^{r_{\rm st}^L}_{r_{t_c}} \tau_+(r,P) dr \, ,
\end{equation}
\end{subequations}
where $t_{t_c} (t_{t_h})$ is the value of the time coordinate at the turning point $r_{t_c} (r_{t_h})$.
Notice that from \eqref{eq:def_tau} we get
\begin{equation}
    \int_a^b \tau_{\pm}(r,P) \, dr = \int_a^b \frac{-P}{f(r) \sqrt{P^2 + f(r) (r/L)^{2(d-1)}}} \, dr \pm r^*(b) \mp r^*(a) \, .
    \label{eq:tau_integral}
\end{equation}
Therefore, by summing the identities~\eqref{eq:identities_CV_case3}, after some algebra we obtain
\begin{align}
\label{eq:time L and R BC}
    t_R-t_L =& -\qty(\int_{r_{t_h}}^{r_{\rm st}^R} + \int_{r_{t_h}}^{r_{t_c}} +\int_{r_{\rm st}^L}^{r_{t_c}})\frac{P}{f(r) \sqrt{P^2 + f(r) (r/L)^{2(d-1)}}} \, dr \, .
\end{align}
We now have two main choices:
\begin{itemize}
    \item $t_R = t_L$, time grows in a symmetric way on the two stretched horizons. 
    In agreement with the time translation symmetry \eqref{eq:time_shift_symmetry}, the volume is time-independent.
    Indeed, all the surfaces with $t_R = t_L$ can be obtained from the $P=0$ solution at $t_R = t_L=0$ by applying the time translation \eqref{eq:time_shift_symmetry}.
    \item $t_R \neq t_L$, the anchoring times are not symmetric.
    The volume is time-dependent. 
    In particular, as argued around \eqref{eq:symmetric_times}, it is not restrictive to choose the antisymmetric configuration $-t_L=t_R=t/2$.
\end{itemize}

\begin{figure}[t!]
    \centering
    \includegraphics[width=0.6\textwidth]{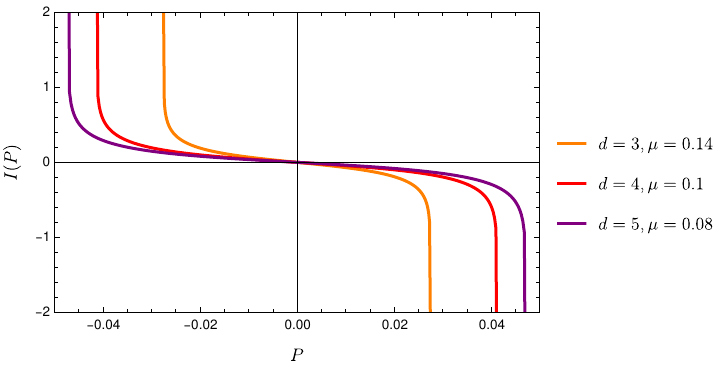}
    \caption{Plot of the integral $I$ in \eqref{eq:integral_deltat} as a function of the conserved momentum $|P| \leq \sqrt{U_{\rm max}}$, for various choices of the dimension $d$ and the mass parameter $\mu$. We fix $L=1$. }
    \label{fig:plot_deltat_case3}
\end{figure}

Before proceeding, we discuss the consistency of the identities \eqref{eq:identities_CV_case3}.
When $t_L=t_R$ and the stretched horizons are located at the same radial coordinate $r_{\rm st}^L = r_{\rm st}^R$, the extremal surfaces in the Penrose diagram are symmetric, therefore the turning points are located at $t_{t_h}=t_{t_c}=0$.
By plugging these conditions and the identity \eqref{eq:tau_integral} inside \eqref{eq:identity2_CV_case3}, we find the following constraint
\beq
I(P) = 0 \, , \qquad
I(P) \equiv \int_{r_{t_h}}^{r_{t_c}} \frac{P}{f(r) \sqrt{P^2 + f(r) (r/L)^{2(d-1)}}} dr \, .
\label{eq:integral_deltat}
\eeq
A priori, for any allowed value of $P$, one has to compute the turning points $r_{t_c}, r_{t_h}$, and then check whether the constraint \eqref{eq:integral_deltat} is satisfied.
This step is performed numerically, for various choices of the spacetime dimensions and of the mass parameter, in Fig.~\ref{fig:plot_deltat_case3}. 
It is clear that the constraint only holds when $P=0$.
This result is consistent with the first bullet below \eqref{eq:time L and R BC}.
In other words, the volume is time-independent, and this case corresponds to the trivial evolution performed with the Killing vectors.

On the contrary, when $t_R=-t_L=t/2$, the extremal surface is \textit{not} symmetric in the Penrose diagram, therefore we are allowed to have $t_{t_c} \ne 0$ and $t_{t_h} \ne 0$.
In this way, there are three unknown variables $t_{t_c}, t_{t_h}, t$ that can be determined by using the three identities \eqref{eq:identities_CV_case3}.
The conserved momentum $P$ is not constrained except for the condition $|P| \leq \sqrt{U_{\rm max}}$, which defines the existence of two turning points.

\begin{figure}[t!]
    \centering
    \subfloat[]{\includegraphics[width=0.48\textwidth]{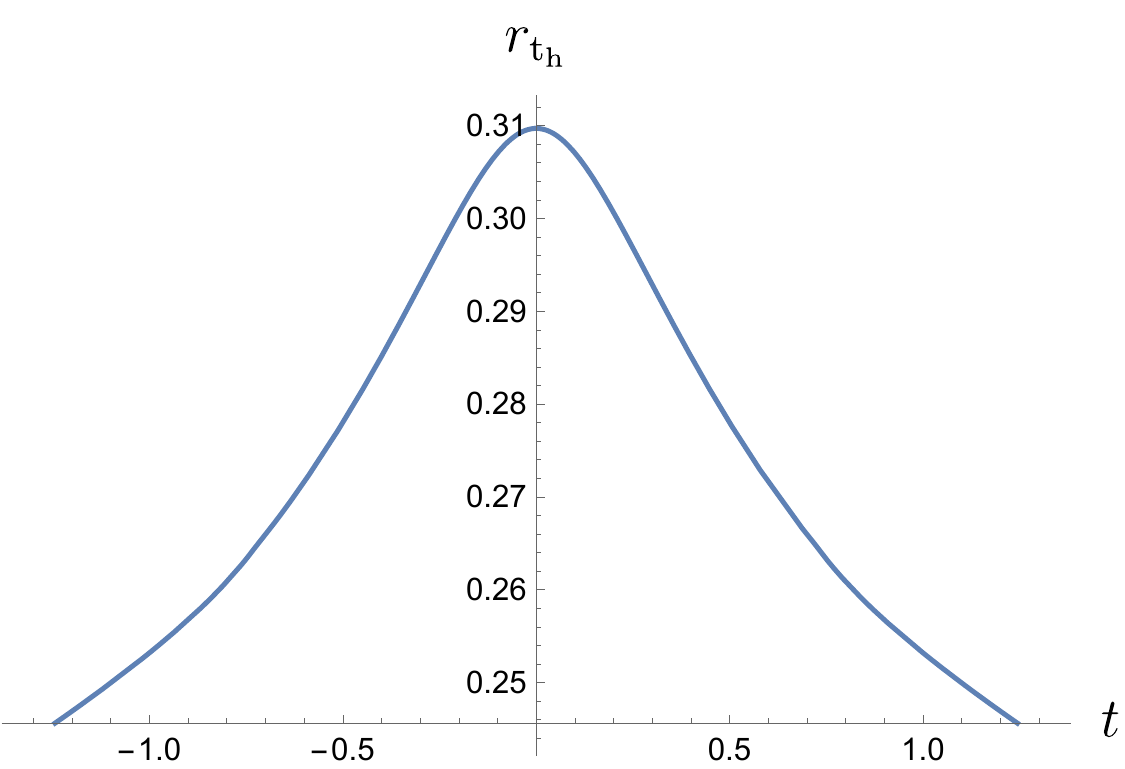}}\hfill\subfloat[]{\includegraphics[width=0.48\textwidth]{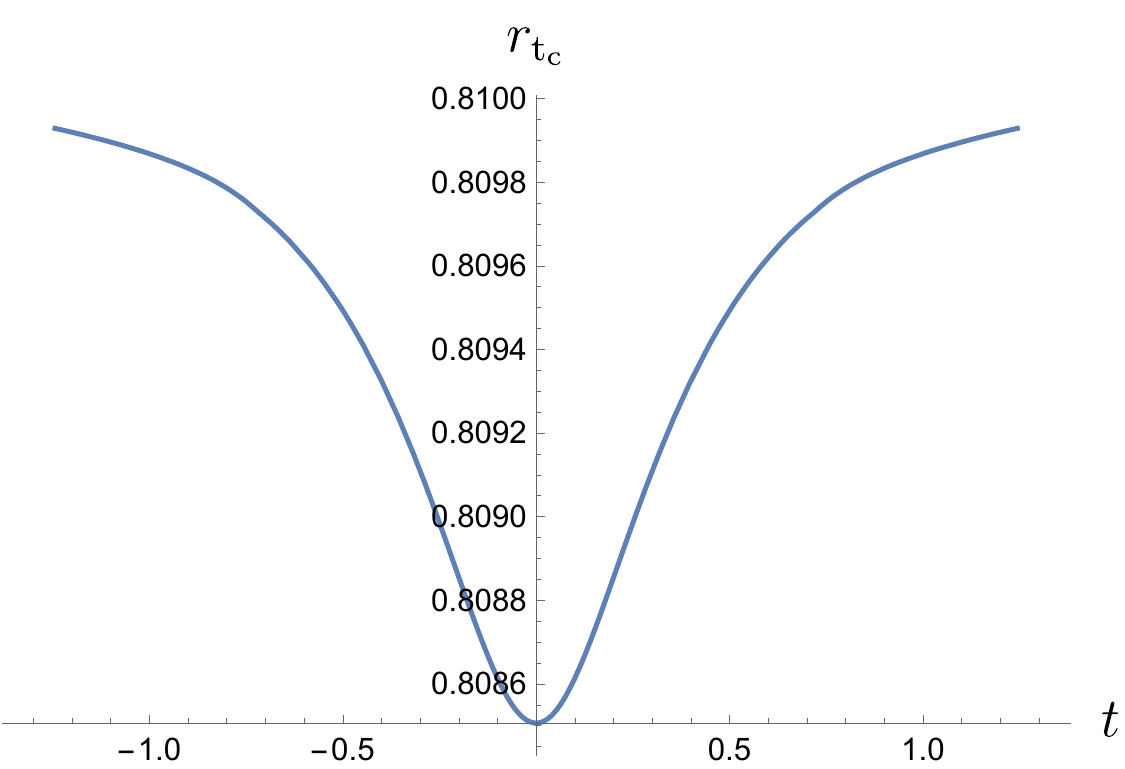}}\\
    \subfloat[]{\includegraphics[width=0.48\textwidth]{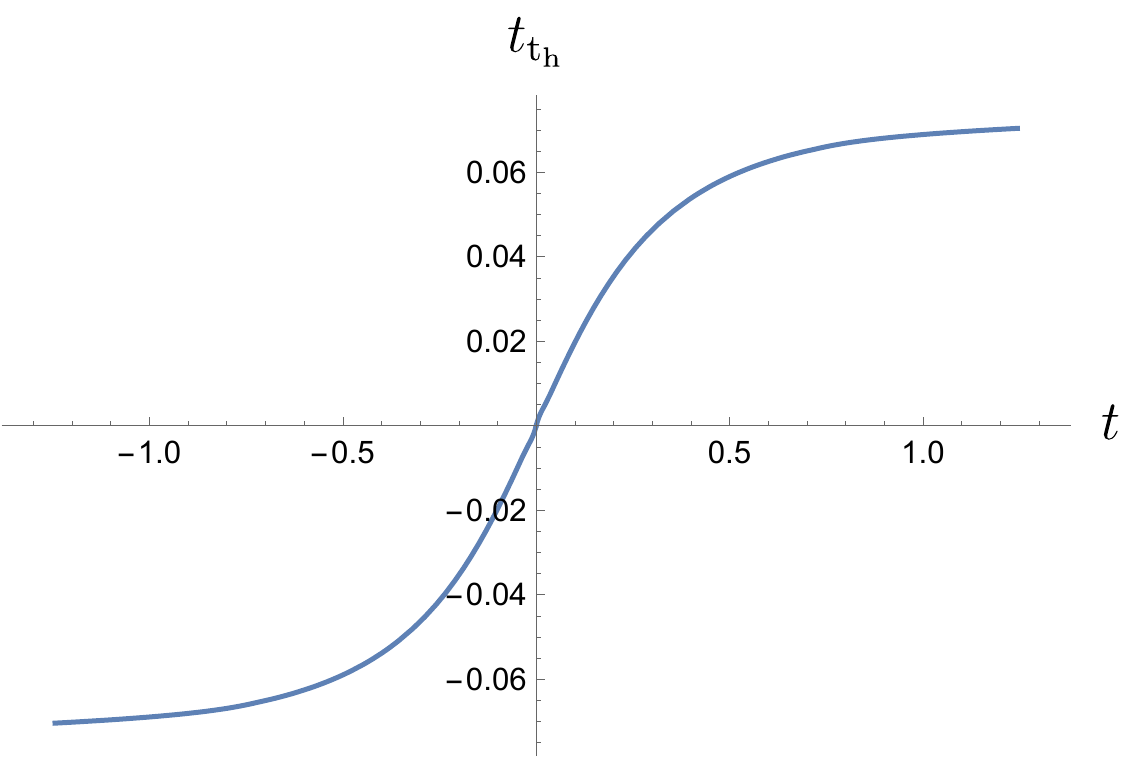}}\hfill\subfloat[]{\includegraphics[width=0.48\textwidth]{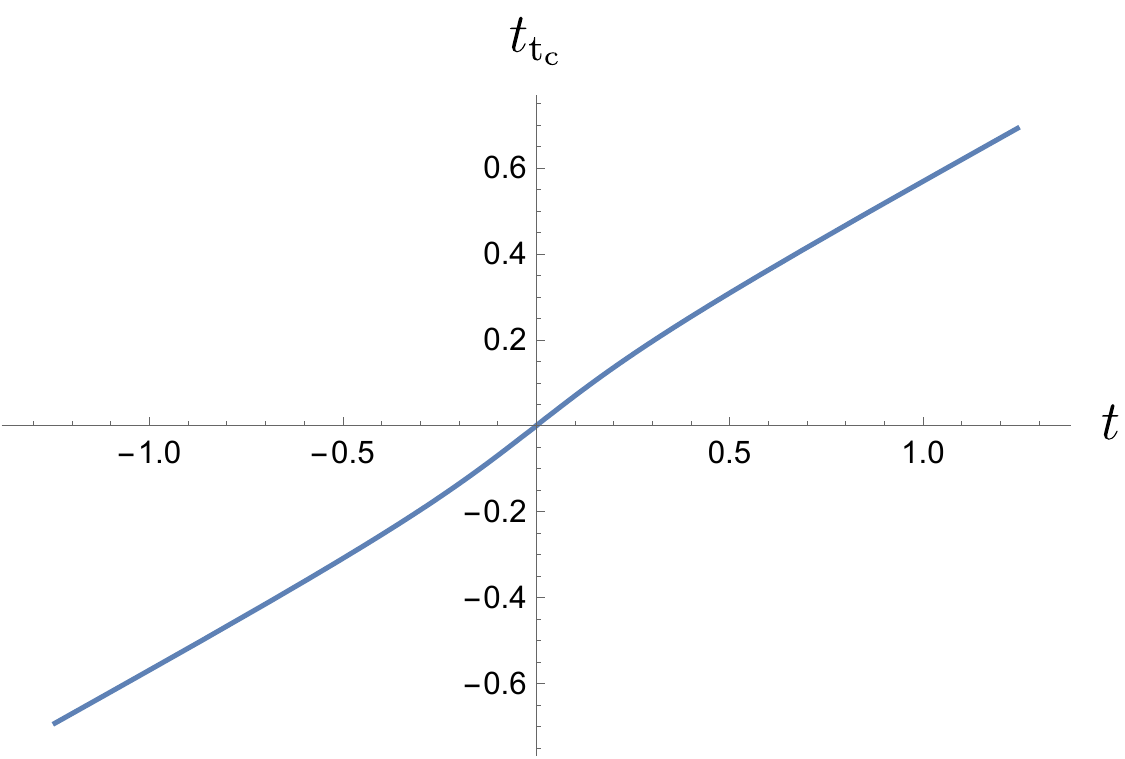}}
    \caption{Plots of the radial and time coordinates of the turning point in the black hole and cosmological patches as functions of the boundary time $t$ in case 3. We set $d=3, \mu=0.14$ and $L=1$.}
    \label{fig:turning point in time}
\end{figure}

Let us focus on the time-dependent (antisymmetric) choice.
Similar manipulations as in cases 1-2, applied to eqs.~\eqref{eq:Vol BC} and \eqref{eq:time L and R BC}, lead to
\beq
     \frac{dV}{dt} = \Omega_{d-1} L^{d-1} P~.
     \label{eq:CVrate_case3}
\eeq
We can understand the evolution of the extremal surfaces, schematically depicted in Fig.~\ref{fig:Ueff_3}, by studying the positions $(r_{t_c}, r_{t_h})$ and the time coordinates $(t_{t_c}, t_{t_h})$ of the turning points as functions of the boundary time. These plots are shown in Fig.~\ref{fig:turning point in time}.
At times $t>0$, we find that the turning point $r_{t_c}$ moves in the past exterior of the cosmological horizon towards the bottom-left corner $C^-$ of the Penrose diagram where the horizons, the singularity, and past infinity $\mathcal{I}^-$ meet. This corner is characterized by the condition $t_{t_c}=\infty$, while the corresponding radial coordinate is ill-defined.
At late times in the evolution, the turning point formally reaches the corner $C^-$ with finite value $\bar{r}_{t_c} > r_c$ of the radial coordinate, such that $U_{\rm eff}(\bar{r}_{t_c})=U_{\rm max}$.
On the other hand, the turning point $r_{t_h}$ moves towards the past singularity inside the interior of the black hole, until it asymptotically approaches a finite positive time $0<t_{t_h} < \infty$ and a final slice at constant radial coordinate $r_{f_h} < r_h$ such that $U_{\rm eff}'(r_{f_h})=0$.

Starting from \eqref{eq:time contribution4} and applying manipulations similar to \eqref{eq:t_Pmax_case2}, we find that the condition $U'(r_{f_h})=0$ implies $t_R (\sqrt{U_{\rm max}}) \rightarrow \infty$.
By imposing the antisymmetry $t_L=-t_R$ in the time evolution, we find that eqs.~\eqref{eq:identity2_CV_case3} and \eqref{eq:time contribution1} are solved when $t_{t_c} \rightarrow \infty$. 
These conditions define the late-time regime of the complexity evolution.
It is difficult to numerically plot the exact shape of the extremal surface at late times. The previous observations suggest that the extremal surface in the limit $t \rightarrow \infty$ becomes disconnected, composed of two parts: the single point at the corner $C^-$, plus a surface at constant radius $r_{f_h}$ in the black hole region.
A priori, it may be possible that the extremal surface remains connected in the strict limit $t \rightarrow \infty$.
However, one can at least rule out that the extremal surface in the inflating patch becomes null and hugs the cosmological horizon, since this would lead to an inconsistent limit of the spacelike surfaces determined by the equations of motion \eqref{eq:classical_motion_AdS}. We reserve the precise study of the extremal surface in the late time limit for future studies.  

\begin{figure}[t!]
    \centering
   \subfloat[]{ \label{subfig:CV_case3}
\includegraphics[width=0.48\textwidth]{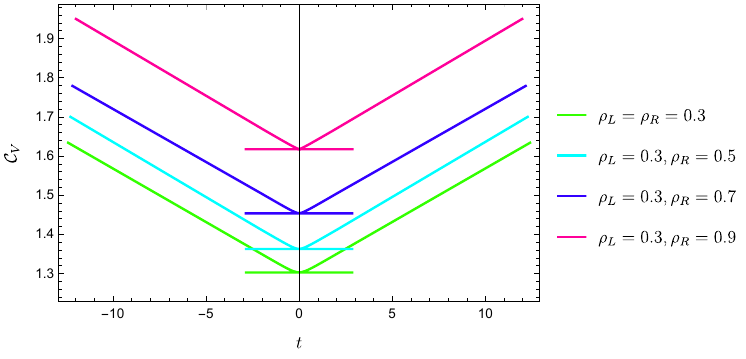}}\hfill \subfloat[]{\label{subfigb:CVrate_case3} 
\includegraphics[width=0.48\textwidth]{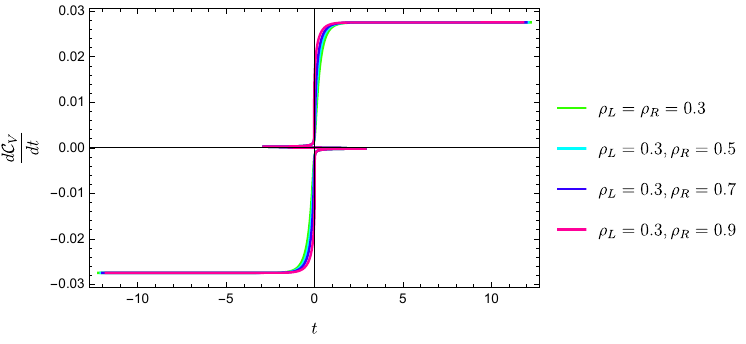}}
    \caption{Time dependence of (a): CV and (b): its rate for the prescription in Fig. \ref{fig:Ueff_3}. Different curves correspond to different values of the left and right stretched horizon parameters.
    It is understood that the complexity function is obtained by taking the maximum of each colored curve for any fixed time $t$.
    We have set $d=3$, $\mu = 0.14$, and $G_N L=1$. }
    \label{fig:volume_case3}
\end{figure}

Nonetheless, we find the rate of growth of the volume by plugging the value of $P=\sqrt{U_{\rm max}}$ inside \eqref{eq:CVrate_case3}. The result reads
\beq
\lim_{t\rightarrow\infty} \frac{dV}{dt}= \Omega_{d-1}\sqrt{-f(r_f)}r_f^{d-1}\, .
\label{CV_latetimes_case3}
\eeq
We conclude that CV conjecture approaches a linear increase at late times in case 3, characterized by the volume measure evaluated on the final slice inside the black hole. 
The full-time dependence of CV and its rate are plotted in Fig.~\ref{fig:volume_case3}.

\subsubsection*{Case 4}

\begin{figure}[t!]
  	\centering
  	\subfloat[]{ \label{subfig:extremal_surface_CV_case4} \includegraphics[width=0.5\textwidth]{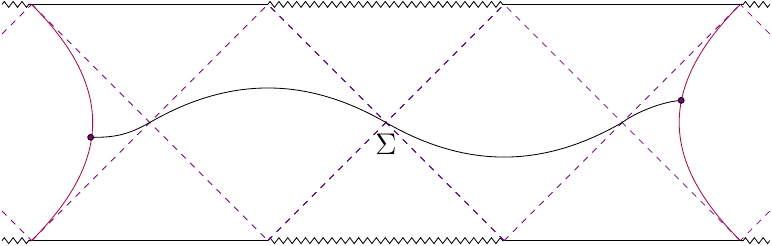}}
    \qquad\subfloat[]{ \label{potential_CV_case4} \includegraphics[width=0.4\textwidth]{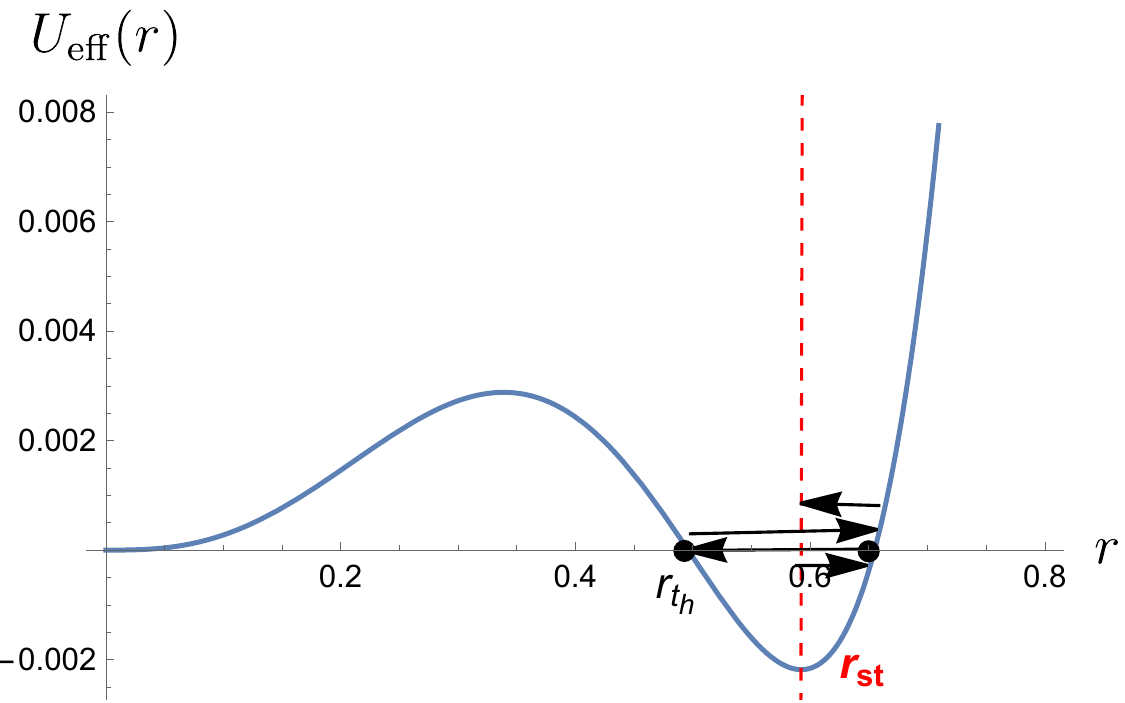}}
  	\caption{(a): The left and right stretched horizons are placed in non-consecutive static patches.
   Given the restriction \eqref{eq:integral_deltat}, only the conserved momentum $P=0$ is allowed for the surface $\Sigma$ depicted in the diagram.
   (b): Qualitative behavior of the effective potential $U_{\rm eff}(r)$. The surface $\Sigma$ connects the two stretched horizons after going through three turning points: $r_{t_c}$, $r_{t_h}$, and $r_{t_c}$ again. 
   In this case, the turning points coincide with the location of the cosmological and black hole horizons, respectively.
    We have set $d=3$, $\mu= 0.187$, and $L=1$.    }
  	\label{fig:Ueff_4}
  \end{figure}

The configuration is illustrated in Fig.~\ref{fig:Ueff_4}.
We locate the left and right stretched horizons at $r_{\rm st}^L = r_{\rm st}^R \equiv r_{\rm st}$ and, following the orientation of the Killing vector in Fig.~\ref{fig:WDW_case_45}, we define the left and right time as $-t_L$ and $t_R$, respectively.
Due to the time translation symmetry \eqref{eq:time_shift_symmetry}, it is not restrictive to consider the symmetric configuration \eqref{eq:symmetric_times}. 
An analysis of the effective potential in Fig.~\ref{fig:Ueff_4} implies that the conserved momentum needs to satisfy $0 \leq P^2 \leq U_{\rm max}$ (where $U_{\rm max}$ is the local maximum of the effective potential) in order to obtain a connected extremal surface anchored to the stretched horizons.    
In this regime, the expressions for the volume and boundary time can be found in a similar manner to eqs.~\eqref{eq:Vol BC} and \eqref{eq:time L and R BC}, but in this case there is an additional turning point in the inflating region of the second patch.
The symmetries of the configuration imply that the turning points in the cosmological regions are located at the same radial coordinate $r_{t_{c1}} = r_{t_{c2}} \equiv r_{t_c}$, therefore we find
\begin{equation}\label{eq:Vol CC B}
    \frac{V(P)}{\Omega_{d-1}} = \frac{2}{L^{d-1}} \qty(\int^{r_{t_c}}_{r_{\rm st}}+\int^{r_{t_c}}_{r_{t_h}}) \frac{r^{2(d-1)}}{\sqrt{P^2 + f(r) (r/L)^{2(d-1)}}} dr \, ,
\end{equation}
In performing the previous steps, we employed a smooth coordinate map in the static patch region that connects different copies of SdS space.
For a fixed $P$, the maximal surface explores the future (past) black hole interior and the past (future) inflating regions. 
Next, we study the difference of the null coordinates $u(v)$ on the extremal surface, moving from the right to the left of the Penrose diagram.
The following relations hold
\begin{subequations}
\label{eq:identities_CV_case4}
\begin{equation}\label{eq:time contribution1_case4}
   t_{t_c} -r^*(r_{t_c}) - t_R +r^*(r_{\rm st}^R)  =
   \int_{r_{\rm st}^R}^{r_{t_c}} \frac{\dot{u}_+}{\dot{r}_+} dr =
   -\int^{r_{\rm st}^R}_{r_{t_c}} \tau_-(r,P) dr \, ,
\end{equation}
\begin{equation}
  t_{t_h} +  r^*(r_{t_h}) -t_{t_c} - r^*(r_{t_c}) =
   \int_{r_{t_c}}^{r_{t_h}} \frac{\dot{v}_-}{\dot{r}_-} dr =
   \int_{r_{t_h}}^{r_{t_c}} \tau_-(r,P) dr \, ,
   \label{eq:identity2_CV_case4}
\end{equation}
\begin{equation}
  t_{t_c} -  r^*(r_{t_c}) -t_{t_h} + r^*(r_{t_h}) =
   \int_{r_{t_h}}^{r_{t_c}} \frac{\dot{u}_+}{\dot{r}_+} dr =
   \int_{r_{t_h}}^{r_{t_c}} \tau_-(r,P) dr \, ,
   \label{eq:identity3_CV_case4}
\end{equation}
\begin{equation}\label{eq:time contribution4_case4}
  -t_L + r^*(r_{\rm st}^L) - t_{t_c} + r^*(r_{t_c}) =
   \int^{r_{\rm st}^L}_{r_{t_c}} \frac{\dot{v}_-}{\dot{r}_-} dr =
    \int^{r_{t_c}}_{r_{\rm st}^L} \tau_-(r,P) dr \, ,
\end{equation}
\end{subequations}
where we momentarily kept $t_L, t_R$ arbitrary.
Since the turning points in the two inflating regions are located at the same radial coordinate, we find that the summation of  eqs.~\eqref{eq:identity2_CV_case4}-\eqref{eq:identity3_CV_case4}
leads to the same constraint obtained in \eqref{eq:integral_deltat}.
This constraint can also be achieved by considering either \eqref{eq:identity2_CV_case4} or \eqref{eq:identity3_CV_case4} separately, in the symmetric setting \eqref{eq:symmetric_times} where $t_{t_c}=t_{t_h}=0$. 
In analogy with our discussion in case 3 of this section, the vanishing of $I(P)$ forces $P=0$, which means that the maximal volume surface would only exist for a single boundary time in the symmetric configuration, namely $t_L=t_R=0$. A possible way to determine an extremal surface for any boundary time is to evolve its endpoints in the antisymmetric way $t_L=-t_R$. However, this leads to the trivial time evolution generated by the Killing vector $\partial_t$. 
This means that the maximal volume slice will remain with the same constant value
\begin{equation}
    V(0)=\frac{2\Omega_{d-1}}{L^{d-1}}\qty(\int_{r_{\rm st}}^{r_{t_c}}+\int_{r_{t_h}}^{r_{t_c}})\frac{r^{d-1}~\rmd r}{\sqrt{f(r)}}~.
\end{equation}
Thus, the choice $t_L=-t_R$ allows for an evolution (although trivial) of the codimension-one observable.

\subsubsection*{Case 5}
\begin{figure}[t!]
  	\centering
\subfloat[]{\includegraphics[width=0.5\textwidth]{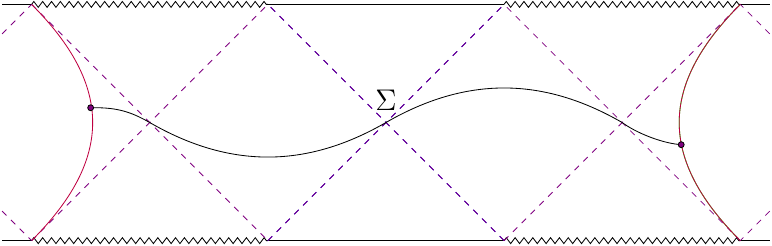}}\qquad\subfloat[]{\includegraphics[width=0.4\textwidth]{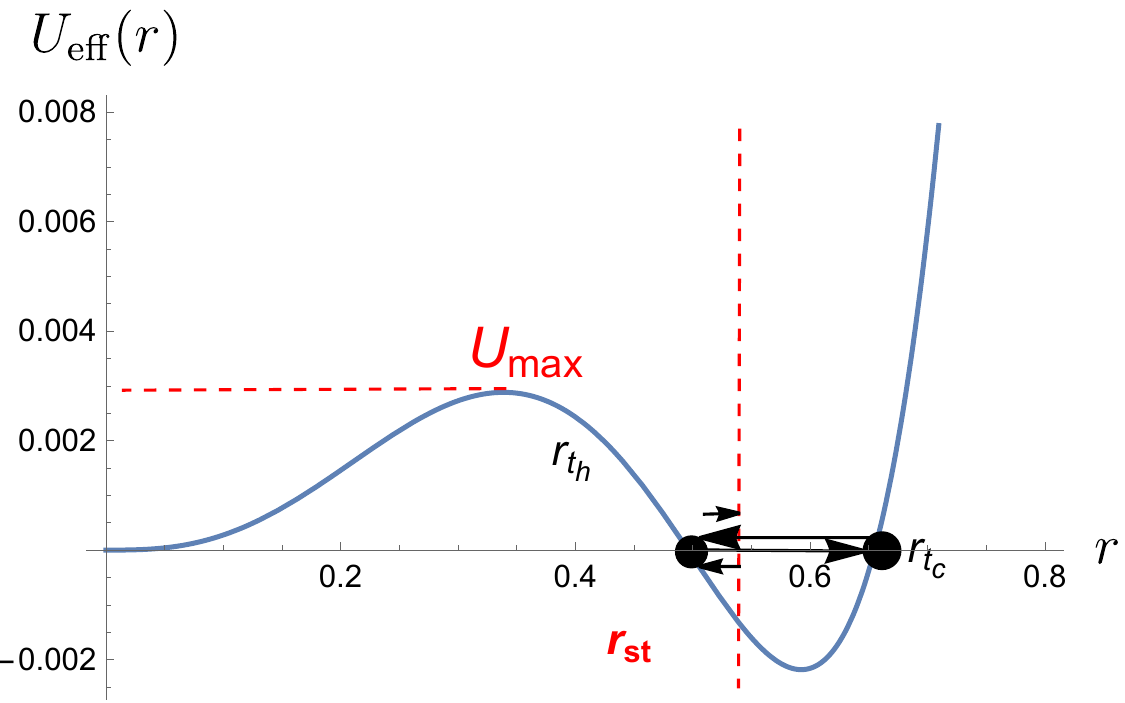}}
  	\caption{(a): The left and right stretched horizons are placed in non-consecutive static patches with a generic boundary time $t_R=-t_L$. 
   Due to the constraint \eqref{eq:integral_deltat}, only the conserved momentum $P=0$ is allowed for the surface $\Sigma$ depicted in the diagram.
    (b): Qualitative behavior of the effective potential $U_{\rm eff}(r)$. The surface $\Sigma$ connects the stretched horizons after going through three turning points: $r_{t_h}$, $r_{t_c}$, and $r_{t_h}$ again. 
   In this case, the turning points coincide with the location of the cosmological and black hole horizons, respectively.
    We have set $d=3$, $\mu= 0.187$, and $L=1$.   }
  	\label{fig:Ueff_5}
  \end{figure}

This configuration is illustrated in Fig.~\ref{fig:Ueff_5}. 
We take $r_{\rm st}^L = r_{\rm st}^R \equiv r_{\rm st}$ and we define the time on the left and right stretched horizons as $t_L$ and $-t_R$, respectively (see Fig.~\ref{fig:WDW_case_45}). Again, it is not restrictive to take the symmetric configuration \eqref{eq:symmetric_times}. Similarly to the previous cases, one can express
\begin{equation}\label{eq:Vol BB C}
    \frac{V(P)}{\Omega_{d-1}} = \frac{2}{L^{d-1}} \qty(\int_{r_{t_h}}^{r_{\rm st}}+\int_{r_{t_h}}^{r_{t_c}}) \frac{r^{2(d-1)}}{\sqrt{P^2 + f(r) (r/L)^{2(d-1)}}} dr \, .
\end{equation}
Furthermore, we obtain several identities by computing the null coordinates at various points along the extremal surface
\begin{subequations}
\label{eq:identities_CV_case5}
\begin{equation}\label{eq:time contribution1_case5}
   t_{t_h} -r^*(r_{t_h}) + t_R +r^*(r_{\rm st}^R)  =
   \int_{r_{\rm st}^R}^{r_{t_h}} \frac{\dot{u}_-}{\dot{r}_-} dr =
   \int^{r_{\rm st}^R}_{r_{t_h}} \tau_+(r,P) dr \, ,
\end{equation}
\begin{equation}
  t_{t_c} +  r^*(r_{t_c}) -t_{t_h} - r^*(r_{t_h}) =
   \int_{r_{t_h}}^{r_{t_c}} \frac{\dot{v}_+}{\dot{r}_+} dr =
   \int_{r_{t_h}}^{r_{t_c}} \tau_+(r,P) dr \, ,
   \label{eq:identity2_CV_case5}
\end{equation}
\begin{equation}
  t_{t_h} -  r^*(r_{t_h}) -t_{t_c} + r^*(r_{t_c}) =
   \int_{r_{t_c}}^{r_{t_h}} \frac{\dot{u}_-}{\dot{r}_-} dr =
   \int_{r_{t_h}}^{r_{t_c}} \tau_+(r,P) dr \, ,
   \label{eq:identity3_CV_case5}
\end{equation}
\begin{equation}\label{eq:time contribution4_case5}
  t_L + r^*(r_{\rm st}^L) - t_{t_h} + r^*(r_{t_h}) =
   \int^{r_{\rm st}^L}_{r_{t_h}} \frac{\dot{v}_+}{\dot{r}_+} dr =
    \int_{r_{t_h}}^{r_{\rm st}^L} \tau_+(r,P) dr \, ,
\end{equation}
\end{subequations}
where we momentarily allowed $t_L, t_R$ to be arbitrary.
By adding eqs.~\eqref{eq:identity2_CV_case5} and \eqref{eq:identity3_CV_case5}, we find again the constraint \eqref{eq:integral_deltat}. 
This implies that $P=0$, therefore only the solution $t_L=t_R=0$  is allowed in the non-trivial case determined by the prescription \ref{rule}.
At arbitrary boundary time, the only allowed configuration is the antisymmetric one (with $t_L=-t_R$) for which the volume is simply given by \eqref{eq:Vol BB C} with $P=0$.

\subsection{Complexity=Anything}\label{sec:CAny CMC}

\subsubsection*{Case 1}
\subsubsection*{Eternal late time growth}
\begin{figure}
    \centering
    \subfloat[]{\includegraphics[width=0.324\textwidth]{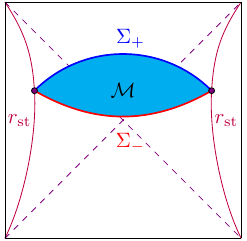}}\hspace{0.5cm}\subfloat[]{\includegraphics[width=0.62\textwidth]{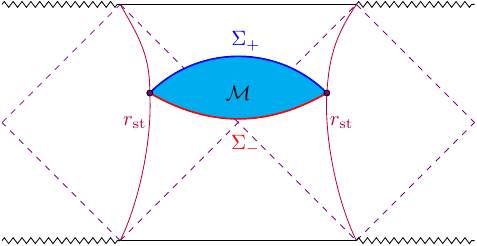}}
    \caption{Implementation of the codimension-one CAny proposals with CMC slices, $\Sigma_+$ (blue) and $\Sigma_-$ (red) in the bulk region $\mathcal{M}$ (cyan), as evaluation regions in the unperturbed (a) dS$_{d+1}$ space, (b) SdS$_{d+1\geq4}$ space. The notation follows our conventions in Fig. \ref{fig:SW SdS}.}
    \label{fig:CAny SdS}
\end{figure}

We will apply the techniques in (\ref{eq:C epsilon inter}) to evaluate (\ref{eq:C epsilon inter}) in the inflating patch of SdS spacetime. See Fig \ref{fig:CAny SdS} for an illustration of the proposal.
We will consider the coordinate $r\in[r_{\rm st},~r_t]$ where $r_{\rm st}\in[0,~r_c]$. Without loss of generality, we consider the case of a symmetric time evolution \eqref{eq:symmetric_times} and we locate the stretched horizons at the same radial coordinate $r_{\rm st} = r^L_{\rm st}=r^R_{\rm st}$, and set $K_{\epsilon}=-\epsilon\abs{K}$.
The codimension-one observable in \eqref{eq:C epsilon inter} can be expressed as
\begin{equation}\label{eq:Anything complexity cod1}
\mathcal{C}^\epsilon=\frac{2\Omega_{d-1} L^{d-2}}{G_N} \int_{{r}_{\rm st}}^{{r}_{t_c}} \frac{a(r)~ {\qty(\frac{r}{L})}^{2(d-1)}  }{\sqrt{-\mathcal{U}(P_{\epsilon},\,{r})}} \, dr \, .
\end{equation}
Meanwhile, the evolution of the boundary time follows from
\begin{equation}\label{eq: t CAny}
\begin{aligned}
t&=\int_{\Sigma_\epsilon} \rmd r\frac{\dot{v}-\dot{r}/{f(r)}}{\sqrt{-\mathcal{U}(P_v^\epsilon,\,r)}}= 2  \int_{{r}_{\rm st}}^{{r}_{t_c}} \frac{ P_{\epsilon}- \epsilon L\frac{\abs{K}}{d} \qty(\frac{r}{L})^d  }{f(r)\,\sqrt{-\mathcal{U}(P_{\epsilon}, r)}} \, dr \, .
\end{aligned}
\end{equation}
This allows to bring the integral (\ref{eq:Anything complexity cod1}) to the form
\begin{equation}\label{eq:CAny late time C1}
    \begin{aligned}
    {\mathcal{C}^\epsilon}=&\frac{\Omega_{d-1} L^{d-2}}{G_N} \, a(r_{t_c})\sqrt{-f(r_{{t_c}})\qty(\frac{r_{{t_c}}}{L})^{2(d-1)}}~t\\
    &+\frac{2\Omega_{d-1} L^{d-2}}{G_N}\int_{r_{\rm st}}^{r_{{t_c}}}  dr \, \frac{a(r)f(r)\,\qty(\frac{r}{L})^{2(d-1)}-\zeta(r)}{f(r)\sqrt{-\mathcal{U}(P_\epsilon,\,r)}} ~,
\end{aligned}
\end{equation} 
where
\begin{equation}
    \zeta(r)=a(r_{t_c})\sqrt{-f(r_{{t_c}})\qty(\frac{r_{\rm t_c}}{L})^{2(d-1)}}\left(P_\epsilon -\frac{\epsilon {L~\abs{K}}}{d} \qty(\frac{r}{L})^{d} \right)~.
\end{equation}
As discussed in \cite{Aguilar-Gutierrez:2023zqm}, one can perform an additional minimization between $\mathcal{C}^{\pm}$ to define a time-reversal invariant observable in asymptotically dS space that avoids a hyperfast growth. For our purposes, it is sufficient to note that at late (early) times, the knowledge of $\mathcal{C}^{-} (\mathcal{C}^+)$ will be sufficient to determine the complexity, while the other surface might not be defined.

We can then straightforwardly take the time derivative,
\begin{align}
   {\dv{\mathcal{C}^\epsilon}{t}}=&\frac{\Omega_{d-1} L^{d-2}}{G_N} \, a(r_{t_c})\sqrt{-f(r_{{t_c}})\qty(\frac{r_{{t_c}}}{L})^{2(d-1)}}\\
    &+\frac{2\Omega_{d-1} L^{d-2}}{G_N}\dv{P_v^\epsilon}{t}\int_{r_{\rm st}}^{r_{{t_c}}}  dr \, \frac{a(r)f(r)\,\qty(\frac{r}{L})^{2(d-1)}-\zeta(r)}{f(r)\sqrt{-\mathcal{U}(P_\epsilon,\,r)}} ~.\nonumber
\end{align}
Thus, in the $t\rightarrow\infty$ regime we then recover a very similar expression to (\ref{eq:Vol late timesAdS}):
\begin{equation}\label{eq:Vol late times}
    \lim_{t\rightarrow\infty}\dv{t} \mathcal{C}^\epsilon \simeq \frac{\Omega_{d-1}L^{d-2}}{G_N}\sqrt{-f(r_{f}) \qty(\frac{r_f}{L})^{2(d-1)}}\,a(r_{f})~\text{ with }r_{f}\equiv\lim_{t\rightarrow \infty}r_{t}~.
\end{equation}
where $r_f$ is a local maximum, with the same constraints as (\ref{eq:extr potentialAdS}), resulting in the same relation as (\ref{eq:rt locationAdS}) (the only difference being the blackening factor). Let us then label the following function:
\begin{equation}\label{eq:rt location}
\begin{aligned}
    W(r_f,\,K_\epsilon)&\equiv 4 r_{f}f\left(r_{f}\right)\left((d-1)f'\left(r_{f}\right)+K_\epsilon^2 r_{f}\right)+4(d-1)^2f\left(r_{f}\right){}^2+r_{f}^2f'\left(r_{f}\right){}^2
\end{aligned}
\end{equation}
When there are no roots to the above function in the range $r_c \leq r_f < \infty$, \eqref{eq:Vol late times} does not hold, and complexity presents instead a hyperfast growth in the inflating patch.
Indeed, the latter behavior happens when $K_{\epsilon}=0$, leading to the CV conjecture studied in Sec. \ref{ssec:CV}.

The novelty carried by the class of holographic proposals \eqref{eq:Volepsilon} is the existence of a range of extrinsic curvatures for the CMC slices  $\abs{K}\geq K_{\rm crit}$ such that the relation \eqref{eq:rt location} admits a turning point located beyond the cosmological horizon, and the hyperfast growth is avoided.
At the special values of the mass parameter $\mu=0$ (pure dS space) and $\mu=\mu_N$ (extremal case), the roots $r_f$ of the function $W(r_f,\,K_\epsilon)$ can be found explicitly
\begin{align}
\label{eq:rf dS}
    \qty(\tfrac{{r}^{\rm (dS)}_f}{L})^2 &= \tfrac{{K_\eps}^2L^2-2 d(d-1)\pm|{K}|L \sqrt{{K_\eps}^2L^2-4(d-1)}}{2( {K_\eps}^2- d^2)}~,\quad\abs{K}L\geq 2\sqrt{d-1}~;\\\label{eq:rf N}
    {r}^{\rm (N)}_f&=\sqrt{\frac{d-2}{d}}L~,\quad\abs{K}L\geq\sqrt{d}~.
\end{align}

\begin{figure}
\centering
\subfloat[]{\label{subfig:CMC_slices_case1} \includegraphics[height=0.3\textwidth]{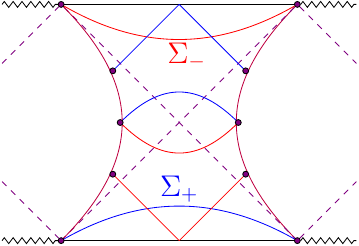}}\hspace{0.75cm}\subfloat[]{\label{subfig:Ueff_CAny_case1}  \includegraphics[height=0.3\textwidth]{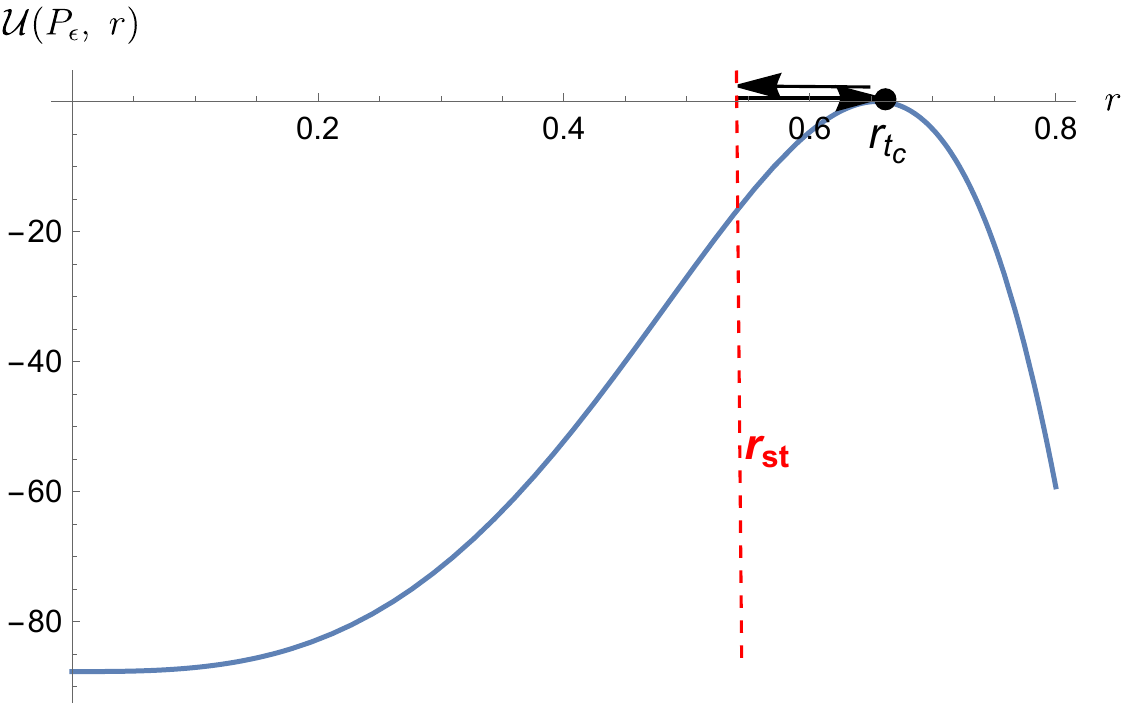}}
    \caption{Case 1: (a) Representation of the profile of the CMC slices $\Sigma_+$ (blue) and $\Sigma_-$ (red) when $\abs{K}\geq K_{\rm crit}$ for SdS space. (b) Effective potential (\ref{eq:U CAny}). A particle coming from $r_{\rm st}$ will be reflected at the turning point $r_{t_c}$ if it exists; otherwise it will fall into timelike infinity $\mathcal{I}^{\pm}$. Numerical values are the same as in Fig.~\ref{fig:Ueff_1}, in addition to $P_\epsilon=-9.37$ and $\abs{K}=100$.  }
    \label{fig:CMC1}
\end{figure}

\begin{figure}
    \centering
   \subfloat[]{\includegraphics[width=0.48\textwidth]{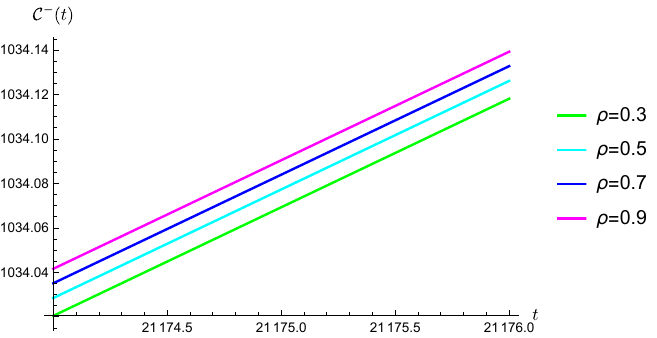}}\hspace{0.25cm}
   \subfloat[]{\includegraphics[width=0.48\textwidth]{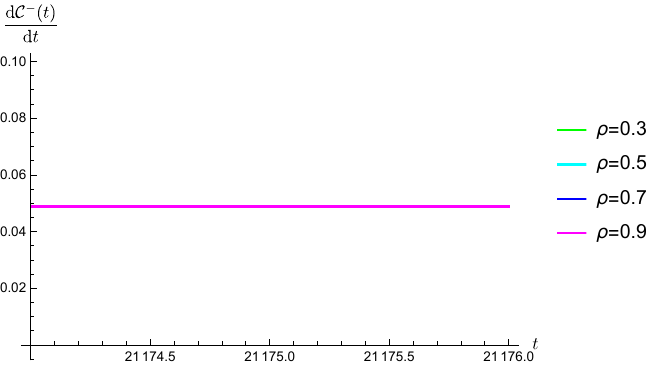}}
    \caption{Case 1: Evaluation of (a) the late-time behavior of $\mathcal{C}^-$ determined by \eqref{eq:CAny late time C1} and of (b) its rate of growth as a function of time. We take $a(r)=1$, $K_-=100/L$, besides the same parameters as in Fig.~\ref{fig:volume_cosmo}. Notice that all the different choices of $\rho$ lead to the same late-time rate of growth.}
    \label{fig:CAny1}
\end{figure}

The evolution of the CMC slices in the SdS background is shown in Fig.~\ref{subfig:CMC_slices_case1}. 
We depict the effective potential $\mathcal{U}$ in Fig.~\ref{subfig:Ueff_CAny_case1} for a specific value of the conserved momentum.
This plot shows that at the corresponding boundary time, determined according to \eqref{eq: t CAny}, there is a single turning point corresponding to the existence of a root of the effective potential.
The late-time behavior of complexity and its rate of growth are depicted in Fig.~\ref{fig:CAny1}, referring to the case $K_{\epsilon} \geq K_{\rm crit}$ and for a trivial function $F=1$ in \eqref{eq:Volepsilon}.
As anticipated below \eqref{eq:rt location}, the main feature is that CAny displays a linear increase with a rate independent of $\rho$, the value that determines the location of the stretched horizons.

Let us now show that for a generic SdS black hole spacetimes ($d\geq3$), there will always be a real root to (\ref{eq:rt location}). One can evaluate $W(r_f,\,K_\epsilon)$ in (\ref{eq:rt location}) with the roots in (\ref{eq:rf dS}, \ref{eq:rf N}) while keeping the mass of the black hole arbitrary. We will denote $m\equiv \mu/\mu_N\in[0,\,1]$, such that we may express:
\begin{align}
    &W\qty({r}^{\rm (dS)}_f,\,2\sqrt{d-1})=\frac{4 m  \left((d-2)^d d^2m-4 (d-2)^2 ((d-1) d)^{d/2}\right)}{(d-2)^{4-d} (d-1)^{d-2} d^{d}}~,\label{eq:HdS}\\
    &W\qty({r}^{\rm (N)}_f,\,K_\epsilon)=\frac{4(1-m)\qty(2K_\epsilon^2(d-2)+d^2(1-m))}{d^2}~.\label{eq:HN}
\end{align}
Notice that (\ref{eq:HdS}) is clearly negative for all $d\geq3$ and $m\in(0,\,1)$, while (\ref{eq:HN}) is positive. Moreover, as we increase $\abs{K}L>2\sqrt{d-1}$, $W\qty({r}^{\rm (dS)}_f,\,K_\epsilon)$ becomes more negative in (\ref{eq:rt location}). Then, according to the \emph{intermediate value theorem}, there will exist at least a real root ${r}_f\in \qty[{r}^{\rm (dS)}_f,\,{r}^{\rm (N)}_f]$ for general SdS$_{d+1}$ space.

On the other hand, since we have allowed $a(r)$ to be an arbitrary function in (\ref{eq:Vol late times}), we see that when $r_f\rightarrow\infty$\footnote{This condition is satisfied in (\ref{eq:rf dS}) when $K_\epsilon=d$ in dS$_2$ and (S)dS$_3$ space.} there would be arbitrary types of late-time growth for $C^\epsilon$ depending on the particular choice of $a(r)$. For instance, the case $a(r)=1$ leads to late-time exponential behavior when $r_f\rightarrow\infty$; meanwhile, having a different degree of divergence in (\ref{eq:Vol late times}) would lead to enhancement or decrease in the late-time growth. For $d = 3$ and higher we find finite $r_{f}$ for $\abs{K} = K_{\rm{crit}}$, which translates to the linear growth.

However, for this to be a valid CAny proposal, we require also a modification, as explained below.

When one evaluates the early time evolution in (\ref{eq:Vol late times}) $t\rightarrow -\infty$, there is a sign flip in $K_\epsilon\rightarrow-K_\epsilon$. As a result, the rate of growth of the CAny observables at early and late times does not coincide for a given CMC slice. The future or past growth would be given by (\ref{eq:Vol late times}), while the other generates hyperfast growth. This is illustrated in Fig. \ref{subfig:CMC_slices_case1}.
\begin{figure}
\centering
\subfloat[]{\includegraphics[width=0.33\textwidth]{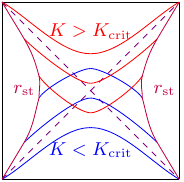}}\hspace{0.2cm}\subfloat[]{\includegraphics[width=0.64\textwidth]{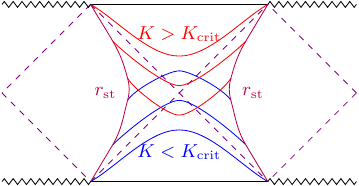}}
    \caption{Time symmetric complexity proposal~\eqref{eq:switchback result} allowing for early- and late-time linear growth in (a) empty ${\rm dS}_{d+1}$ and (b) ${\rm SdS}_{d+1\geq4}$. The CMC with ${K}_\epsilon < -K_{\rm crit}$ dominates in the past, shown in blue; while the CMC with ${K}_\epsilon > K_{\rm crit}$ dominates in the future.}
    \label{fig:timesymmetric}
\end{figure}

If we want to assign a Nielsen unitary complexity~\cite{Nielsen1,Chapman:2021jbh} interpretation to this setting, then the complexity of a unitary is the same as its inverse. This implies time asymmetric quantities in time-symmetric setups either do not capture (this type of) complexity or the considered time evolution is not unitary. The same occurs for CAny proposals on CMC slices even for AdS Schwarzschild with the location of the early/late turning point not being time-reflection symmetric. Importantly, the switchback effect is not respected in this case, as one requires a cancellation between early and late-time contributions to the complexity growth. We will make this more explicit below.

\subsubsection{The switchback effect}
We will study the set of observables (\ref{eq:Volepsilon}, \ref{eq:regions CMC}) in the multiple shockwave geometry (\ref{eq:metric SW alpha}), as displayed in Fig. \ref{fig:many SWs}.

\begin{figure}
    \centering
    \subfloat[]{\includegraphics[width=0.32\textwidth]{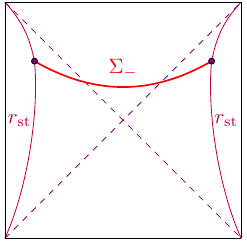}}\hfill\subfloat[]{\includegraphics[width=0.32\textwidth]{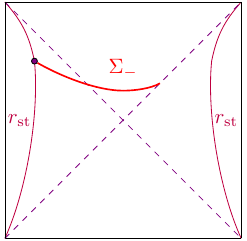}}\hfill\subfloat[]{\includegraphics[width=0.32\textwidth]{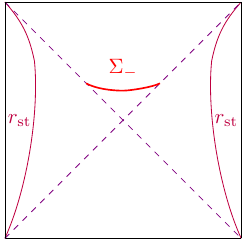}}
    \caption{Different configurations of the extremal complexity surface $\Sigma_\epsilon$ appearing in (\ref{eq:multiple SW}) and illustrated for $\epsilon=-$ in SdS$_3$ space. (a) $\mathcal{C}^-(t_L,\,t_R)$; (b) $\mathcal{C}^-(V_R,~t_L)$; and (c) $\mathcal{C}^-(V_R,~U_L)$.}
    \label{fig:anchoring}
\end{figure}

Accounting for the sign of the shift in the backreacted metric (\ref{eq:metric SW alpha}), the functional (\ref{eq:Volepsilon}) has an additive property under these insertions in the strong shockwave limit \cite{Belin:2021bga,Belin:2022xmt},
\begin{equation}\label{eq:multiple SW}
\begin{aligned}
    {\mathcal{C}}^\epsilon(t_L,\,t_R)=&\mathcal{C}^\epsilon(t_R,\,V_1)+\mathcal{C}^\epsilon(V_1-\alpha_1,\,U_2)+\dots\\
    &+\mathcal{C}^\epsilon(U_{n-1}+\alpha_{n-1},\,V_{n})+\mathcal{C}^\epsilon(V_{n}-\alpha_{n},\,t_L)
\end{aligned}
\end{equation}
where all the contributions $\mathcal{C}^\epsilon(\cdot,\,\cdot)$ follow the same definition (\ref{eq:Volepsilon}), but $\Sigma_\epsilon$ is anchored between endpoints that are located either on the left/right cosmological horizon, $r_c$, or on the stretched horizon $r_{\rm st}$. The different cases are illustrated in Fig. \ref{fig:anchoring}.
The procedure to evaluate (\ref{eq:multiple SW}) follows straightforwardly as we have described in Sec. \ref{sec:CAny CMCAdS}, with the modification $\alpha_i\rightarrow-\alpha_i$, $k=1$, setting $r_h\rightarrow r_c$, $r_{\rm bdy}\rightarrow r_{\rm st}$ and inverting the integration ranges. 

The same procedure as in Sec. \ref{sec:CAny CMCAdS} then leads to the late-time contributions:
\begin{align}
{\mathcal{C}^\epsilon}(t_R,~V_L)&=\frac{\Omega_{d-1}L^{d-1}}{G_N}P_{\infty}^\epsilon\log{V_L\rme^{t_R/L}}~,\\
{\mathcal{C}^\epsilon}(V_R,~U_L)&=\frac{\Omega_{d-1}L^{d-1}}{G_N}P_{\infty}^\epsilon\log{U_LV_R}~,\\
{\mathcal{C}^\epsilon}(V_R,~t_L)&=\frac{\Omega_{d-1}L^{d-1}}{G_N}P^\epsilon_{\infty}\log{\rme^{t_L/L}V_R}~,
\end{align}
with $P_\infty^\epsilon$ in (\ref{eq:P infty epsilon}); and the early time contribution is given by
\begin{equation}
{\mathcal{C}^\epsilon}(V_L,\,U_R)=\frac{\Omega_{d-1}L^{d-1}}{G_N}P_{-\infty}^\epsilon\log{V_LU_R}~,
\end{equation}
with $P_{-\infty}^\epsilon$ in (\ref{eq:P -infty epsilon}). 

As mentioned above, for $t\rightarrow -\infty$, there is a sign flip in $K_\epsilon\rightarrow-K_\epsilon$, and $r_I$ is then a solution to (\ref{eq:extr potentialAdS}, \ref{eq:rt location}) with the appropriate modification of $K_\epsilon$, the CMC slices that display late time growth in the far past/future also lead to hyperfast growth in the future/past respectively, which means that there is no maximal turning point in either case. This implies \textit{no switchback effect} if we evaluate the same observable $\mathcal{C}^\epsilon$ at all times.

Let us then consider a protocol where we evaluate (\ref{eq:C epsilon inter}) over different CMC slices in the past and future, such that there are always solutions $r_f$ and $r_I$ with respect to the stretch horizon evolution. (\ref{eq:multiple SW}) then transforms into
\begin{equation}\label{eq:Cepsilon tot}
\begin{aligned}
{\mathcal{C}^\epsilon}(t_L,\,t_R)\simeq& \frac{\Omega_{d-1}L^{d-1}}{G_N}\Bigl[P_{\infty}^\epsilon \log(V_1\rme^{t_R/L})+P_{-\infty}^\epsilon\log(V_1-\alpha_1)U_2\\
&+P_{\infty}^\epsilon\log(U_2+\alpha_2)V_3+\dots+P_\infty^\epsilon\log((V_n-\alpha_n)\rme^{t_L/L})\Bigr]~.
\end{aligned}
\end{equation}
Extremizing (\ref{eq:Cepsilon tot}) with respect to an arbitrary interception point ($V_i$, $U_i$), just as in (\ref{eq:cond SW1}) leads us to (\ref{eq:cond SW2}) with $\alpha_i\rightarrow-\alpha_i$. Replacing the interception points into (\ref{eq:Cepsilon tot}) generates:
\begin{equation}\label{eq:Intermediate switchback}
{\mathcal{C}^\epsilon}(t_L,\,t_R)\simeq\frac{\Omega_{d-1}L^{d-2}}{G_N}\qty(P_{\infty}^\epsilon (t_R+t_L)+(P_{\infty}^\epsilon+P_{-\infty}^\epsilon)\qty(\sum_{k=1}^nt_k-nt^{(c)}_*))~,
\end{equation}
where $t^{(c)}_*$ is given in (\ref{eq:scrambling cosmo}), and we again discard constant subleading terms of $P_{+\infty}^\epsilon$, $P_{-\infty}^\epsilon$.

Importantly, it was noticed in Sec. \ref{sec:CAny CMCAdS} the CAny proposals with a generic functional $F[\dots]$ in (\ref{eq:C CMC}) for an AdS black hole background only satisfy the switchback effect when the rate of growth in the past and future are the same. This means that for (\ref{eq:C CMC}) to obey the definition of holographic complexity in \cite{Belin:2021bga,Belin:2022xmt}, we require
\begin{equation}\label{eq:P epsilon past future}
P^\epsilon_{+\infty}=P^\epsilon_{-\infty}~.
\end{equation}
In that case, the evaluation of (\ref{eq:Intermediate switchback}) reduces to
\begin{equation}\label{eq:switchback result}
    \mathcal{C}^\epsilon(t_L,\,t_R)\propto\abs{t_R+t_1}+\abs{t_2-t_1}+\dots+\abs{t_n-t_L}-2nt^{(c)}_*~,
\end{equation}
where the term $-2n t^{(c)}_*$ appears due to cancellation in the complexity growth due to early and late time perturbations.

\begin{figure}
    \centering
    \subfloat[]{\includegraphics[width=0.325\textwidth]{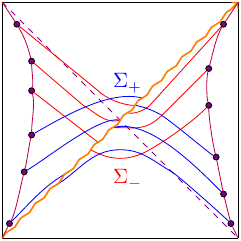}}\hspace{0.1cm}
        \subfloat[]{\includegraphics[width=0.63\textwidth]{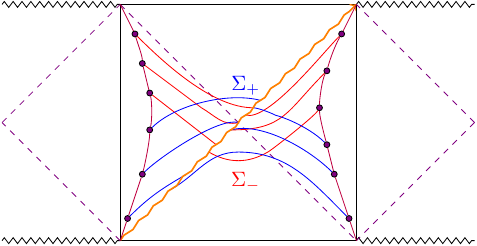}}
    \caption{Representative CMC slices anchored to the stretched horizon (fuchsia) in a single shockwave geometry in (a) SdS$_3$ space; (b) SdS$_{d+1\geq4}$ space. We employ the discontinuous Kruskal coordinates $\tilde{U}$, $\tilde{V}$ in (\ref{eq:tilde metric}) in the Penrose diagram to facilitate the representation of the CMC slices.}
    \label{fig:SW CAny SdS}
\end{figure}

Notice that a possible way to satisfy (\ref{eq:P epsilon past future}) in SdS space can be obtained by setting $K_-=-K_+$ and selecting a complexity proposal $\mathcal{C}$ as\footnote{However, instead of minimization, one might as well perform a maximization over the CMC slices, or an averaging, as either of those would satisfy (\ref{eq:P epsilon past future}) in AdS black holes; although that would reproduce the hyperfast growth in SdS space.}
\begin{equation}\label{eq:C min}
    \mathcal{C}=\min_t\qty(\mathcal{C}^+(t),\,\mathcal{C}^-(t))~.
\end{equation}
See Fig. \ref{fig:SW CAny SdS} for an illustration of the evolution of the CMC slices in the shockwave geometry.
Notice that this protocol is a further enlargement of the space of CAny proposals, as one needs (\ref{eq:P epsilon past future}) to be respected even for planar AdS black holes, which is not the case for a general functional $F[\dots]$ in (\ref{eq:C CMC}). Moreover, (\ref{eq:C min}) might also be advantageous when considering more complicated black holes in AdS space. A potential subtlety with this generalization is that the complexity growth rate (\ref{eq:Vol late times}) might become discontinuous at the time when the change of CMC slice occurs. However, this is in principle allowed in the definition of holographic complexity proposals \cite{Belin:2021bga,Belin:2022xmt}.

We close the subsection with a few remarks. First, the result (\ref{eq:switchback result}) reproduces the same type of behavior as the switchback effect for AdS black holes, at least for the CAny proposals with early and late-time linear growth in SdS space. Second, as we previously, there are fine-tuned situations where $\mathcal{C}^\epsilon$ can have any type of early and late-time growth behavior for SdS$_{3}$. It might be interesting to study the modifications in the switchback in those cases. Lastly, the switchback effect has also been recovered in a different and more explicit analysis for particular asymptotically dS backgrounds \cite{Anegawa:2023dad,Baiguera:2023tpt,Baiguera:2024xju}, hinting at the possibility that this is a universal phenomenon in shockwave geometries.

\subsubsection*{Other cases}
\begin{figure}
    \centering
    \includegraphics[width=\textwidth]{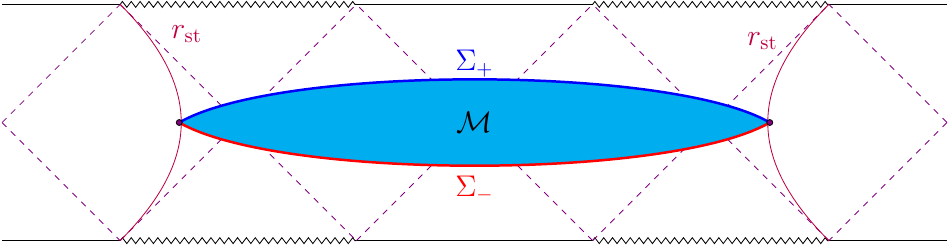}
\caption{Proposal for evaluating the volume of CMC slices for SdS$_{d+1}$ black holes. $\mathcal{M}$ (cyan) is the bulk region bounded by the slices $\Sigma_-$ and $\Sigma_+$ (red and blue, respectively), where $\mathcal{C}^{-}$ and $\mathcal{C}^{+}$ in \eqref{eq:Volepsilon} are evaluated. 
In the picture, we anchor the bulk region to the black hole stretched horizons (fuchsia lines) at particular locations (purple dots).
Similar considerations apply to the cosmological stretched horizons. The precise profile of the $\Sigma_\epsilon$ slices is determined by the extremization of \eqref{eq:regions CMC}.}
    \label{fig:ProposaldS}
\end{figure}

Having evaluated Case 1, we are interested in the various other configurations of the stretched horizons presented in Sec. \ref{ssec:stretched_horizons}, as illustrated in Fig. \ref{fig:ProposaldS}. Unlike in the previous case, we leave a detailed analysis of the switchback effect for these different configurations for future directions.

\subsubsection*{Case 2}

We consider again a symmetric configuration of the boundary times \eqref{eq:symmetric_times} and of the stretched horizons $r_{\rm st}^L = r_{\rm st}^R$.
The evaluation of complexity and of the boundary time is very similar to eqs.~\eqref{eq:Anything complexity cod1} and \eqref{eq: t CAny}:
\begin{align}\label{eq:CAny 2}
\mathcal{C}^\epsilon&=\frac{2\Omega_{d-1} L^{d-2}}{G_N} \int^{{r}_{\rm st}}_{{r}_{t_h}} \frac{ {\qty(\frac{r}{L})}^{2(d-1)}}{\sqrt{-\mathcal{U}(P_v^{\epsilon},\,{r})}} \, dr \, ,\\
t&= -2\int^{{r}_{\rm st}}_{{r}_{t_h}}\frac{\left(P_{\epsilon} - \epsilon L\frac{\abs{K}}{d} \qty(\frac{r}{L})^d \right)}{f(r)\,\sqrt{-\mathcal{U}(P_{\epsilon}, r)}} \, dr \, .
\label{eq:t CAny 2}
\end{align}
After similar manipulations, we find that the rate of growth at late times is formally given by the same expression \eqref{eq:Vol late times} with $r_f$ satisfying \eqref{eq:rt location}.
However, the difference with the previous case is that the radial coordinate $r_f$ of the turning point generating the late time linear growth now lies inside the interval $[0,~r_{h}]$ (with $r_{\rm st}>r_h$). 
One can now find a solution for $r_f$ in this range even when $K_\epsilon=0$, as obtained in case 2 of Sec.~\ref{ssec:CV_dS}, and in contrast to case 1 for the CAny proposal.
This result is consistent with the behavior of black holes in asymptotically AdS space \cite{Belin:2022xmt,Jorstad:2023kmq}.

\begin{figure}
\centering
\subfloat[]{\includegraphics[height=0.3\textwidth]{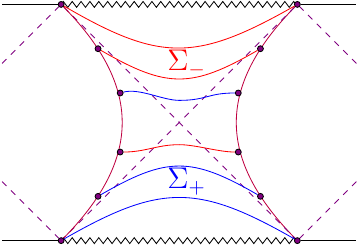}}\hspace{0.75cm}\subfloat[]{\includegraphics[height=0.3\textwidth]{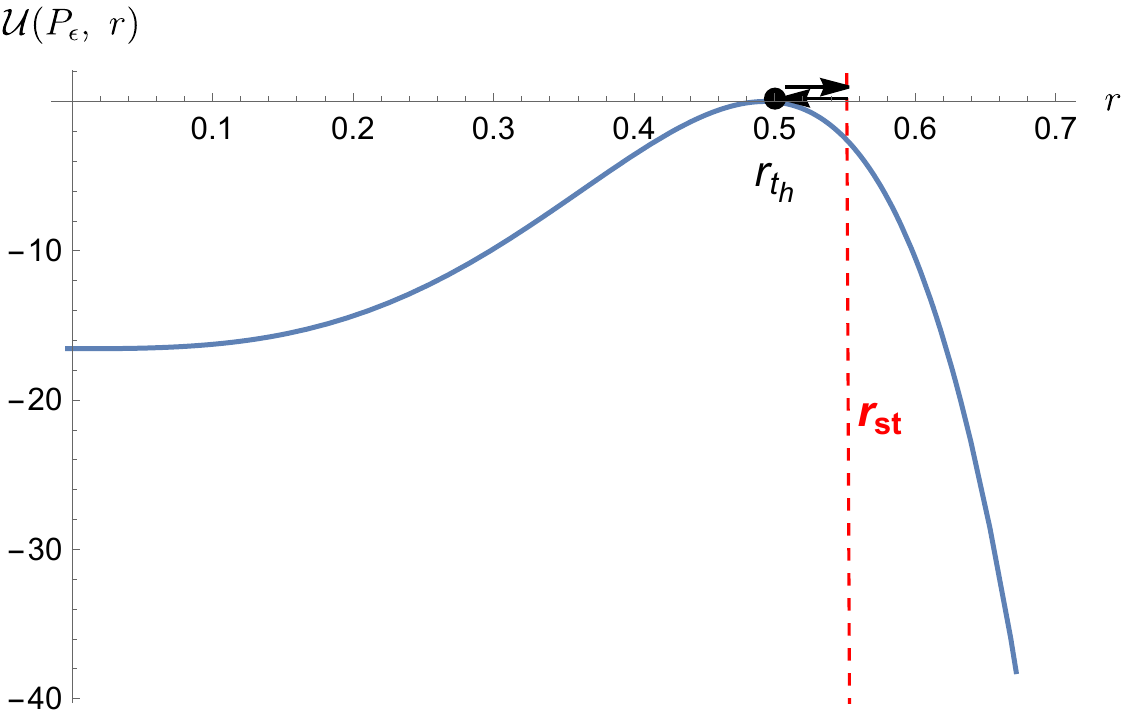}}
    \caption{Case 2: (a) Representation of the profile of the CMC slices for $\Sigma_-$ (red) and $\Sigma_+$ (blue) for SdS space with a pair of black hole stretched horizons. (b) Effective potential (\ref{eq:U CAny}). A particle moving in the potential coming from $r_{\rm st}$ will be reflected at the turning point $r_{t_h}$. The numerical parameters are chosen to be the same as in Fig. \ref{fig:CMC1}, except for $P_\epsilon=-4.07$.}
    \label{fig:CMC2}
\end{figure}

\begin{figure}
    \centering
    \subfloat[]{\includegraphics[width=0.48\textwidth]{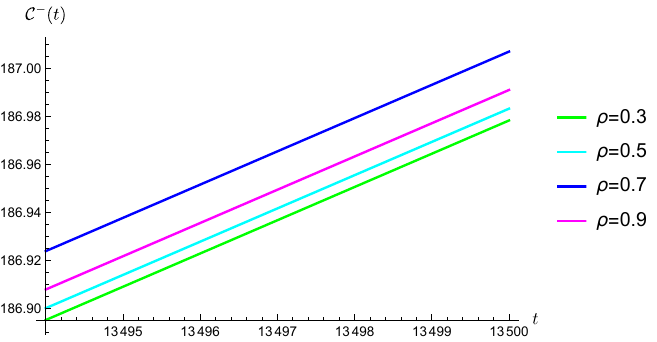}}\hspace{0.2cm}\subfloat[]{\includegraphics[width=0.48\textwidth]{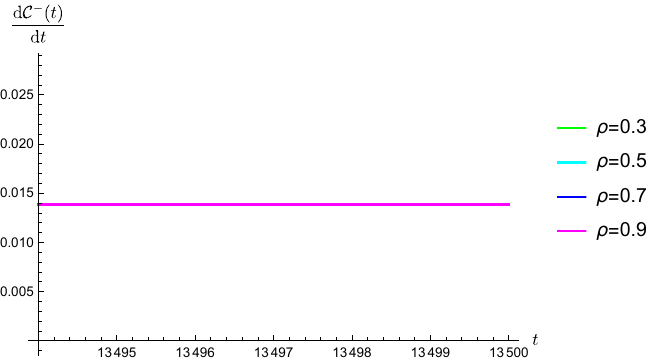}}
    \caption{Case 2: late-time behavior of (a) $\mathcal{C}^\epsilon$ in \eqref{eq:CAny 2} and (b) its rate of growth. The same parameters as in Fig.~\ref{fig:CAny1} have been used, including $a(r)=1$ and $K_-=100/L$.}
    \label{fig:CAny2}
\end{figure}

The CMC slices in the SdS black hole background and a plot of the effective potential for a specific value of $P_{\epsilon}$ are shown in Fig.~\ref{fig:CMC2}, while the corresponding plot for the complexity growth at late times is shown in Fig.~\ref{fig:CAny2}.


\subsubsection*{Case 3}
We consider the case of a cosmological stretched horizon located at $r_{\rm st}^L$ on the left side, and a black hole stretched horizon at $r_{\rm st}^R$ on the right side of the Penrose diagram. 
The general CAny observable defined in \eqref{eq:C epsilon inter} becomes
    \begin{align}\label{eq:Cepsilon car2}
    \mathcal{C}^\epsilon&=\frac{\Omega_{d-1} L^{d-2}}{G_N}\qty(\int_{r^{\rm L}_{\rm st}}^{r_{\rm t_c}}+  \int^{r_{\rm t_c}}_{r_{\rm  t_h}}+\int^{r^{R}_{\rm st}}_{r_{\rm t_h}})\frac{\qty(\frac{r}{L})^{2(d-1)}a(r)}{\sqrt{-\mathcal{U}(P_{\epsilon},\,r)}} \, \rmd {r} ~.
\end{align}
Next, we follow similar steps to case 3 in Sec.~\ref{ssec:CV_dS} to derive an expression for the boundary times. 
Namely, we define the anchoring times at the left and right stretched horizons as $-t_L$ and $-t_R$, and we obtain the same \eqref{eq:identities_CV_case3} {except for} the modification $\tau_\pm\rightarrow\tau_\pm^\epsilon$ in (\ref{eq:def_tau}), where
\begin{equation}
    \tau_{\pm}^\epsilon(P_\epsilon,~r) \equiv \frac{-\qty(P_\epsilon-\epsilon L\frac{\abs{K}}{d}\qty(\frac{r}{L})^d )\pm \sqrt{-\mathcal{U}(P_\epsilon,~r)}}{f(r) \sqrt{-\mathcal{U}(P_\epsilon,~r)}} \, .
    \label{eq:def_taup}
\end{equation}
After this replacement, we can directly add the various contributions to find
\begin{align}
    t_R-t_L &= \qty(\int_{r_{\rm st}^L}^{r_{t_c}} + \int_{r_{\rm t_h}}^{r_{t_c}} + \int_{r_{t_h}}^{r_{\rm st}^R} )\frac{P_{\epsilon} - \epsilon L\frac{\abs{K}}{d} \qty(\frac{r}{L})^d }{f(r) \sqrt{-\mathcal{U}(P_{\epsilon}, r)}} \, \rmd r \, ,\label{eq: t CMC 3}\\
   t_{t_c}-t_{t_h} &=-I(P_\epsilon,~K_\epsilon),\quad I(P_\epsilon,~K_\epsilon)=\int_{r_{t_h}}^{r_{t_c}}\frac{P_{\epsilon} - \epsilon L\frac{\abs{K}}{d} \qty(\frac{r}{L})^d }{f(r) \sqrt{-\mathcal{U}(P_{\epsilon}, r)}} \, \rmd r \, .\label{eq:time dif turning points}
\end{align}
When $\abs{K}\geq K_{\rm crit}$, one can numerically show that there do not exist two turning points $r_{t_c}$ and $r_{t_h}$ at fixed boundary time, thus corresponding to disconnected extremal surfaces $\Sigma_\epsilon$. 
For this reason, we will restrict the following analysis to the case $\abs{K} < K_{\rm crit}$.

Similar to case 3 in Sec. \ref{ssec:CV_dS}, we have two possibilities:
\begin{itemize}
    \item In the symmetric configuration $t_L=t_R$, the only solution to \eqref{eq: t CMC 3} is characterized by the conditions $P_{\epsilon}=K_{\epsilon}=0$.
Therefore, the CAny proposal is reduced to CV, and the same analysis performed in Sec. \ref{ssec:CV_dS} follows.
In particular, the previous result is consistent with the constraint \eqref{eq:time dif turning points}, and leads to a time-independent complexity observable \eqref{eq:Cepsilon car2}. 
\item In contrast, when we consider (without loss of generality) the relation $t_R=-t_L$ dictated by prescription \ref{rule}, we can have a non-trivial time evolution with non-vanishing $t_{t_c}, t_{t_h}$. We will consider this choice in the remainder of the section.
\end{itemize} 

\begin{figure}
\centering
\subfloat[]{\includegraphics[height=0.25\textwidth]{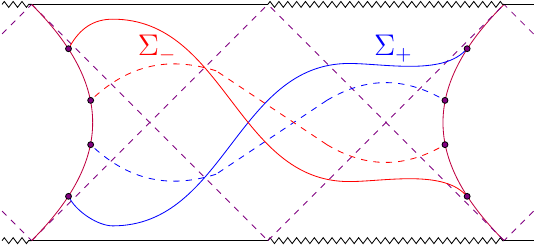}}\hspace{0.2cm}\subfloat[]{\includegraphics[height=0.25\textwidth]{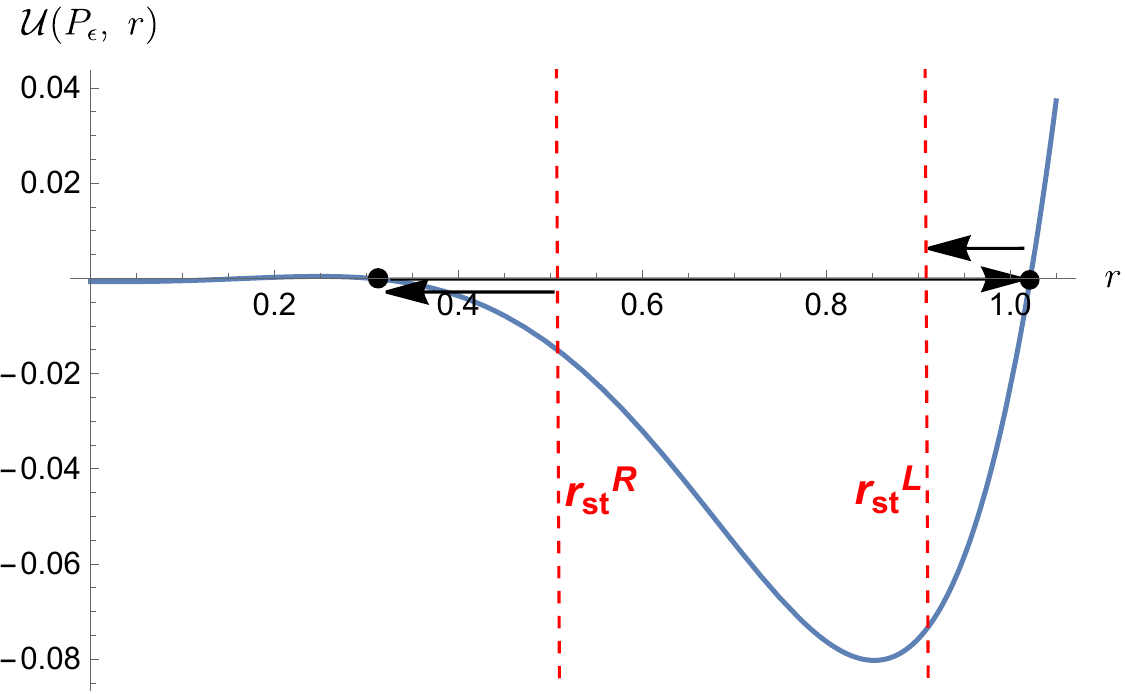}}
    \caption{Case 3: (a) Representation of the profile of the CMC slices for $\Sigma_-$ (red) and $\Sigma_+$ (blue) when $\abs{K}\leq K_{\rm crit}$ for a SdS black hole, intersecting the stretched horizons at the green dots. {Dashed and solid lines represent different boundary times along the evolution.} (b) Effective potential (\ref{eq:U CAny}). A particle moving in the potential coming from $r_{\rm st}^R$ will be reflected at the turning points and reach $r_{\rm st}^L$. The numerical parameters are chosen to be the same as in Fig. \ref{fig:CMC2}, except for $\abs{K}=\sqrt{3}$, and $P_\epsilon=-0.027$.}
    \label{fig:CMC3}
\end{figure}

We proceed to evaluate \eqref{eq:Cepsilon car2} carefully. Since under the gauge fixing \eqref{gauge} we have $P_{\epsilon} - \epsilon L\frac{\abs{K}}{d}=-\dot{r}-f(r)\dot{u}= -f(r) \dot{t}$, the sign of this combination will be determined by the orientation of the Killing vectors, as we remarked in footnote \ref{fnt:Sign of P}. 
The CMC slices corresponding to this regime are shown in Fig.~\ref{fig:CMC3}.

We can now take care of the integrand (\ref{eq:Cepsilon car2}) at the turning points. 
At late times, \eqref{eq:rt location} admits a turning point $r_{t_h}$ inside the black hole region but does not admit a turning point $r_{t_c}$ outside the cosmological horizon.
For this reason, we can conveniently re-express the complexity as
\begin{equation}\label{eq:Cepsilon close form BC}
\begin{aligned}
    &\frac{G_N~\mathcal{C}^\epsilon}{\Omega_{d-1} L^{d-2}}=\sqrt{-f(r_{t_h})\qty(\frac{r_{t_h}}{L})^{2(d-1)}}a(r_{t_h})\qty(\int^{r^{\rm R}_{\rm st}}_{r_{\rm t_h}}+  \int^{r_{\rm t_c}}_{r_{\rm  t_h}}+\int^{r_{\rm t_c}}_{r_{\rm st}^{\rm L}})\tfrac{\left(P_{\epsilon}- \epsilon L\frac{\abs{K}}{d} \qty(\frac{r}{L})^d \right)d r}{f(r)\sqrt{-\mathcal{U}(P_{\epsilon}, r)}}\\
&+\qty(\int^{r^{R}_{\rm st}}_{r_{\rm t_h}} + \int^{r_{\rm t_c}}_{r_{\rm  t_h}}+\int^{r_{\rm t_c}}_{r_{\rm st}^{\rm L}})\tfrac{f(r)\,\qty(\frac{r}{L})^{2(d-1)}a(r)-\sqrt{-f(r_{t_h})\qty(\frac{r_{t_h}}{L})^{2(d-1)}}a(r_{t_h})\left(P_{\epsilon} -\frac{\abs{K}}{d} \qty(\frac{r}{L})^d \right)}{f(r)\sqrt{-\mathcal{U}(P_{\epsilon},\,r)}}\rmd r~.
\end{aligned}
\end{equation}
By denoting with $r_{f_h}$ the local maximum of the effective potential $\mathcal{U}$ approached by the CMC slice at late times, we obtain that the time derivative of \eqref{eq:Cepsilon close form BC} reads
\begin{equation}\label{eq:t inft dCdt CAny3}
    \lim_{{t}\rightarrow\infty}\dv{\mathcal{C}^\epsilon}{{t}}=\frac{\Omega_{d-1} L^{d-2}}{G_N}\sqrt{-f(r_{f_h})\qty(\frac{r_{f_h}}{L})^{2(d-1)}}a(r_{f_h})~.
\end{equation}
The existence of a final slice at a constant radius approached by the CMC slices is responsible for the linear growth at late times.
Notice that the result for $K_\epsilon < K_{\rm crit}$ agrees with the analysis of case 3 in Sec.~\ref{ssec:CV_dS} when  $K_\epsilon=0$ and $a(r)=1$. 

\begin{figure}
    \centering
    \subfloat[]{\includegraphics[width=0.48\textwidth]{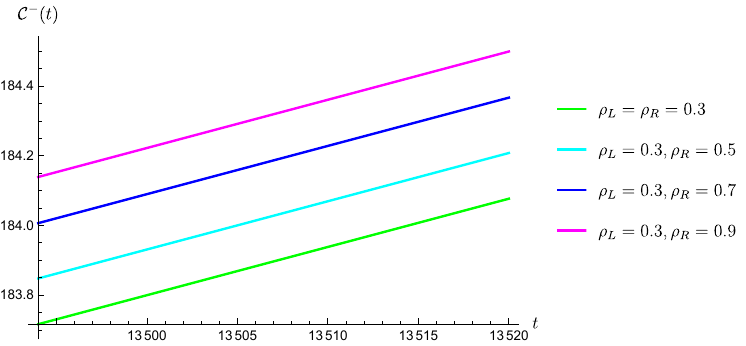}}\hspace{0.2cm}\subfloat[]{\includegraphics[width=0.48\textwidth]{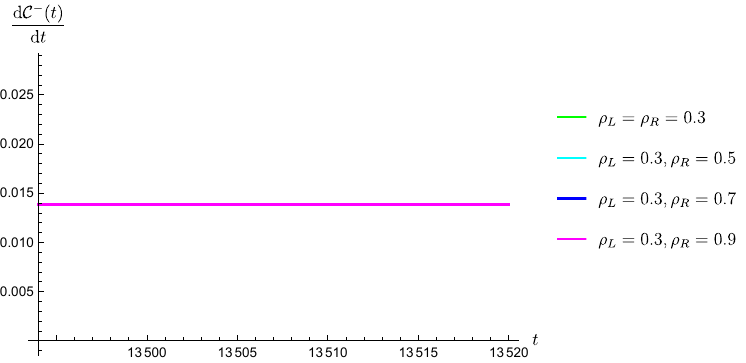}}
    \caption{Case 3: plot of (a) $\mathcal{C}^-$ in \eqref{eq:Cepsilon close form BC} and (b) its rate of growth. We use the same parameters as in Fig. \ref{fig:CAny1}, and various choices of $\rho_{\rm L}$ and $\rho_{\rm R}$, as indicated in the legend of the figure.}
    \label{fig:CAny3}
\end{figure}

\subsubsection*{Case 4}

\begin{figure}
    \centering
    \includegraphics[width=0.7\textwidth]{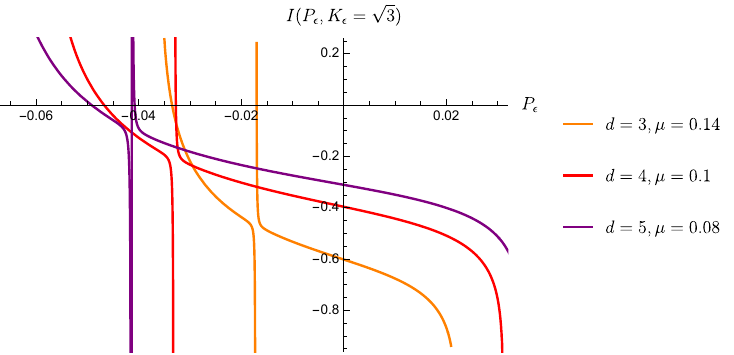}
    \caption{Plot of the integral $I(P_\epsilon,~K)$ in (\ref{eq:time dif turning points}) as a function of the conserved momentum $P_\epsilon$ for fixed $\abs{K}=\sqrt{3} < K_{\rm crit}$ and for various $d, \mu$. We fix $L=1$ and $\epsilon=-1$. The function only vanishes for particular values of $P_\epsilon$.}
    \label{fig:CAny2ndInt}
\end{figure}

We assume that the cosmological stretched horizons are located at the same radial coordinate $r_{\rm st}^L=r_{\rm st}^R \equiv r_{\rm st}$. Without loss of generality, we also consider the symmetric configuration \eqref{eq:symmetric_times} of the boundary times.
In each copy of the extended SdS$^n$ geometry, all connected surfaces $\Sigma_\epsilon$ must at least pass through a turning point behind the black hole horizon and one beyond the cosmological horizon. 
Specializing to the case $n=2$, we have a single turning point $r_{t_h}$ in the interior of the black hole, and two turning points outside the cosmological horizon, located at the same radial coordinate $r_{t_c}$ by symmetry.
When $|K_{\epsilon}| \geq K_{\rm crit}$, there are no connected surfaces.
In the other regime $|K_{\epsilon}| <K_{\rm crit}$, we perform manipulations similar to case 3 to get
\begin{align}\label{eq:Cepsilon car}
    \mathcal{C}^\epsilon&=\frac{2\Omega_{d-1} L^{d-2}}{G_N}\qty(\int_{r_{\rm st}}^{r_{\rm t_c}}+ \int^{r_{t_c}}_{r_{\rm t_h}})\frac{\qty(\frac{r}{L})^{2(d-1)}a(r)}{\sqrt{-\mathcal{U}(P_{\epsilon},\,r)}}d r~,
    \end{align}
\begin{equation}\label{eq:t CMC close form}
    t_R+t_L= 2\qty(\int_{r_{\rm st}}^{r_{\rm t_c}}+ \int^{r_{t_c}}_{r_{\rm t_h}})\frac{\left(P_{\epsilon} - \epsilon L\frac{\abs{K}}{d} \qty(\frac{r}{L})^d \right)d r}{f(r)\,\sqrt{-\mathcal{U}(P_{\epsilon}, r)}} ~,
\end{equation}
together with the constraint (\ref{eq:time dif turning points}). 
Now, there are two main possibilities:
\begin{itemize}
 \item When $t_R=-t_L$, the condition \eqref{eq:time dif turning points}, together with \eqref{eq:t CMC close form}, implies that
 $P_{\epsilon}=K_{\epsilon}=0$, thus leading to the time-independent evolution of CV already analyzed in Sec. \ref{ssec:CV_dS}.  
\item When we consider (without loss of generality) the time-dependent case $t_R= t_L$ proposed by the prescription \ref{rule},  
symmetry arguments imply that the turning points are located at $t_{t_c}=t_{t_h}=0$. In this case, one has to numerically check whether the constraint \eqref{eq:time dif turning points}, with vanishing left-hand-side, is satisfied.
The plots of the integral $I(P_{\epsilon}, K_{\epsilon})$ are depicted in Fig.~\ref{fig:CAny2ndInt} for the case $|K_{\epsilon}| < K_{\rm crit}$.
One can clearly see that the constraint is only satisfied for a specific value of $P_{\epsilon}$. Therefore, holographic complexity would only be defined for a single boundary time.     
\end{itemize}

In summary, CMC surfaces connecting the two stretched horizons only exist at a fixed symmetric boundary time.
Alternatively, we can have solutions defined at arbitrary (but antisymmetric) boundary times by evolving the trivial surface at constant $t=0$ with the time isometry.

\subsubsection*{Case 5}
Analogous to the evaluation of case 5 in Sec. \ref{ssec:CV_dS}, we consider a symmetric location of the black hole stretched horizons $r_{\rm st}^{\rm L}=r_{\rm st}^{\rm R}=r_{\rm st}$. We find that
    \begin{align}\label{eq:Cepsilon car3}
    \mathcal{C}^\epsilon&=\frac{2\Omega_{d-1} L^{d-2}}{G_N}\qty(\int^{r_{\rm st}}_{r_{\rm t_h}}+\int^{r_{\rm t_c}}_{r_{\rm  t_h}})\frac{\qty(\frac{r}{L})^{2(d-1)}a(r)}{\sqrt{-\mathcal{U}(P_{\epsilon},\,r)}}d r~,\\
 t_R+t_L&= -2\qty(\int^{r_{\rm st}}_{r_{\rm t_h}}+\int^{r_{\rm t_c}}_{r_{\rm  t_h}})\frac{\left(P_{\epsilon} - \epsilon L\frac{\abs{K}}{d} \qty(\frac{r}{L})^d \right)d r}{f(r)\,\sqrt{-\mathcal{U}(P_{\epsilon}, r)}} ~,   \label{eq:t CAny case5}
\end{align}
together with the constraint \eqref{eq:time dif turning points}.
The conclusions are the same as case 4: either the CMC slices degenerate to the volume case with a trivial evolution at antisymmetric times $t_L= -t_R$, or they exist at a single boundary time, such that $t_L=t_R$ and the constraint \eqref{eq:time dif turning points} is satisfied.

\section{Discussion}
\label{sec:ConclusionsSdSn}
In this work, we studied the time dependence of holographic complexity in the extended SdS black hole geometry for several configurations of the stretched horizons.
The results are summarized in table~\ref{tab:resultsSdSn}.

In case 1, which investigates the cosmological patch of the geometry, complexity is characterized by a hyperfast growth, extending previous studies in empty dS space \cite{Jorstad:2022mls}.
The only exception to this behavior is represented by a class of codimension-one CAny observables, which present an eternal evolution with a linear increase at late times, if the extrinsic curvature on CMC slices where they are evaluated is larger than a critical value \cite{Aguilar-Gutierrez:2023zqm}.
These results may suggest that the hyperfast growth could be used as a way to discriminate whether a complexity conjecture is well-behaved or not.
In case 2, where only the black hole patch of the geometry is considered, we achieve the characteristic linear growth for all the complexity conjectures, in agreement with the behavior of AdS black holes (\eg see \cite{Carmi:2017jqz,Belin:2021bga,Belin:2022xmt}).
This seems a robust feature that holds independently of the asymptotics of the spacetime where complexity is evaluated.
Cases 3-5 are novel set-ups that access both the interior of the black hole and the exterior of the cosmological horizon.
In the case of codimension-zero proposals, all the complexity observables are time-independent. This happens because the WDW patch always reaches timelike infinities $\mathcal{I}^{\pm}$ and the singularities. 
In particular, the portion of WDW patch emerging from the past of the Penrose diagram is always compensated by the disappearance of a region of equal size into the future part of the geometry. Remarkably, this feature happens separately on the left and right sides of the diagram, therefore it is not a consequence of any synchronization of the boundary times on the two stretched horizons.

For codimension-one observables, the situation is more peculiar.
Our main result is that in the volume case, \textit{connected extremal surfaces anchored at symmetric times to the stretched horizons only exist when $t_L=t_R=0$}.
In case 3, the isometries of the geometry allow for a non-trivial evolution of complexity that approaches linear growth at late times. 
In cases 4-5, the isometries imply that only the trivial evolution is possible, leading to a time-independent result. 
Similar features appear for the codimension-one CAny proposal.
We conclude that the extended SdS black hole represents an arena where the location of the stretched horizons distinguishes among different behaviors for the holographic proposals.

It is interesting to compare our results on codimension-one extremal surfaces with the recent study of spacelike geodesics connecting static sphere observers in SdS space \cite{Faruk:2023uzs}, in connection to our case 3.  
When the boundary times are symmetric ($t_L=t_R$), they found that there exists a small window where these spacelike geodesics exist, that connect the future (past) of the inflating patch with the past (future) of the black hole patch.
In empty dS space, the only symmetric spacelike geodesics connecting antipodal static patch observers require $t_{L}=t_R=0$, \eg see \cite{Chapman:2021eyy,Galante:2022nhj}.
These findings resonate with our results, except that we only have the time $t_L=t_R=0$ allowed, instead of a small boundary time interval.
These discrepancies can be probably attributed to the different dimensionality between the geometric objects that we studied, compared to the analysis in the literature.
In the case of geodesics, it is interesting to notice that more general configurations are allowed by considering curves with a complex length \cite{Chapman:2022mqd,Aalsma:2022eru}.
This suggests that we may consider complex maximal volume to find a non-trivial time evolution for symmetric configurations in multiple copies of the SdS geometry.

We also studied the appearance of the switchback effect in asymptotically dS spacetimes in the late (and early time) evolution of the codimension-one CAny observables under shockwave insertions. We picked a set of observables that are evaluated in CMC slices of different curvature in the past and future boundaries. We proved that under a weakly gravitating regime, the CAny observables show a reduction in the complexity growth due to cancellations of the energy perturbations. We also explicitly verified one of the predictions in \cite{Aguilar-Gutierrez:2023zqm}, namely that a time-reversal symmetric protocol would be necessary for the switchback effect to occur. Moreover, our findings show a great similarity with respect to the behavior of CAny proposals for AdS black holes under the switchback effect. 

On the other hand, as we found in Fig. \ref{fig:CAny SdS}, the CAny observables generically reach a terminal turning $r_f$ (as well as a time-reversal version) determined by the choice of the CMC slices through (\ref{eq:rt location}). This implies that the CAny proposals in the thesis do not prove the whole cosmological patch of SdS spacetime. However, we would expect that any notion of static patch holography should also encode the degrees of freedom of the inflating region, similar to investigations in asymptotically AdS space \cite{Jorstad:2023kmq}. Nevertheless, one can probe more of the geometry outside the cosmological horizon using the alternating shockwave geometry. It seems that adding perturbations in the stretched horizon reveals more information, even when its explicit localization is irrelevant.

\subsubsection*{Outlook}
Next, let us discuss some further developments in the study of gravitational observables.

One of the motivations for this work was to relate our results with the coarse-graining of information for general observables in quantum cosmology.
According to static patch holography, we expect that the information
about the geometry in the static patch can be retrieved from observables anchored at the stretched horizon, either near the black hole or the cosmological horizons.
On the other hand, in the quantum cosmology context, one usually considers a meta-observer that lives close to a spacelike surface (such as a reheating surface, see the horizontal purple lines in Fig.~\ref{fig:extSdS}).
One may expect that the observables anchored to the stretched horizons should give the same information to which this meta-observer has access. 
If this is true, then one would conclude that the encoding of information in the extended SdS geometry is redundant.
This is because the meta-observer living on a single spacelike slice close to $\mathcal{I}^{\pm}$ would have access to the information contained in the other copies of the extended SdS background as well.
Our results imply that if the observer were to measure gravitational probes such as those in case 3, they would be able to extract different types of dynamical information with codimension-one observables, while they would not find any time evolution from the codimension-zero observables that we studied.
In general, we expect that other codimension-zero observables in the CAny class \cite{Belin:2022xmt} could allow for non-trivial dynamics, as long as the bulk region where complexity is evaluated does not reach both $\mathcal{I}^\pm$ and the singularities of the black hole. It might also be fruitful to study if introducing shockwave perturbations in the geometry affects the coarse-graining of information found in \cite{Aguilar-Gutierrez:2021bns}. We leave this analysis for future studies.

Second, as emphasized in Sec. \ref{sec:CAny CMC}, the switchback effect is a crucial ingredient of all the complexity proposals.
While this property has already been analyzed in asymptotically dS space in many instances \cite{Baiguera:2023tpt,Anegawa:2023wrk,Aguilar-Gutierrez:2023pnn},
it would be interesting to extend the analysis to the observables considered in this work, beyond our presentation in Sec. \ref{sec:CAny CMC} (case 1).
In particular, a natural step would be to investigate whether the time-independent observables in cases 3-5 would develop non-trivial dynamics after the insertion of a shockwave.

Third, we plan to compute the holographic complexity conjectures in lower-dimensional gravity.
Two-dimensional empty dS space arises from the dimensional reduction of near-extremal SdS black holes in higher dimensions. Explicitly, when $r_{c,h}\rightarrow r_N$ defined in \eqref{eq:critical_mass_SdS}, the geometry is well-known to become dS$_2\times$S$^{d-1}$ (\eg see \cite{Anninos:2012qw})
    \begin{equation}
        \rmd s^2=-\qty( 1-\tfrac{\rho^2}{L_2^2})\rmd\tau^2+\qty( 1-\tfrac{\rho^2}{L_2^2})^{-1}\rmd \rho^2+r_N^2\rmd\Omega_{d-1}^2~,
    \end{equation}
where $L_2$ is the length scale of dS$_2$, and $\rho$ is a radial coordinate measuring the finite separation between the cosmological and black hole horizons. There are different reasons to be interested in this regime. 
First, let us consider the complexity rate obtained at late times for the CAny observables in cases 1-3 (see table \ref{tab:resultsSdSn} and references therein).
The results always depend on the location $r_{f_i}$ of a final slice which is approached by the extremal surface at late times. In the near-extremal limit, the solution for $r_f$ in \eqref{eq:rt location} has been previously found to coincide with the Nariai radius $r_N$ \cite{Aguilar-Gutierrez:2023zqm}.
While in the general case the turning points in the cosmological and black hole region (if they both exist) satisfy $r_{t_c} > r_{t_h}$, the naive near-extremal limit $\mu \rightarrow \mu_N$ at late times would imply that $r_{t_c} \rightarrow r_{t_h} \rightarrow r_N $. According to this limit, we would conclude that the growth rate at late times reads
\begin{equation}
    \lim_{t\rightarrow\infty}\dv{t} \mathcal{C}^\epsilon =0~,
    \label{eq:extremal_limit}
\end{equation}
for cases 1-3, while this equation is already true without taking the near-extremal limit in cases 4-5.
The Lloyd bound on complexity for asymptotically AdS black holes is conjectured to take the form $d\mathcal{C}/dt \sim TS$ \cite{Susskind:2014rva,Brown:2015lvg}. 
The result \eqref{eq:extremal_limit} is usually associated with the vanishing temperature of extremal black holes. 
While the naive Hawking temperature associated with the SdS solution (see \eqref{eq:hor_cosm_temp_SdS}) vanishes in the limit $r_{c,~h} \rightarrow r_N$, a careful analysis of the surface gravity experienced by a static patch observer moving along a worldline shows that the temperature should be non-zero (see footnote \ref{foot:extremal_temp}) \cite{Bousso:1996au,Morvan:2022ybp}. 
If we assume that the Lloyd bound also holds in SdS space, then we expect that a non-vanishing $d\mathcal{C}/dt\sim TS$ should be reproduced using the dS$_2\times$S$^{d-1}$ geometry.

Another reason to be interested in a two-dimensional empty dS case is that one might be able to propose modifications of the CAny conjectures based on our higher dimensional results, in a similar fashion as the refinement of CV considered in \cite{Anegawa:2023wrk}.
In the case of the observables in the extended SdS$^n$ with $n>1$, the two-dimensional quantities analog to cases 4-5 are Jackiw-Teitelbom (JT) gravity multiverse models \cite{Aguilar-Gutierrez:2021bns,Levine:2022wos}.
Moreover, in two dimensions it is possible to consider centaur geometries where patches of dS space are glued to asymptotic AdS regions, which have a standard timelike boundary where AdS/CFT duality applies \cite{Anninos:2017hhn,Anninos:2018svg,Anninos:2020cwo,Anninos:2022hqo,Aguilar-Gutierrez:2023odp,Galante:2023uyf}. This might allow for a clearer interpretation of the properties studied in this work in the context of dS$_2$ space, appearing from near extremal limits near horizon limits of black hole geometries
\cite{Maldacena:2019cbz,Castro:2022cuo}. It would be interesting to study holographic complexity for multiple copies of this geometry and check whether the time-independence of the codimension-zero proposals in cases 3-5 and codimension-one proposals in 4-5 is lifted. 
In particular, the case of a single copy was studied in \cite{Chapman:2021eyy}, where no hyperfast growth arises. 

Fourth, the covariant objects computed in this work can be interpreted as proper measurements of computational complexity only if we find a dual quantum state and a corresponding circuit that exhibits the same features. 
One may then hope to refine these toy models using Nielsen or Krylov definitions of complexity in quantum mechanics \cite{Nielsen1,Nielsen2,Nielsen3,Parker:2018yvk}. 
We present work in this direction in Ch. \ref{ch:ComplexityDSSYK} for the DSSYK model. 
Concretely, one should analyze quantum circuit observables that display perturbations on the stretched horizon that can be related to an epidemic type of growth given the insertion of operators, as overviewed in Sec. \ref{ssec:QC switchback}.

Finally, another interesting and challenging task would be to include quantum corrections in the evaluation of holographic complexity, as recently uncovered in JT gravity \cite{Carrasco:2023fcj}. 
A useful test ground for these corrections is provided by double holography, where quantum backreaction can be tracked in an exactly solvable regime \cite{Emparan:2021hyr,Chen:2023tpi}. A brane-world version of the extended SdS black hole was recently proposed in \cite{Aguilar-Gutierrez:2023zoi}, and a SdS$_3$ black hole with quantum corrections has appeared in \cite{Emparan:2022ijy}. It would be interesting to incorporate the switchback effect to characterize perturbations in a double holographic setting that has a clear CFT dual. This effective theory might be further modified by adding intrinsic gravity theories on the brane, leading to a more intricate holographic complexity evolution \cite{Hernandez:2020nem,Aguilar-Gutierrez:2023ccv,Aguilar-Gutierrez:2023zoi}.
Therefore, it might be useful to test whether the complexity proposals explored in our work display either a similar or a radically different behavior compared to these models.

\cleardoublepage

\part{Complexity in the double-scaled Sachdev–Ye–Kitaev model}\label{part:3}
  \graphicspath{{\chaptersdir/DSSYK/image/}}%
  \chapter{The double-scaled Sachdev-Ye-Kitaev model and de Sitter holography}\label{ch:DSSYK}
This chapter reviews a recent holographic correspondence between the double-scaled SYK model (DSSYK) model, Liouville-de Sitter CFT (LdS$_2$), and SdS$_3$ space. We begin summarizing these recent developments in connection with static patch holography. We will proceed by reviewing the chord Hilbert space description of the DSSYK model, mainly based on \cite{Aguilar-Gutierrez:2024rka}. This formulation will allow us to define notions of complexity for SdS$_3$ space from proper quantum information definitions of complexity in the DSSYK model in the next chapter.

\section{A quantum mechanical dual to 3D Schwarzschild-de Sitter space?}\label{sec:3D SdS holo}
So far, we have motivated the stretched horizon as an anchoring region where a holographic theory dual to dS space might live, based on \cite{Susskind:2021dfc}, and we have employed this hypothesis to evaluate holographic complexity proposals associated with dS space. We will study an explicit quantum theory presumed to be dual to SdS$_3$ space.

The Sachdev-Ye-Kitaev (SYK) model plays an important role as a toy model in nuclear physics \cite{French:1970ztu,Bohigas:1971vpj}, condensed matter \cite{Sachdev_1993,Sachdev:2010um} (see \cite{Chowdhury:2021qpy} for a review), and especially for us, high energy physics \cite{kitaevTalks,Cotler:2016fpe,Maldacena:2016upp,Saad:2018bqo,Maldacena:2018lmt,Jensen:2016pah,Polchinski:2016xgd}. It describes $N$ Majorana fermions in $(0+1)$-dimensions (which has a Hilbert space dimension $2^{N/2}$) with all to all $p$ body interactions governed by the hermitian Hamiltonian
\begin{equation}\label{eq:DDSSYK Hamiltonian}
    H=\rmi^{p/2}\sum_{1\leq i_1<\dots<i_p\leq N}J_{i_1,\dots,~i_p}\psi_{i_1}\dots\psi_{i_p}~,
\end{equation}
where 
\begin{equation}\label{eq:Majo fer}
    \qty{\psi_{i},~\psi_{j}}=2\delta_{ij}~,
\end{equation}
with $i=1,\dots N$; and the coupling constants $J_{i_1\dots i_p}$ obey the following Gaussian distribution
\begin{equation}\label{eq:GEA J}
\expval{(J_{i_1\dots i_p})}=0~,\quad\expval{(J_{i_1\dots i_p})^2}=\begin{pmatrix}
        N\\
        p
    \end{pmatrix}^{-1}\frac{J^2N}{2p^2}~.
\end{equation}
Some of its remarkable properties are enumerated below:
\begin{itemize}
    \item In the large-$N$ limit and low-temperature regime, the model becomes solvable in the sense that its thermal partition function, as well as the 2 and 4-point correlation functions of the fermion operators $\psi_{i}$ can be found analytically for any number $p\in\mathbb{N}$ \cite{kitaevTalks,Maldacena:2016hyu}.
    \item In the large $N$ regime and low temperatures, it possesses a (near) conformal symmetry (which we denote nCFT$_1$) dual to a near-AdS$_2$ black hole spacetime \cite{Maldacena:2016upp}, which provides an explicit example of the AdS$_2$/CFT$_1$ correspondence.
    \item It is a maximally chaotic system, meaning that its Lyapunov exponent (as measured by OTOCs) reaches the maximal bound conjectured \cite{Maldacena:2015waa,Murthy:2019fgs} $\kappa_L = 2\pi/\beta $ with $\beta$ being the inverse temperature of the system, which is also expected for black holes in Einstein gravity \cite{Maldacena:2015waa}.
\end{itemize}
To study the properties of this model in an analytically solvable regime that does not rely on low energies, the double scaling limit
\begin{equation}\label{eq:double scaling}
    N,~ p \rightarrow \infty~,\quad q\equiv\rme^{-\lambda} \equiv \rme^{-\frac{2p^2}{N}}~ \text{fixed}~,
\end{equation}
has been scrutinized in the literature, see e.g. \cite{kitaevTalks,Maldacena:2016hyu,Cotler:2016fpe,Berkooz:2018qkz,Berkooz:2018jqr}. In this regime, the partition function and the 2 and 4-point correlation functions can also be solved exactly \cite{Berkooz:2018jqr}, at any energy scale. Another remarkable advantage is that there exist many models in the same universality class as the DSSYK model, which means that their ensemble averaged observables can be solved using the same type of counting techniques as what we discuss in this chapter. The reader is referred to \cite{Erdos:2014zgc,Berkooz:2018qkz} for more details of this universality class, which includes the hypercube model of Parisi \cite{Parisi_1994}, as well as recently studied examples of double scaled integrable-chaotic models \cite{Berkooz:2024evs,Berkooz:2024ofm,Almheiri:2024xtw}, and double scaled chaotic bosonic models \cite{Jia:2024tii}.

Intriguingly, the DSSYK model has a \emph{maximal entropy state} \cite{Lin:2022rbf}, which is one of the main characteristics of dS space associated with the entropy given by (\ref{eq:GH entropy}) \cite{Chandrasekaran:2022cip}.\footnote{Different systems share this characteristic, they can be elegantly studied in algebraic quantum field theory with von Neumann algebras \cite{Chandrasekaran:2022cip,Lin:2022rbf,Jensen:2023yxy,Kudler-Flam:2023qfl,Witten:2023qsv,Witten:2023xze,Seo:2022pqj,Gomez:2023wrq,Gomez:2023upk,Gomez:2023tkr,Gomez:2023jbg,Aguilar-Gutierrez:2023odp,Basteiro:2024cuh,Xu:2024hoc}. We will not enter into the details about this area. The reader is referred to \cite{Papadodimas:2013jku,Jefferson:2018ksk,Leutheusser:2021frk,Witten:2021unn} for early work on von Neumann algebras in quantum gravity, and \cite{Witten:2018zxz,Witten:2021jzq,Sorce:2023fdx,Casini:2022rlv} for recent reviews.} Recently, there have been several exciting developments in dS holography based on the DSSYK model in the series of works \cite{Narovlansky:2023lfz,Verlinde:2024znh,Verlinde:2024zrh} (see also \cite{Blommaert:2023opb,Milekhin:2023bjv,Almheiri:2024xtw,Blommaert:2024ydx}). It has been argued that a pair of DSSYK models can have a dual interpretation in terms of (1+2)-dimensional (non-rotating) SdS$_3$ space.\footnote{It has been recently shown \cite{Blommaert:2024ydx} that the DSSYK model without additional constraints is dual to a dilaton-gravity theory of the form
\begin{equation}\label{eq:Sin Phi}
    I_{\rm dilaton-gravity}=\frac{1}{\abs{\log q}}\int\rmd^2x\sqrt{-g}\qty[\Phi\mathcal{R}-2\sin\Phi]+I_{\rm bdry}~.
\end{equation}
with $I_{\rm bdry}$ an appropriate boundary action. We instead consider the proposal \cite{Narovlansky:2023lfz,Verlinde:2024znh,Verlinde:2024zrh} where the pair of DSSYK models have additional constraints (\ref{eq:phy states},\ref{eq:phy ops}). We leave a detailed study of complexity in (\ref{eq:Sin Phi}) and its connections with the DSSYK model for future studies (see Ch. \ref{ch:Conclusion} for further comments).} {We will briefly review the different sides of the correspondence below.}

On the bulk side, SdS$_3$ space (\ref{eq:spherically symmetric metric}, \ref{eq:asympt_dS}) is locally isomorphic to dS$_3$ space; however, the term $M$ modifies the periodicity of the angular coordinate, which we denote $\Phi$ (i.e. $\rmd\Phi=d\Omega_{1}$ in (\ref{eq:spherically symmetric metric})) by
\begin{equation}
    \Phi\sim\Phi+2\pi(1-\alpha)~,\quad\alpha\equiv1-\sqrt{1-8G_NM}~.
\end{equation}
\cite{Verlinde:2024znh} proposed to identify a holonomy variable measuring the conical deficit angle, $2\pi\alpha$, produced by matter sources along the poles of the sphere, with the Hamiltonian for SdS$_3$ space. They studied the canonical quantization of this proposal in the Chern-Simons (CS) formulation of SdS$_3$ space (see e.g. \cite{Witten:1988hc,Witten:1998qj,Castro:2011xb,Castro:2023dxp,Castro:2023bvo}) which turns out to take the same form of a pair of DSSYK models, subject to physical constraints.

On the quantum mechanical side of the correspondence, each DSSYK model is described by Hamiltonians $H^{\rm L/R}$ of the form (\ref{eq:DDSSYK Hamiltonian}) where L and R are labeling the different theories. It was argued in \cite{Narovlansky:2023lfz} that the doubled DSSYK system (\ref{eq:DDSSYK Hamiltonian}) can describe the same correlators as dS$_3$ space by gauging discrete symmetries of quantum gravity (specifically charge-parity and time reversal invariance \cite{Harlow:2023hjb}), and imposing a Hamiltonian constraint on the physical states (i.e. gauge and diffeomorphism invariant states) of the system,
    \begin{equation}\label{eq:phy states}
        (H^{\rm L}-H^{\rm R})\ket{\chi_{\rm phys}}=0~,
    \end{equation}
which translates to the requirement that for the physical operators acting on the system,
    \begin{equation}\label{eq:phy ops}
        [H^{\rm L}-H^{\rm R},~\mathcal{O}_{\rm phys}]=0~.
    \end{equation}
Interestingly, the maximal entropy state corresponds to an energy eigenstate 
\begin{equation}\label{eq:ref energy state}
    \ket{\mathbb{E}_0}\equiv\ket{E_0^{\rm L},~E_0^{\rm R}}~.
\end{equation}
Under these considerations, it was found in \cite{Verlinde:2024znh} that one can develop a dictionary between the doubled DSSYK model and (2+1)-dimensional SdS$_3$ space, even away from the $G_N\rightarrow0$ regime previously employed in \cite{Narovlansky:2023lfz}. The holographic dictionary so far has succeeded in matching partition functions, correlators, and quasinormal modes of dS space. The state in (\ref{eq:ref energy state}) has been identified with the maximal entropy state of dS space, $\ket{\psi_{\rm dS}}$. {According to the interpretation in \cite{Verlinde:2024znh}, the microscopic theory dual might be located along the worldline of the observers, or on the cosmological horizon, given that $E_0$ corresponds to the maximum of the density of states $\rme^{S(E)}$.}

It might be surprising for the reader that there is a duality between 3D gravity (SdS$_3$) and a quantum mechanical theory (DSSYK), in contrast, for instance, to the holographic dictionary between (nearly)-AdS$_2$ space \cite{Maldacena:2016upp}, described by Jackiw–Teitelboim (JT) gravity \cite{JACKIW1985343,TEITELBOIM198341}, with the triple scaling limit of the SYK model (i.e. $\lambda\ll1$ and energies $E/J\ll1$) \cite{Sachdev:2010um,Maldacena:2016hyu}. In \cite{Verlinde:2024zrh}, it was shown that there is an alternative procedure in the CS quantization of SdS$_3$, depending on the order when the physical constraints are imposed. This results in a third member of the correspondence, a two-dimensional gravity theory that will be referred to as Liouville-de Sitter (LdS) in the remainder of the chapter. This theory in Lorentzian signature is defined in terms of two space-like Liouville-CFT$_2$,\footnote{The word ``spacelike" is used when $\phi\in\mathbb{R}$ and the kinetic term in (\ref{eq:action LdS}) has positive sign; while a negative sign would instead be referred to as ``timelike". See \cite{Nakayama:2004vk,Teschner:2001rv,Vargas:2017swx} for reviews on Liouville-CFTs, and \cite{Polyakov:1981rd} for initial work in this area.} as
\begin{equation}\label{eq:action LdS}
    \begin{aligned}
        I_{\rm LdS}=&I_{\rm sL}[\phi_+]+I_{\rm sL}[\phi_-]~,\\
        I_{\rm sL}[\phi_\pm]=&\frac{1}{4\pi}\int_{\Sigma}\rmd^2\sigma~\sqrt{\abs{h}}\qty[h^{\mu\nu}\partial_\mu\phi_\pm{\partial_\nu}\phi_\pm+Q_\pm \mathcal{R}_h\phi_\pm+\mu_{\rm B}\rme^{2b_\pm\phi_\pm}]\\
        &+\frac{1}{2\pi}\int_{\partial\Sigma}\rmd\tau~ \abs{h}^{1/4}(Q_\pm k+\mu_B\rme^{b_\pm\phi_\pm})
    \end{aligned}
\end{equation}
where $\Sigma$ is the boundary manifold (such that $\partial\Sigma$ corresponds to the geodesic of S or N pole worldline observer in SdS$_3$ space \cite{Verlinde:2024zrh}); $\mu_B$ is called the boundary cosmological constant, which parametrizes the boundary conditions of the theory; $Q_\pm=b_\pm+b_\pm^{-1}$ is the background charge; $h_{ij}$ the boundary metric; $\mathcal{R}_h$ its scalar curvature; $k$ the boundary curvature; $\tau$ is a time-like coordinate along $\partial{\Sigma}$, and $b_\pm\in\mathbb{C}$ are constants which obey $b_+=(b_-)^*$ and $b_+^2\in\rmi\mathbb{R}$. Dirichlet boundary conditions along $\partial\Sigma$ are imposed and the value of $\mu_B$ is fixed \footnote{In terms of the Liouville CFT literature, $\partial\Sigma$ corresponds to a Fateev-Zamolodchikov-Zamolodchikov-Teschner (FZZT) brane \cite{Fateev:2000ik,Teschner:2000md}. The reader is referred to \cite{Chatterjee:2024phq} for a recent and rigorous introduction to the topic.}

Upon quantization, the central charge of the $\pm$ sectors is complex, while the complete theory has a real central charge, given by
\begin{equation}
    c_\pm=1+Q_\pm^2~,\quad c_{+}+c_-=26~.
\end{equation}
It was found in \cite{Verlinde:2024zrh} that the correlation functions of physical operators (see Sec. \ref{sec:Krylov}) in this theory agree with those in the doubled DSSYK model, which together with the original description of SdS$_3$ space, provide compelling evidence for a holographic triality.

Below, we will review in more detail the basic theoretical aspects of the DSSYK model that will be applied in the next chapter.

\section{The double-scaled Sachdev-Ye-Kitaev model}
\subsection{Chord diagrams}\label{ssec:Chords}
In the double scaling limit (\ref{eq:double scaling}) one can obtain a statistical description of the system by studying the ensemble-averaged moments of the Hamiltonian in (\ref{eq:DDSSYK Hamiltonian}), denoted as $\expval{\Tr[H^k]}_J$, where the averaging refers to (\ref{eq:GEA J}), and the trace is taken with respect to the Hilbert space of the Majorana fermions $\psi_i$. We will briefly explain the techniques developed in \cite{Erd_s_2014,Cotler:2016fpe,Berkooz:2018qkz,Berkooz:2018jqr} to evaluate these quantities using (\ref{eq:DDSSYK Hamiltonian}),
\begin{equation}\label{eq:sum moments}
    \expval{H^k}_J=\rmi^{kp/2}\sum_{I_1,\dots, I_k}\expval{J_{I_1}\dots J_{I_k}}\Tr\qty(\psi_{I_1}\dots\psi_{I_k})~,
\end{equation}
where the subindex $I\equiv\qty{i_1,\dots i_p}$ is a collective index of size $p$ (which we denote $\abs{I}=p$), and similarly $\psi_I=\psi_{i_1}\dots\psi_{i_p}$ is a string of fermionic operators. Since the couplings $J_I$ obey a Gaussian distribution (\ref{eq:GEA J}), one can express (\ref{eq:sum moments}) as a sum of all pairwise contractions of $J_I$ times the appropriate trace. We now perform the ensemble averaging (\ref{eq:GEA J}), which requires that the operator strings $\Psi_I$ appear twice in the trace (\ref{eq:sum moments})
\begin{equation}\label{eq:sum moments2}
    \expval{H^k}_J=\rmi^{kp/2}\frac{J^k}{\lambda^{k/2}}\begin{pmatrix}
        N\\
        p
    \end{pmatrix}^{-k/2}\sum_{\text{pairings}}\sum_{I_1,\dots, I_{k/2}}\Tr\qty(\Psi_{I_1}\Psi_{I_2}\dots\Psi_{I_1}\dots)~,
\end{equation}
where $\sum_{\text{pairings}}$ indicates that only the strings $\Psi_I$ with the same indices are contracted.

Diagrammatically, we will represent the trace in (\ref{eq:sum moments2}) by a circle where we place $k$ nodes, each one representing the strings $\Psi_I$.\footnote{One needs to perform a Wick rotation $t\rightarrow \rmi t^{(\rm E)}$ to represent the strings $\Psi_I$ in the circle, whose periodicity is fixed by the inverse temperature $\beta$ when we work in the canonical ensemble.} The $k/2$ pairwise contractions, required by Wick's theorem, are represented by lines (called \textit{chords}) connecting the vertices. This is called a \textit{chord diagram}; see Fig. \ref{fig:SlicedChordDiagram} for an example.

Moreover, since we are considering Majorana fermions, using (\ref{eq:Majo fer}) we have that
\begin{equation}\label{eq:annhil Psi}
    \Psi_I\Psi_I=\rmi^p~,
\end{equation}
which simplifies the evaluation of the trace by noticing from (\ref{eq:Majo fer}) that 
\begin{equation}\label{eq:comm Psi}
    \Psi_I\Psi_{I'}=(-1)^{\abs{I\cap I'}}\Psi_{I'}\Psi_I~,
\end{equation}
where ${\abs{I\cap I'}}$ is the number of indices that appear both in $I$ and $I'$.

There are two key results shown rigorously in \cite{Berkooz:2018jqr} that allow for simplifications in the double scaling limit (\ref{eq:double scaling}): (i) ${\abs{I\cap I'}}$ obeys a \textit{Poisson probability distribution}
\begin{equation}\label{eq:poisson distr}
    \text{Pr}(\abs{I\cap I'},~m)=\frac{\lambda_P^m\rme^{-\lambda_P}}{m!}~,
\end{equation}
where $m$ is the number of sites in the intersection $\abs{I\cap I'}$, and $\lambda_P=p^2/N$ is the mean value of the distribution. (ii) ${\abs{I_l\cap I_m\cap I_n}}=0~\forall~l\neq m\neq n$. 

The factor $\begin{pmatrix}
        N\\
        p
    \end{pmatrix}^{-k/2}\sum_{I_1\dots,I_{k/2}}\Tr(\Psi_{I_1}\Psi_{I_2}\dots\Psi_{I_1}\dots)$ in (\ref{eq:sum moments2}) represents the sum of probabilities of appearances of a specific pairing $\Psi_{I_1}\Psi_{I_2}\dots\Psi_{I_1}$. Then, the factor produced by each Hamiltonian interception (i.e. when we arrange $\Tr(\Psi_{I_1}\Psi_{I_2}\dots\Psi_{I_1}\dots)$ into an ordered the set of strings as $\Tr(\Psi_{I_1}\Psi_{I_1}\dots\Psi_{I_{k/2}}\Psi_{I_{k/2}})$) can be evaluated as the sum of probabilities for m sites of intersections (\ref{eq:poisson distr}) with the factors $(-1)$ coming from the commutation between the strings in (\ref{eq:sum moments2}). Each Hamiltonian interception then contributes by a factor:
\begin{equation}\label{eq:H cap H factor}
   \rme^{-p^2/N}\sum_{m=0}^\infty(-1)^m\frac{(p^2/N)^m}{k!}\equiv q~,
\end{equation}
where $q=\rme^{-\lambda}$. Lastly, once we have ordered all strings in a pairwise matter (\ref{eq:comm Psi}) simplifies the factor $\rmi^{kp/2}$ in (\ref{eq:sum moments2}), leading to:
\begin{equation}\label{eq:chord tech}
    \expval{\Tr(H^k)}_J=\frac{J^{2k}}{\lambda^k}\sum_{\text{chord diagrams}}q^{\text{number of }(H\cap H)}~.
\end{equation}
where $q=\rme^{-\lambda}$, and \# is a shorthand for ``number of", and $H\cap H$ denotes Hamiltonian interceptions. This means that there will be a relative weight $q^n$ when any given chord intercepts with $n$ other chords.

\subsection{Chord Hilbert space without matter}\label{ssec:Chord Hilbert}
\begin{figure}[t!]
    \centering
    \subfloat[]{\includegraphics[width=0.35\textwidth]{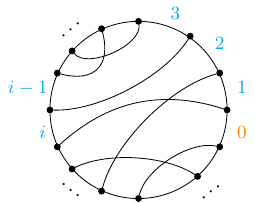}}\hspace{0.5cm}\subfloat[]{\includegraphics[width=0.6\textwidth]{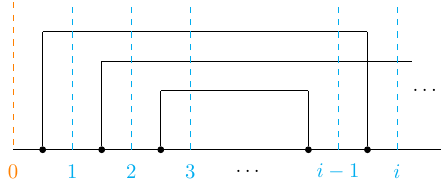}}
    \caption{(a) Example of disk chord diagram, where we label the different levels (cyan) before each vertex (black dot), where $0$ (orange) represents the level where we will cut the diagram. (b) The chord diagram is sliced open (each level is represented with a dashed line). Each chord is a Wick contraction between the nodes (black dots) corresponding to the Hamiltonians in (\ref{eq:chord tech}), which can then end on the subsequent levels.}
    \label{fig:SlicedChordDiagram}
\end{figure}

We will now provide a brief overview (mostly based on \cite{Berkooz:2018jqr,Rabinovici:2023yex}) of how to use (\ref{eq:chord tech}) to evaluate amplitudes that only involve the Hamiltonian moments. First, consider slicing open the chord diagram at any chosen point, so that the total number of nodes (which depends on how many closed or open chords we consider) lie on a line rather than a circle, as shown in Fig. \ref{fig:SlicedChordDiagram}. This represents a transition from a state without chords, to one with $k$ chords, which transitions back to one without chords.

\begin{figure}[t!]
    \centering
    \subfloat[]{\label{subfig:ClosedOpen}\includegraphics[width=0.35\textwidth]{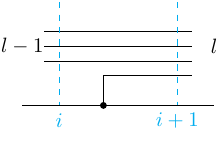}}\hspace{0.75cm}\subfloat[]{\label{subfig:OpenClosed}\includegraphics[width=0.35\textwidth]{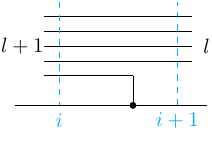}}
    \caption{Two ways to end with $l$ open chords after vertex $i+1$. (a) $l-1$ open chords before vertex $i$. (b) $l+1$ open chords before vertex $i$.}
    \label{fig:OpenClosedChords}
\end{figure}

Let $v_{l}^{(i)}$ denote the sum of chord diagrams with $l$ open chords at the $i$-th vertex in the sliced amplitude, and starting at the zeroth vertex. For a generic vertex $i$ with $l$ open chords immediately after it, one has 2 possibilities (Fig. \ref{fig:OpenClosedChords}): (a) that $l-1$ of them were open at level $i-1$ and one chord opens just before the vertex $i$, and (b) that $l+1$ of them were open at level $i-1$ and one chord is closed at vertex $i$. In the latter case, the chord that closes off might cross with any of the other $l$ open chords. 
Considering the factor (\ref{eq:chord tech}), the recursion relation for the total number of \emph{involuntary interceptions} at a given vertex becomes
\begin{equation}\label{eq:relation v}
    v_{l}^{(i+1)}=\frac{J}{\sqrt{\lambda}}\qty(v_{l-1}^{(i)}+\sum_{j=1}^lq^jv_{l+1}^{(i)})\equiv Tv_{l}^{(i)}~,
\end{equation}
where in the last line we have introduced the so-called ``\emph{transfer matrix}", $T$. 

We can associate a Hilbert space with this construction. Consider the stage $i=0$ where there are no open chords, so that we can define a corresponding vacuum state $v^{(0)}_0=\ket{0}$. By applying the recursion relation (\ref{eq:relation v}) we have that at stage $i=k$ (corresponding to the power of the Hamiltonian moment (\ref{eq:chord tech})) we have applied the transfer matrix $k$ times on the vacuum state of the ensemble theory $\ket{0}$, which has no open chords since it's connected to stage $0$,
\begin{equation}\label{eq:state closed chord}
    v^{(k)}_0=T^k\ket{0}~.
\end{equation}
The Hamiltonian moments in (\ref{eq:chord tech}) can be then expressed by projecting the final state, $v^{(k)}_0$, onto the initial one, $v^{(0)}_0$, which means
\begin{equation}\label{eq:momemts transfer}
    \expval{\Tr(H^k)}_J=\bra{0}T^{k}\ket{0}~,\quad \forall k~\text{even}~.
\end{equation}
We then notice that the transfer matrix acts as the Hamiltonian (up to a sign) in the ensemble-averaged theory. We would like to associate an auxiliary \emph{chord Hilbert space} $\mathcal{H}=\bigoplus_{n=0}^{\infty}\mathbb{C}\tilde{\ket{n}}$, where $n$ represents the number of open chords and the tilde means that that the states have not yet been normalized\footnote{Notice that $\mathcal{H}$ is not the same as the physical Hilbert space of the Majorana fermions $\psi_i$ in (\ref{eq:DDSSYK Hamiltonian}) since the states $\qty{\ket{n}}$ are defined in the ensemble-averaged theory (corresponding to the \textit{bulk} Hilbert space), so the left and right-hand sides of (\ref{eq:momemts transfer}) are described by different Hilbert spaces.}. In terms of the sum of chord diagrams (\ref{eq:state closed chord}, \ref{eq:relation v}), the $\qty{\tilde{\ket{n}}}$ states are defined as
\begin{equation}\label{eq:solu T}
    v^{(i)}_n=T^i\tilde{\ket{n}}~.
\end{equation}
The inner product between states in this Hilbert space can be defined from the Hamiltonian moments (\ref{eq:momemts transfer}) by inserting a completeness relation $\mathbbm{1}=\sum_n\tilde{\ket{n}}\tilde{\bra{n}}$, such that
\begin{equation}\label{eq:completeness in trace}
    \bra{0}T^{a+b}\ket{0}=\sum_{mn}\bra{\tilde{m}}\ket{\tilde{n}}\bra{0}T^a\tilde{\ket{m}}\tilde{\bra{n}}T^b\ket{0}~,
\end{equation}
From (\ref{eq:relation v}, \ref{eq:solu T}) and (\ref{eq:completeness in trace}), one recovers
\begin{equation}
    \bra{\tilde{m}}\ket{\tilde{n}}=\delta_{n,m}\prod_{i=1}^n\frac{1-q^i}{1-q}=\delta_{n,m}\frac{(q;~q)_n}{(1-q)^n}~,
\end{equation}
where $(a;~q)_n$ is the q-Pochhammer symbol:
\begin{equation}\label{eq:q-Pochhammer}
    (a;~q)_n\equiv\prod_{k=0}^{n-1}(1-aq^k)~,\quad (a_0,\dots, a_N;~q)_n=\prod_{i=1}^N(a_i;~q)~.
\end{equation}
However, (\ref{eq:relation v}) and (\ref{eq:solu T}) implies that $T$ would not be symmetric on this basis. We can perform a Lanczos algorithm to find an orthonormalized chord basis $\qty{\ket{n}}$ (i.e. $\bra{n}\ket{m}=\delta_{nm}$), which is related to the original chord basis $\tilde{\ket{n}}$ by (see Sec. \ref{ssec:def Krylov}):
\begin{equation}
    \ket{n}\equiv b_n^{-1}\tilde{\ket{n}},\quad \ket{n+1}=\frac{1}{b_{n+1}}\qty(T\ket{n}-b_n\ket{n-1})~,
\end{equation}
where $b_0=1$, and the recursion relation follows from (\ref{eq:relation v}). The coefficients $b_n$ are determined explicitly through the Lanczos algorithm with the Hamiltonian moments (\ref{eq:momemts transfer}) as
\begin{equation}
    b_n=\sqrt{\frac{1-q^n}{1-q}}~.
\end{equation}
The transfer matrix in the $\qty{\ket{n}}$ basis is then
\begin{align}
    T\ket{n}&=\frac{J}{\sqrt{\lambda}}\qty(\sqrt{\frac{1-q^n}{1-q}}\ket{n-1}+\sqrt{\frac{1-q^{n+1}}{1-q}}\ket{n+1})~.\label{eq:Transfer matrix}
\end{align}
Moreover, we can define the following operators:
\begin{equation}\label{eq: Oscillators 1}
    A\ket{n}=\sqrt{\frac{1-q^n}{1-q}}\ket{n-1}~,\quad A^\dagger\ket{n}=\sqrt{\frac{1-q^{n+1}}{1-q}}\ket{n+1}~,
\end{equation}
which obey the q-deformed commutation relation:
\begin{equation}\label{eq: Oscillators 2}
    \qty[A,~A^\dagger]_q\equiv AA^\dagger-qA^\dagger A=1~.
\end{equation}
$T$ can be then described in terms of a q-deformed harmonic oscillator as:\footnote{This model has a quantum group symmetry, $SU_q(1,~1)$, which is defined as a deformation of the Lie group SU(1,~1), whose generators ($\mathscr{D}$, $\mathscr{E}$, $\mathscr{F}$) obey q-deformed Lie algebra, denoted $U_q(su(1,~1))$:
\begin{equation}\label{eq:Qalgebra}
\begin{aligned}
    \qty[\mathscr{D},~\mathscr{E}]_{1}=\mathscr{E},\quad \qty[\mathscr{D},~\mathscr{F}]_{1}=-\mathscr{F},\quad \qty[\mathscr{E},~\mathscr{F}]_{1}=\frac{q^{\mathscr{D}}-q^{-\mathscr{D}}}{q-q^{-1}}~.
\end{aligned}
\end{equation}
Let $\mathscr{K}\equiv q^{\mathscr{D}}$. The algebra (\ref{eq: Oscillators 2}) can be obtained as a contraction of (\ref{eq:Qalgebra}):
\begin{equation}
\begin{aligned}
    A=\lim_{\varepsilon\rightarrow0}(q-q^{-1})^{1/2}\mathscr{E}~\varepsilon,\quad A^\dagger=\lim_{\varepsilon\rightarrow0}(q-q^{-1})^{1/2}\mathscr{F}~\varepsilon~,\quad q^{-\hat{n}}=\lim_{\varepsilon\rightarrow0}\mathscr{K}~\varepsilon~.
\end{aligned}
\end{equation}
See \cite{Ohtsuki:2002ud} for a pedagogical introduction to quantum groups and \cite{Berkooz:2018jqr,Blommaert:2023opb,Lin:2023trc,Berkooz:2022mfk} for more details about the quantum group of the DSSYK model. It has been conjectured that the dual geometry of the DSSYK model for finite $q$ can be described by a particle moving on a non-commutative deformation of AdS$_3$ space \cite{Berkooz:2022mfk,Blommaert:2023opb}, or on the quantum group manifold of SU$_q$(1,~1) \cite{Blommaert:2023opb,Blommaert:2023wad}; and possibly in other non-commutative geometries \cite{Almheiri:2024ayc}.}
\begin{equation}\label{eq:H DDSSYK momnetum}
\begin{aligned}
T&=\frac{J}{\sqrt{\lambda}}(A+A^\dagger)~.
\end{aligned}
\end{equation}
For our later discussion, it is convenient to introduce the chord number operator $\hat{n}$ and its conjugate momentum, $\hat{p}$:
\begin{equation}\label{eq:Normalized ops}
A=\rme^{\rmi \hat{p}}\sqrt{\frac{1-q^{\hat{n}}}{1-q}},\quad A^\dagger=\sqrt{\frac{1-q^{\hat{n}}}{1-q}}\rme^{-\rmi \hat{p}}~,
\end{equation}
which obey the relations
\begin{equation}\label{eq:chord numb op}
\begin{aligned}
    &\qty[\hat{n},~\rme^{\rmi \hat{p}}]_1=\rme^{\rmi \hat{p}}~,\quad\qty[\hat{n},~\rme^{-\rmi \hat{p}}]_1=-\rme^{-\rmi \hat{p}}~,\quad\hat{n}\ket{n}=n\ket{n}~.
    \end{aligned}
\end{equation}
The transfer matrix takes the form
\begin{equation}\label{eq:H DDSSYK momnetum2}
\begin{aligned}
T=&\frac{J}{\sqrt{\lambda(1-q)}}\qty(\rme^{\rmi \hat{p}}\sqrt{1-q^{\hat{n}}}+\sqrt{1-q^{\hat{n}}}\rme^{-\rmi \hat{p}})~.
\end{aligned}
\end{equation}
The reader can recognize that this operator corresponds to the Hamiltonian of a scattering problem.

As we have mentioned, there is an ambiguity in assigning the relative sign between the transfer matrix and the Hamiltonian (\ref{eq:momemts transfer}). We will consider $T=-H$ for the Hamiltonian to be bounded. Let us now construct a basis of states $\ket{\theta}$ where $T$ is diagonal, which can be then related to the eigenvalues of the Hamiltonian (\ref{eq:eigenval DSSYK}) as
\begin{equation}\label{eq:transfer in theta basis}
    T\ket{\theta}=-E(\theta)\ket{\theta}~.
\end{equation}
We can now combine the recursion relation (\ref{eq:Transfer matrix}) with the diagonalization above (\ref{eq:transfer in theta basis}) by defining a representation of the inner product between these two bases, $\psi_n(\theta,~q)\equiv\bra{\theta}\ket{n}$, which then obeys the relation:
\begin{equation}\label{eq:recursion}
    E(\theta)\psi_n(\theta,~q)=\frac{J}{\sqrt{\lambda}}\qty(\sqrt{\frac{1-q^n}{1-q}}\psi_{n-1}(\theta,~q)+\sqrt{\frac{1-q^{n+1}}{1-q}}\psi_{n+1}(\theta,~q))~,
\end{equation}
where we impose as initial condition $\bra{\theta}\ket{0}=1$ (where $\ket{0}$ is constructed as a normalized vacuum state). The solution to the recursion relation (\ref{eq:recursion}) are the q-Hermite polynomials \cite{Berkooz:2018jqr,Berkooz:2018qkz}:
\begin{equation}\label{eq: proj E0 theta}
    \psi_n(\theta,~q)=\bra{\theta}\ket{n}=\frac{H_n(\cos\theta|q)}{\sqrt{(q;~q)_n}}~,
\end{equation}
\begin{equation}\label{eq:eigenval DSSYK}
    E(\theta)=-\frac{2J\cos(\theta)}{\sqrt{\lambda(1-q)}}~.
\end{equation}
The q-Hermite polynomials can be expanded as:
\begin{equation}\label{eq:H_n def}
    H_n(\cos\theta|q)=\sum_{k=0}^n\begin{bmatrix}
        n\\
        k
    \end{bmatrix}_q\rme^{\rmi(n-2k)\theta}~,\quad
    \begin{bmatrix}
        n\\
        k
    \end{bmatrix}_{q}\equiv\frac{(q;~q)_{n}}{(q;~q)_{n-k}(q;~q)_{k}}~.
\end{equation}
Notice that since $E(\theta)$ is a continuous spectrum, its eigenstates cannot be part of a Hilbert space (they are not square integrable). In fact, since the q-Hermite polynomials obey the orthogonality relations:
\begin{equation}
    \int_0^\pi\frac{\rmd\theta}{2\pi}(q,~\rme^{\pm\rmi\theta};~q)_\infty H_m(\cos\theta|q)H_n(\cos\theta|q)=\delta_{m,n}(q;~q)_n~,
\end{equation}
where we introduced the notation $(q,~\rme^{\pm 2 \rmi \theta};q)_\infty=(q,~\rme^{2 \rmi \theta};q)_\infty(q,~\rme^{-2 \rmi \theta};q)_\infty$, then it follows that we can represent the $\ket{\theta}$ basis as
\begin{align}
\bra{\theta}\ket{\theta_0}&=\frac{2\pi}{\mu(\theta)}\delta(\theta-\theta_0)~,\quad\mu(\theta)=(q,~\rme^{\pm 2 \rmi \theta};q)_\infty\ ,\quad\label{eq:norm theta}\\
    1&=\int_0^\pi\frac{\rmd\theta}{2\pi}\mu(\theta)\ket{\theta}\bra{\theta}~.\label{eq:identity theta}
\end{align}
\subsubsection{Semiclassical thermodynamics}
To illustrate the techniques above, the partition function of the DSSYK model on the disk (see Fig. \ref{fig:SlicedChordDiagram}) can be expressed using the moments (\ref{eq:momemts transfer}) and identity (\ref{eq:identity theta}) as:
\begin{equation}
\begin{aligned}\label{eq:partition DSSYK}
        \expval{\Tr\qty(\rme^{-\beta H})}_J&=\bra{0}\rme^{-\beta T}\ket{0}=\int_0^\pi  \frac{d\theta}{2 \pi} \mu(\theta)~\rme^{-\beta E(\theta)}\\
        &=\frac{\sqrt{1-q}}{J\beta}\sum_{\nu=0}^\infty(-1)^pq^{\frac{\nu(\nu+1)}{2}}(2\nu+1)I_{2\nu+1}\qty(\frac{2J\beta}{\sqrt{1-q}})~,
\end{aligned}
\end{equation}
where $I_\nu(x)$ is the modified Bessel function of the first kind. In fact, one can include 1-loop corrections to the partition function (\ref{eq:partition DSSYK}) \cite{Goel:2023svz}, which, in the semiclassical limit (i.e. $q\rightarrow1$) takes the form
\begin{equation}\label{eq:partition function 1loop}
\begin{aligned}
    \eval{Z(\beta)}_{q\rightarrow1}=\int\rmd E(\theta)\rme^{S(\theta)-\beta E(\theta)}~,\quad S(\theta)=2\frac{\pi\theta-\theta^2}{\lambda}~.
\end{aligned}
\end{equation}
where $S(\theta)$ corresponds to the entropy in the microcanonical ensemble. One can use the above relation to derive the microcanonical inverse temperature of the saddle point solution
\begin{equation}\label{eq:beta DSSYK}
    \beta=\dv{S(\theta)}{E(\theta)}=\frac{2\pi-4\theta}{J\sin\theta}\sqrt{\frac{1-q}{\lambda}}~.
\end{equation}
Notice that the maximal entropy is reached when $\theta=\pi/2+2n\pi$. The existence of maximal entropy state is one of the similarities with dS space that motivated the proposal in \cite{Narovlansky:2023lfz,Verlinde:2024znh,Verlinde:2024zrh}, as we explained in Sec. \ref{sec:3D SdS holo}. However, notice that the microcanonical temperature would be infinite. The interpretation in \cite{Narovlansky:2023lfz,Verlinde:2024znh,Verlinde:2024zrh,Blommaert:2024ydx} is that there are two types of temperatures to describe dS physics, one describing a Boltzmann distribution for matter in the maximal entropy state, which would correspond to (\ref{eq:beta DSSYK}) in the DSSYK model, and a temperature measuring the decay of correlation functions with static patch time respect to a worldline observer (previously referred to as tomperature \cite{Lin:2022nss}) which remains finite. We will discuss the temperature associated with correlation functions in the DSSYK model in Sec. \ref{sec:matter DSSYK}.

\subsection{Special limits}
To conclude our discussion of general aspects. We briefly comment on the special limits of the DSSYK model where one reproduces the same statistical distribution as certain systems described below. In particular, the triple scaling limit allows one to study the low energy physics of the DSSSYK model, which is a well-known and perhaps, the simplest explicit example of the AdS/CFT correspondence, which we will explore in more detail in Ch. \ref{ch:ComplexityDSSYK}.

\paragraph{Random matrix theory}
Consider $q=0$, the probability density function in (\ref{eq:identity theta}) in this limit can be expressed
\begin{equation}
    \frac{1}{2\pi}\mu(\theta)=\frac{2}{\pi}\sin\theta~.
\end{equation}
This result coincides with the mean number of nonzero elements per row of real, symmetric, random $N\times N$ matrix theories (\gls{RMTs}) in the $N\gg1$ limit, called the Wigner semicircle law \cite{Eynard:2015aea}. The reason this should be expected is that $\lambda\rightarrow\infty$, so the number of all-to-all body interactions becomes much larger than the number of fermions, which makes the theory completely non-local, such that its density of states resembles that of a RMT universality class (whose precise ensemble is determined by the values of $N$ and $p$ \cite{Berkooz:2020fvm}).

\paragraph{SYK$_2$}
Consider:
\begin{equation}
    \lambda\rightarrow0~,\quad n\ll \lambda^{-1}~.
\end{equation}
In this regime, the probability density function with 1-loop corrections (\ref{eq:partition function 1loop}) can be expressed as
\begin{equation}
    \frac{1}{2\pi}\mu(\theta)=\frac{1}{\pi}\exp\qty({\lambda^{-1}\qty(\frac{\pi^2}{6}-2\qty(\theta-\frac{\pi}{2})^2)})~.
\end{equation}
This reproduces the same density of states obeyed by the large-N SYK$_2$ model (i.e. a Gaussian distribution) \cite{Garcia-Garcia:2018fns} (i.e. (\ref{eq:H DDSSYK momnetum}) for $p=2$ and $N\rightarrow\infty$), even though we started from a double scaling limit. This is one of the simplest solvable examples of the SYK model; its two-point correlation functions are reported in \cite{Maldacena:2016hyu}.

\paragraph{Triple scaling limit}
As we will discuss in Ch. \ref{ch:ComplexityDSSYK}, there is a special regime, called the \emph{triple scaling limit}, where one takes an additional step by considering:
\begin{equation}\label{eq:triple scaling}
    \lambda\rightarrow0~,\quad \ell\equiv n\lambda\rightarrow\infty~,\quad\frac{\rme^{-\ell}}{4\lambda^2}=\rme^{-\tilde{\ell}}~\text{fixed}~.
\end{equation}
This limit translates to the low-energy behavior for the Hamiltonian of the system (\ref{eq:H DDSSYK momnetum2}), which takes the form:
\begin{equation}\label{eq:H TSSYK}
    H_{\rm TSSYK}-E_0=2J\lambda\qty(\frac{\hat{p}^2}{2\lambda^2}+2\rme^{-\tilde{\ell}})+\mathcal{O}(\lambda^2)
\end{equation}
where $H_{\rm TSSYK}=-T_{\rm TSSYK}$ and $E_0=-\frac{2J}{\lambda}$ is just a constant shift. 

Moreover, the probability distribution (\ref{eq:identity theta}) in the (\ref{eq:triple scaling}) regime transforms to
\begin{equation}\label{eq:prob triple}
\frac{1}{2\pi}\mu(\theta)=4\sqrt{\frac{2}{\lambda\pi}}\rme^{-2\lambda^{-1}\qty(\pi^2+\qty(\theta-\frac{\pi}{2})^2)}\sin\theta\sinh\frac{2\pi\theta}{\lambda}\sinh\frac{2\pi(\pi-\theta)}{\lambda}+\mathcal{O}(\lambda)~.
\end{equation}
Both the Hamiltonian (\ref{eq:H TSSYK}) and the spectral density (\ref{eq:prob triple}) in this limit can be exactly mapped to the ADM Hamiltonian and the energy density of JT gravity \cite{Berkooz:2018jqr}. This relation provides an explicit example of bulk emergence from primitive structure (the chord diagrams, instead of $\psi_i$ and $J_I$) \cite{Almheiri:2024xtw}. Here, the chord number states $\ket{n}$ have a bulk interpretation as wormhole states connecting the asymptotic boundaries of an eternal AdS$_2$ black hole with a fixed (renormalized and discrete) geodesic length. More details, including a connection with Liouville quantum mechanics, can be found in \cite{Lin:2022rbf,Rabinovici:2023yex}.

To conclude our discussion of the chord Hilbert space without matter operators, we introduce a doubled Hilbert space description for the chord number states $\qty{\ket{n}}$. This allows a connection between the DSSYK model and tensor networks, as recently argued by \cite{Okuyama:2024yya}. This description will be useful to study the spread complexity in the doubled DSSYK model in Ch. \ref{ch:ComplexityDSSYK}.

\subsection{The doubled Hilbert space}\label{ssec:HxH and MERA}

Let us denote $X$ as an arbitrary operator acting on the Hilbert space $\mathcal{H}$ of one of the DSSYK models. We introduce its state representation in a doubled Hilbert space $\mathcal{H}\otimes\mathcal{H}$ in the chord number basis as
\begin{equation}
\begin{aligned}
\hat{X}&=\sum_{m,\,n}X_{nm}\ket{m}\bra{n}~,\\
    |{X})&=\sum_{m,~n}X_{nm}\ket{m,~n}~,\label{eq:Op as state Oku}
    \end{aligned}
\end{equation}
where $X_{nm}\equiv\bra{m}X\ket{n}$, $\ket{m,~n}\equiv\ket{m}\otimes\ket{n}$. Notice that the inner product between operators $({Y}|{X})$ is determined by the chord basis elements.

Given that the chord number basis is orthonormal, one can then represent the identity operator as
\begin{equation}
    |\mathbbm{1})=\sum_{n=0}^\infty \ket{n,~n}=\mathcal{E}\ket{0,~0}
\end{equation}
where
\begin{equation}\label{eq:entangler}
    \mathcal{E}=\sum_n\frac{(A^\dagger\otimes A^\dagger)^n}{(q;~q)_{n}}~.
\end{equation}
Manifestly, $\mathcal{E}$ maximally entangles the vacuum state $\ket{0,~0}$. It was pointed out \cite{Okuyama:2024yya} that $\mathcal{E}$ is reminiscent of an entangler operator in a MERA network \cite{Vidal:2007hda,Vidal:2008zz}. 

\subsection{Adding matter operators}\label{sec:matter DSSYK}
So far, our review has been focused on the counting rules for the Hamiltonian moments (\ref{eq:chord tech}). We can easily generalize the previous arguments by introducing matter operators (also called matter chords) that have the form:
\begin{align}\label{eq:matter ops DSSYK}
\mathcal{O}_\Delta(t)&=\rmi^{\frac{p'}{2}}\sum_{i_1,\dots~,i_{p'}}K_{i_1\dots i_{p'}}\psi_{i_1}(t)\dots\psi_{i_{p'}}(t)~.
    \end{align}
    Here $K_{i_1\dots i_{p'}}$ are Gaussian random couplings, independent of $J_{i_1\dots i_{p}}$, and $\Delta\equiv p'/p$, meaning that
    \begin{equation}\label{eq:ensemble av operators}
        \expval{K_{I}}=0~,\quad \expval{K_{I}J_{I'}}=0~,\quad \expval{(K_{I})^2}=\frac{\mathcal{K}^2}{\Delta^2\lambda}\begin{pmatrix}
        N\\
        p'
    \end{pmatrix}^{-1}~.
    \end{equation}
    where $\mathcal{K}$ is a constant. Now, one has to account for the $H$ and $\mathcal{O}$-nodes where we perform the Wick contractions, and average over the random couplings $J_{I}$ and $K_{I'}$.

Next, we would like to evaluate the n-point correlation functions of the form
\begin{equation}\label{eq:matter corr}
\begin{aligned}
    &\expval{\Tr(H^{k_1}\mathcal{O}_\Delta(t_1)\dots H^{k_n}\mathcal{O}_\Delta(t_n)H^{k_{n+1}})}_{J,~K}\\
    &=\sum_{\qty{I}}\expval{J_{I_1}\dots J_{I_{k_1}}K_{I'_1}\dots}\Tr\qty(\Psi_{I_1}\dots\Psi_{I_{k_1}}\Psi_{I'_{1}}\dots)~.
\end{aligned}
\end{equation}
As in Sec. \ref{ssec:Chords}, we need to perform an ensemble averaging with (\ref{eq:GEA J}, \ref{eq:ensemble av operators}), and given that they are Gaussian, and apply pairwise Wick contractions of the strings $\Psi_I$ between operators of the same type (i.e. $H^{k}$ or $\mathcal{O}_\Delta(t)$). 

\begin{figure}
    \centering
    \subfloat[]{\includegraphics[width=0.29\textwidth]{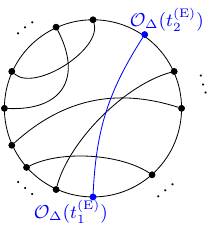}}\hfill\subfloat[]{\includegraphics[width=0.69\textwidth]{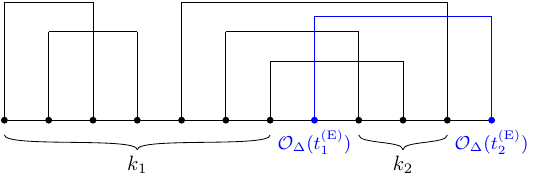}}
    \caption{Chord diagram for the correlation function (\ref{eq:correlator chord}) when $n=2$, which is represented as (a) the thermal circle, (b) a line where one of the operators $\mathcal{O}_\Delta$ appear in the rightmost end, and $k_1$, $k_2$ are the number of Hamiltonian insertions. The matter chords (blue) have been Wick rotated $t_i\rightarrow t_i^{(\rm E)}$.}
    \label{fig:MatterChord}
\end{figure}

To represent the matter chords diagrammatically, the difference is that when we locate the strings $\Psi_{I'}$ corresponding to an operator $\mathcal{O}_\Delta(t_1)$ with a coupling $K_{I'}$, we must label it differently from those of the Hamiltonian moments since it has a different weight (i.e. the factor $q^\Delta$). This is illustrated in Fig. \ref{fig:MatterChord} for a 2-point correlation function.

Let us now derive the appropriate factors to evaluate the correlator (\ref{eq:matter corr}). The procedure can be done just as in the single type of chord case (Sec. \ref{ssec:Chords}) but since $\Psi_I$ and $\Psi_{I'}$ can now have different sizes, meaning $\abs{I}=p$ while $\abs{I'}=p'$ then the Poisson distribution in (\ref{eq:poisson distr}) has a mean value $\lambda_P=\frac{p p'}{N}=\frac{\Delta p^2}{N}$ for the interception between Hamiltonian and matter chords ($H\cap \mathcal{O}$), while we have $\lambda_P=\frac{p^2}{N}$ and $\lambda_P=\frac{\Delta^2{p}^2}{N}$ for $H\cap H$ and $\mathcal{O}\cap \mathcal{O}$ respectively. Meanwhile, we will get factors $(-1)^{\abs{I\cap I'}}$ when commuting $\Psi_I$ and $\Psi_{I'}$ (\ref{eq:comm Psi}). This leads to the rules:
    \begin{itemize}
        \item Each $H\cap \mathcal{O}$ contributes a factor:
        \begin{equation}\label{eq:H cap O factor}
   \rme^{-\Delta p^2/N}\sum_{m=0}^\infty(-1)^m\frac{(\Delta p^2/N)^m}{k!}= q^\Delta~.
\end{equation}
        \item For each $\mathcal{O}\cap \mathcal{O}$
        \begin{equation}\label{eq:O cap O factor}
   \rme^{-\Delta^2p^2/N}\sum_{m=0}^\infty(-1)^m\frac{(\Delta^2p^2/N)^m}{k!}= q^{\Delta^2}~.
\end{equation}
    \end{itemize}
    Therefore, we can combine the results (\ref{eq:H cap H factor}, \ref{eq:H cap O factor}, \ref{eq:O cap O factor}) into (\ref{eq:matter corr})
\begin{equation}\label{eq:correlator chord}
\begin{aligned}
    &\expval{\Tr(H^{k_1}\mathcal{O}_\Delta(t_1)\dots H^{k_n}\mathcal{O}_\Delta(t_n)H^{k_{n+1}})}_{J,~K}\\
    &=\frac{J^{k_1+\dots+k_{n+1}}\mathcal{K}^n}{\Delta^n\lambda^{(k_1+\dots+k_{n+1}+n)/2}}\sum_{\rm chord~diagrams}q^{\#(H\cap H)}q^{\Delta\#(H\cap\mathcal{O})}q^{\Delta^2\#(\mathcal{O}\cap \mathcal{O})}~.
\end{aligned}
\end{equation}
To simplify the evaluations, we assume there are no intersections of the form $(\mathcal{O}\cap\mathcal{O})$, which can be justified e.g. by considering bulk-free fields dual to the matter operator $\mathcal{O}_\Delta$.

One can deduce the rules for evaluating chord diagrams, similar to Sec. \ref{eq:correlator chord}. Let us specialize for the 2-point correlation functions, meaning $n=2$ in (\ref{eq:correlator chord}), as this will be used in Sec. \ref{sec:Krylov}. To facilitate the process, we cut the chord diagram in a way that one of the matter operators ends on the rightmost edge, as illustrated in Fig. \ref{fig:MatterChord}. We see that for the first $k_1$ Hamiltonian interceptions we have exactly the same contribution as in the case without matter operators. However, the open chords crossing the first $\mathcal{O}_\Delta$ factor will generate an additional $(q^{\Delta})^{\#(H\cap \mathcal{O})}$ given (\ref{eq:correlator chord}). The last step is to pair the remaining $k_2$ Hamiltonian insertions, which would produce the same $T^{k_2}\ket{0}$ factor. 

Therefore, the action of a pair of operators $\mathcal{O}_\Delta$ in the chord Hilbert $\mathcal{H}$ can be expressed as\footnote{We have set $\frac{\mathcal{K}}{\Delta\sqrt{\lambda}}=1$ in this expression {to normalize $\expval{\tr(\mathcal{O}_{\Delta}^2)}_{K}=1$, following the convention in \cite{Berkooz:2018jqr,Berkooz:2018qkz}. The parameter $\mathcal{K}$ can be reintroduced by dimensional analysis.}}
\begin{equation}\label{eq:exp val correlators}
    \expval{\Tr\qty(H^{k_1}\mathcal{O}_\Delta H^{k_2}\mathcal{O}_\Delta )}_J=\bra{0}T^{k_1}q^{\Delta \hat{n}}T^{k_2}\ket{0}~.
\end{equation}
We will then associate correlation functions in the auxiliary Hilbert space as
\begin{equation}\label{eq:correlator DSSYK}
    \bra{0}\mathcal{O}_\Delta\rme^{\beta_1T}\mathcal{O}_\Delta\rme^{\beta_2T}\ket{0}=\int\prod_{i=1,~2}\frac{\rmd\theta_i}{2\pi}\mu(\theta_i)\rme^{-\beta_i E(\theta_i)}\bra{\theta_1}q^{\Delta\hat{n}}\ket{\theta_2}~,
\end{equation}
where (\ref{eq:identity theta}, \ref{eq:exp val correlators}) were used. By inserting the identify (\ref{eq:identity theta}) on the left hand side, one can associate \cite{Berkooz:2018jqr}
\begin{equation}\label{eq:1point O delta}
    \bra{\theta_1}\mathcal{O}_\Delta\ket{\theta_2}=\sqrt{\bra{\theta_1}q^{\Delta\hat{n}}\ket{\theta_2}}=\sqrt{\frac{(q^{2\Delta};q)_\infty}{(q^{\Delta}\rme^{\rmi(\pm\theta-1\pm\theta_2)};q)_\infty}}~.
\end{equation}
where we have used the identity (\ref{eq:identity theta}) and the so-called n-orthogonality relation of the q-Hermite polynomials \cite{Berkooz:2018jqr}:
\begin{equation}
\sum_{n}\frac{q^{\Delta n}}{(q;q)_n}H_n(\cos\theta|q)H_n(\cos\theta'|q)=\frac{(q^{2\Delta};q)_\infty}{(q^{\Delta}\rme^{\rmi(\pm\theta\pm\theta')};q)_\infty}~.
\end{equation}
The relation (\ref{eq:1point O delta}) can be used to derive correlation functions in the doubled DSSYK model introduced at the start of the section, as we will discuss in Sec. \ref{sec:Krylov}.

To conclude our review of the DSSYK model, we remark that the auxiliary Hilbert space of the DSSYK model can be generalized to include states with an arbitrary number of matter operators, as recently developed in \cite{Lin:2022rbf,Lin:2023trc,Xu:2024hoc,Okuyama:2024gsn}. This will not be required for our evaluations in the next chapter, so we will not discuss it. Nevertheless, one can extend our arguments in the following chapter using the formalism of these recent developments, as we comment in Sec. \ref{sec:discu}.

\cleardoublepage
  \graphicspath{{\chaptersdir/ComplexityDSSYK/image/}}%
  \chapter{Complexity in de Sitter space from the double scaled SYK model}\label{ch:ComplexityDSSYK}
This chapter is mostly a reprint of \cite{Aguilar-Gutierrez:2024nau}. Our aim is to define complexity in dS space from microscopic principles. Based on recent developments pointing towards a correspondence between a pair of double-scaled Sachdev–Ye–Kitaev (DSSYK) models/ 2D Liouville-de Sitter (LdS$_2$) field theory/ 3D Schwarzschild de Sitter (SdS$_3$) space in \cite{Narovlansky:2023lfz,Verlinde:2024znh,Verlinde:2024zrh}, we study concrete complexity proposals in the microscopic models and their dual descriptions. First, we examine the \textit{spread complexity} of the maximal entropy state of the doubled DSSYK model. We show that it counts the number of entangled chord states in its doubled Hilbert space. We interpret spread complexity in terms of a time difference between antipodal observers in SdS$_3$ space, and a boundary time difference of the dual LdS$_2$ CFTs. This provides a new connection between entanglement and geometry in dS space. Second, \textit{Krylov complexity}, which describes operator growth, is computed for physical operators on all sides of the correspondence. Their late time evolution behaves as expected for chaotic systems. Later, we define the \textit{query complexity} in the LdS$_2$ model as the number of steps in an algorithm computing n-point correlation functions of boundary operators of the corresponding antipodal points in SdS$_3$ space. We interpret query complexity as the number of matter operator chord insertions in a cylinder amplitude in the DSSYK, and the number of junctions of Wilson lines between antipodal static patch observers in SdS$_3$ space. Finally, we evaluate a specific proposal of \textit{Nielsen complexity} for the DSSYK model and comment on its possible dual manifestations.

We will now apply our understanding of complexity in quantum mechanical and quantum field theory systems, reviewed in Ch. \ref{sec:QC proposals} to the doubled DSSYK model and the LdS$_2$ CFT, in relation to the dS$_3$ space holographic proposal in Sec. \ref{sec:3D SdS holo}. 

\subsection*{Main question: How to define complexity in dS space?}
An exciting possibility from this recent holographic framework in terms of the DSSYK model is to study notions of quantum information theory for SdS$_3$ space and the Liouville CFT side from the quantum mechanical dual description. Concretely, we ask:
\begin{quote}
\emph{Can the doubled DSSYK model provide first principles to properly define quantum information-theoretic notions of complexity in dS space?}
\end{quote}
As we have emphasized in Ch. \ref{ch:QComplexity}, there is a large variety of notions of complexity for quantum systems, which we would like to compare with the holographic complexity proposals studied in Ch. \ref{ch:ComplexitydS}. In particular, whether the microscopic notion of complexity display a relation similar to the hyperfast growth (\ref{eq:hyperfast}) or the eternal growth discovered for the proposals in Sec. \ref{sec:CAny CMC}. Given that the previous studies have considered different gravitational observables without a clear holographic description in terms of complexity, our work aims to examine some microscopic notions of complexity and interpret their bulk description from the dS holographic dictionary based on the series of works in \cite{Narovlansky:2023lfz,Verlinde:2024zrh,Verlinde:2024znh}, and compare their evolution with the previous holographic complexity proposals.

\subsection*{Outline of the chapter}
The main purpose of our work is to study concrete notions of complexity in dS space based on the microscopic dual theories identified in \cite{Narovlansky:2023lfz,Verlinde:2024znh,Verlinde:2024zrh}, as well as to develop its geometrical interpretation in the bulk. Concretely, we study the definitions of spread, Krylov, and query complexity on all sides of the dS correspondence, and a particular proposal of Nielsen complexity in the DSSYK model. The main results of our work are summarized in Table \ref{tab:results}.
\begin{table}[t!]   
\begin{center}    
\begin{tabular}  {|C{1.6cm}|C{3cm}|C{3cm}|C{3cm}|} \hline  
Proposal & Doubled DSSYK model & LdS$_2$ CFT & SdS$_3$ space\\ \hline
Spread complexity, $\mathcal{C}_{\rm S}$& Number of entangled chord states (\ref{eq:C S DDSSYK}) for the state (\ref{eq:ref energy state})& Boundary time difference (\ref{eq:CS LdS})& Static patch time difference (\ref{eq:relation y and spread})\\ \hline
Krylov complexity, $\mathcal{C}_{\rm K}$& Exponential growth (\ref{eq:late time CK}) of physical operators in (\ref{eq:O phys DDSSYK}) &Exponential growth (\ref{eq:late time CK}) of physical operators (\ref{eq:O phys LdS})& Exponential growth (\ref{eq:late time CK}) of physical operators (\ref{eq:O phys SdS})\\ \hline
Query complexity, $\mathcal{C}_{\rm Q}$& Number of matter chord insertions (\ref{eq:CQ DDSSYK}) & Number of fusions (\ref{eq:Query n fusions}) & Number of junctions of Wilson lines (\ref{eq:CAny})\\ \hline
Nielsen complexity, $\mathcal{C}_{\rm N}$& Distance (\ref{eq:CN after minimization}) between $\mathbbm{1}$ and a unitary (\ref{eq:Nielsen intermediate}) in its group manifold & ??? & ???\\ \hline
\end{tabular}   
\caption{Different quantum complexity proposals (spread, Krylov, query, and Nielsen complexity) in this work and their interpretation for each side of the doubled DSSYK model/LdS$_2$ CFT/SdS$_3$ space correspondence. For comments about holographic duals to Nielsen complexity, see Sec \ref{sec:comments Nielsen}.}
\label{tab:results}
\end{center}
\end{table}

First, using the doubled Hilbert space formalism of \cite{Okuyama:2024yya} we provide a natural interpretation of spread complexity in the doubled DSSYK model as a map that counts the number of entangled chord states, which are projected onto the maximal entropy state (\ref{eq:ref energy state}). In this formalism, there is a description of the DSSYK model reminiscent of a \textit{multi-scale entanglement renormalization ansatz} ({MERA}) network \cite{Vidal:2007hda,Vidal:2008zz}\footnote{The reader is referred to App. \ref{app:MERA} for a brief introduction to the MERA network.}, as noticed by \cite{Okuyama:2024yya}. We use a known entry in the holographic dictionary relating chord number with a geodesic length\footnote{Note however, that this is not in contradiction with \cite{Aguilar-Gutierrez:2023nyk}, where it was found that spread and Krylov complexity cannot represent distances in metric spaces, as a geodesic length between two points is not the same as their distance.} measuring static patch time difference between the N and S poles in SdS$_3$ space \cite{Verlinde:2024znh}, to interpret boost symmetries in SdS$_3$ space in terms of spread complexity in the doubled DSSYK model. Since spread complexity counts entangled chord states in the doubled Hilbert space, our study provides a connection between entanglement and geometry similar to the ER=EPR conjecture \cite{Maldacena:2013xja}.

Secondly, we study the notion of Krylov complexity for physical operators. To our convenience, the correlation functions of physical operators in the maximal entropy state have been computed and matched by \cite{Narovlansky:2023lfz,Verlinde:2024znh,Verlinde:2024zrh}. Since Krylov complexity is completely determined in terms of the correlation functions of the chosen initial operator (see Sec. \ref{ssec:def Krylov}), our results describe the same physical operator growth for the doubled DSSYK model/LdS$_2$ CFT/SdS$_3$ space. We show that this complexity proposal displays an exponential growth behavior with respect to physical time, as expected for maximally chaotic systems. Its evolution is quite similar to previous studies on the Krylov complexity of the DSSYK model \cite{Parker:2018yvk,Bhattacharjee:2022ave}. Some details differ with respect to previous studies, given that we evaluate Krylov complexity for physical operators obeying the constraint (\ref{eq:phy ops}) instead of the $\psi_i$ fields themselves.

Later, we study the notion of query complexity for the LdS$_2$ CFT, originally proposed by \cite{Chen:2020nlj} for vacuum CFT$_2$ states dual to AdS$_3$ space. In this context, query complexity is the number of steps taken in an algorithm that reproduces multipoint correlators based on fusion rules in the CFT. However, our implementation of the type of correlation functions differs substantially from \cite{Chen:2020nlj} in that we consider pairwise contractions between matter operators inserted on the north and south poles of SdS$_3$ space (corresponding to the boundaries of the LdS$_2$ CFTs), instead of operators within a single AdS global time slice. Moreover, using the dictionary in \cite{Narovlansky:2023lfz,Verlinde:2024znh,Verlinde:2024zrh}, the protocol can be interpreted in terms of the doubled DSSYK model as counting the number of connections of matter operator chords in a cylinder amplitude, where the ends of the cylinder correspond to the pairs of DSSYK theories. Like in spread complexity, the correlation functions can be expressed in terms of MERA entangler and disentangler operators between the pairs of DSSYK theories. Lastly, we argue that query complexity in the bulk theory corresponds to the number of junctions of Wilson lines connecting the north and south poles of SdS$_3$ space in its CS formulation, which computes the correlation functions, and we describe how this approach relates to the holographic complexity proposals in asymptotically dS space \cite{Jorstad:2022mls,Auzzi:2023qbm,Anegawa:2023wrk,Anegawa:2023dad,Baiguera:2023tpt,Aguilar-Gutierrez:2023zqm,Aguilar-Gutierrez:2023tic,Aguilar-Gutierrez:2023pnn,Aguilar-Gutierrez:2024rka}.

Finally, motivated by the connections between computational complexity and holography, we study a particular notion of Nielsen operator complexity in the DSSYK model. We adopt a measure of complexity that is invariant under unitary transformations and time reversal, which allows for a tractable evaluation. We find a linear time growth, as expected for generic chaotic systems (see e.g. \cite{Chapman:2021jbh}.) Although, in contrast with the other proposals, we do not identify a holographic dual description, we notice that this type of evolution is compatible with certain holographic complexity proposals in asymptotically dS spacetimes \cite{Aguilar-Gutierrez:2023zqm,Aguilar-Gutierrez:2023tic,Aguilar-Gutierrez:2023pnn,Aguilar-Gutierrez:2024rka}. We then evaluate the low-energy limit of Nielsen complexity in the DSSYK, which is appropriate for describing JT gravity.

In summary, in Sec. \ref{sec:Spread} we evaluate the spread complexity of the maximal entropy state (\ref{eq:ref energy state}) in the Hilbert space of the doubled DSSYK model and study its holographic manifestations. Sec. \ref{sec:Krylov} is dedicated to deriving the Krylov complexity for the doubled DSSYK model, LdS$_2$ CFT and SdS$_3$ space using the known correlation functions in all sides of the correspondence \cite{Verlinde:2024znh,Verlinde:2024zrh}, and analyzing its late time behavior, which follows that of a maximally chaotic system. In Sec. \ref{sec:Query}, we use the proposal for query complexity, originally developed in \cite{Chen:2020nlj}, to define complexity in the LdS$_2$ CFT, and study its manifestation in terms of chord diagrams in the DSSYK model, and bulk invariant quantities in SdS$_3$ space. Sec. \ref{sec:comments Nielsen} contains new results on Nielsen's geometric approach to operator complexity in the DSSYK model, and some comments about its relation with dS holography, and JT gravity. Finally, Sec. \ref{sec:discu} contains a summary of the findings and some outlook questions.

\section{Towards spread complexity in de Sitter space}\label{sec:Spread}
In this section, we first study $\mathcal{C}_{\rm S}$ in the doubled Hilbert space description of the DSSYK model in terms of entanglers in a MERA network. Our arguments at this point are not restrained to dS holography, and they can be applied to the triple-scaled SYK/JT gravity correspondence. Afterward, we show that $\mathcal{C}_{\rm S}$ with $\ket{\mathbb{E}_0}$ as the reference state reproduces a geodesic distance in SdS$_3$ space from the holographic dictionary in \cite{Verlinde:2024znh}.

Importantly for us, and as noticed in \cite{Lin:2022rbf,Rabinovici:2023yex}, the Krylov basis of the DSSYK model is given by the chord number of states, i.e. 
\begin{equation}\label{eq:Krylov basis DSSYK}
\ket{K_n}=\ket{n}~,\quad b_n=-J\sqrt{\frac{1-q^n}{\lambda(1-q)}}~,
\end{equation}
since the Hamiltonian (i.e. the transfer matrix (\ref{eq:Transfer matrix}) up to a sign) becomes tridiagonal in this basis. Therefore, spread complexity is related to the chord number operator through the relation
\begin{equation}\label{eq:spread chord number}
    \mathcal{C}_{\rm S}(t)=\bra{\phi(t)}\hat{n}\ket{\phi(t)}~.
\end{equation}
This allowed \cite{Rabinovici:2023yex} to identify the spread complexity of the $\ket{\phi(0)}=\ket{0}$ state with a wormhole length in the JT gravity dual description to the DSSYK model.

We start defining the operator
\begin{equation}\label{eq:N bb}
\begin{aligned}
    {\mathbb{N}}&\equiv \frac{1}{2}\qty(\hat{n}\otimes\mathbbm{1}+\mathbbm{1}\otimes\hat{n})~,\\
    |{\mathbb{N}})&=\sum_nn\ket{n,~n}~.
\end{aligned}
\end{equation}
Using the identification (\ref{eq:Krylov basis DSSYK}), we can then express the spread complexity of a time-evolved state $\phi(t)$ (\ref{eq:spread complexity}) in the doubled Hilbert space formalism as
\begin{equation}
    \mathcal{C}_{\rm S}(t)=\bra{\phi(t),~\phi^*(t)}{\mathbb{N}})~.
\end{equation}
Expanding the evolved state in its Krylov basis as $\ket{\phi(t)}=\sum_n\phi_n(t)\ket{n}$, and employing (\ref{eq:entangler}):
\begin{equation}\label{eq:spread as counting gates}
\begin{aligned}
    \mathcal{C}_{\rm S}(t)&=\bra{0,~0}\mathcal{O}_{\phi(t)}^\dagger~\mathbb{N}~\mathcal{E}\ket{0,~0}~,
\end{aligned}
\end{equation}
where we have defined
\begin{equation}\label{eq:general OP MERA}
    \mathcal{O}_{\phi(t)}\equiv\sum_{n,~m}\frac{\phi_n(t)\qty(A^\dagger)^n}{\sqrt{(q;~q)_n}}\otimes\frac{\phi_m(t)\qty(A^\dagger)^m}{\sqrt{(q;~q)_m}}
\end{equation}
It can be seen in (\ref{eq:spread as counting gates}) that spread complexity has a natural interpretation as a map from an operator ($\mathbb{N}~\mathcal{E}$) that counts entangled chord states in the state $\ket{\phi(t),~\phi^*(t)}$, which has been prepared from the vacuum through the map $\mathcal{O}_{\phi(t)}\ket{0,~0}$.

\subsubsection*{Interpretation for the double-scaled Sachdev-Ye-Kitaev model}\label{sssec:Application spread}
\begin{figure}[t!]
    \centering
    \includegraphics[width=0.3\textwidth]{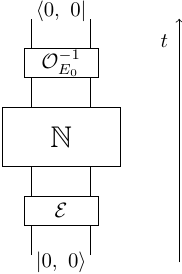}
    \caption{Illustration of spread complexity $\mathcal{C}_{\rm S}(t)$ (\ref{eq:spread complexity}) in $\mathcal{H}\otimes\mathcal{H}$. It maps the operator $\hat{\mathbb{N}}$ that counts the number of entangled chord states through $\mathcal{E}$, and are projected onto the state $\mathcal{O}_{E_0}\ket{0,~0}$, where $\mathcal{O}_{E_0}$ is defined in (\ref{eq:O E0}). See Sec. \ref{sec:discu} for comments on the DSSYK model and tensor networks.}
    \label{fig:spread MERA}
\end{figure}

We now specialize the previous discussion to the doubled DSSYK model \cite{Narovlansky:2023lfz,Verlinde:2024znh,Verlinde:2024zrh}. We take as reference state $\ket{\mathbb{E}_0}$ (\ref{eq:ref energy state}), which to the $\ket{\psi_{\rm dS}}$ in the bulk. We can express this state using the identity (\ref{eq:identity theta}) as:
\begin{equation}
    \ket{E_0}=\sum_n\frac{H_n(\cos\theta_0|q)}{\sqrt{(q;~q)_n}}\ket{n}~.
\end{equation}
(\ref{eq:spread as counting gates}) then simplifies to
\begin{equation}\label{eq:C S DDSSYK}
    \boxed{\mathcal{C}_{\rm S}(t)=\bra{0,~0}\mathcal{O}^{\dagger}_{E_0}\mathbb{N}~\mathcal{E}\ket{0,~0}~,}
\end{equation}
where
\begin{equation}\label{eq:O E0}
    \mathcal{O}_{E_0}=\sum_{n,~m}\frac{H_n(\cos\theta_0|q)\qty(A^\dagger)^n}{{(q;~q)_n}}\otimes\frac{H_m(\cos\theta_0|q)\qty(A^\dagger)^m}{{(q;~q)_m}}~.
\end{equation}
Notice that the time dependence has dropped out in (\ref{eq:C S DDSSYK}) due to $\ket{E_0}$ being an eigenstate of $T$. One can further simply (\ref{eq:O E0}) using the following identify (see e.g. \cite{Okuyama:2023byh}):
\begin{equation}
    \sum_n\frac{H_n(\cos\theta|q)t^n}{(q;~q)_n}=\frac{1}{(t~\rme^{\pm \rmi\theta};~q)_\infty}~,
\end{equation}
such that (\ref{eq:O E0}) becomes
\begin{equation}
    \mathcal{O}_{E(\theta)}=\frac{1}{(A^\dagger~\rme^{\pm \rmi\theta};~q)_\infty}\otimes \frac{1}{(A^\dagger~\rme^{\pm \rmi\theta};~q)_\infty}~.
\end{equation}
From the expressions above, we notice that $\mathcal{C}_{\rm S}$ manifestly counts the chord states that have been entangled through the gate $\mathcal{E}$, which are then projected to the maximal entropy state $\ket{\mathbb{E}_0}$, which is created from the vacuum $\ket{0,~0}$ by acting with $\mathcal{O}_{E_0}$. See Fig. \ref{fig:spread MERA} for an illustration.

\begin{figure}
    \centering
    \includegraphics[width=0.75\textwidth]{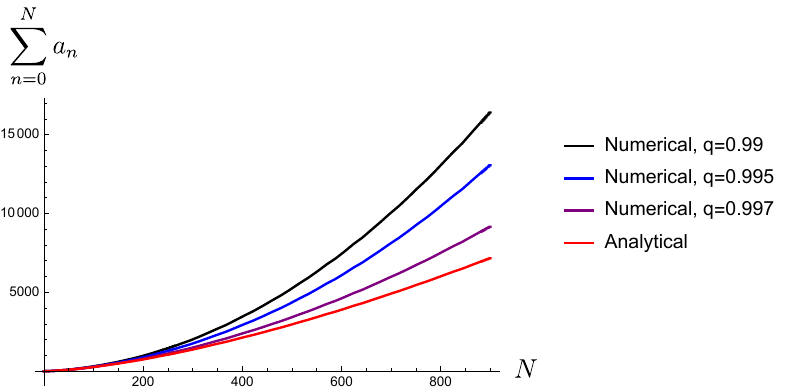}
    \caption{Evaluation of the series in (\ref{eq:towards CS exp}) with $q=0.99$ (black), $q=0.994$ (blue), $q=0.998$ (purple) and its analytic approximation in (\ref{eq:semiclassical spread}) (red), where, instead of the infinite summation, we have included $N$ as the upper limit of summation. Here, $a_n=n\frac{\qty(H_n(\cos\theta_0|q))^2}{\qty(q;~q)_n}$ in the former case; and $a_n=\frac{2^n\pi}{(n-1)!\Gamma\qty(\frac{1-n}{2})^2}$ in the latter. We observe that the analytic approximation lower bounds the numerical ones, and they diverge as we increase the upper bound $N$.}
    \label{fig:spread_approx}
\end{figure}

We now perform the evaluation of (\ref{eq:C S DDSSYK}) explicitly using (\ref{eq: proj E0 theta}):
\begin{equation}\label{eq:towards CS exp}
    \mathcal{C}_{\rm S}=\sum_nn\frac{\qty(H_n(\cos\theta_0|q))^2}{\qty(q;~q)_n}~.
\end{equation}
In the $\lambda\rightarrow0$ regime, one can approximate \cite{Tang:2023ocr}
\begin{equation}
    H_n(x|q)\simeq\sqrt{\frac{\lambda}{2}}H_n\qty(x\sqrt{\frac{2}{\lambda}})~,
\end{equation}
where $H_n(x)$ is the Hermite polynomial of degree $n$. Moreover, $\lim_{\lambda\rightarrow0}(q;q)_n=\lambda^nn!$ from the definition (\ref{eq:q-Pochhammer}). Thus, in the semiclassical regime ($\lambda\rightarrow0$), and considering the maximal entropy state in (\ref{eq:towards CS exp}), $\theta_0=\pi/2$, one then recovers
\begin{equation}\label{eq:semiclassical spread}
    \mathcal{C}_{\rm S}\simeq\sum_{n=0}^\infty\frac{2^n\pi}{(n-1)!\Gamma\qty(\frac{1-n}{2})^2}~.
\end{equation}
This is a diverging series, given that $\lim_{n\rightarrow\infty}\frac{2^n\pi}{(n-1)!\Gamma\qty(\frac{1-n}{2})^2}\neq0$. We have confirmed the approximation bounds from below (\ref{eq:towards CS exp}) when $q$ is close to $1$, as illustrated in Fig. \ref{fig:spread_approx}. 
This divergence implies that the chord number operator needs to be renormalized in the semiclassical limit. We will discuss this point in the following subsection, guided by the holographic dictionary.

We conclude this subsection with a few remarks:
\begin{itemize}
    \item \textbf{Microcanonical ensemble}: Notice that there is a straightforward extension, by switching from a canonical ensemble to a microcanonical one where we consider
\begin{equation}\label{eq:microcanonical state}
    \rho_{\rm dS}=\frac{1}{N}\sum_E\ket{E}\bra{E}~,
\end{equation}
centered around $E=E_0$ to evaluate the spread complexity \cite{Alishahiha:2022anw} as
\begin{equation}\label{eq:spread micro}
    \mathcal{C}_{\rm S}=\frac{1}{N}\sum_{E,~n}n\bra{E,~E}\ket{n,~n}=\frac{1}{N}\sum_{E}\bra{0,~0}\mathcal{O}_{E(\theta)}^{\dagger}~\mathbb{N}~\mathcal{E}\ket{0,~0}~.
\end{equation}
\item \textbf{Triple scaling limit}: In this regime, one might choose $\ket{n=0}$ as the reference state, representing the canonical ensemble thermofield doubled (TFD) state in the infinite temperature limit \cite{Rabinovici:2023yex}. In this case, (\ref{eq:general OP MERA}) adopts a simple expression in terms of Hamiltonian evolution (through the transfer matrix $T$) in each $\mathcal{H}$, which we denote:
\begin{equation}\label{eq:triple scaling op mera}
    \mathcal{O}_{T}(t)\equiv\rme^{\rmi t T}\otimes\rme^{-\rmi t T}~.
\end{equation}
\end{itemize}
We will now move on to study the dual interpretation of the spread complexity for the state $\ket{\mathbb{E}_0}$, and especially its time independence.

\subsection{Dual interpretations}
\begin{figure}[t!]
    \centering
    \includegraphics[width=0.5\textwidth]{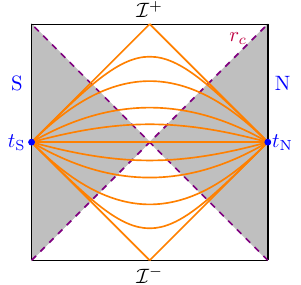}
    \caption{Geodesic curves (orange) joining antipodal static patch observers (S and N, blue) in SdS$_3$ space, and probing the region outside the cosmological horizon ($r_c$, purple dashed lines). $\mathcal{C}_{\rm S}$ measures the time difference between the antipodal observers, which has been fixed to $t_{\rm N}=t_{\rm S}$ (blue dots) in the diagram.}
    \label{fig:dS geo}
\end{figure}

We would like to translate the above results using the bulk dictionary developed in \cite{Verlinde:2024znh}, where a bulk phase space variable $z$ was related to the chord number operator $\hat{n}$ though 
\begin{align}
    \rme^{-8\pi {G_N}(\hat{n}-1/2)/{\ell_{\rm dS}}}=-\rme^{-2z}~.\label{eq:classical phase}
\end{align}
Here $z$ is an operator whose expectation value on the $\ket{\psi_{\rm dS}}$ state corresponds to a (regularized) geodesic length in SdS$_3$ space measuring the static patch time difference between the antipodal observers (i.e. $\expval{z}_{\rm dS}=t_{\rm N}-t_{\rm S}$). Geometrically, $\expval{z}_{\rm dS}$ and the deficit angle in SdS$_3$ space determine the geodesic lengths connecting the antipodal observers, as shown in Fig. \ref{fig:dS geo}.

Using the dictionary, we take expectation values in $\ket{\psi_{\rm dS}}$ for the bulk side of (\ref{eq:classical phase}), and $\ket{E_0}$ on the chord number operator leads to:
\begin{equation}\label{eq:relation y and spread}
    \boxed{2\expval{z}_{\rm dS}=\frac{8\pi G_N}{\ell_{\rm dS}}\mathcal{C}_{\rm S}+\qty(\rmi\pi-\frac{8\pi G_N}{\ell_{\rm dS}})\bra{E_0}\ket{E_0}~.}
\end{equation}
The factor $\rmi\pi$ is related to the integration of the Wilson defining the holonomy variable which is involved in the identification of $z$ with a time difference \cite{Verlinde:2024znh}, and it can be shifted away. 

Next, we would like to interpret the relation (\ref{eq:norm theta}). We have that $\bra{E_0}\ket{E_0}\rightarrow\infty$, and as we have noticed in (\ref{eq:semiclassical spread}) $\mathcal{C}_{\rm S}$ also diverges in the $q\rightarrow1$ limit (i.e. when $\ket{E_0}$ corresponds to the pure dS state). We need to renormalize both of these terms in (\ref{eq:relation y and spread}). For instance, the dS space boost isometries give us the freedom to set $t_{\rm N}=t_{\rm S}$. This can be motivated on the doubled DSSYK side from the Hamiltonian constraint in physical states (\ref{eq:phy states}) which correspond to synchronizing the clocks (physical time) for the left and right (L/R) systems \cite{Narovlansky:2023lfz}. We will then normalize $\bra{E_0}\ket{E_0}_{q\rightarrow1}=\eval{\mathcal{C}_{\rm S}}_{q\rightarrow1}$ in (\ref{eq:relation y and spread}) such that
\begin{equation}
    \lim_{q\rightarrow1}\Re(\expval{z}_{\rm dS})=0~.\label{eq:redef E_0}
\end{equation}
We, therefore, conclude that the boost symmetries in SdS$_3$ space can be interpreted in terms of renormalization in the spread complexity of the $\ket{E_0^{\rm L},~E_0^{\rm R}}$ state in the DSSYK model. There are a few remarks about the above analysis:
\begin{itemize}
\item \textbf{Time independence}: The fact that the spread complexity for this state does not depend on the value of $t_{\rm N}$ or $t_{\rm S}$, might be interpreted in the bulk description with the total entropy perceived by the N and S pole observers being a time-independent constant (\ref{eq:GH entropy}). Given that $\mathcal{C}_{\rm S}$ is a constant, the precise bulk dictionary allowed us to set the time difference to vanish, which is related to the freedom in fixing the time difference between the observers by boost symmetry.

\item \textbf{Relation with entanglement in the doubled DSSYK model}: As we commented in Sec. \ref{sssec:Application spread}, spread complexity counts the number of entangled chord states pairs in the maximal entropy state of the DSSYK model. Given that the spread complexity of the doubled DSSYK model is dual a geodesic length in the SdS$_3$ space (measuring a time difference), we find agreement with previous studies \cite{Lashkari:2013koa,Faulkner:2013ica,Swingle:2014uza,Faulkner:2017tkh,Haehl:2017sot,Agon:2020mvu,Agon:2021tia,VanRaamsdonk:2009ar,VanRaamsdonk:2010pw,Bianchi:2012ev,Maldacena:2013xja,Balasubramanian:2014sra} suggesting that entanglement builds spacetime, and in this case, spread complexity provides a measure of this relation.

\item \textbf{Connections with JT gravity}: In the triple scaled SYK model/ JT gravity correspondence, spread complexity is identified with a geodesic length between asymptotic AdS$_2$ boundaries \cite{Rabinovici:2023yex}, suggesting a relation with the CV conjecture \cite{Susskind:2014rva}. Our observations in the dS holographic context share some similarities. The spread complexity for the $\ket{E_0}$ state has a geodesic length interpretation, although in terms of a time-like coordinate difference between antipodal observers; and, the bulk has two space-like dimensions more than the theory where spread complexity is evaluated.

\item \textbf{Liouville-dS CFT}: As we mentioned in the introduction, the LdS$_2$ CFT, described by the field $\phi_\pm$, is located in a disk region $\Sigma$, whose boundary describes the time-like geodesic of a worldline observer. Since the N/S pole static patch time in SdS$_3$ space, $t$, is identified with the boundary time along $\partial\Sigma$, $\tau$ \cite{Verlinde:2024zrh}, $\mathcal{C}_{\rm S}$ can also be identified with a proper time difference between the L/R LdS$_2$ CFTs on the maximal entropy state, that is
\begin{equation}\label{eq:CS LdS}
    \tau_{\rm L}-\tau_{\rm R}\propto \mathcal{C}_{\rm S}~.
\end{equation}
Since the spread complexity counts the number of entangled modes in the doubled DSSYK model, this suggests there is entanglement between the pairs of CFTs.
\end{itemize}

\section{Towards Krylov complexity in de Sitter space}\label{sec:Krylov}
This section investigates the evolution of Krylov complexity for physical operators $\mathcal{O}^{\rm phys}_\Delta$ (i.e. those obeying (\ref{eq:phy ops})) in the different sides of the correspondence, which are shown explicitly below:
\begin{itemize}
    \item Doubled DSSYK model:
    \begin{equation}
        \begin{aligned}\label{eq:O phys DDSSYK}
        &\mathcal{O}^{\rm phys}_\Delta(\tau)=\int\rmd t~O_{1-\Delta}^{\rm L}(t)O_{\Delta}^{\rm R}(\tau-t)~.
    \end{aligned}
    \end{equation}
    where $\mathcal{O}^{\rm L/R}_\Delta(\tau)$ are shown in (\ref{eq:matter ops DSSYK}). 
\item LdS$_2$ CFT:
\begin{equation}
\begin{aligned}\label{eq:O phys LdS}
        \mathcal{O}^{\rm phys}_\Delta(\tau)&=\int\rmd t~V_{1-\Delta}^-(t)~V_{\Delta}^+(\tau-t)~,\\
    V^\pm_\Delta&=\rme^{b_\pm\Delta \phi_{\pm}}~,
\end{aligned}
\end{equation}
where $V^\pm_\Delta(\tau)$ are boundary vertex operators, parametrized by the proper time $\tau$ in $\partial\Sigma$ (\ref{eq:action LdS}).
\item SdS$_3$ space: Scalar fields with conformal weight $\Delta$, 
\begin{equation}\label{eq:O phys SdS}
    \mathcal{O}^{\rm phys}_\Delta(x)=\phi_\Delta(x)~.
\end{equation}
\end{itemize}
We consider $|\mathcal{O}^{\rm phys}_\Delta)$ as the initial operator in the Lanczos algorithm to calculate their Krylov complexity. The first amplitude in (\ref{eq:2pnt correlator Krylov}) is determined by the 2-point correlation function:
\begin{equation}\label{eq:2pnt correlator Verlinde}
\begin{aligned}
    \varphi_0(\tau)&=\frac{(O^{\rm phys}_\Delta(0)|O^{\rm phys}_\Delta(\tau))}{(O^{\rm phys}_\Delta(0)|O^{\rm phys}_\Delta(0))}~.
\end{aligned}
\end{equation}
Meanwhile the case of SdS$_3$ space, we take $\tau=\tau(x_1,~x_2)$ as the proper time between the insertion of the fields $\phi_\Delta(x_1)$ and $\phi_\Delta(x_2)$ on time-like separated points $x_1$, $x_2$.

The 2-point correlation function of the physical operators in (\ref{eq:2pnt correlator Verlinde}) has been computed for the state $\ket{\mathbb{E}_0}$ in the doubled DSSYK model based on the result (\ref{eq:1point O delta}), and matched with the correlators of its presumed holographic duals \cite{Verlinde:2024znh,Verlinde:2024zrh}. Explicitly, using (\ref{eq:1point O delta}) and (\ref{eq:O phys DDSSYK}) in (\ref{eq:2pnt correlator Verlinde}) leads to:
\begin{equation}\label{eq:auto correlator dS}
\begin{aligned}
    \varphi_0(\tau)&=\mathcal{N}^{-1}\int_0^\pi\rmd\theta_1\frac{\mu(\theta_1)\rme^{-\rmi\tau\,E(\theta_1)}}{(q^{1-\Delta}\rme^{\pm\rmi\theta_0\pm\rmi\theta_1};~q)_\infty(q^{\Delta}\rme^{\pm\rmi\theta_0\pm\rmi\theta_1};~q)_\infty}~,\\
    \mathcal{N}&\equiv\int_0^\pi\rmd\theta_1\frac{\mu(\theta_1)}{(q^{1-\Delta}\rme^{\pm\rmi\theta_0\pm\rmi\theta_1};~q)_\infty(q^{\Delta}\rme^{\pm\rmi\theta_0\pm\rmi\theta_1};~q)_\infty}~,
\end{aligned}
\end{equation}
where $q\in[0,~1]$, and $\theta_0=\frac{\pi}{2}$ (corresponding to $E_0=0$ for the maximal entropy state). Given that we are considering Hermitian physical operators (\ref{eq:O phys DDSSYK}-\ref{eq:O phys SdS}), only the even moments in (\ref{eq:2pnt correlator Krylov}) will contribute to $\mathcal{C}_{\rm K}$. The moments, determined from (\ref{eq:auto correlator dS}), are
\begin{equation}
    m_{2n}=\mathcal{N}^{-1}\int_0^\pi\rmd\theta_1\frac{\mu(\theta_1)\rm(E(\theta_1))^{2n}}{(q^{1-\Delta}\rme^{\pm\rmi\theta_0\pm\rmi\theta_1};~q)_\infty(q^{\Delta}\rme^{\pm\rmi\theta_0\pm\rmi\theta_1};~q)_\infty}~.
\end{equation}
In principle, one can proceed to evaluate the Krylov complexity (\ref{eq:Krylov complexity}) exactly. To simplify the evaluation of the amplitudes $\varphi_n(\tau)$, we will work in $q\rightarrow1$ limit (i.e. $G_N\rightarrow0$),
\begin{align}
    \varphi_0(\tau)&=\frac{\sinh(\nu \tau)}{\nu\sinh(\tau)}~,\label{eq: correlator massive s q->1}\\
    m_{2n}&=\frac{2}{\nu}\sum_{k=0}^\infty\nu^{2n-2k+1}\frac{(1-2^{2k-1}){\rm B}_{2k}}{(2k)!(2n-2k+1)!}~,
\end{align}
where ${\rm B}_{n}$ are the Bernoulli numbers; $\tau$ has been rescaled by $\ell_{\rm dS}$ to make it dimensionless; and $\nu\equiv2\Delta-1$, which is related to the scalar particle's mass in SdS$_3$ space through $m^2\ell^2_{\rm dS}=4\Delta(1-\Delta)$, i.e.
\begin{equation}\label{eq:nu scalar field SdS3}
\nu=\sqrt{1-m^2\ell^2_{\rm dS}}    ~.
\end{equation}
Notice that $\nu\in[0,~1]$ when $m\ell_{\rm dS}\leq1$ and $\nu\in\rmi\mathbb{R}$ otherwise. Motivated by the duality with the doubled DSSYK model, where $\Delta\in[0,~1]$ (\ref{eq:O phys DDSSYK}), we will consider $m^2\ell^2_{\rm dS}\leq1$ in the evaluation. The Lanczos coefficients can be determined through the algorithm (\ref{eq:Alt Lanczos}), leading to
\begin{equation}\label{eq:Lanczos}
    b_n=n\sqrt{\frac{n^2-\nu^2}{4n^2-1}}~.
\end{equation}
Notice that the growth of $b_n$ is linear in $n$ for $n\gg1$, so that it satisfies Carleman's condition \cite{carleman1926fonctions}. The linear growth in the coefficients is generically found in chaotic systems \cite{Hashimoto:2023swv}, although it is also sensitive to the choice of the initial operator \cite{Espanol:2022cqr,Rabinovici:2022beu}.

We can now compute the amplitudes $\varphi_n(t)$ through the recursion relation (\ref{eq:sch eq K operator}). One can check that the amplitude for arbitrary $\nu$ and $n$ in the early and late time regime take the form:
\begin{equation}    
\varphi_n(\tau)=\frac{\tau^n\prod_{k=1}^n\sqrt{k^2-\nu^2}}{(2n-1)!!\sqrt{2n+1}}+\mathcal{O}(\tau^{n+2})~,\quad n\geq1~,
\end{equation}
\begin{equation}    
\varphi_n(\tau)=\rme^{(\nu-1)\tau}\frac{\sqrt{2n+1}}{\nu}\prod_{k=1}^n\frac{k-\nu}{\sqrt{k^2-\nu^2}}+\mathcal{O}(\rme^{-(1+\nu)\tau})~,\quad n\geq1~.
\end{equation}
Moreover, there is a particular non-trivial value, $\nu=1/2$, for which we find a closed form relation for the amplitudes $\varphi_n$ and the corresponding Krylov complexity (\ref{eq:Krylov complexity}) (see also \cite{Caputa:2021sib}):
\begin{align}
    \nu=1/2~:\quad {\varphi_n(\tau)}&=\sech(\frac{\tau}{2})\tanh^n\qty(\frac{\tau}{2})~,\quad b_n=n/2~,\\ {\mathcal{C}_{\rm K}(\tau)}&=\sinh^2\qty(\frac{\tau}{2})~.
\end{align}
Meanwhile, for $\nu\neq1/2$, we can still find the late time behavior of $\mathcal{C}_{\rm K}$ using a result shown in \cite{Parker:2018yvk}. Assuming smoothness of the Lanczos coefficients $b_n$ with $n$ for a local operator, it was shown that if $b_n=\frac{\lambda_{\rm K}}{2} n+\mathcal{O}(1)$ ($\lambda_{\rm K}\in\mathbb{R}$) for $n\gg1$, then $\mathcal{C}_{\rm K}(\tau)$ grows exponentially, with $\lambda_{\rm K}$ being the exponent. In our case, given that $b_n$ in (\ref{eq:Lanczos}) is smooth, and $\lambda_{\rm K}=1$; we conclude that
\begin{equation}\label{eq:late time CK}
    \boxed{\mathcal{C}_{\rm K}(\tau\gg1)\propto \rme^{\tau}~,\quad\forall\nu\in[0,~1)~.}
\end{equation}

\subsection{Dual interpretations}\label{ssec:Bound OTOC}
The late-time exponential growth in $C_{\rm K}(\tau)$ was conjectured to be universally displayed by maximally chaotic systems in \cite{Parker:2018yvk}, so our results are consistent with the expectation that the DSSYK model is a maximally chaotic system \cite{Berkooz:2018jqr}, and with previous studies finding exponential growth of the Krylov complexity \cite{Parker:2018yvk,Bhattacharjee:2022ave}, albeit for the $\psi_i(\tau)$ operators in (\ref{eq:DDSSYK Hamiltonian}).\footnote{A first connection between Krylov complexity and dS holography appeared in \cite{Bhattacharjee:2022ave}. A ``cosmic time" scale appears in the exponent of $\mathcal{C}_{\rm K}$ which is equivalent to a rescaling $\tau\rightarrow p\tau$ in the correlation function of $\psi_i(\tau)$. In the double scaling limit (\ref{eq:double scaling}), the enhanced growth of $\mathcal{C}_{\rm K}$ was associated with hyperfast scrambling in the DSSYK model conjectured in \cite{Susskind:2021esx}.} The same holds for the other two members in the dS holographic proposal of \cite{Narovlansky:2023lfz,Verlinde:2024znh,Verlinde:2024zrh} since our calculations employ the known correlation functions of the physical operators in (\ref{eq:O phys DDSSYK} - \ref{eq:O phys SdS}). Moreover, according to the conjecture in \cite{Parker:2018yvk}, the exponent in (\ref{eq:late time CK}) corresponds to the maximal Lyapunov exponent measured by OTOCs \cite{Maldacena:2015waa,Murthy:2019fgs}. Our results are in \textbf{agreement} with the conjecture since the physical temperature identified in the correlator (\ref{eq: correlator massive s q->1}) corresponds to $\beta_{\rm dS}=2\pi$ \cite{Narovlansky:2023lfz} (while its Boltzmann temperature, appearing in the thermal density matrix $\rho_{\rm dS}=Z^{-1}\rme^{-\beta H}$ is conjectured to be infinite in \cite{Rahman:2023pgt}). 

Moreover, notice that the late-time exponential growth reproduces the evolution CAny observables evaluated in CMC slices in Sec. \ref{sec:CAny CMC} when $\abs{K}=K_{\rm crit}$ for $d=1$ or $2$, and $F\qty[\dots]=1$. However, this agreement might be a coincidence as Krylov complexity is a measure of operator growth, while (\ref{eq:C CMC}) is interpreted as state complexity, which only depends on the background metric (i.e. without matter, unlike the bulk interpretation of Krylov complexity in this section).

Lastly, we notice that $\mathcal{C}_{\rm K}=\,$const. (no operator spread) in the critical case $\nu=1$, given that the input correlator (\ref{eq: correlator massive s q->1}) corresponds to a scalar propagating on a time-like trajectory, or equivalently, no operator insertion in the doubled DSSYK and LdS$_2$ duals (\ref{eq:O phys DDSSYK}, \ref{eq:O phys LdS} respectively).

As a side comment, motivated by the recent discussions about spread and Krylov complexities for density matrices \cite{Caputa:2024vrn}, one might study the Krylov complexity of the density matrix $\rho_{\rm dS}$ in (\ref{eq:spread micro}) and compare with the features of the spread complexity encountered in Sec. \ref{sec:Spread}. Given that the evolution is controlled by the Liouville-von Neumann equation
\begin{equation}
    \partial_t\rho(t)=-\rmi\mathcal{L}\rho(t)~,
\end{equation}
with $|\rho(t=0))=|\rho_{\rm dS})$, it follows that the density matrix does not evolve in static patch time, i.e. $|\rho(t))=|\rho_{\rm dS})$, given that $\ket{E}$ are energy eigenstates. The Krylov complexity for $\rho_{\rm dS}$ is then trivial, in contrast to the spread complexity in (\ref{eq:spread micro}).

\section{Towards query complexity in de Sitter space}\label{sec:Query}
In this section, we formulate an algorithm computing correlation functions for LdS$_2$ CFTs building on Sec. \ref{ssec:query}, and also guided by the diagrammatic structure of n-point correlation functions in the cylinder amplitude of the DSSYK model. We make contact with SdS$_3$ through its CS formulation, which is a SL$(2,~\mathbb{C})$ topological field theory, with $k\rightarrow\rmi\kappa$ and $\kappa\in\mathbb{R}$ in (\ref{eq:CS AdS3}).

In terms of the LdS$_2$ theory, it was argued in \cite{Verlinde:2024zrh} that the natural vacuum state dual to $\ket{\mathbb{E}_0}$ and $\ket{\psi_{\rm dS}}$, which we denote $\ket{s_0}$, corresponds to a FZZT brane with $\mu_{\rm B}=0$ in (\ref{eq:action LdS}) (equivalent to setting $E_0=0$ for the maximal entropy state). For higher energy states, let us consider the region $\Sigma$ where the LdS$_2$ CFT is defined (\ref{eq:action LdS}), and insert physical operators $\mathcal{O}_\Delta^{\rm phys}(\tau_1)$ and $\mathcal{O}_\Delta^{\rm phys}(\tau_0)$ on the vacuum state $\ket{s_0}$. Let us denote a segment $s_1\in\partial\Sigma$ where $\partial s_1=\qty{\tau_0,~\tau_1}$. Then, excited states can be represented as
\begin{equation}\label{eq: momentum eigenstates LdS}
\begin{aligned}
    \ket{s_1}=\hat{W}(\tau_0,~\tau_1)\ket{s_0}~,\quad\hat{W}(\tau_0,~\tau_1)\equiv\mathcal{O}_\Delta^{\rm phys}(\tau_1)\mathcal{O}_\Delta^{\rm phys}(\tau_0)~,
\end{aligned}
\end{equation}
where $\hat{W}$ is a CFT operator, whose expectation value on state $\ket{s_0}$ can be expressed in the bulk in terms of Wilson lines
\begin{equation}\label{eq:momentum eigenstates LdS}
\begin{aligned}
    &\bra{s_0}\hat{W}(\tau_0,~\tau_1)\ket{s_0}\\
    &=\bra{\psi_{\rm dS}}\tr_{R_j}\qty(P\exp(-\int_{\gamma(\tau_0,\tau_1)} \mathcal{A})P\exp(-\int_{\gamma(\tau_0,\tau_1)} \bar{\mathcal{A}}))\ket{\psi_{\rm dS}}~,
\end{aligned}
\end{equation}
with $R_j$ a representation of $SL(2,~\mathbb{C})$; $j=1/2+\rmi~s_0$ is the spin; and ${\gamma}(\tau_0,~\tau_1)$ is a path between the insertion times ${\tau_0},~{\tau_1}$ along $\partial\Sigma$.

We will be considering the same definitions for the algorithm computing the arbitrary n-point correlation functions as we explained in Sec. \ref{ssec:query}. However, there will be key differences in its implementation with respect to the proposal in \cite{Chen:2020nlj}, which are intrinsically connected to the dual theories. We previously discussed that the query complexity of a vacuum CFT trivializes if several operators are inserted in the exact same location in pure AdS space at a constant time slice. In contrast, for the LdS CFT case, if one were to remove $\partial\Sigma$ (corresponding to the worldline of the static patch observers in SdS$_3$ space), this potentially eliminates the degrees of freedom of the doubled DSSYK model. However, one can instead insert the physical operators of the LdS CFT at \textit{the same spatial location but at different proper times} $\tau_i$. 

\begin{figure}
    \centering
    \subfloat[]{\includegraphics[width=0.6\textwidth]{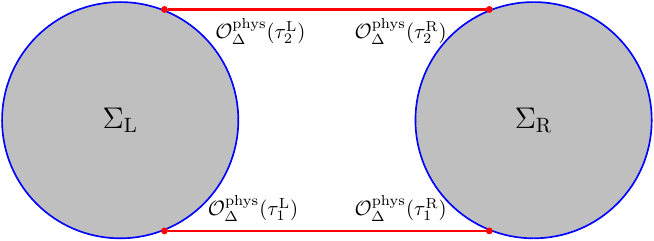}}\\
    \subfloat[]{\includegraphics[width=0.6\textwidth]{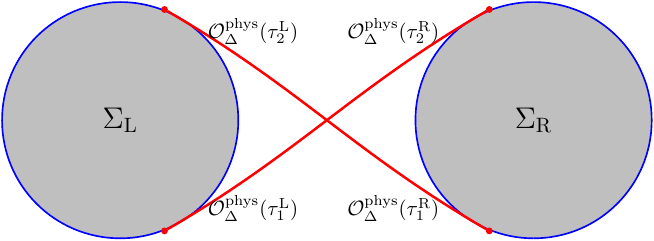}}
    \caption{Pair of disks $\Sigma_{\rm L/R}$ where some operators $\mathcal{O}_{\Delta}^{\rm phys}(\tau^{\rm L/R}_i)$ are inserted (red dots) in $\partial\Sigma_{\rm L/R}$. We illustrate two out of all symmetric pairwise {contractions} between the matter operators on the L/R boundaries (red lines). \textit{Above}: $\mathcal{O}_{\Delta}^{\rm phys}(\tau_{1,~2}^{\rm R})$ and $\mathcal{O}_{\Delta}^{\rm phys}(\tau_{1,~2}^{\rm L})$ are contracted. \textit{Below}: $\mathcal{O}_{\Delta}^{\rm phys}(\tau_1^{\rm R/L})$ with $\mathcal{O}_{\Delta}^{\rm phys}(\tau_2^{\rm L/R})$.}
    \label{fig:LR_LdS}
\end{figure}

Moreover, counting the number of fusions for matter operators can be simplified substantially using cylinder amplitudes in the DSSYK model \cite{Okuyama:2023yat} (see Fig. \ref{fig:Correlators_DDSSYK}). Guided by the correspondence with the doubled DSSYK model, we choose to study the correlators where the matter fields are pairwise connected between the L/R LdS CFTs (corresponding to the N and S poles), as illustrated in Fig. \ref{fig:LR_LdS}.
Then, we will formulate the same type of CFT algorithm that we discussed in Sec. \ref{ssec:query}, in terms of the fusion algebra for the states $\qty{\ket{s}}$, but we consider operators from both the L and R side LdS CFTs. This means that the map (\ref{eq:fusions}) takes any number of incoming representations of the states (\ref{eq:momentum eigenstates LdS}) at an overlapping time (e.g. $\tau^{\rm L}_i$), and generates an additional one. For instance, in the case of two incoming matter chords and one outgoing:
\begin{equation}\label{eq:junction rule Liouville}
\begin{aligned}
    &\mathcal{J}(W(\tau^{\rm L}_{i},~\tau^{\rm R}_j)\ket{s_0},~W(\tau^{\rm L}_{i},~\tau^{\rm R}_{k})\ket{s_0})\\
    &=\int\rmd\tau^{\rm R}_l~c(\tau^{\rm L}_i,~\tau^{\rm R}_j,~\tau^{\rm R}_k,~\tau^{\rm R}_l)W(\tau^{\rm L}_{i},~\tau^{\rm R}_l)\ket{s_0}~,
\end{aligned}
\end{equation}
where $c(\tau^{\rm L}_i,~\tau^{\rm R}_j,~\tau^{\rm R}_k,~\tau^{\rm R}_l)$ corresponds to a conformal kinematical factor that obeys the deformability rule in Sec. \ref{ssec:query}, which is determined from OPE data of the representations in (\ref{eq:junction rule Liouville}). Then, correlation functions on a given multiplet representation of SL($2,~\mathbb{C}$) appear from contracting operators according to (\ref{eq:junction rule Liouville}) in the relevant multiplet representation. We will now adopt the definition for query complexity in the LdS CFT as the number of applications of fusion rules (\ref{eq:Query n fusions}) using (\ref{eq:junction rule Liouville}) iteratively, and study its dual interpretation.

\subsection{Dual interpretations}
\begin{figure}
    \centering
    \includegraphics[width=\textwidth]{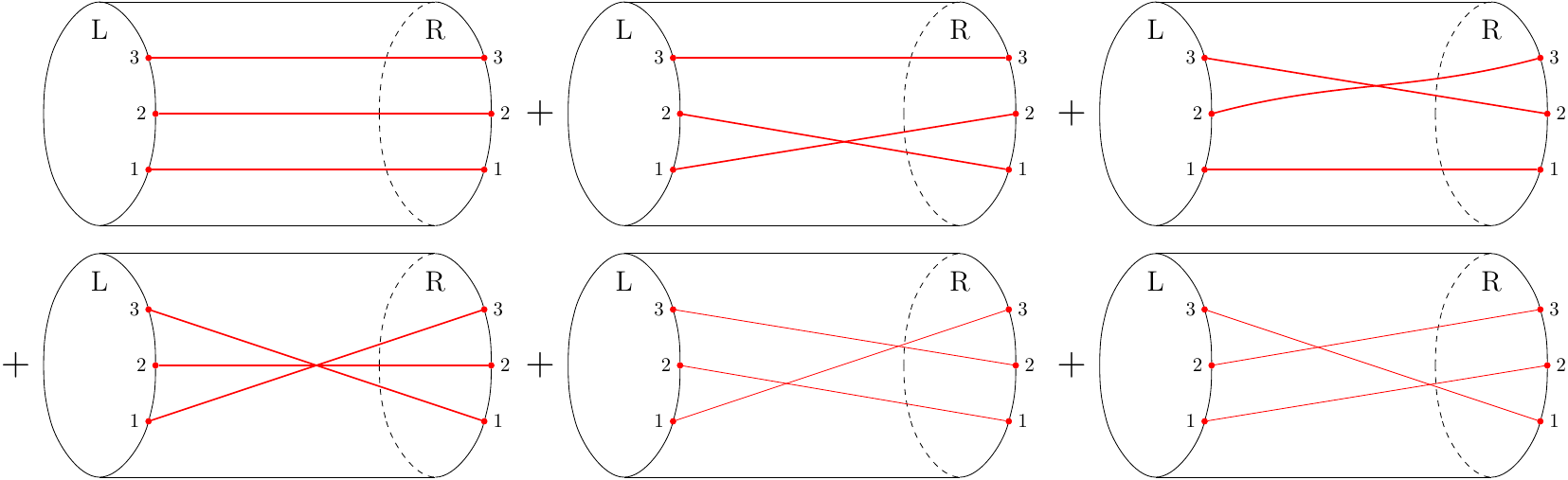}
    \caption{$S_{k=3}$ symmetric matter operator chords (red lines) in the cylinder amplitude of the doubled DSSYK model.}
    \label{fig:Correlators_DDSSYK}
\end{figure}

Our motivation for constructing the algorithm computing correlation functions between the L/R LdS CFTs is the wormhole amplitude describing the matter operator chords extending between two disks, which in our context corresponds to the pair of DSSYK models (L/R). See Fig. \ref{fig:Correlators_DDSSYK} for a representation of this system.
Based on the two-matrix model formulation of the DSSYK model \cite{Jafferis:2022wez,Jafferis:2022uhu}, an arbitrary ($2k$)-point function in the L/R edges of the cylinder is given by \cite{Okuyama:2023yat}
\begin{equation}
    \begin{aligned}\label{eq:matter correlators query}
    &\expval{\tr(\prod_{i=1}^k\rme^{\rmi t_i^{\rm L}T}\mathcal{O}^{\rm phys}_\Delta(t^{\rm L}_i))\tr(\prod_{i=1}^k\rme^{\rmi t_i^{\rm R}T}\mathcal{O}^{\rm phys}_\Delta(t^{\rm R}_i))}_{\rm cyl}\\
    &=\sum_{\sigma\in S_k}\tr_{\mathcal{H}}\qty(\prod_{i=0}^k\rme^{\rmi\qty(t^{\rm L}_{i}+t^{\rm R}_{\sigma(i)})T}q^{\Delta \hat{n}})~.
\end{aligned}
\end{equation}
Here $S_k$ represents the symmetric group, and $\mathcal{H}$ is the chord Hilbert space. 

If the dS holographic dictionary \cite{Verlinde:2024zrh} holds, the ($2k$)-point function (\ref{eq:matter correlators query}) corresponds to a ($2k$)-point function in the LdS$_2$ CFT, which can be computed using the rule (\ref{eq:junction rule Liouville}). It can be seen, for instance in Fig. \ref{fig:LR_LdS}, that to compute the $(2k)$-point correlator in the LdS$_2$ CFT one needs a total of ($2k$)-junctions between pairwise contractions of matter operators $\mathcal{O}_\Delta^{\rm phys}(\tau_i^{\rm L})$ and $\mathcal{O}_\Delta^{\rm phys}(\tau_j^{\rm R})$ in $\Sigma_{\rm L/R}$ respectively. The same can be seen in the cylinder diagram of the DSSYK model (Fig. \ref{fig:Correlators_DDSSYK}). 

If we sum over an $S_k$ orbit, one recovers an object (i.e. the $(2k)$-point correlator) that lives on the trivial representation of $S_k$.\footnote{This is similar to the s-wave reduction of the vacuum CFT in Sec. \ref{ssec:query}.} {Moreover, lower order-point correlation functions of the L/R LdS$_2$ CFTs can be computed by adding a trivial representation in the junction rule (\ref{eq:junction rule Liouville}). From the perspective of the DSSYK model, instead of summing an $S_k$ orbit for $k$ fixed to compute the ($2k$)-point function (\ref{eq:matter correlators query}), we have to evaluate over all the $(2k)$-point correlation functions for $k\leq k_{\rm max}$, such that lower than $k_{\rm max}$-point correlators are included in the DSSYK dual of query complexity.} Thus, the CFT definition in (\ref{eq:Query n fusions}) corresponds to a sum of matter insertions $k\leq k_{\rm max}$ in the doubled DSSYK model, resulting in:
\begin{equation}\label{eq:CQ DDSSYK}
    \boxed{\mathcal{C}_{\rm Q}= k_{\rm max}(k_{\rm max}+1)/2~,}
\end{equation}
where the number of insertions in the edges of the cylinder determines the number of $S_{k\leq k_{\rm max}}$ combinations of chord diagrams in (\ref{eq:matter correlators query}).

\begin{figure}
    \centering
    \includegraphics[width=0.7\textwidth]{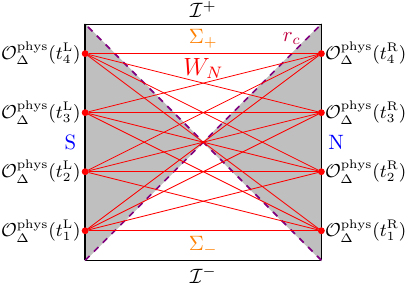}
    \caption{Wilson line network (labeled $W_N$, red lines) for a few operator insertions (red dots) along the static patch worldlines of SdS$_3$ space following the algorithm computing $S_{k\leq k_{\rm max}}$ symmetric ($2k$)-point correlation functions in the (\ref{eq:matter correlators query}), which is illustrated here for a fixed number, $k=4$. The future (past) surface where the network ends (starts) is labeled $\Sigma_+$ ($\Sigma_-$). Notice that there are $k$ number of junctions for every vertex.}
    \label{fig:SdS Wilson Network}
\end{figure}

Meanwhile, from the bulk perspective, query complexity counts the number of junctions of Wilson lines connecting the N to S patches, as shown in Fig. \ref{fig:SdS Wilson Network}.
In contrast to the case presented in Sec. \ref{ssec:query} there is no static time slice connecting both the N and S poles and, in principle, query complexity in this protocol needs to incorporate general local geometric invariants in the bulk spacetime of the form
\begin{equation}\label{eq:CAny}
    \boxed{\mathcal{C}_{\rm Q}=\int_{W_N}\rmd^{3}x~ \mathcal{F}[g_{\mu\nu}]+\sum_{\varepsilon=\pm}\int_{\Sigma_\varepsilon}\rmd^{2}\sigma~ \mathcal{G}_\varepsilon[g_{\mu\nu},~X_\varepsilon^\mu]~,}
\end{equation}
where $W_N$ corresponds to the Wilson line network manifold (see Fig. \ref{fig:SdS Wilson Network}); $\Sigma_{\pm}$ are codimension-one space-like slices on the future and past of $W_N$ which represent cutoff surfaces of the network near $\mathcal{I}^\pm$; $\sigma_i$ ($i=1,~2$) are coordinates on the $\Sigma_{\pm}$; while $\mathcal{F}[g_{\mu\nu}]$ is a scalar functional of the 3D bulk curvature invariants involving the metric $g_{\mu\nu}$; and similarly for $\mathcal{G}_{\pm}[g_{\mu\nu},~X_\pm^\mu]$, which are local invariant functionals constructed from the bulk metric and the embedding functions $X_\pm^\mu(\sigma_i)$ of $\Sigma_\pm$ respectively (e.g. the extrinsic curvature and torsion encountered in Sec. \ref{ssec:query}). {(\ref{eq:CAny}) is part of the family of codimension-zero CAny conjectures \cite{Belin:2021bga,Belin:2022xmt} for SdS$_3$ space \cite{Jorstad:2022mls,Aguilar-Gutierrez:2023zqm,Aguilar-Gutierrez:2023pnn,Aguilar-Gutierrez:2024rka}. In this case, the complexity surfaces would be anchored to the N and S worldline observers. Similar to footnote \ref{fn:Powers K T}, we can further restrict the form of $\mathcal{F}[\dots]$ and $\mathcal{G}_{\pm}[\dots]$ by requiring $\mathcal{C}_{\rm Q}$ to be well-behaved under the constraint that the Wilson line network is not smooth at the location of the junctions (i.e. where the operators $\mathcal{O}(x_i)$ are located in Fig. \ref{fig:SdS Wilson Network}). This requirement would then allow us to have a concrete holographic dual of query complexity, which we leave for future work.}

\section{Nielsen complexity in the double-scaled Sachdev-Ye-Kitaev model}\label{sec:comments Nielsen}
In this section, we investigate Nielsen's geometric approach \cite{Nielsen1,Nielsen2,Nielsen3} in the DSSYK model using the bi-invariant proposal in (\ref{eq:biinvariant complexity}). It reproduces the linear growth expected for the CAny holographic complexity proposals in AdS black hole backgrounds \cite{Belin:2021bga,Belin:2022xmt}, and in asymptotically dS spacetimes \cite{Aguilar-Gutierrez:2023zqm,Aguilar-Gutierrez:2023pnn,Aguilar-Gutierrez:2024rka}.

We begin evaluating (\ref{eq:biinvariant complexity}). In the context of the DSSYK model, the most natural choice for the Hilbert space where unitaries (\ref{eq:unitaries}) can act, and evaluate the traces, is over the auxiliary chord Hilbert space $\mathcal{H}$. In that case, we can use $H=-T$ and (\ref{eq:identity theta}), to express (\ref{eq:unitaries}) as
\begin{align}
  x(t)=\rme^{-\rmi V}~,\quad  V=\int_0^\pi\frac{\rmd\theta}{2\pi}\mu(\theta)\qty(tE(\theta)+2\pi K_n)\ket{\theta}\bra{\theta}~.\label{eq:Nielsen intermediate}
\end{align}
(\ref{eq:biinvariant complexity}) with (\ref{eq:Nielsen intermediate}) then transforms into
\begin{equation}\label{eq:complicated CN}
    \mathcal{C}_{\rm N}(t)=\min_{\qty{K_n:~\sum_{n=0}^\infty K_n=0}}\sqrt{\sum_{m=0}^\infty{\int_0^\pi\frac{\rmd\theta}{2\pi}\mu(\theta)\qty(tE(\theta)+2\pi K_n)^2\frac{\qty(H_m(\cos\theta|q))^2~}{(q;~q)_m}}}~.
\end{equation}
We can perform the minimization above noticing that since $T$ is traceless, then it follows that $K_n=0$. (\ref{eq:complicated CN}) becomes:
\begin{equation}\label{eq:CN after minimization}
\begin{aligned}
\mathcal{C}_{\rm N}(t)&=t\sqrt{\sum_{m=0}^\infty{\int_0^\pi\frac{\rmd\theta}{2\pi}\mu(\theta)E(\theta)^2\frac{\qty(H_m(\cos\theta|q))^2~}{(q;~q)_m}}}\\
    &=t\sqrt{\bra{0,~0}\mathcal{E}^{\dagger}(T\otimes T)\mathcal{E}\ket{0,~0}}~,
\end{aligned}
\end{equation}
where in the second line, we have used the symmetry of $T$ in the orthonormal chord basis (\ref{eq:Transfer matrix}), and we reintroduced the entangler operator (\ref{eq:entangler}) in the doubled Hilbert space. Thus, the particular definition of Nielsen complexity (\ref{eq:biinvariant complexity}) in the DSSYK model measures the vacuum expectation value of an operator entangling chords in a doubled Hilbert space (such as for the doubled DSSYK model), with a similar structure to spread complexity in (\ref{eq:spread as counting gates}).

In the context of dS holography, the choice of bi-invariant $\mathcal{C}_{\rm N}(t)$ would have a corresponding dual observable exhibiting a static patch time linear growth in SdS$_3$ space. However, we \textit{have not} identified such observable with the dS holographic dictionary. Nevertheless, we hope this is a first step towards developing this side of the dictionary. Moreover, notice that the evolution in (\ref{eq:CN after minimization}) is by the codimension-one CAny proposals in Sec. \ref{sec:CAny CMC} in asymptotically dS spacetimes \cite{Aguilar-Gutierrez:2023zqm}. However, there is a lot of freedom in the definition of Nielsen complexity (\ref{eq:Nielsen General}), so in principle, there could be other proposals where the hyperfast growth (\ref{eq:hyperfast}) could be reproduced instead \cite{Jorstad:2022mls}.

\subsection{Jackiw–Teitelboim gravity regime}
Next, we would like to evaluate (\ref{eq:complicated CN}) in the semiclassical limit, where $\lambda\rightarrow0$ (i.e. $q\rightarrow1$). The evaluation is still quite involved, but there is an analytically tractable limit, where the dominating terms in the sum are $m\sim\mathcal{O}(1/\lambda)$ (i.e. the triple scaling limit). Under these considerations, we can approximate the integral in (\ref{eq:complicated CN}) as \cite{Okuyama:2023yat}
\begin{align}
   &\int_0^\pi\frac{\rmd\theta}{2\pi}\mu(\theta)\tilde{G}(E(\theta))\frac{\qty(H_m(\cos\theta|q))^2}{(q;~q)_m}\simeq\int_0^\pi\frac{\rmd\theta}{2\pi}\int_0^{2\pi}\frac{\rmd\Phi}{2\pi}\rme^{-\lambda \tilde{F}(\theta,~\Phi)}\tilde{G}(E(\theta))~,\label{eq:int DiLog}
\end{align}
where $\tilde{G}(E(\theta))=t^2E(\theta)^2$ in our case, while
\begin{equation}
    \tilde{F}(\theta,~\Phi)\equiv\rmi\lambda m\Phi+\text{Li}_2(\rme^{2\rmi\Phi})-2\text{Li}_2(\rme^{\rmi\Phi})+\sum_{\varepsilon=\pm}\qty[\text{Li}_2(\rme^{\varepsilon2\rmi\theta})-\text{Li}_2(\rme^{\rmi\Phi+\varepsilon2\rmi\theta})]~,
\end{equation}
and Li$_2(x)=\sum_{k=1}^\infty x^k/k^2$ is the dilogarithm function. 

We now evaluate (\ref{eq:int DiLog}) with a saddle point approximation, which satisfies the conditions:
\begin{equation}
    \eval{\partial_{\theta}\tilde{F}(\theta,~\Phi)}_{\theta=\theta_S,~\Phi=\Phi_S}=\eval{\partial_{\Phi}\tilde{F}(\theta,~\Phi)}_{\theta=\theta_S,~\Phi=\Phi_S}=0~,
\end{equation}
resulting in $\sin^2\theta_S=q^m$, $\Phi_S=0$. Thus, combining (\ref{eq:complicated CN}) with (\ref{eq:int DiLog}) in the saddle point solution ($\theta_S$, $\Phi_S$), one has:
\begin{equation}\label{eq:Nielsen JT limit t}
    \mathcal{C}_{\rm N}(t)\simeq t\sqrt{\frac{J^2}{\lambda}\sum_{m=0}^\infty\frac{1-q^m}{1-q}}~.
\end{equation}
Note that the infinite sum above is divergent for $t\neq0$. This is expected for complexity without a regulating surface \cite{Chen:2020nlj} in the bulk of an asymptotically AdS spacetime, and the corresponding divergences in the triple-scaled regime of the SYK model \cite{Rabinovici:2023yex}. In this case, one can instead consider that the cutoff can be implemented by truncating the series to a finite number of terms.

As a remark, while the definition of $\mathcal{C}_{\rm N}$ in (\ref{eq:biinvariant complexity}) generically reproduces a late time linear growth; as we have emphasized below (\ref{eq:complicated CN}), many other possible behaviors could be recovered by properly choosing the cost function. For instance, one could use a different symmetry principle with respect to (\ref{eq:f unitary}, \ref{eq:f time reverse}). Also notice that while (\ref{eq:complicated CN}) is valid for the doubled DSSYK model, the condition (\ref{eq:int DiLog}) leading to (\ref{eq:Nielsen JT limit t}) assumes that $\lambda\rightarrow0$ and $m\sim\mathcal{O}(1/\lambda)$. In the holographic context, this regime is appropriate for studying JT gravity (see e.g. \cite{Lin:2022rbf}) instead of dS space.

\section{Discussion}\label{sec:discu}
In \emph{summary}, we studied concrete notions of complexity in the context of the holographic correspondence between the doubled DSSYK model, LdS$_2$ CFT, and SdS$_3$ space. The main results are shown in Table \ref{tab:results}. First, we showed that the \textit{spread complexity} in the doubled DSSYK model can be expressed as the number of entangled chord states in its maximal entropy state. We interpreted boost symmetries fixing the time difference between antipodal observers in SdS$_3$ space in terms of a renormalization condition of the spread complexity in the maximal entropy state. This leads to a connection between entanglement, geometry, and complexity in dS holography. Second, we used the correlation functions in the doubled DSSYK model, the LdS$_2$ CFT, and SdS$_3$ space to calculate the respective \textit{Krylov complexity} on all sides of the correspondence, and we showed they display the exponential time growth expected for maximally chaotic systems, with the expected maximal Lyapunov exponent. Later, we introduced the concept of \textit{query complexity} for the LdS$_2$ CFT, which counts the number of steps in an algorithm computing multipoint correlators between antipodal static patches. We described query complexity in terms of matter chord diagrams in a cylinder geometry in the doubled DSSYK model and a network of Wilson lines in SdS$_3$ space. Here, we recognized a connection with the CAny proposals \cite{Belin:2021bga,Belin:2022xmt} in dS space. The geometric invariant terms involved in query complexity can be further constrained by demanding regularity on the network. Finally, we evaluated the \textit{Nielsen operator complexity} of the DSSYK model for a specific proposal where linear time growth is recovered. Although we did not identify its precise dual observable in the other sides of the correspondence in this latter proposal, it shares the late time behavior of certain holographic complexity conjectures in asymptotically dS spacetimes \cite{Aguilar-Gutierrez:2023zqm,Aguilar-Gutierrez:2023tic,Aguilar-Gutierrez:2023pnn,Aguilar-Gutierrez:2024rka}. 

Based on all these approaches, the doubled DSSYK model \cite{Narovlansky:2023lfz,Verlinde:2024znh,Verlinde:2024zrh} is a promising arena to develop complexity, and perhaps other quantum information-theoretic notions, in low-dimensional dS space holography. While several elements of the correspondence need to be developed further, we hope this work provides a step forward. We conclude with some questions left for future work.

\subsubsection*{Gibbons-Hawking entropy and thermodynamics} 
As we noticed in Sec. \ref{sec:Spread}, the spread complexity for the maximal entropy state is time-independent, which we related to the boost isometries allowing us to fix the time-difference between antipodal observers in SdS$_3$ space. Moreover, this observable has a natural interpretation in terms of counting entangled chord states. Perhaps, the time-independence of spread complexity in this case is a microscopic manifestation of the Gibbons-Hawking entropy being constant in time from the perspective of a worldline static patch observer. It would be interesting to study how the thermodynamic properties of the bulk theory are encoded in the explicit microscopic models and whether they can be manifested in the complexity proposals. 

For instance, there is a conjectured relation between the temperature, entropy and holographic complexity of AdS black holes \cite{Brown:2015bva,Brown:2015lvg} based on the Lloyd bound \cite{lloyd2000ultimate} (recently also hinted for SdS$_{d+1\geq4}$ black holes in \cite{Aguilar-Gutierrez:2024rka}):
\begin{equation}
    \text{Late~times:}\quad\dv{\mathcal{C}}{t}\sim TS~.
\end{equation}
It would be interesting to learn if a similar type of relation can be found in the dS holographic approach for any of the complexity proposals in our work.

We comment more about important connections between thermodynamics and complexity for the DSSYK model in Ch. \ref{ch:Conclusion}.

\subsection*{Complexity in tensor networks}
We have found that the spread and Nielsen complexity can be expressed in terms of entangler/disentangler operators (see (\ref{eq:C S DDSSYK}, \ref{eq:CN after minimization}) respectively) acting on the doubled Hilbert space of the DSSYK model \cite{Okuyama:2024yya}, which is reminiscent of a MERA network.
However, to take a step further and associate this type of tensor network to dS$_3$ space\footnote{See \cite{Beny:2011vh,Bao:2017qmt,Bao:2017iye,Niermann:2021wco,Cao:2023gkw} for previous approaches to tensor networks and quantum circuit models of dS space.}, one would need to include a universal gate set of both (dis)entanglers and isometries in a hierarchical order that determines the causal structure of the MERA network \cite{Vidal:2008zz,Beny:2011vh}. This is currently not present (at least not manifestly) in any of the proposals. It would be interesting to pursue this quantum circuit description of dS space (see related work in \cite{Beny:2011vh}) emerging from the doubled DSSYK model.

Meanwhile, on the 3D gravity side, there have been recent exciting developments in constructing tensor networks based on the CS formulation of 3D gravity \cite{Chen:2024unp,Akers:2024wab}, which connects to our discussion on introducing Wilson line networks to study query complexity. Perhaps this can be more naturally studied within the Fubini-Study distance approach to holographic complexity in \cite{Erdmenger:2022lov}.

\subsection*{Late time linear, or hyperfast growth?}
We studied a very particular notion of Nielsen complexity for the DSSYK model, which we showed grows linearly in time and it can be described in terms of entangler operators (\ref{eq:CN after minimization}). Given the universal scaling of computational complexity with system size, which motivated the CAny conjectures \cite{Belin:2021bga,Belin:2022xmt}, it would be useful to learn if the characteristics that we have encountered for the pair of DSSYK models are also generic for more intricate Nielsen complexity proposals, in view of related recent studies for bipartite multiparticle quantum systems in \cite{Baiguera:2023bhm}. It could be beneficial to show if the evolution of some of these definitions in the DSSYK model can be matched with the hyperfast growth of certain holographic complexity proposals (\ref{eq:hyperfast}) in asymptotically dS spacetimes \cite{Susskind:2021esx,Chapman:2021eyy,Jorstad:2022mls,Galante:2022nhj,Auzzi:2023qbm,Anegawa:2023wrk,Anegawa:2023dad,Baiguera:2023tpt,Aguilar-Gutierrez:2024rka}.

\subsection*{Generalization of Krylov complexity}
Lastly, there has been a recent construction of multiparticle wavefunctions in the DSSYK model with matter \cite{Xu:2024hoc}. It should be possible to extend our study considering the wavefunctions in the evaluation of spread complexity (\ref{eq:spread complexity}) with a 1-particle insertion, although, perhaps the holographic dictionary might need to be extended in order to associate a bulk description in these cases. Spread complexity for these states could then be associated with the Krylov complexity obtained through 2-point correlation functions (\ref{eq:Krylov complexity}) of a given matter operator. It would be very interesting to study how our results for Krylov and spread complexity would be related to each other in this proposal, as the connection between these definitions has been recently scrutinized in \cite{Caputa:2024vrn}.

\cleardoublepage

\part{Epilogue}\label{part:4}
  \graphicspath{{\chaptersdir/Conclusion/image/}}%
  \chapter{Conclusions and outlook}\label{ch:Conclusion}
We have studied two approaches to define different complexity proposals in asymptotically dS spacetimes from the perspective of static patch holography. Let us reflect on the main findings, their connections, and some future directions.

We first took a bottom-up approach in Ch. \ref{ch:ComplexitydS} to study several codimension-zero (CA, CV2.0) and codimension-one (CV, CAny) holographic complexity proposals, defined for AdS black holes, in an extended SdS spacetime background, which we viewed as a toy model of eternal inflation. We probed single or multiple inflating and black hole patches, to generalize previous studies that were limited to the inflating patch of dS space. We proposed a set of complexity observables (codimension-one CAny observables evaluated on CMC slices) that have very different behavior compared to the others. When limited to the inflating patch, we showed that hyperfast growth in holographic complexity found in the CV, CA, CV2.0 proposals is avoided by a set of CAny proposals, which instead evolve eternally. Moreover, the same switchback effect found in AdS black holes can be recovered for our set of proposals. Interestingly, once we include both black hole and inflating patches, the codimension-zero, and codimension-one are highly constrained. The codimension-zero proposals will always be time-independent, while the codimension-one can only be time-dependent for a particular choice of time flow along the stretched horizon.

In contrast, we took a microscopic perspective in Ch. \ref{ch:ComplexityDSSYK}, where we considered the DSSYK model as the putative quantum mechanical dual theory to SdS$_3$ space, and to a LdS$_2$ CFT, based on the relevant findings in \cite{Narovlansky:2023lfz,Verlinde:2024znh,Verlinde:2024zrh}. We studied the notions of spread, Krylov, query, and Nielsen complexity in the microscopic model, and we used some elements of the holographic dictionary in \cite{Narovlansky:2023lfz,Verlinde:2024znh,Verlinde:2024zrh} to obtain a dual interpretation for each side of the putative correspondence. In particular, we found that spread complexity in the DSSYK model is related to geodesic lengths in SdS$_3$ space that measure time differences between antipodal observers. Krylov complexity instead represents operator growth on all the different sides of the correspondence. Query complexity in the LdS$_2$ CFT has a bulk description in terms of a CAny observable. Lastly, a particular choice of Nielsen complexity has the time growth as the codimension-one CAny proposals on CMC slices at late times, connecting to our work in Ch. \ref{ch:ComplexitydS}. While this correspondence is still in early phases, we hope our work can spark future developments in this area.

\section{Outlook}
We now conclude with some future directions. We will focus on directions that connect the bottom-up and first principles approaches to define complexity in the bulk in Ch. \ref{ch:ComplexitydS} and \ref{ch:ComplexityDSSYK} respectively.\footnote{The reader is referred to the outlook in Sec. \ref{sec:ConclusionsSdSn} and \ref{sec:discu} for future directions in the previous chapters.}

A clear limitation in Ch. \ref{ch:ComplexityDSSYK} is that the discussion is very special to lower dimensions. Finding a holographic dual to dS space in higher dimensions remains an important problem (see recent progress in \cite{Bobev:2022lcc}); we hope that the techniques employed here can be improved to study more intricate cases.

\subsection*{Non-hermitian dS holography}
As mentioned in Ch. \ref{ch:introduction}, one of the simplifying assumptions through the thesis has been considering the experience of a static patch observer in terms of a dual unitary quantum system,  as motivated by the cosmological central dogma in static patch holography. However, in reality, they have access to a subregion of spacetime, which in principle is better described by an open quantum system. For instance, dS space is known to undergo vacuum decay due to bubble nucleation (see e.g. \cite{Kashyap:2015lva}). We have argued that at time scales smaller than the dS radius these effects should not be very relevant for the static patch observer. However, since we are interested in the late time regime for defining holographic complexity, one should, in principle, account for this, once the unitary case is well understood.

To model these effects on the quantum theory side, a reasonable improvement would be to construct a non-Hermitian Hamiltonian in a DSSYK approach to dS holography. There are different ways of incorporating non-Hermitian terms in (\ref{eq:DDSSYK Hamiltonian}) while preserving the chord techniques. For instance, by having multiple DSSYK Hamiltonians (\ref{eq:DDSSYK Hamiltonian}) with a different number of fermion interactions (see \cite{Anninos:2022qgy,Anninos:2020cwo} for the original proposal, motivated from thermodynamic considerations):
\begin{equation}\label{eq:non unitary}
    H=H^{(p)}_{\rm DSSYK}+\sum_i\lambda_i H^{(q_i)}_{\rm DSSYK}~,
\end{equation}
where the superscript ($p$, $q_i$) denotes the number of fermion interactions in (\ref{eq:DDSSYK Hamiltonian}), with $\lambda_i\in\mathbb{C}$, and $q_i\rightarrow\infty$ in the double scaling limit (while keeping $q_i<p$ for the non-unitary term to be relevant). A similar type of interpolated model has recently appeared in \cite{Berkooz:2024evs,Berkooz:2024ofm}. 
It has been previously studied in \cite{Garcia-Garcia:2020ttf} that by having a (non-doubled scaled) system of uncoupled SYK models whose couplings $J_{i_1\dots i_p}\in\mathbb{C}$ are complex conjugates of each other, it produces a mass gap in the energy spectrum of the system that is similar to the Maldacena-Qi Euclidean wormhole in JT gravity, leaking out information to a region of spacetime that the observers would not have access to. A similar effect for a non-Hermitian SYK model was noticed to be reminiscent of dissipation effects in dS$_2$ space \cite{Garcia-Garcia:2022adg}. 

It would be useful to study the double scaling limit for these models, where we might separate the real and imaginary parts of the Hamiltonian and derive the chord diagram rules with two types of chords. Spread and Krylov complexity have been useful probes for these effects in several previous studies \cite{Bhattacharjee:2023uwx,Bhattacharya:2023zqt,Bhattacharjee:2022lzy,Garcia-Garcia:2018fns,Bhattacharya:2023yec}, so we expect that the techniques developed in the unitary setting during our work can be adapted for these more intricate scenarios.

Lastly, non-unitary evolution is important for modeling measurement-induced dynamics. This has been recently studied in holographic systems by \cite{Antonini:2023aza,Antonini:2022lmg}, and using Krylov complexity in \cite{Bhattacharya:2023yec}. It could be interesting to incorporate these effects in dS holography and probe them with the complexity proposals of Ch. \ref{ch:ComplexityDSSYK}.

\subsection*{Holographic dictionary for the stretched horizon and $G_N/L$}
Our study of holographic complexity in Ch. \ref{ch:ComplexitydS} relies on the existence of a stretched horizon, somewhere in the static patch that is close to the cosmological or black hole horizons of SdS space, where gravity is non-dynamical. However, our study in Ch. \ref{ch:ComplexityDSSYK} is based on a model where there is no such stretched horizon, instead, it relies on Dirichlet boundaries along the static patch worldline observer. It is an interesting question whether these different approaches can be unified. As we mentioned in Ch. \ref{ch:introduction}, \ref{ch:ComplexitydS}, in AdS/CFT there are the so-called $\TT+\Lambda_2$-deformations \cite{Gorbenko:2018oov,Lewkowycz:2019xse,Shyam:2021ciy,Coleman:2021nor,Silverstein:2022dfj,Batra:2024kjl}, where a timelike Dirichlet boundary in dS$_3$ space appears by performing a irrelevant deformation of a CFT holographic to empty AdS$_3$ space or a BTZ black hole. It is possible to perform these deformations in quantum mechanical theories \cite{Gross:2019ach,Gross:2019uxi}. Preliminary results indicate that the location of the stretched horizon in the SdS$_3$ space is reproduced from the coupling constant of the irrelevant deformation in the DSSYK model, and it can be pushed all the way to the worldline observer. This will allow for a more clear interpretation of the complexity observables from the holographic dictionary.

An important lesson from $\TT+\Lambda_2$-deformations \cite{Batra:2024kjl} is that the dual theory to dS$_3$ space in this construction has a finite number of degrees of freedom, and so is the Gibbons-Hawking entropy, i.e. $G_N/L$ is always finite. This observation strongly connects with a debate in the DSSYK literature \cite{Rahman:2023pgt}, on the holographic dictionary for the parameter $G_N/L$ should scale as $N$ \cite{Susskind:2021dfc,Rahman:2022jsf,Rahman:2023pgt,Rahman:2024vyg}, or $G_N/L\sim \abs{\log q}$ as in the bulk dual models to the DSSYK model \cite{Blommaert:2023opb,Narovlansky:2023lfz,Blommaert:2024ydx,Milekhin:2023bjv}. The latter case seems to be counter-intuitive with respect to usual holography, where we would expect it to be related to the number of degrees of the system $N$ Majorana fermions. In the double scaling limit (where $N\xrightarrow{}\infty$ while keeping $q$ finite), the Gibbons-Hawking entropy $\propto N$ would be divergent in the former proposals, while in the explicit examples \cite{Blommaert:2023opb,Narovlansky:2023lfz,Blommaert:2024ydx,Milekhin:2023bjv}, it remains finite. From the viewpoint of the author, the apparent discrepancy arises from the fact that the double scaling theory describes an ensemble averaging of the fermions, so degrees of freedom are not really captured by the factor $N$, but rather, the parameter $\lambda=\abs{\log q}$ which controls the density of states of the DSSYK model (i.e. $\mu(\theta)/2\pi$ in (\ref{eq:identity theta})). We expect the finiteness of $G_N/L$ from the DSSYK model can be related to the finiteness of Gibbons-Hawking entropy of SdS$_3$ space in the model of \cite{Narovlansky:2023lfz,Verlinde:2024znh,Verlinde:2024zrh}.

\subsection*{Well-posed initial value problem and thermal stability of dS space}
Throughout the thesis, we have evaluated complexity observables anchored to a Dirichlet time-like boundary in the static patch of asymptotically dS spacetimes: the stretched horizon in Ch. \ref{ch:ComplexitydS}, and, possibly, the worldline static patch observer in Ch. \ref{ch:ComplexityDSSYK}. It has been recently shown in \cite{Anninos:2024wpy} that dS$_{4}$ space with Dirichlet timelike boundaries is not a well-posed initial value problem. Meanwhile, it is also well known that dS$_3$ space with Dirichlet time-like boundaries is thermodynamically unstable \cite{Svesko:2022txo,Banihashemi:2022jys,Banihashemi:2022htw,Anninos:2024wpy}. This would mean that the complexity observables should be evaluated with different boundary conditions that are well-posed and where the inflating patch remains thermodynamically stable (such as conformal boundary conditions \cite{Anninos:2024wpy}). 

On the bulk side, it would be a substantial improvement to revisit the holographic complexity proposals studied in Ch. \ref{ch:ComplexitydS} considering conformal boundary conditions. On the DSSYK side, it is, therefore, an important problem to study these modifications in the holographic framework developed by \cite{Narovlansky:2023lfz,Verlinde:2024zrh,Verlinde:2024znh}. On the other hand, it has been found \cite{Blommaert:2023opb,Blommaert:2024ydx} that dS$_2$ space geometries interpolating with AdS$_2$ space with timelike boundaries can be obtained from the DSSYK model. Our preliminary studies have found that the particular model in \cite{Blommaert:2023opb,Blommaert:2024ydx} is thermodynamically stable. One can in principle repeat our same type of analysis as in Ch. \ref{ch:ComplexityDSSYK} to search for a sharp definition of complexity in dS$_2$ space from a stable quantum mechanical dual theory. 
Moreover, since the alternative approach of \cite{Blommaert:2023opb,Blommaert:2023wad,Blommaert:2024ydx} reproduces a well-posed solution, it could be useful to implement in our future study of a gravitational dual of a DSSYK model that incorporates non-Hermitian terms (\ref{eq:non unitary}).

In connection to the above, it has been recently discussed by \cite{Milekhin:2024vbb} that a very fast type of diffusion of matter process (coined superdiffusion) should be reproduced by the holographic dual of (S)dS space. It would be very interesting to study if these effects may appear in four-point functions of the DSSYK model in the approaches to dS holography of \cite{Narovlansky:2023lfz,Verlinde:2024znh,Verlinde:2024zrh}; \cite{Blommaert:2023opb,Blommaert:2023wad,Blommaert:2024ydx}; some recent works \cite{Milekhin:2023bjv}, and \cite{Almheiri:2024xtw}; or our future non-unitary proposal.

\subsection*{Switchback effect in microscopic models}
The switchback effect plays a fundamental role in the definition of the holographic complexity proposals in asymptotically AdS spacetimes \cite{Belin:2021bga,Belin:2022xmt}. In the microscopic approach for defining complexity in dS space, Ch. \ref{ch:ComplexityDSSYK}, we have focused on the evolution of different complexity proposals in the maximal entropy state of the DSSYK model, and its correlation functions. Since our study in Sec. \ref{sec:CAny CMC} indicates that there is a set of geometric observables that display the same switchback effect as the dual quantum circuit model to an AdS black hole, it would be important to confirm if this type of effect can also be recovered for some of the different definitions we studied in \ref{ch:ComplexityDSSYK}, at least when probed with alternating localized energy pulses.

We expect that the shockwave analysis can be carried out by considering six-point operator insertions in the thermal circle, Fig. \ref{fig:MatterChord}, based on a similar analysis of shockwaves performed in the high-temperature regime of the DSSYK model \cite{Almheiri:2024xtw} (previously considered by \cite{Gao:2023gta}). If only some of the complexity proposals in Ch. \ref{ch:ComplexityDSSYK} can display this type of behavior, it might help us to constrain the possible entries duals to the codimension-one CAny proposal in Sec. \ref{sec:CAny CMC}.

\subsection*{Causal connectivity of dS space}
Another important question in static patch holography is how does the causal connectivity of dS space arise from entanglement between holographic quantum mechanical dual theories \cite{Susskind:2021esx,Susskind:2022dfz,Franken:2023pni,Franken:2023jas,Franken:2024ruw,Rondeau:2024ftb} on two static patches? In geometric terms, this question means: if each of the quantum mechanical theories on the stretched horizons describes a static patch, how do they get connected with the intermediate inflating regions (see e.g. Fig. \ref{fig:StrechedHorizon})? This question has not been addressed with explicit microscopic models so far. Since our findings show that holographic complexity is able to probe the geometry between any pair of stretched horizons, and that spread complexity measures their entanglement in the doubled Hilbert space, we would like to understand better how precisely the entanglement entropy between the pair of DSSYK models allows to reconstruct a portion of spacetime. Related progress in this direction has been performed in \cite{Tang:2023ocr}.

\instructionsconclusions

\cleardoublepage

\appendix

\chapter{Details on the complexity=action proposal}\label{app:details_CA}

In this appendix, we present additional technical details on the computation of the boundary action $I_{\rm bdy}$ in eq.~\eqref{eq:bdy_action} for various configurations of the WDW patch.

\section{Case 1}
\label{app:ssec:case1}

Assuming that $t_{\infty} > 0$, we focus on the regime at intermediate times $-t_{\infty} \leq t \leq t_{\infty}$ when the WDW patch takes the shape in Fig.~\ref{fig:WDW_evo1}.
In this setting all the codimension-one boundaries are null, therefore there are no GHY terms.
In case 1, the special positions of the WDW patch $r_{\pm}$ are defined in eq.~\eqref{eq:umax_vmin_WDWpatch}.

\paragraph{Joint terms.}
The geometry contains four null-null joints: the top and bottom vertices of the WDW patch, plus two joints located at the stretched horizons.
The one-forms orthogonal to the null boundaries of the WDW patch are given by
\beq
\begin{aligned}
&  \mathrm{TL}: \qquad
k_{\mu}^{\rm TL} dx^{\mu} = - \alpha \le du + \frac{2}{f(r)} dr \ri\Big|_{v= -t_L + r^*(r_{\rm st}^L)} \\
&  \mathrm{BL}: \qquad
k_{\mu}^{\rm BL} dx^{\mu} = 
 \beta  du|_{u = -t_L - r^*(r_{\rm st}^L)} 
 \\
&  \mathrm{TR}: \qquad
k_{\mu}^{\rm TR} dx^{\mu} = 
 \beta'  du|_{u = t_R - r^*(r_{\rm st}^R)}  \\
&  \mathrm{BR}: \qquad
k_{\mu}^{\rm BR} dx^{\mu} = - \alpha' \le du + \frac{2}{f(r)} dr \ri\Big|_{v= t_R + r^*(r_{\rm st}^R)}\\
\end{aligned}
\label{eq:list_null_normals_joints}
\eeq
where R denotes right, L left, T top, B bottom.
The corresponding joint contributions read
\begin{subequations}
\beq
\begin{aligned}
    I_{\mathcal{J}}^{\rm st,L} &= - \frac{\Omega_{d-1}}{8 \pi G_N} (r_{\rm st}^L)^{d-1} \log \left| \frac{\alpha \beta}{f(r_{\rm st}^L)} \right| \, ,\\
I_{\mathcal{J}}^{\rm st,R} &= - \frac{\Omega_{d-1}}{8 \pi G_N} (r_{\rm st}^R)^{d-1} \log \left| \frac{\alpha' \beta'}{f(r_{\rm st}^R)} \right| \, ,
\end{aligned}
\label{eq:joints_stretched_horizons}
\eeq
\beq
\begin{aligned}
I_{\mathcal{J}}^{\rm T} &= \frac{\Omega_{d-1}}{8 \pi G_N} (r_{+})^{d-1} \log \left| \frac{\alpha \beta'}{f(r_+)}  \right| \, , \\
I_{\mathcal{J}}^{\rm B} &= \frac{\Omega_{d-1}}{8 \pi G_N} (r_{-})^{d-1} \log \left| \frac{\alpha' \beta}{f(r_-)}  \right| \, . 
\end{aligned}
\eeq    
\label{eq:joints_case1}
\end{subequations}

\paragraph{Contributions on null boundaries.}
The parametrization \eqref{eq:list_null_normals_joints} is affine, therefore the codimension-one boundary terms in eq.~\eqref{eq:null_codim1_term} vanish. 
The counterterm \eqref{eq:counterterm_action} is non-trivial.
To evaluate the latter, we compute the expansion parameter of the null geodesics delimiting the WDW patch
\beq
\begin{aligned}
&  \mathrm{TL}: \qquad
\Theta^{\rm TL}  =  - \frac{\alpha (d-1)}{r} \, , \qquad
&  \mathrm{BL}: \qquad
\Theta^{\rm BL}  = - \frac{\beta (d-1)}{r} \, ,  \\
&  \mathrm{TR}: \qquad
\Theta^{\rm TR}  =  - \frac{\beta' (d-1)}{r}  \, , \qquad
&  \mathrm{BR}: \qquad
\Theta^{\rm BR}  = - \frac{\alpha' (d-1)}{r} \, . \\
\end{aligned}
\eeq
Correspondingly, the null counterterm evaluated at each null boundary reads
\begin{subequations}
\beq
\begin{aligned}
I_{\rm ct}^{\rm TL} = -\frac{\Omega_{d-1}}{8\pi G_N} &\left\lbrace (r_{+})^{d-1} \left[ \log \left| \frac{\alpha \ell_{\rm ct} (d-1)}{r_{+}} \right| + \frac{1}{d-1} \right]\right.\\
&\left.- (r_{\rm st}^L)^{d-1} \left[ \log \left| \frac{\alpha \ell_{\rm ct} (d-1)}{r_{\rm st}^L} \right| + \frac{1}{d-1} \right]  \right\rbrace \, ,    
\end{aligned}
\eeq
\beq
\begin{aligned}
    I_{\rm ct}^{\rm BL} = -\frac{\Omega_{d-1}}{8\pi G_N} &\left\lbrace (r_{-})^{d-1} \left[ \log \left| \frac{\beta \ell_{\rm ct} (d-1)}{r_{-}} \right| + \frac{1}{d-1} \right]\right. \\
&\left.- (r_{\rm st}^L)^{d-1} \left[ \log \left| \frac{\beta \ell_{\rm ct} (d-1)}{r_{\rm st}^L} \right| + \frac{1}{d-1} \right]  \right\rbrace \, ,
\end{aligned}
\eeq
\beq
\begin{aligned}
I_{\rm ct}^{\rm TR} = -\frac{\Omega_{d-1}}{8\pi G_N} &\left\lbrace (r_{+})^{d-1} \left[ \log \left| \frac{\beta' \ell_{\rm ct} (d-1)}{r_{+}} \right| + \frac{1}{d-1} \right] \right.\\
&\left.- (r_{\rm st}^R)^{d-1} \left[ \log \left| \frac{\beta' \ell_{\rm ct} (d-1)}{r_{\rm st}^R} \right| + \frac{1}{d-1} \right]  \right\rbrace \, ,    
\end{aligned}
\eeq
\beq
\begin{aligned}
I_{\rm ct}^{\rm BR} = -\frac{\Omega_{d-1}}{8\pi G_N} &\left\lbrace (r_{-})^{d-1} \left[ \log \left| \frac{\alpha' \ell_{\rm ct} (d-1)}{r_{-}} \right| + \frac{1}{d-1} \right] \right.\\
&\left.- (r_{\rm st}^R)^{d-1} \left[ \log \left| \frac{\alpha' \ell_{\rm ct} (d-1)}{r_{\rm st}^R} \right| + \frac{1}{d-1} \right]  \right\rbrace \, .
\end{aligned}
\eeq
\label{eq:Ict_case1}
\end{subequations}

\paragraph{Total boundary action.}
Finally, the boundary action \eqref{eq:bdy_action} is given by the summation of eqs.~\eqref{eq:joints_case1} and \eqref{eq:Ict_case1}. In the symmetric case $r_{\rm st}^L = r_{\rm st}^R$, and considering $-t_{\infty} \leq t \leq t_{\infty}$, the result reads
\beq
\begin{aligned}
   I_{\rm bdy} (t) = \frac{\Omega_{d-1}}{8 \pi G_N} &\left\lbrace  
   - 2 (r_{\rm st})^{d-1} \left[ \log \left| \frac{(r_{\rm st})^2}{f(r_{\rm st}) \ell_{\rm ct}^2 (d-1)^2} \right| - \frac{2}{d-1} \right]
 \right. \\
& \left.  + (r_{+})^{d-1} \left[ \log \left| \frac{(r_{+})^2}{f (r_{+}) \ell_{\rm ct}^2 (d-1)^2} \right| - \frac{2}{d-1} \right] \right.\\
&\left.+ (r_{-})^{d-1} \left[ \log \left| \frac{(r_{-})^2}{f (r_{-}) \ell_{\rm ct}^2 (d-1)^2} \right| - \frac{2}{d-1} \right]
\right\rbrace \, ,
\end{aligned}
\label{eq:bdy_action_case1}
\eeq
where $r_{\pm}$ were defined in eq.~\eqref{eq:umax_vmin_WDWpatch}, and we assumed that the stretched horizons on the two sides of the geometry are located at the same radial coordinate, \ie $r^L_{\rm st} =r^{R}_{\rm st}$.

\section{Case 2}
\label{app:ssec:case2}

When the stretched horizons are determined according to case 2 in section \ref{ssec:stretched_horizons}, we need to study the early and late time regimes to find the characteristic linear growth of complexity.
Depending on the sign of the critical time $t_0$, the time dependence of the WDW patch can follow two different evolutions, represented by the Penrose diagrams in Fig.~\ref{fig:WDW_evo1} or \ref{fig:WDW_evo2}.
The formal expression of CA at early and late times is universal, while it differs at intermediate times between the two cases.

\paragraph{Evolution with critical time $t_0>0$ (see Fig.~\ref{fig:WDW_evo1}).}
The analytic expression of the boundary terms at intermediate times $-t_0 \leq t \leq t_0$ is given by the same analytic formula reported in eq.~\eqref{eq:bdy_action_case1}, with the only difference that the locations of the joints $r_{\pm}$ are now defined by eq.~\eqref{eq:umax_vmin_WDWpatch_case2}.

\vskip 2mm
\noindent
At times $t>t_0$, the configuration of the WDW patch is given by Fig.~\ref{fig:WDW_evo1}.
The main novelty is the existence of GHY terms due to the intersection of the WDW patch with the spacelike cutoff surface at $r=r_{\rm min}$.
The corresponding induced metric reads
\beq
\begin{aligned}
    \sqrt{h} &= r^{d-1} \sqrt{f (r)}\Big|_{r=r_{\rm min}} \, ,\\
K &= \sqrt{f(r)} \partial_r \sqrt{h} = \frac{1}{2r \sqrt{f(r)}} \left[ 2(d-1) f(r) + r f'(r) \right]\Big|_{r= r_{\rm min}} \, ,
\end{aligned}
\label{eq:induced_metric_GHY}
\eeq
which gives the GHY contribution
\beq
\begin{aligned}
I_{\rm GHY} & = - \frac{\Omega_{d-1}}{16 \pi G_N} (r_{\rm min})^{d-2} \le  t_R + t_L + r^*(r_{\rm st}^R) + r^*(r_{\rm st}^L) - 2 r^* (r_{\rm min})  \ri \\
& \times  \left[ 2(d-1) f(r) + r f'(r) \right]\Big|_{r= r_{\rm min}}  \, .
\end{aligned}
\eeq
There are some special choices that simplify the previous result. First, we consider the symmetric time evolution \eqref{eq:symmetric_times}.
Second, we locate the stretched horizons at the same radial coordinate $r^R_{\rm st} = r^L_{\rm st}$.
Finally, we take the limit $r_{\rm min} \rightarrow 0$.
Altogether, these choices lead to 
\beq
\lim_{r_{\rm min} \rightarrow 0} I_{\rm GHY}  =  \frac{\Omega_{d-1}}{8 \pi G_N} \mu d \le  t + 2 r^*(r_{\rm st}) - 2 r^* (0)  \ri \, ,
\eeq
where $\mu$ is the mass parameter entering the blackening factor in eq.~\eqref{eq:asympt_dS}.\footnote{\label{fn:AdS CA App}The result above can be trivially adapted for the AdS black hole, studied in Sec. \ref{ssec:CAAdS}, using the metric (\ref{eq:metric SAdS})
\beq
\lim_{r_{\rm min} \rightarrow 0} I_{\rm GHY}  =  \begin{cases}
    \frac{\Omega_{k,1}}{8 \pi G_N} \qty(\frac{C}{L^2} +k)\le  t + 2 r^*(r_{\rm bdy}) - 2 r^* (0)  \ri&d=2 \, ,\\
    \frac{\Omega_{k,d-1}}{8 \pi G_N} \frac{C~d}{2L^2} \le  t + 2 r^*(r_{\rm bdy}) - 2 r^* (0)  \ri&d\geq3
\end{cases}
\eeq}
Next, let us evaluate the joints.
The expressions in eq.~\eqref{eq:joints_case1} are also valid in this setting, except for the top vertex $I_{\mathcal{J}}^{\rm T}$, which now moved behind the cutoff surface at $r=r_{\rm min}$.
There are instead two new joints arising from the intersection of the future boundaries of the WDW patch with the cutoff.
The one-form normal to such a spacelike surface is given by
\beq
t_{\mu} dx^{\mu} = - \frac{dr}{\sqrt{-f(r)}}\Big|_{r = r_{\rm min}} \, . 
\eeq
This allows to determine the joints
\beq
\begin{aligned}
    I_{\mathcal{J}}^{\rm TL} &= \frac{\Omega_{d-1}}{8 \pi G_N} (r_{\rm min})^{d-1} \log \left| \frac{\alpha}{f(r_{\rm min})}  \right| \, , \\
I_{\mathcal{J}}^{\rm TR} &= \frac{\Omega_{d-1}}{8 \pi G_N} (r_{\rm min})^{d-1} \log \left| \frac{\beta'}{f(r_{\rm min})}  \right| \, . 
\end{aligned}
\label{eq:joints_case2_app}
\eeq
These expressions vanish in the limit $r_{\rm min} \rightarrow 0$.

Finally, the counterterms \eqref{eq:Ict_case1} referring to the top-left and top-right null boundaries of the WDW patch get modified as
\begin{subequations}
\beq
I_{\rm ct}^{\rm TL} = \frac{\Omega_{d-1}}{8\pi G_N} 
(r_{\rm st})^{d-1} \left[ \log \left| \frac{\alpha \ell_{\rm ct} (d-1)}{r_{\rm st}} \right| + \frac{1}{d-1} \right]  \, ,
\eeq
\beq
I_{\rm ct}^{\rm TR} = \frac{\Omega_{d-1}}{8\pi G_N} 
 (r_{\rm st})^{d-1} \left[ \log \left| \frac{\beta' \ell_{\rm ct} (d-1)}{r_{\rm st}} \right| + \frac{1}{d-1} \right] \, ,
\eeq
\label{eq:Ict_case2_late_times}
\end{subequations}
where we already considered the symmetric case $r^R_{\rm st} = r^L_{\rm st}$ and performed the limit $r_{\rm min} \rightarrow 0$.
The contributions arising from the past boundaries in eq.~\eqref{eq:Ict_case1} are unchanged.

Summing all the boundary terms determined above according to eq.~\eqref{eq:bdy_action}, we obtain
\beq
\begin{aligned}
 I_{\rm bdy} (t>t_0) = \frac{\Omega_{d-1}}{8 \pi G_N} & \left\lbrace 
   \mu d \le  t + 2 r^*(r_{\rm st}) - 2 r^* (0)  \ri 
 \right.  \\
  &\left. - 2 (r_{\rm st})^{d-1} \left[ \log \left| \frac{(r_{\rm st})^2}{f_2(r_{\rm st}) \ell_{\rm ct}^2 (d-1)^2} \right| - \frac{2}{d-1} \right]
 \right. \\
 &\left.  + (r_{-})^{d-1} \left[ \log \left| \frac{(r_{-})^2}{f (r_{-}) \ell_{\rm ct}^2 (d-1)^2} \right| - \frac{2}{d-1} \right]
\right\rbrace \, .
\end{aligned}
\label{eq:bdy_action_case2_latetimes}
\eeq

\vskip 2mm
\noindent
The computation of CA at times $t< t_0$ is very similar as the case of late times, with the roles of $r_+$ and $r_-$ exchanged.
We directly report the result
\beq
\begin{aligned}
 I_{\rm bdy} (t<-t_0) = \frac{\Omega_{d-1}}{8 \pi G_N} & \left\lbrace  
  \mu d \le  -t + 2 r^*(r_{\rm st}) - 2 r^* (0)  \ri \right.
  \\
  &\left.- 2 (r_{\rm st})^{d-1} \left[ \log \left| \frac{(r_{\rm st})^2}{f_2(r_{\rm st}) \ell_{\rm ct}^2 (d-1)^2} \right| - \frac{2}{d-1} \right]
 \right. \\
& \left.  + (r_{+})^{d-1} \left[ \log \left| \frac{(r_{+})^2}{f (r_{+}) \ell_{\rm ct}^2 (d-1)^2} \right| - \frac{2}{d-1} \right] 
\right\rbrace \, .
\end{aligned}
\label{eq:bdy_action_case2_earlytimes}
\eeq

\paragraph{Evolution with critical time $t_0<0$ (see Fig.~\ref{fig:WDW_evo2}).}
Since we already argued that early and late times contribute to CA with the same expressions \eqref{eq:bdy_action_case2_earlytimes} and \eqref{eq:bdy_action_case2_latetimes} computed above, it is sufficient to focus on the intermediate time regime $t_0 \leq t \leq -t_0$.

There are two GHY terms, coming from either the past or the future cutoff surfaces near the singularities.
The sum of these contributions reads
\beq
\begin{aligned}
I_{\rm GHY}^{\rm T} + I_{\rm GHY}^{\rm B}  = \frac{\Omega_{d-1}}{8 \pi G_N} (r_{\rm min})^{d-2}&
\le r^*(r_{\rm st}^L) + r^*(r_{\rm st}^R) - 2 r^*(r_{\rm min})   \ri \\
&\left[ 2(d-1) f(r_{\rm min}) + r f'(r_{\rm min}) \right] \, .
\end{aligned}
\label{eq:IGHY_case2_app}
\eeq
Next, we consider the joints.
At the stretched horizons we have the usual contributions in eq.~\eqref{eq:joints_stretched_horizons}.
The intersections with the singularities lead to vanishing results when performing the limit $r_{\rm min} \rightarrow 0$, similarly to the analog computation performed in eq.~\eqref{eq:joints_case2_app}.
Finally, the counterterms include contributions similar to eq.~\eqref{eq:Ict_case2_late_times}, and their sum reads
\beq
\begin{aligned}
I_{\rm ct} = \frac{\Omega_{d-1}}{8\pi G_N} & \left\lbrace 
(r_{\rm st}^L)^{d-1} \left[ \log \left| \frac{\alpha \ell_{\rm ct} (d-1)}{r_{\rm st}^L} \right| + \frac{1}{d-1} \right] \right.\\
&\left.  + (r_{\rm st}^R)^{d-1} \left[ \log \left| \frac{\beta' \ell_{\rm ct} (d-1)}{r_{\rm st}^R} \right| + \frac{1}{d-1} \right] \right. \\
& \left.  + (r_{\rm st}^L)^{d-1} \left[ \log \left| \frac{\beta \ell_{\rm ct} (d-1)}{r_{\rm st}^L} \right| + \frac{1}{d-1} \right]\right.\\
& \left.+ (r_{\rm st}^R)^{d-1} \left[ \log \left| \frac{\alpha' \ell_{\rm ct} (d-1)}{r_{\rm st}^R} \right| + \frac{1}{d-1} \right]
\right\rbrace \, .
\end{aligned}
\label{eq:Ict_case2_app}
\eeq
The relevant observation is that the sum of boundary terms \eqref{eq:IGHY_case2_app} and \eqref{eq:Ict_case2_app} is time-independent, \ie 
\beq
\frac{d I_{\rm bdy}}{dt} (t_0 \leq t \leq -t_0) = 0 \, .
\label{eq:rate_bdy_app_case2}
\eeq

\section{Cases 3-5}
\label{app:ssec:case3}

In case 3, the top and bottom joints of the WDW patch always lie behind the cutoff surfaces, see Fig.~\ref{fig:WDW_BH_stretched}.
In the following, we will focus on the left side of the diagram.

\paragraph{GHY terms.}
There are two codimension-one spacelike surfaces, corresponding to the intersection of the WDW patch with the cutoff surfaces located near the singularities.
The induced metric is given by eq.~\eqref{eq:induced_metric_GHY}, and the corresponding GHY terms sum to
\beq
\begin{aligned}
I_{\rm GHY}^{\rm L} =  &\frac{\Omega_{d-1}}{16\pi G_N} (r_{\rm max})^{d-2}
\left[ 2(d-1) f(r_{\rm max}) + r f'(r_{\rm max}) \right]\\
&\times\le  \int_{-\infty}^{-t_L-r^*(r_{\rm max})} du  + \int_{-t_L+r^*(r_{\rm max})}^{\infty} dv  \ri  
\, ,    
\end{aligned}
\eeq
where we expressed the result as an integral over $u(v)$ for convenience.
The important observation is that the time dependence on $t_L$ cancels when summing the two terms.

\paragraph{Joint terms.}
The right joint evaluated at the stretched horizons has the same formal expression $I_{\mathcal{J}}^{\rm st, L}$ computed in eq.~\eqref{eq:joints_stretched_horizons}.
The intersections between the cutoff surfaces and the WDW patch generate two more joints on the left region, given by  
\beq
\begin{aligned}
 &   I_{\mathcal{J}}^{\rm TL} =  \frac{\Omega_{d-1}}{8 \pi G_N} (r_{\rm max})^{d-1} \log \left| \frac{\alpha}{\sqrt{f(r_{\rm max})}} \right| \, ,\\
&I_{\mathcal{J}}^{\rm BL} =  \frac{\Omega_{d-1}}{8 \pi G_N} (r_{\rm max})^{d-1} \log \left| \frac{\beta}{\sqrt{f(r_{\rm max})}} \right| \, .
\end{aligned}
\eeq

\paragraph{Counterterm.}
The parametrization \eqref{eq:list_null_normals_joints} is affine, therefore there are no codimension-one boundary terms for the null surfaces except for the counterterms.
The null counterterms read
\begin{subequations}
\beq
\begin{aligned}
    I_{\rm ct}^{\rm TL} = -\frac{\Omega_{d-1}}{8\pi G_N} &\left\lbrace (r_{\rm max})^{d-1} \left[ \log \left| \frac{\alpha \ell_{\rm ct} (d-1)}{r_{\rm max}} \right| + \frac{1}{d-1} \right] \right.\\
    &\left.- (r_{\rm st})^{d-1} \left[ \log \left| \frac{\alpha \ell_{\rm ct} (d-1)}{r_{\rm st}} \right| + \frac{1}{d-1} \right]  \right\rbrace \, ,
\end{aligned}
\eeq
\beq
\begin{aligned}
I_{\rm ct}^{\rm BL} = -\frac{\Omega_{d-1}}{8\pi G_N} &\left\lbrace (r_{\rm max})^{d-1} \left[ \log \left| \frac{\beta \ell_{\rm ct} (d-1)}{r_{\rm max}} \right| + \frac{1}{d-1} \right] \right.\\
&\left.- (r_{\rm st})^{d-1} \left[ \log \left| \frac{\beta \ell_{\rm ct} (d-1)}{r_{\rm st}} \right| + \frac{1}{d-1} \right]  \right\rbrace \, .
\end{aligned}
\eeq
\end{subequations}
Combining with the joints, the normalization of null normals cancels from the action.

\paragraph{Total sum.}
The sum of all the boundary terms \eqref{eq:bdy_action} on the left side of the diagram is time-independent, without the need for any particular assumption on the boundary times $t_L, t_R$.
An analog computation reveals that the right side is also independently time-independent, leading to the result
\beq
\frac{d I_{\rm bdy}}{dt_L} = \frac{dI_{\rm bdy}}{dt_R}=0 \, .
\label{eq:rate_CA_bdy_case3}
\eeq

\vskip 2mm
\noindent
Finally, let us point out that cases 4 and 5 do not present additional new terms, instead they always reduce to contributions similar to other cases.

\instructionsappendices
\cleardoublepage
\chapter{Multi-scale entanglement renormalization ansatz tensor networks}\label{app:MERA}
This appendix provides a brief introduction to the MERA tensor network, which we have mentioned in Chs. \ref{ch:DSSYK}, \ref{ch:ComplexityDSSYK}.

First, let us describe a lattice with $N$ sites, where each site has an associated vector and an associated Hilbert space. The Hilbert space of the lattice is normally spanned by the tensor product of local basis states $\ket{j_1\dots j_N}$, which amounts to a description of the wavefunction in a total Hilbert space of dimension $d^N$, given by
\begin{equation}
    \ket{\Psi}=\sum_{j_1\dots j_=1}^dT_{j_1\dots j_N}\ket{j_1\dots j_N}~,
\end{equation}
where $T_{j_1\dots j_N}$ are referred to as probability amplitudes.

\begin{figure}
    \centering
    \includegraphics[width=0.9\textwidth]{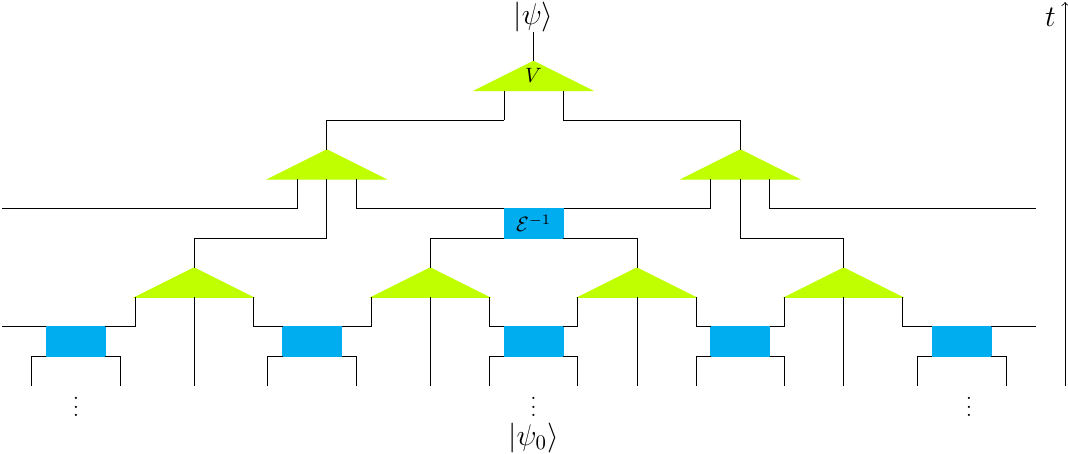}
    \caption{Example of a MERA tensor network generating state $\ket{\psi}$ from $\ket{\psi_0}$. The green triangles represent the isometry operators, and the cyan rectangles the disentanglers $\mathcal{E}^{-1}$. The time flow is represented by the black arrow.}
    \label{fig:MERA example}
\end{figure}

The \textbf{MERA} network is a hierarchical tensor network where a one-dimensional state takes up a decomposition $T_{j_1\dots j_N}$ in terms of a set of tensors organized in a two-dimensional layered graph. It takes an initial state as input, such as a two-body product state $\ket{\psi_0}=\ket{0,~0}$, and outputs a state $\ket{\psi}$. An example of the network is shown in Fig. \ref{fig:MERA example}.

Let \{$j_1$, $j_2$, \dots\} label state contractions (referred to as ``wires'') in the network and \{$i_1$, $i_2$, \dots\} outgoing states. Each site of the graph represents a tensor and those may divided in
\begin{itemize}
    \item \textbf{Isometries}: They combine incoming wires into a single outgoing one. We represent the isometry, $V$ as
    \begin{equation}
        \sum_{j_1,~j_2}V_{j_1j_2}^{i_1}\ket{j_1,~j_2}=\ket{i_1}~, \quad V^\dagger V=\mathbbm{1}~,
    \end{equation}
    this turns state $\ket{\psi_j}$ of $N_j$ wires into state $\ket{\psi_{j+1}}$ of $N_{j+1}=N_j/2$ wires. Notice that $V$ cannot be unitary because the output dimension must be larger than the input dimension. The time-reversed process is also possible, where the output dimension is smaller than the input.
    
    \item \textbf{Entanglers/disentanglers}: These operations do not alter input and output dimensions, instead the entanglers and disentanglers, which we denote by $\mathcal{E}$ and $\mathcal{E}^{-1}$ respectively, increase or decrease the entanglement between nearest neighbors. They act on states as
    \begin{equation}
        \sum_{j_1,~j_2}\mathcal{E}_{j_1j_2}^{i_1i_2}\ket{j_1,~j_2}=\ket{i_1,~i_2}~.
    \end{equation}
The original construction of the MERA networks \cite{Vidal:2008zz} considers $\mathcal{E}$ and $\mathcal{E}^{-1}$ are usually assumed to be unitary operators. 
\end{itemize}
These operations are applied by a sequence of states $\qty{\ket{\psi_0},\dots,\ket{\psi}}$. If the MERA tensor network is homogeneous (scale-invariant), the entangler/disentangler and isometries determine the full graph. This sequence of operations determines a class of real space coarse-graining transformations known as entanglement renormalization in quantum many-body systems \cite{Evenbly:2007hxg,Vidal:2008zz,Giovannetti_2009,Evenbly_2009,Vidal:2009cbi,Beny:2011vh}. The network maps a given density matrix to a compressed state (or enlarged state, depending on the time orientation). It has been argued that this process allows for the efficient evaluation of local expectation values \cite{Vidal:2007hda}.

\instructionsappendices
\cleardoublepage
\backmatter

\includebibliography
\bibliographystyle{utphys}
\bibliography{allpapers}
\instructionsbibliography

\makebackcoverXII

\end{document}